\documentclass{aastex631} 

\begin{document}
\title{Performance of SK-Gd's Upgraded Real-time Supernova Monitoring System}

\correspondingauthor{Y.~Kashiwagi}
\email{kasiwagi@km.icrr.u-tokyo.ac.jp}

\newcommand{\AFFicrr}{\affiliation{Kamioka Observatory, Institute for Cosmic Ray Research, University of Tokyo, Kamioka, Gifu 506-1205, Japan}}
\newcommand{\AFFkashiwa}{\affiliation{Research Center for Cosmic Neutrinos, Institute for Cosmic Ray Research, University of Tokyo, Kashiwa, Chiba 277-8582, Japan}}
\newcommand{\AFFipmu}{\affiliation{Kavli Institute for the Physics and
Mathematics of the Universe (WPI), The University of Tokyo Institutes for Advanced Study,
University of Tokyo, Kashiwa, Chiba 277-8583, Japan }}
\newcommand{\AFFmad}{\affiliation{Department of Theoretical Physics, University Autonoma Madrid, 28049 Madrid, Spain}}
\newcommand{\AFFubc}{\affiliation{Department of Physics and Astronomy, University of British Columbia, Vancouver, BC, V6T1Z4, Canada}}
\newcommand{\AFFbu}{\affiliation{Department of Physics, Boston University, Boston, MA 02215, USA}}
\newcommand{\AFFuci}{\affiliation{Department of Physics and Astronomy, University of California, Irvine, Irvine, CA 92697-4575, USA }}
\newcommand{\AFFcsu}{\affiliation{Department of Physics, California State University, Dominguez Hills, Carson, CA 90747, USA}}
\newcommand{\AFFcnm}{\affiliation{Institute for Universe and Elementary Particles, Chonnam National University, Gwangju 61186, Korea}}
\newcommand{\AFFduke}{\affiliation{Department of Physics, Duke University, Durham NC 27708, USA}}
\newcommand{\AFFgifu}{\affiliation{Department of Physics, Gifu University, Gifu, Gifu 501-1193, Japan}}
\newcommand{\AFFgist}{\affiliation{GIST College, Gwangju Institute of Science and Technology, Gwangju 500-712, Korea}}
\newcommand{\AFFuh}{\affiliation{Department of Physics and Astronomy, University of Hawaii, Honolulu, HI 96822, USA}}
\newcommand{\AFFicl}{\affiliation{Department of Physics, Imperial College London , London, SW7 2AZ, United Kingdom }}
\newcommand{\AFFkek}{\affiliation{High Energy Accelerator Research Organization (KEK), Tsukuba, Ibaraki 305-0801, Japan }}
\newcommand{\AFFkobe}{\affiliation{Department of Physics, Kobe University, Kobe, Hyogo 657-8501, Japan}}
\newcommand{\AFFkyoto}{\affiliation{Department of Physics, Kyoto University, Kyoto, Kyoto 606-8502, Japan}}
\newcommand{\AFFliv}{\affiliation{Department of Physics, University of Liverpool, Liverpool, L69 7ZE, United Kingdom}}
\newcommand{\AFFmiyagi}{\affiliation{Department of Physics, Miyagi University of Education, Sendai, Miyagi 980-0845, Japan}}
\newcommand{\AFFnagoya}{\affiliation{Institute for Space-Earth Environmental Research, Nagoya University, Nagoya, Aichi 464-8602, Japan}}
\newcommand{\AFFkmi}{\affiliation{Kobayashi-Maskawa Institute for the Origin of Particles and the Universe, Nagoya University, Nagoya, Aichi 464-8602, Japan}}
\newcommand{\AFFpol}{\affiliation{National Centre For Nuclear Research, 02-093 Warsaw, Poland}}
\newcommand{\AFFsuny}{\affiliation{Department of Physics and Astronomy, State University of New York at Stony Brook, NY 11794-3800, USA}}
\newcommand{\AFFokayama}{\affiliation{Department of Physics, Okayama University, Okayama, Okayama 700-8530, Japan }}
\newcommand{\AFFosaka}{\affiliation{Department of Physics, Osaka University, Toyonaka, Osaka 560-0043, Japan}}
\newcommand{\AFFox}{\affiliation{Department of Physics, Oxford University, Oxford, OX1 3PU, United Kingdom}}
\newcommand{\AFFqmul}{\affiliation{School of Physics and Astronomy, Queen Mary University of London, London, E1 4NS, United Kingdom}}
\newcommand{\AFFregina}{\affiliation{Department of Physics, University of Regina, 3737 Wascana Parkway, Regina, SK, S4SOA2, Canada}}
\newcommand{\AFFseoul}{\affiliation{Department of Physics, Seoul National University, Seoul 151-742, Korea}}
\newcommand{\AFFsheff}{\affiliation{Department of Physics and Astronomy, University of Sheffield, S3 7RH, Sheffield, United Kingdom}}
\newcommand{\AFFshizuokasc}{\affiliation{Department of Informatics in
Social Welfare, Shizuoka University of Welfare, Yaizu, Shizuoka, 425-8611, Japan}}
\newcommand{\AFFstfc}{\affiliation{STFC, Rutherford Appleton Laboratory, Harwell Oxford, and Daresbury Laboratory, Warrington, OX11 0QX, United Kingdom}}
\newcommand{\AFFskk}{\affiliation{Department of Physics, Sungkyunkwan University, Suwon 440-746, Korea}}
\newcommand{\AFFtodai}{\affiliation{Department of Physics, University of Tokyo, Bunkyo, Tokyo 113-0033, Japan }}
\newcommand{\AFFtit}{\affiliation{Department of Physics,Tokyo Institute of Technology, Meguro, Tokyo 152-8551, Japan }}
\newcommand{\AFFtus}{\affiliation{Department of Physics, Faculty of Science and Technology, Tokyo University of Science, Noda, Chiba 278-8510, Japan }}
\newcommand{\AFFtriumf}{\affiliation{TRIUMF, 4004 Wesbrook Mall, Vancouver, BC, V6T2A3, Canada }}
\newcommand{\AFFtokai}{\affiliation{Department of Physics, Tokai University, Hiratsuka, Kanagawa 259-1292, Japan}}
\newcommand{\AFFtsinghua}{\affiliation{Department of Engineering Physics, Tsinghua University, Beijing, 100084, China}}
\newcommand{\AFFynu}{\affiliation{Department of Physics, Yokohama National University, Yokohama, Kanagawa, 240-8501, Japan}}
\newcommand{\AFFllr}{\affiliation{Ecole Polytechnique, IN2P3-CNRS, Laboratoire Leprince-Ringuet, F-91120 Palaiseau, France }}
\newcommand{\AFFbari}{\affiliation{ Dipartimento Interuniversitario di Fisica, INFN Sezione di Bari and Universit\`a e Politecnico di Bari, I-70125, Bari, Italy}}
\newcommand{\AFFnapoli}{\affiliation{Dipartimento di Fisica, INFN Sezione di Napoli and Universit\`a di Napoli, I-80126, Napoli, Italy}}
\newcommand{\AFFroma}{\affiliation{INFN Sezione di Roma and Universit\`a di Roma ``La Sapienza'', I-00185, Roma, Italy}}
\newcommand{\AFFpadova}{\affiliation{Dipartimento di Fisica, INFN Sezione di Padova and Universit\`a di Padova, I-35131, Padova, Italy}}
\newcommand{\AFFkeio}{\affiliation{Department of Physics, Keio University, Yokohama, Kanagawa, 223-8522, Japan}}
\newcommand{\AFFwinnipeg}{\affiliation{Department of Physics, University of Winnipeg, MB R3J 3L8, Canada }}
\newcommand{\AFFkcl}{\affiliation{Department of Physics, King's College London, London, WC2R 2LS, UK }}
\newcommand{\AFFwarwick}{\affiliation{Department of Physics, University of Warwick, Coventry, CV4 7AL, UK }}
\newcommand{\AFFral}{\affiliation{Rutherford Appleton Laboratory, Harwell, Oxford, OX11 0QX, UK }}
\newcommand{\AFFwu}{\affiliation{Faculty of Physics, University of Warsaw, Warsaw, 02-093, Poland }}
\newcommand{\AFFbcit}{\affiliation{Department of Physics, British Columbia Institute of Technology, Burnaby, BC, V5G 3H2, Canada }}
\newcommand{\AFFtohoku}{\affiliation{Department of Physics, Faculty of Science, Tohoku University, Sendai, Miyagi, 980-8578, Japan }}
\newcommand{\AFFicise}{\affiliation{Institute For Interdisciplinary Research in Science and Education, ICISE, Quy Nhon, 55121, Vietnam }}
\newcommand{\AFFilance}{\affiliation{ILANCE, CNRS - University of Tokyo International Research Laboratory, Kashiwa, Chiba 277-8582, Japan}}
\newcommand{\AFFibs}{\affiliation{Institute for Basic Science (IBS), Daejeon, 34126, Korea}}
\newcommand{\AFFglasgow}{\affiliation{School of Physics and Astronomy, University of Glasgow, Glasgow, Scotland, G12 8QQ, United Kingdom}}
\newcommand{\AFFoecu}{\affiliation{Media Communication Center, Osaka Electro-Communication University, Neyagawa, Osaka, 572-8530, Japan}}

\AFFicrr
\AFFkashiwa
\AFFmad
\AFFbu
\AFFbcit
\AFFuci
\AFFcsu
\AFFcnm
\AFFduke
\AFFllr
\AFFgifu
\AFFgist
\AFFglasgow
\AFFuh
\AFFibs
\AFFicise
\AFFicl
\AFFbari
\AFFnapoli
\AFFpadova
\AFFroma
\AFFilance
\AFFkeio
\AFFkek
\AFFkcl
\AFFkobe
\AFFkyoto
\AFFliv
\AFFmiyagi
\AFFnagoya
\AFFkmi
\AFFpol
\AFFsuny
\AFFokayama
\AFFoecu
\AFFox
\AFFral
\AFFseoul
\AFFsheff
\AFFshizuokasc
\AFFstfc
\AFFskk
\AFFtohoku
\AFFtokai
\AFFtodai
\AFFipmu
\AFFtit
\AFFtus
\AFFtriumf
\AFFtsinghua
\AFFwu
\AFFwarwick
\AFFwinnipeg
\AFFynu

\author[0000-0002-3926-8333]{Y.~Kashiwagi}
\AFFicrr
\author{K.~Abe}
\AFFicrr
\AFFipmu
\author[0000-0001-9555-6033]{C.~Bronner}
\AFFicrr
\author[0000-0002-8683-5038]{Y.~Hayato}
\AFFicrr
\AFFipmu
\author[0000-0003-1229-9452]{K.~Hiraide}
\AFFicrr
\AFFipmu
\author[0000-0002-8766-3629]{K.~Hosokawa}
\AFFicrr
\author[0000-0002-7791-5044]{K.~Ieki}
\author[0000-0002-4177-5828]{M.~Ikeda}
\AFFicrr
\AFFipmu
\author{J.~Kameda}
\AFFicrr
\AFFipmu
\author{Y.~Kanemura}
\author{R.~Kaneshima}
\AFFicrr
% ...


\collaboration{300}{(The Super-Kamiokande Collaboration)}


\begin{abstract}
Among multi-messenger observations of the next galactic core-collapse supernova, Super-Kamiokande (SK) plays a critical role in detecting the emitted supernova neutrinos, determining the direction to the supernova (SN), and notifying the astronomical community of these observations in advance of the optical signal.  
On 2022, SK has increased the gadolinium dissolved in its water target (SK-Gd) and has achieved a Gd concentration of 0.033\%, resulting in enhanced neutron detection capability, which in turn enables more accurate determination of the supernova direction. 
Accordingly, SK-Gd's real-time supernova monitoring system~\citep{abe2016snwatch} has been upgraded.
SK\_SN Notice, a warning system that works together with this monitoring system, was released on December 13, 2021, and is available through GCN Notices~\citep{barthelmy2000grb}.
When the monitoring system detects an SN-like burst of events, SK\_SN Notice will automatically distribute an alarm with the reconstructed direction to the supernova candidate within a few minutes.
In this paper, we present a systematic study of SK-Gd's response to a simulated galactic SN. 
Assuming a supernova situated at 10~kpc, neutrino fluxes from six supernova models are used to characterize SK-Gd's pointing accuracy using the same tools as the online monitoring system. 
The pointing accuracy is found to vary from 3--7$^\circ$ depending on the models. 
However, if the supernova is closer than 10~kpc, SK\_SN Notice can issue an alarm with three-degree accuracy, which will benefit follow-up observations by optical telescopes with large fields of view.
\end{abstract}



\section{Introduction} \label{sec:Introduction}
The Super-Kamiokande (SK) detector~\citep{FUKUDA2003SK} is a 50,000~m$^3$ water Cherenkov detector with a fiducial volume of 22.5k~m$^3$, consisting of a cylindrical stainless steel tank (39.3~m in diameter and 41.4~m in height) lined with by photo-multiplier tubes (PMTs).  It has an optically separated concentric structure with an inner detector (ID) covered with 11,129 PMTs in 50~cm diameter and an outer detector (OD) with 1,885 outward-facing 20~cm PMTs.
SK observes neutrinos from various sources, searches for proton decay, and searches for exotic particles such as dark matter.  One of the vital roles of SK is to detect core-collapse supernova (SN) neutrinos\footnote{Hereafter, the word ``supernova'' or ``SN'' in this paper always means a core-collapse supernova.}, to determine the direction of the SN, and to issue an alarm to astronomical observatories for multi-messenger observations of the event.  For accurate pointing to the SN, it is essential to extract as many elastic scattering (ES) events ($\nu + \mathrm{e^-} \to \nu + \mathrm{e^-}$) as possible, since the outgoing electron's direction tracks that of the incoming neutrino.  
These events must be separated from inverse beta decay (IBD) events ($\bar{\nu}_\mathrm{e} + \mathrm{p} \to \mathrm{e^+ + n}$), that dominate the event rate but whose positron is weakly correlated with the neutrino direction. 
This can be achieved by tagging the IBD interactions via their emitted neutron, which produces delayed gamma ray signals after capturing on nuclei in SK and can be used to form a coincidence with the positron. 
Prior to 2020 neutron capture on hydrogen was used to identify such neutrons with roughly 25\% efficiency, but thereafter gadolinium has been 
dissolved in the detector water to improve the tagging efficiency (SK-Gd project)~\citep{abe2022firstGdLoading,abe2024second}.

\begin{figure}[htb!]
    \centering
    \includegraphics[width=0.6\textwidth]{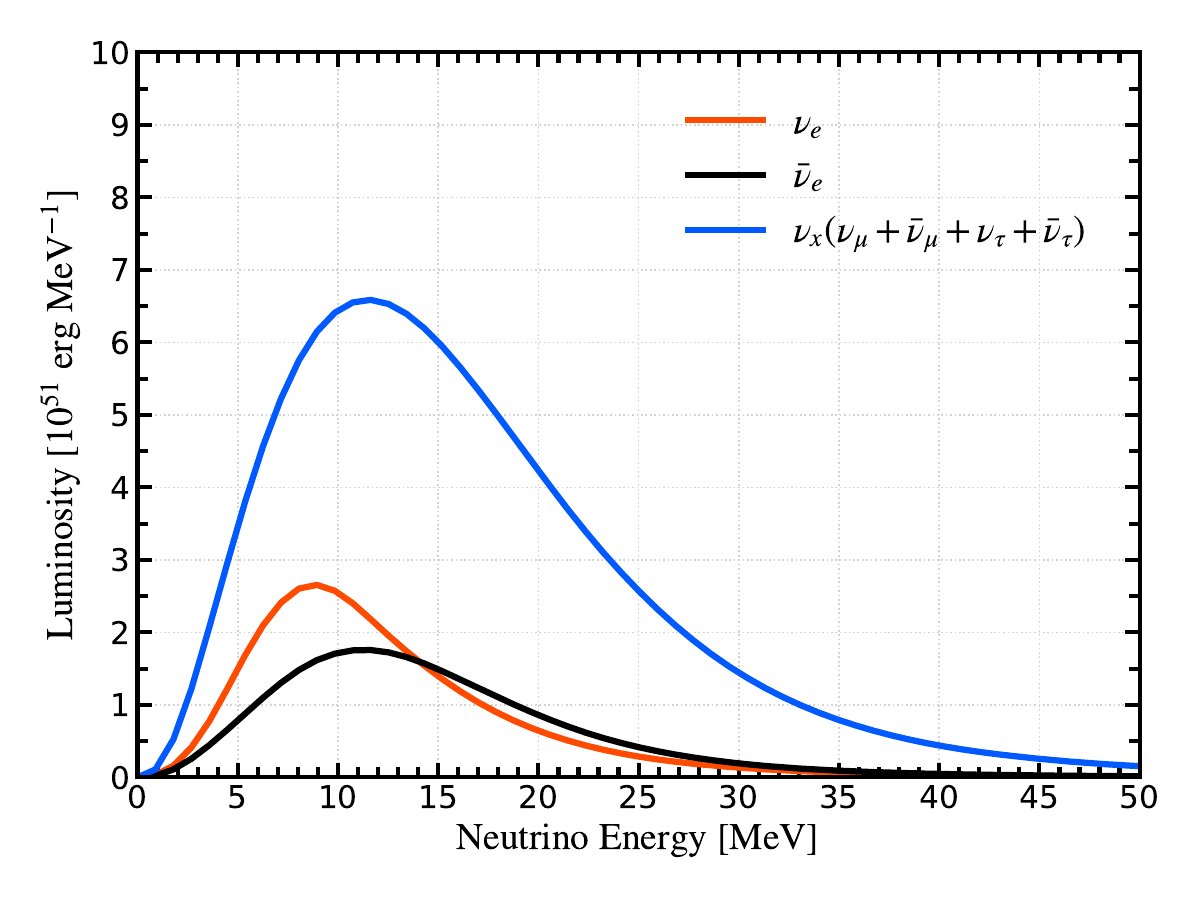}
    \caption{SN neutrino energy spectrum of the Nakazato model (see Section~\ref{subsubsec:NakazatoModel}) for each flavor.  The red, black, and blue lines represent $\nu_\mathrm{e}$, $\bar\nu_\mathrm{e}$, and $\nu_x (=\nu_\mu+\nu_\tau+\bar\nu_\mu+\bar\nu_\tau)$, respectively.}
    \label{fig:TypicalNeutrinoEnergySpectra}
\end{figure}

The observational importance of neutrinos from SN burst neutrinos lies in the fact that they rarely interact with matter before reaching a terrestrial detector and therefore carry information from the dying star at the moment of their production. 
 Figure~\ref{fig:TypicalNeutrinoEnergySpectra} shows a typical SN neutrino energy spectrum; all the neutrino flavors are emitted, with energies less than ${\sim}$30~MeV.
Ultimately, SN neutrinos carry 99\% of the energy released by the star's gravitational collapse.
The neutrino emission process starts shortly before the core's collapse for stars heavier than $10 M_\odot$ where $M_\odot$ is the total mass of the Sun when the star's silicon shell begins to burn (pre-SN neutrinos~\citep{2004AcPPB..35.1981O}).
When an iron core is formed in the innermost part of the star and fusion stops, gravity induces electron capture ($\mathrm{p} + \mathrm{e}^- \to  \mathrm{n} + \mathrm{\nu_{e}}$) and photodisintegration of iron nuclei initiating the core's collapse.
The star collapses from within and when the matter at the star's center exceeds nuclear density, the core rebounds against itself (core bounce). 
Within ${\sim}$ 0.1~s after the core bounce there is a sharp increase in the number $\nu_\mathrm{e}$ (neutronization burst), 
whose observed luminosity typically reaches ${\sim}10^{53}~\mathrm{erg~s^{-1}}$ (c.f. Figure~\ref{fig:TimeEvolutionLumiAndEave} below). 
The shock wave generated at the core bounce propagates outward but loses energy due to photodisintegration (stagnation of the shock wave). 
Electromagnetic signatures of the burst become visible when this shock wave regains energy by some mechanism and reaches the photosphere near the stellar surface.
Typically, it takes a few hours to a day to blow off the outer stellar layer and produce electromagnetic radiation\footnote{The time becomes shorter if the collapsing star is a Wolf-Rayet star with no hydrogen layer.}.
While shock revival is considered to be caused by neutrino heating (e.g., \cite{1985ApJ...295...14B} and \cite{2001A&A...368..527J}), the details of the heating mechanism are still unknown.
Several scenarios have been proposed;  
hydrodynamical mechanisms such as convection (e.g., \cite{herant1994convection}) and the standing-accretion-shock instability (SASI, e.g., \cite{marek2009delayed}), acoustic oscillations (e.g., \cite{burrows2006acoustic}), and magnetic fields (e.g., \cite{kuroda2021impact}).
Neutrinos are expected to carry information about the explosion mechanism, making their detection during the next SN burst. 
However, SN burst neutrinos have not been observed since their first detection in 1987~\citep{PhysRevLett.58.1494, ALEXEYEV1988209}.

To understand SK's response to SN in advance and evaluate the performance of SK-Gd's SN monitoring system, it is crucial to simulate SN bursts using SN models. 
This paper aims to understand the SK's response to galactic SN systematically and pointing accuracy (angular resolution) since the start of SK-Gd. 
We studied SK's response to a simulated SN burst located at a distance of 10~kpc and pointing accuracy for six SN models considering neutrino oscillations.

This paper is organized as follows.  Section~\ref{sec:SK-Gd} summarizes the overview of SK-Gd.  In Section~\ref{sec:SNWATCHupdate}, we describe the SK-Gd's SN monitoring system.  Section~\ref{sec:Simulations} explains the SN models used in this paper, simulations for event generation, and detector simulation.  
In Section~\ref{sec:Results-truth}, we show the simulated SN interaction events for a simulated SN neutrino burst located at 10~kpc using several SN neutrino emission models and their SK-Gd's response.
The performance of the SK-Gd's real-time SN monitoring system including pointing accuracy is presented in Section~\ref{sec:Results-SNWATCH} before concluding in Section~\ref{sec:Summary}.

\section{SK-G\lowercase{d}} \label{sec:SK-Gd}
In 2020, ultra-pure gadolinium sulfate octahydrate ($\mathrm{Gd_2 (SO_4)_3\cdot 8H_2O}$) was added to SK's ultra-pure water target marking the start of 
the SK-Gd phase of operations~\cite{abe2022firstGdLoading}.
The motivation, as originally posed in \cite{beacom2004antineutrino}, is to enhance SK's ability to detect neutrons due to Gd's 
large neutron capture cross section, 254000~barn for $^{157}$Gd compared to 0.33~barn for proton, followed by the deexcitation 
emission of 8~MeV in gamma rays, which is easier to detect than the 2.2~MeV emitted following neutron capture on a proton. 
During the initial Gd loading, from July to August in 2020, a total concentration of 0.011\% Gd was added to the detector, resulting in 
a neutron capture efficiency of 50\%. 
From May to July 2022 the concentration was increased to 0.033\%, resulting in a 75\% capture efficiency~\citep{abe2022firstGdLoading,abe2024second}.
This increased efficiency is expected to improve SK's sensitivity to the diffuse supernova neutrino background (DSNB) as well as its accuracy for locating a nearby SNe using the neutrino signal.

\begin{figure}[htb!]
    \centering
    \includegraphics[width=0.4\textwidth]{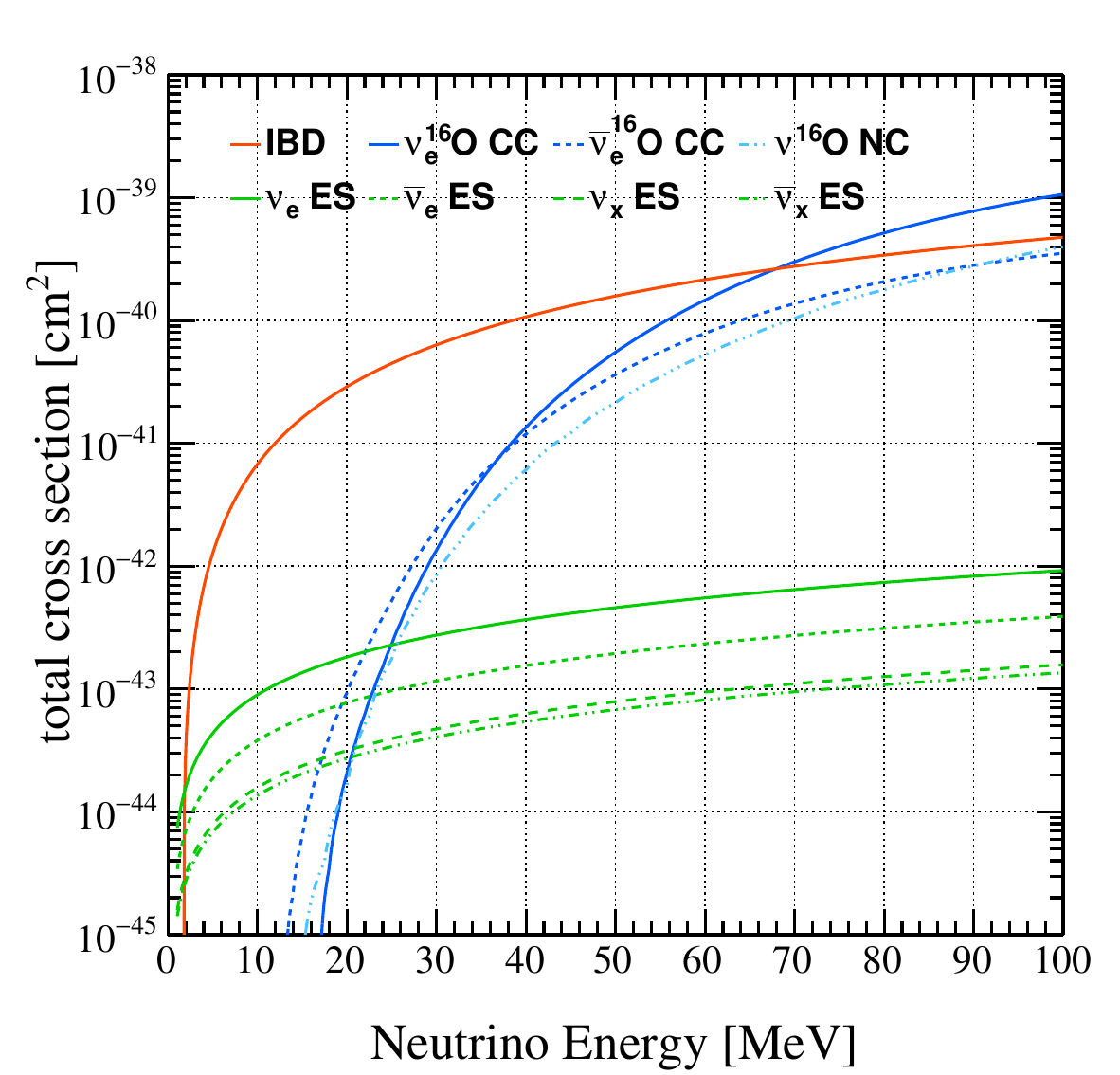}
    \caption{Cross sections for neutrino interactions on water as a function of neutrino energy. 
    The solid red represents IBD and the green lines represent ES for $\nu_\mathrm{e}$ (solid), $\bar{\nu}_\mathrm{e}$ (dotted), $\nu_x$ (dashed), and $\bar{\nu}_x$ (dot-dashed).  The solid blue and dashed blue lines stand for $^{16}$O~CC interactions of $\nu_\mathrm{e}$ and $\overline{\nu}_\mathrm{e}$, respectively.  The dot-dashed light blue line indicates $^{16}$O~NC interactions. Cross sections are calculated according to  \cite{strumia2003precise} for IBD, \cite{bahcall1995solar} for ES, \cite{suzuki2018oxycc} and \cite{nakazato2018charged} for $^{16}$O~CC interactions, and \cite{langanke1996oxync} and \cite{kolbe2002oxync} for $^{16}$O~NC interactions.}
    \label{fig:SNnuCrossSections}
\end{figure}

There are four channels for SN neutrino interaction in water Cherenkov detectors: inverse beta decay (IBD), elastic scattering (ES), charged current interactions on oxygen (hereafter denoted $^{16}$O~CC), such as 
\begin{eqnarray}
    \nu_\mathrm{e} + {}^{16}\mathrm{O} &\to& \mathrm{e^-} + {}^{16}\mathrm{F}\\
    \overline{\nu}_\mathrm{e} + {}^{16}\mathrm{O} &\to& \mathrm{e^+} + {}^{16}\mathrm{N},
\end{eqnarray}
and neutral current interactions on oxygen (hereafter $^{16}$O~NC) such as 
\begin{eqnarray}
    \nu + {}^{16}\mathrm{O} &\to& \nu + {}^{16}\mathrm{O}^* \\
    ^{16}\mathrm{O}^* &\to& ^{15}\mathrm{N} + \mathrm{p} + \gamma \label{eq:OxyNCwithProton}\\
    ^{16}\mathrm{O}^* &\to& ^{15}\mathrm{O} + \mathrm{n} + \gamma
    \label{eq:OxyNCwithNeutron}
\end{eqnarray}
\citep{kolbe2002oxync}.
Figure~\ref{fig:SNnuCrossSections} shows the relevant cross section as a function of neutrino energy.  
As can be seen in the figure, IBD interactions dominate in the energy region below ${\sim} 30$~MeV, accounting for about 90\% of SN neutrino events in SK.
The outgoing positron from these interactions is isotropically emitted, providing little correlation with the incoming neutrino direction. 
In contrast, ES interactions account for only about 5\% of the expected events in SK, but as a forward scattering process, ES interactions provide a tight 
correlation with the neutrino direction.
At around ${\sim} 20$~MeV the inelastic interactions on oxygen, $^{16}$O~CC and $^{16}$O~NC, overtake the ES cross section. 
About 30\% of $^{16}$O~CC interactions are expected to be observed with a delayed neutron capture signal, because the resulting excited nucleus ($^{16}$F and $^{16}$N) may emit one or two neutrons together with gamma rays (typically 5-9~MeV)
Further, the reaction~(\ref{eq:OxyNCwithNeutron}) is also expected to be accompanied by neutron capture.  
In both cases, these neutron captures may cause these interactions to be misidentified as IBD.

\section{Real-time supernova monitoring system} \label{sec:SNWATCHupdate}
\begin{figure}[htb!]
    \centering
    \includegraphics[width=0.5\textwidth]{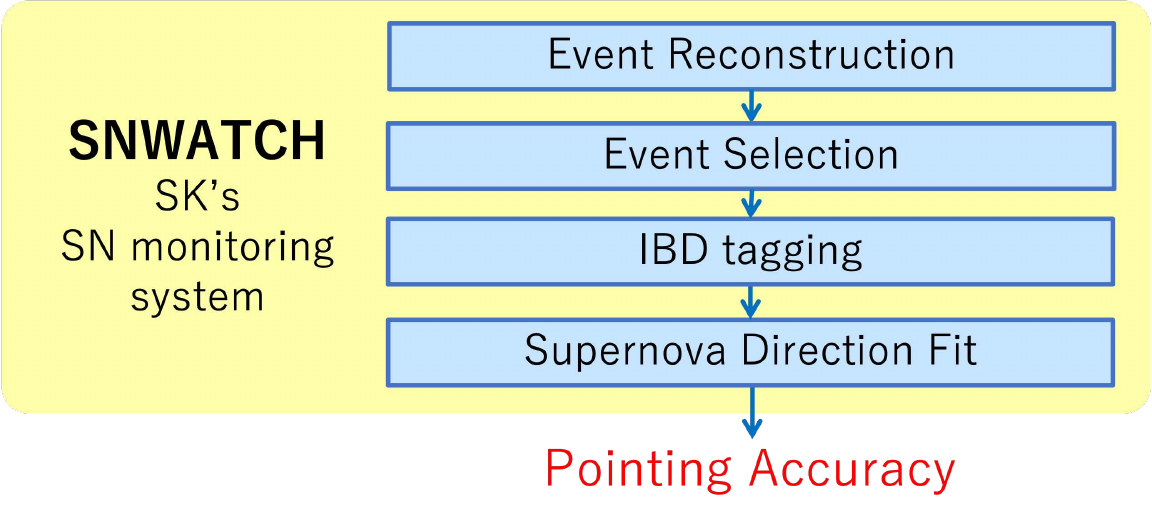}
    \caption{Overview of SK's real-time SN monitoring system, SNWATCH. When an SN-like event burst is detected, SNWATCH performs event reconstruction, event selection, and IBD tagging (see Section~\ref{subsec:EventReconstruction}), before applying SN direction fit (see Section~\ref{subsec:DirectionFitUpdate}). The resulting reconstructed SN direction is announced with the expected pointing accuracy to astronomical networks (see Section~\ref{subsec:SNWarnWithDirInfo}).}
    \label{fig:SNWATCH_system}
\end{figure}

SK's real-time SN monitoring system, SNWATCH \citep{abe2016snwatch}, monitors events in the detector to detect SN-like event bursts. 
Upon SNWATCH's detecting such a burst, SK\_SN Notice, SK's SN warning system working together with SNWATCH, issues a prompt warning to astronomical networks as the first alarm of an SN-like event occurrence, SNWATCH determines its direction, and then SK\_SN Notice broadcasts an announcement of an SN-burst-like detection together with this reconstructed direction and the expected pointing accuracy to astronomical networks.
Figure~\ref{fig:SNWATCH_system} outlines the flow of SNWATCH.
To announce the reconstructed SN direction with the best possible pointing accuracy, SNWATCH needs to identify every interaction channel to extract ES events' SN direction sensitivity as much as possible.  
However, there is a trade-off between accuracy and the time it takes to issue the alarm. Prioritizing accuracy would increase the time to the alarm issue, making it impossible to fulfill the role of SK, which is to detect neutrinos at the very early stages of an SN to enable observations of the optical burst from the beginning to the end. Therefore, to reduce the time to the alert as much as possible, SNWATCH prioritizes identifying ``IBD-like'' events  for extracting ``ES-like'' events at the expense of pointing accuracy.
To identify ``IBD-like'' events, SNWATCH uses delayed coincidence between IBD events and the following neutron capture events on proton and Gd.
Identifying a positron emission event in an IBD event and the following neutron capture event is called IBD tagging.
In this section, we describe the flow of SNWATCH: current event reconstruction, selection, and IBD tagging in  Section~\ref{subsec:EventReconstruction}, the SN direction determination updated to use information from SK-Gd in Section~\ref{subsec:DirectionFitUpdate}, its pointing accuracy and the alarm issue in Section~\ref{subsec:SNWarnWithDirInfo}.

\subsection{Event Reconstruction, Selection, and IBD tagging}\label{subsec:EventReconstruction}
SNWATCH uses a custom online version of the standard SK event reconstruction program \citep{abe2022SKIVsolar, abe2016SKIVsolarPRD} to identify events and reconstruct their vertex position, direction, and total energy.
It differs from the standard program due to the need for a fast real-time reconstruction: it uses a simpler and faster muon reconstruction algorithm. It uses preliminary calibration values to compute PMT hit times and charges. 
This program reconstructs every event detected at the SN-like event bursts, including SN neutrino interactions (IBD, ES, $^{16}$O~CC, and $^{16}$O~NC) and neutron capture events. 
After reconstruction, SNWATCH concentrates on identifying ``IBD-like'' events as fast as possible to extract ``ES-like'' events using IBD tagging, as explained in Section~\ref{sec:SNWATCHupdate}. 
To utilize delayed coincidence in performing IBD tagging, SNWATCH divides reconstructed events according to their reconstructed energy into two samples: ``prompt'' candidates, such as IBD positrons or ES electrons, and ``delayed'' candidates, i.e., ``neutron capture'' candidates. 
We call this event selection.
The conditions for ``prompt'' candidates and ``delayed'' candidates are listed in Table~\ref{tab:EventSelectionConditions}.
In the selection of ``prompt'' candidates, three software triggers that store the data within the time window from $-5$ to $+35~\mathrm{\mu}$s around the triggered time are relevant: low energy (LE) trigger and high energy (HE) trigger with the threshold of ID PMT hits at 49 and 52, respectively, and OD trigger whose threshold is set at 22 OD PMT hits.

\begin{table}[htb!]
    \centering
    \caption{Conditions for ``prompt'' candidates and ``delayed'' candidates.  $E$ is the reconstructed energy.  $g^{2}_{t}$ and $g^{2}_{p}$ represent the PMT timing goodness and the PMT hit pattern goodness, respectively. 
    N/S ratio is the ratio between the number of low-charge PMT hits (below single photo-electron level) and the total number of PMT hits~\citep{Hosaka2006SK-I}.
    LE trigger and HE trigger represent software triggers with ID PMT hit thresholds of 49 and 52, respectively, and OD trigger is set at 22 OD PMT hits. $d_{wall}$ is the distance between the reconstructed vertex and the inner detector walls.}
    \label{tab:EventSelectionConditions}
    \begin{tabular}{c|c}\hline
        conditions for ``prompt'' candidates & conditions for ``delayed'' candidates \\\hline
        $E>7$ MeV & $E<10$ MeV\\
        $g^{2}_{t} \geq 0.4$ & $g^{2}_{t} - g^{2}_{p} > 0$\\
        Number of PMT hit $<500$ & Within the fiducial volume\\
         N/S $\leq 0.4$ & Not a ``prompt'' candidate\\
        $d_{wall} > 200$ cm & \\
        LE- or HE-triggered event & \\
        Not an OD triggered event  & \\\hline
    \end{tabular}
\end{table}

\begin{figure}[htb!]
    \centering
    \includegraphics[width=0.4\textwidth]{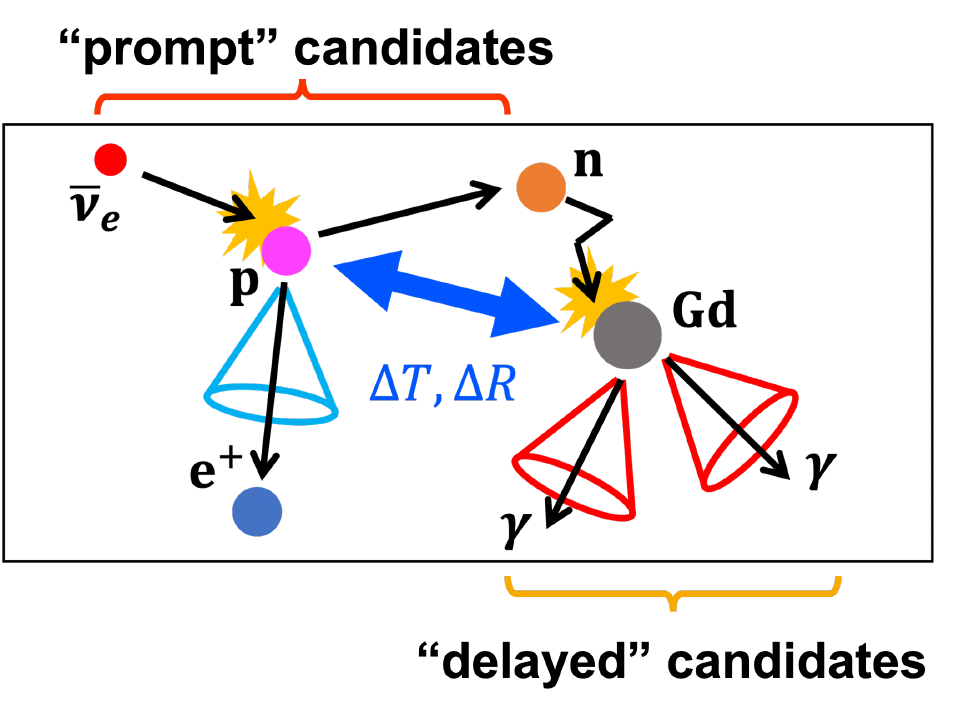}
    \caption{Schematic illustration of IBD tagging algorithm used in SNWATCH. Every possible pairs of ``prompt'' and ``delayed'' candidates with time difference $\Delta T <$ 500~$\mathrm{\mu s}$ and spacial difference $\Delta R <$ 300~cm are tested, and the pair with the smallest $\Delta R$ is selected as an IBD candidate.}
    \label{fig:NeutronTaggingInSNwatch}
\end{figure}

Next, SNWATCH searches for IBD-like interactions, pairing ``prompt'' candidates with ``delayed'' candidates based on differences in their trigger times and the spatial separation of their reconstructed vertices. 
This process is called IBD tagging and is illustrated in Figure~\ref{fig:NeutronTaggingInSNwatch}.
The algorithm was designed to be fast and simple.
Each possible pair of ``prompt'' and ``delayed'' candidates with a trigger time difference $\Delta T <$~500~$\mathrm{\mu s}$ and a vertex separation of  $\Delta R <$~300~$\mathrm{cm}$ are tested and the pair with the smallest spatial distance is selected as an IBD candidate. 
This algorithm provides a tagging efficiency $46.86 \pm 0.04 \%$ of IBD interactions and results in a $98.82 \pm 0.01 \%$ sample 
purity as is discussed below in Section~\ref{subsec:TaggingPerformance} and Table~\ref{tab:IBDtaggingPerformance}. 
Hereafter, we label  ``prompt'' events that have been tagged as an IBD candidate as ``IBD-like'' 
and otherwise as ``ES-like.''
Separating these two event samples improves SNWATCH's accuracy for determining the direction to an SN as described in the following Section \ref{subsec:DirectionFitUpdate}).

\subsection{SN Direction Fit Algorithm} \label{subsec:DirectionFitUpdate}
Though the neutrino direction itself cannot be known directly, the detected recoil electrons of ES can be utilized to reconstruct the SN direction since the outgoing lepton preserves the neutrino's direction, as mentioned in Section~\ref{sec:SK-Gd}. Accordingly, it is essential to collect as many ES events as possible. 
Since ES-like events have high ES purity, separating them from IBD-like events is effective in improving the accuracy of the SN direction determination.

SNWATCH uses a maximum-likelihood (ML) fit to extract the SN direction. 
An unbinned likelihood taken over all events
\begin{equation}
    \mathcal{L}=\exp\left( {\sum_{k,r} N_{r,k}}\right)\prod_i L_i
    \label{eq:likelifhood}
\end{equation}
is constructed. Here, $N_{r, k}$ is the number of events of the interaction $r$ in the $k$-th energy bin.
We note that the index $r$ and $k$ are the same as those described in detail in Section~3.1 of \cite{abe2016snwatch}.
The index $r$ stands for the type of neutrino interaction, i.e., $r=$~IBD, ES of $\bar{\nu}_\mathrm{e}$, ES of other neutrino flavors, or $^{16}$O~CC interactions\footnote{The division of the four interactions represented by the index $r$, introduced in \cite{abe2016snwatch}, does not correspond to the division of the four interaction channels described in Section~\ref{sec:SK-Gd}, but the likelihood in the current SNWATCH still use this division of $r$.}. 
The index $k$ running from 1 to 5 represents the bin of the reconstructed energy $E$ of the ``prompt'' candidates in the unit of MeV: 7~$\leq E <$~10, 10~$\leq E <$~15, 15~$\leq E <$~22, 22~$\leq E <$~35, and 35~$\leq E <$~50, respectively.

\begin{figure}[htb!]
    \centering
    \includegraphics[width=0.4\textwidth]{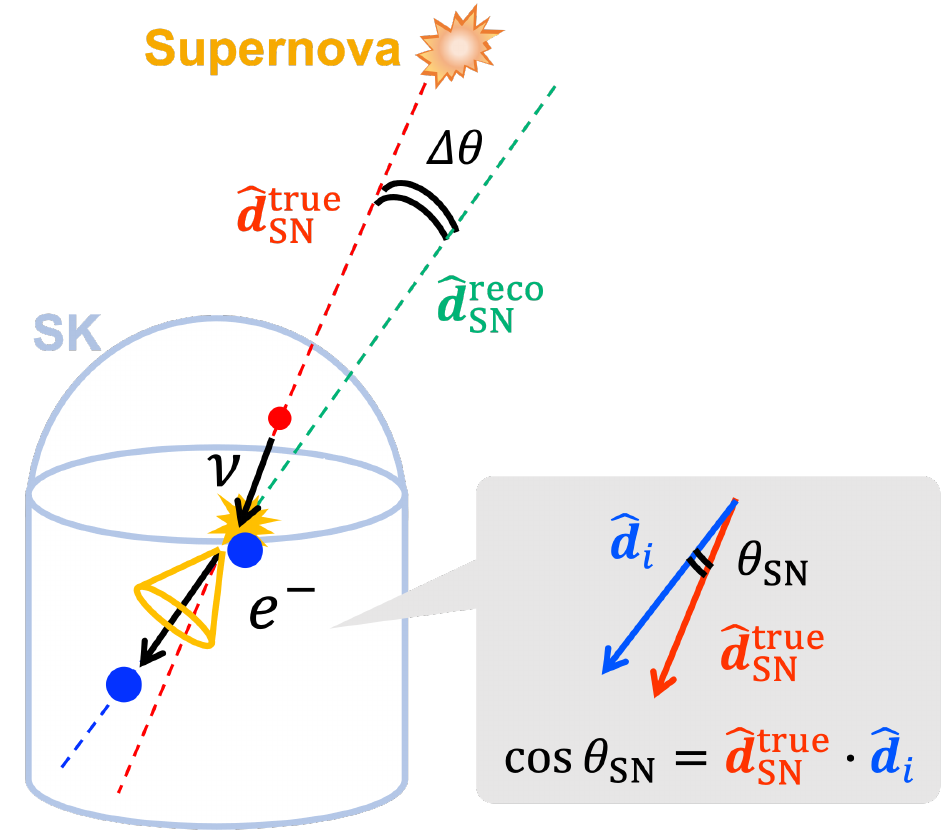}
    \caption{Schematic illustration of $\theta_{\mathrm{SN}}$ and $\Delta\theta$. $\theta_{\mathrm{SN}}$ is the angle between $i$-th event direction in SK $\hat{d}_i$ (the blue arrow in the gray box) and the true SN direction $\hat{d}_\mathrm{SN}^\mathrm{true}$ (${\sim}$~SN neutrino arrival direction)  (the red arrow in the gray box).  An ES event is illustrated as an example.  The direction of the recoil electron corresponds to $\hat{d}_i$.  By applying maximum likelihood fit to the distribution of $\cos\theta_{\mathrm{SN}}$, the reconstructed SN direction $\hat{d}_\mathrm{SN}^\mathrm{reco}$ (the green dashed line) that we would like to determine is obtained in 1~set of MC.  $\Delta\theta$ is the angle difference between $\hat{d}_\mathrm{SN}^\mathrm{reco}$ and $\hat{d}_\mathrm{SN}^\mathrm{true}$, and used to estimate pointing accuracy (see Section \ref{subsec:SNWarnWithDirInfo}).}
    \label{fig:thetaSN}
\end{figure}

$L_i$ in equation~(\ref{eq:likelifhood}) is a likelihood function for $i$-th event, defined as
\begin{equation}
    L_i=\sum_r N_{r,k} t_r(f_i) p_r(E_i, \hat{d}_i; \hat{d}_\mathrm{SN}^\mathrm{reco}).
\end{equation}
Compared to the likelihood function described in~\cite{abe2016snwatch}, a term $t_r(f_i)$ has been added.
This term is defined according to the IBD flag $f_i$ of \textit{i}-th event and the reaction $r$ as: 
\begin{eqnarray}
        t_{r=\mathrm{IBD}}(f_{i})&=&\left\{ 
    \begin{array}{cc}
        \mathrm{IBD~tagging~efficiency} & (f_i = \mathrm{TRUE})\\
        \mathrm{1-(IBD~tagging~efficiency)} & (f_i = \mathrm{FALSE})
    \end{array} \right. \\
    t_{r\neq\mathrm{IBD}}(f_{i})&=&\left\{ 
    \begin{array}{cc}
        0 & (f_i = \mathrm{TRUE})\\
        1 & (f_i = \mathrm{FALSE}).
    \end{array} \right.
\end{eqnarray}
$p_r(E_i, \hat{d}_i; \hat{d}_\mathrm{SN}^\mathrm{reco})$ indicates a probability density function (PDF) for interaction $r$ as a function of the energy $E_i$ and an inner-product of $\hat{d}_\mathrm{SN}^\mathrm{reco}\cdot\hat{d}_i = \cos\theta_\mathrm{SN}^\mathrm{reco}$.
$E_i$ is the total electron energy of $i$-th event and uniquely identifies the index $k$. $\hat{d}_i$ is the reconstructed direction of the \textit{i}-th event and $\hat{d}_\mathrm{SN}^\mathrm{reco}$ is the SN direction we would like to determine. 
In the determination of the PDF, SNWATCH utilizes the SK Monte-Carlo (MC) simulations (described in Section \ref{subsec:EventGeneration}): The generated MC samples are divided into one-MeV bins from 7 to 35~MeV and a combined energy bin greater than 35~MeV. Next, using the known true SN direction $\hat{d}_\mathrm{SN}^{\mathrm{true}}$ illustrated in Figure~\ref{fig:thetaSN} and a model function, $\cos\theta_\mathrm{SN}=\hat{d}_\mathrm{SN}^{\mathrm{true}}\cdot\hat{d}_{i}^{\mathrm{MC}}$ distribution of the generated MC samples is fitted, where $\hat{d}_{i}^{\mathrm{MC}}$ indicates the $i$-th event direction of a generated MC sample.

\subsection{Supernova Warning with Direction Information}\label{subsec:SNWarnWithDirInfo}
SNWATCH's alarm is based on two variables: the number of events in a 20~s time window opened behind the time of each reconstructed event ($N_\mathrm{cluster}$) and a parameter that characterizes the spatial distribution of vertices ($D$).
SNWATCH issues a golden alarm when $N_\mathrm{cluster}\geq60$ and $D=3$, indicating the vertices are uniformly distributed (``volume-like'') (see ~\cite{abe2016snwatch} for details).
The alarm then provides information about the burst candidate, including the number of observed neutrinos, the duration of the neutrino burst, the GPS-based time stamp of the beginning of the burst, and the estimated SN direction with its uncertainty (i.e., pointing accuracy) in the equatorial coordinates.

Pointing accuracy in SNWATCH indicates the performance of determining the SN direction. 
To estimate the pointing accuracy in an SN model-independent way, SNWATCH uses a 15$\times$15 resolution matrix with each matrix element containing 3000~MC samples as described in \cite{abe2016snwatch}. 
SK-Gd aims to realize the SN direction pointing with the accuracy of ${\sim}3^\circ$ for SN bursts located at 10~kpc, which allows the follow-up observation with large telescopes such as Subaru and LSST (Large Synoptic Survey Telescope)~\citep{nakamura2016multimessenger}.

An improved SN warning system at SK, SK\_SN Notice,\footnote{\url{https://gcn.nasa.gov/missions/sksn}} 
started on 2021 December 13 and its alarms can now be received through GCN (General Coordinates Network) Notices\footnote{\url{https://gcn.nasa.gov}}~\citep{barthelmy2000grb} and are generated in a machine-readable format.  
If an SN signal is sufficiently large, an alarm will be automatically published within some minutes of detecting the neutrino burst. 
Signals with a lower significance generate an alarm after undergoing a cross check by analyzers within $\lesssim$ 1~h.  
SK\_SN Notice can be received through the same framework as other GCN Notices; gamma-ray bursts, gravitational waves, and high-energy neutrino alarms.  
A dummy (test) alarm is published as a test on the first day of every month.

Based on SNWATCH's alarm, SK also issues warnings on several networks, currently being SNEWS 1.0 (SuperNova Early Warning System 1.0)~\citep{antonioli2004snews1.0, scholberg2008snews1.0}, IAU CBAT (International Astronomical Union Central Bureau for Astronomical Telegrams)\footnote{\url{http://www.cbat.eps.harvard.edu/}}, and ATEL (The Astronomer's telegram)\footnote{\url{http://www.astronomerstelegram.org/}}~\citep{rutledge1998atel}. SNWATCH has been running since 1996, and no golden alarm has been sent so far.  

The time between the detection and the alarm issue is limited by the processing time of (1)~event reconstruction, (2)~SN direction fit, and (3)~announcement.  So far, (1) and (3) have reduced to less than one minute; however, (2) still takes ${\sim}$~5~minutes for an SN located at 10~kpc with the current maximum likelihood fitter. 
A new direction fitter is under development for faster alarm (within less than 1~minute from the SN neutrino burst detection in SK).

\section{Simulations}\label{sec:Simulations}
\begin{figure}[ht!]
    \centering
    \includegraphics[width=0.5\textwidth]{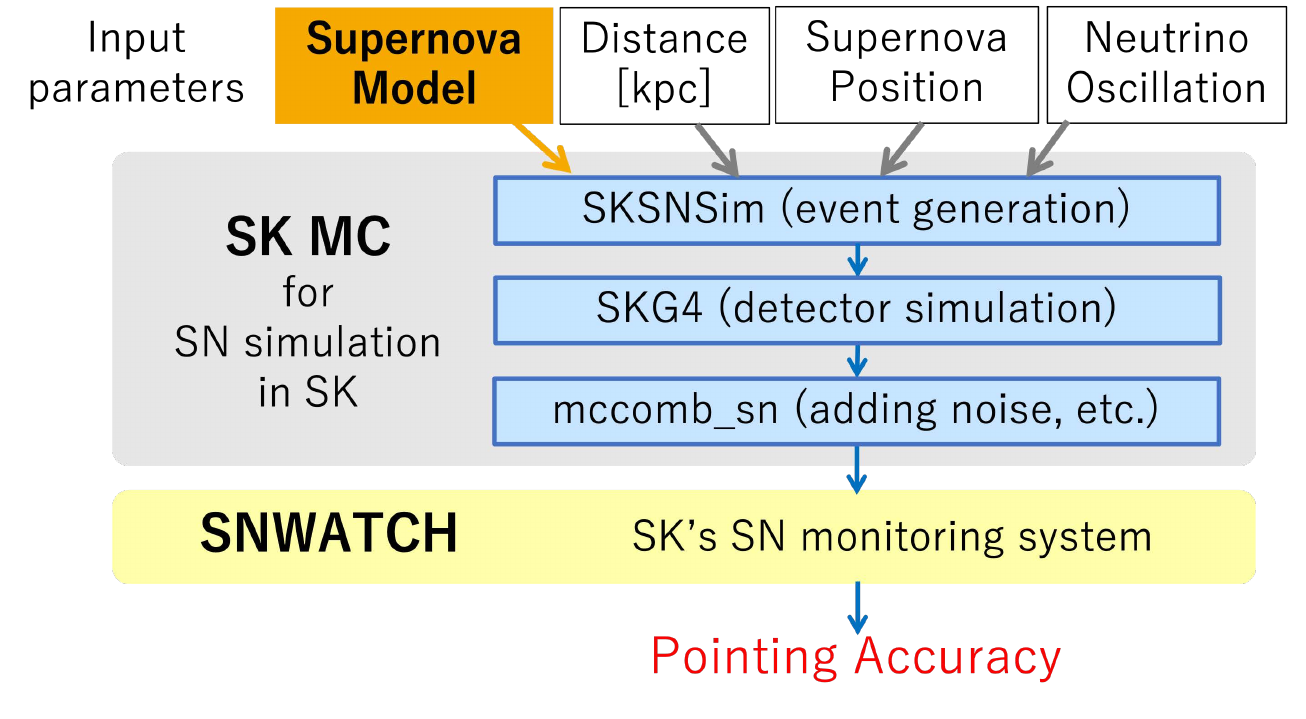}
    \caption{Overview of SN simulation in SK.  An SN neutrino emission data produced by an SN model (Section~\ref{subsec:SNmodelsSummary}), a distance to the SN, a position of the SN, and a neutrino oscillation scenario (no oscillation, oscillation with normal mass ordering (NMO), or oscillation with inverted mass ordering (IMO)) are input into the SK's event generator, SKSNSim (Section~\ref{subsec:EventGeneration}).  The simulated SN neutrino events in SK-Gd are generated with SKSNSim.  Then they are processed with SK's detector simulation, SKG4, and mccomb\_sn (Section~\ref{subsec:DetectorSimulation}).  
    SNWATCH processes the events before applying SN direction fit as described in Section~\ref{sec:SNWATCHupdate}.
    Pointing accuracy is estimated by repeating these sets of simulations 3000~times per an SN model, an SN distance, an SN position, and an oscillation type.}
\label{fig:SKSimulationOverview}
\end{figure}

Since the rate of core-collapse SNe in the galaxy is low, the only way to study the SK detector response to an SN burst is to use simulations.
Figure~\ref{fig:SKSimulationOverview} shows the flow of the SK SN simulation.
First, we choose an SN model (see Section~\ref{subsec:SNmodelsSummary}), the distance to and the galactic coordinates of the SN, and  neutrino oscillation parameters as input to our event generator, SKSNSim (see Section~\ref{subsec:EventGeneration}).  
The output of SKSNSim is then given to the SK detector simulation, SKG4, before noise from random trigger data in the actual detector is added to the events via the mccomb\_sn process. 
It additionally applies software triggers (see Section~\ref{subsec:DetectorSimulation}) to produce a simulated set of events with the same characteristics as the SK data.  
Finally, SNWATCH reconstructs these events, determines the direction of the SN, and estimates the pointing accuracy (explained in Section~\ref{sec:SNWATCHupdate}).
Note that this offline analysis uses the same software as the online version SNWATCH.
The details of these processes are described below.

\subsection{Supernova models}\label{subsec:SNmodelsSummary}
Since SN burst simulations consume considerable computational resources, models often simulate only the first ${\sim}$~1~s post bounce (e.g., \cite{marek2009delayed}).  
However, the typical SN neutrino emission timescale is known to be ${\sim}$~10~s based on observations of SN1987A, indicating that long-time simulations are also necessary.  
To overcome computational resource limitations, SN burst neutrino emission simulations impose simplified assumptions and approximations on the physics involved, such as employing one-dimensional spherically symmetric simulations. 
In many one-dimensional simulations the shock wave does not revive and results in a failed explosion.
Such models therefore assume shock revival mechanisms or artificially enhance neutrino reaction rates in order to produce a successful explosion (see Section~\ref{subsubsec:FischerModel}, for example).  
Changing these mechanisms, the progenitor mass and equation of state (EoS) contribute to a diversity of SN models.  

\begin{figure}[htb!]
\gridline{
    \fig{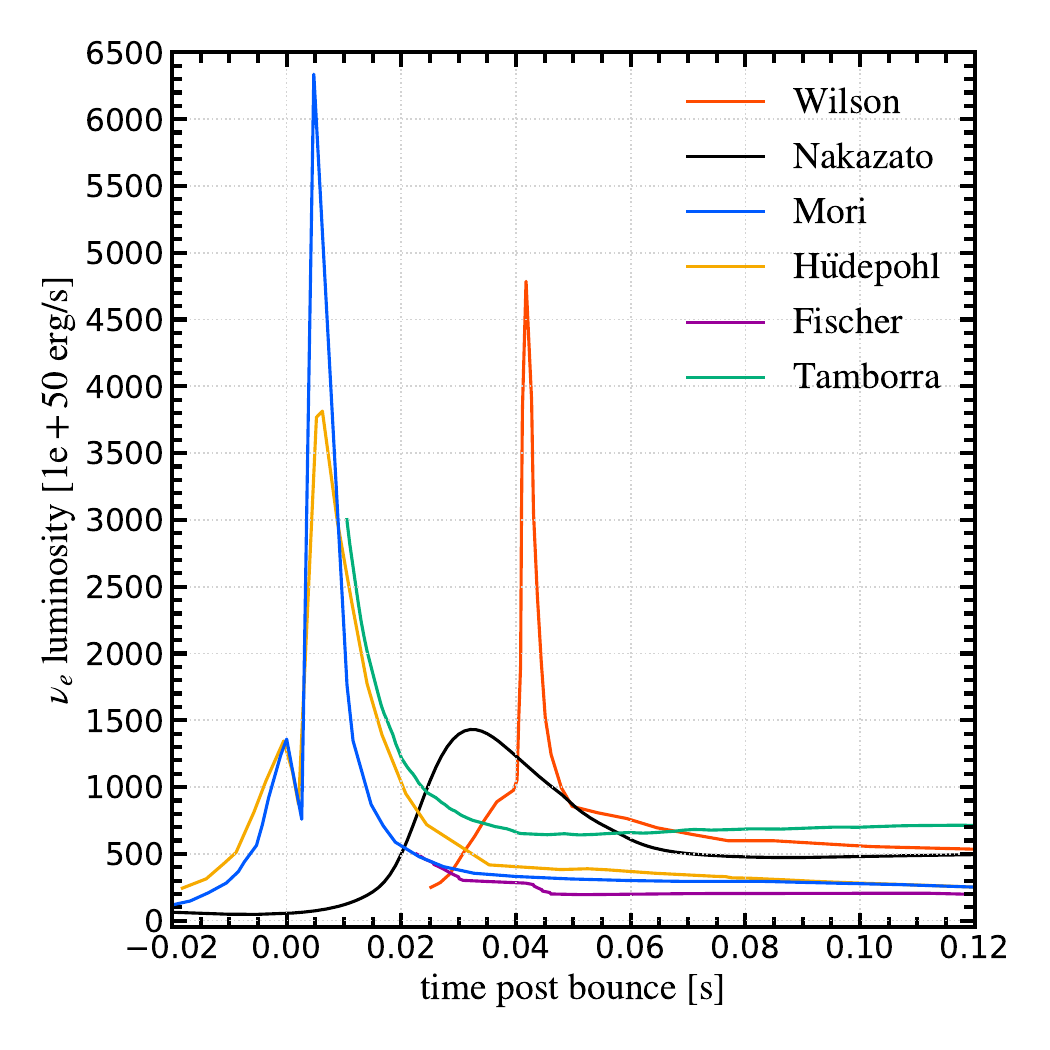}{0.33\textwidth}{(a) $\nu_\mathrm{e}$ luminosity}
    \fig{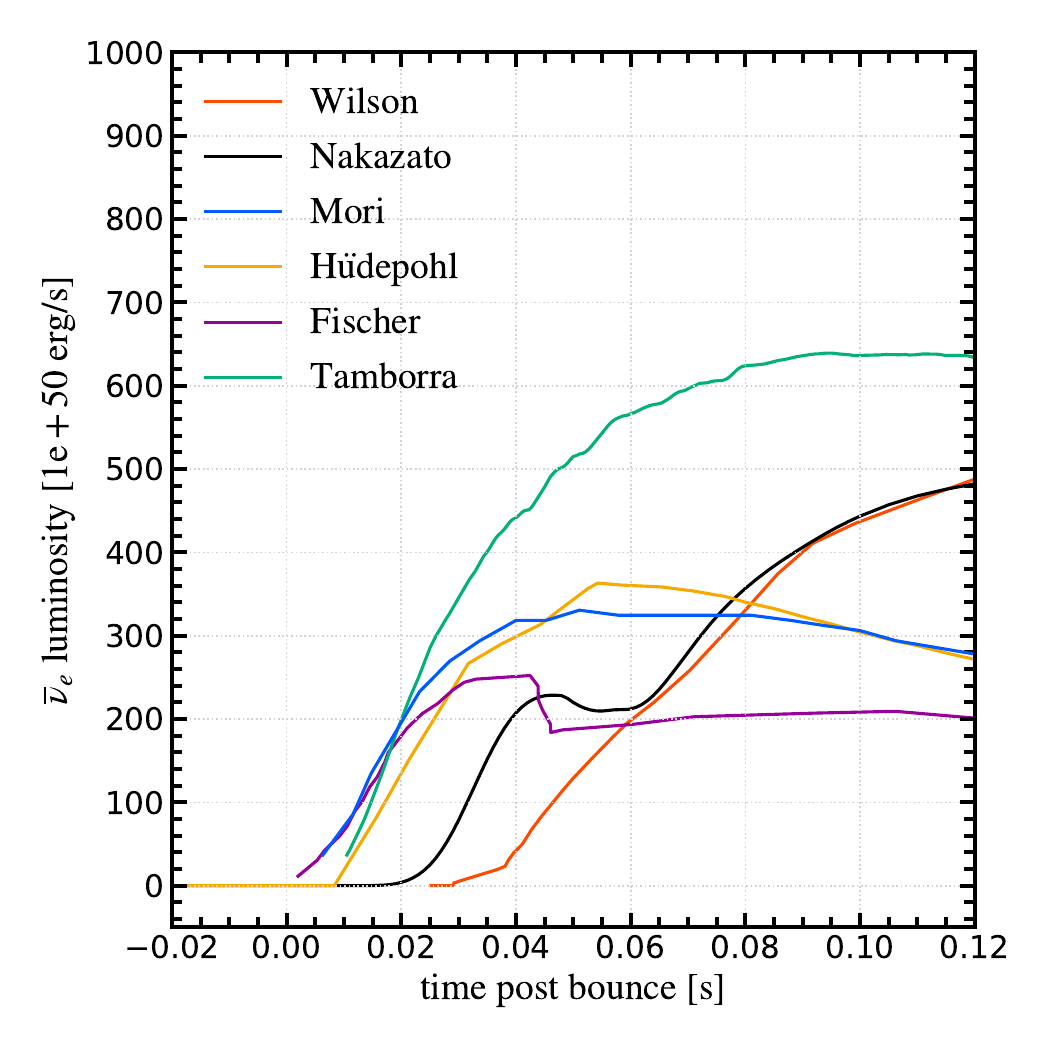}{0.33\textwidth}{(b) $\bar{\nu}_\mathrm{e}$ luminosity}
    \fig{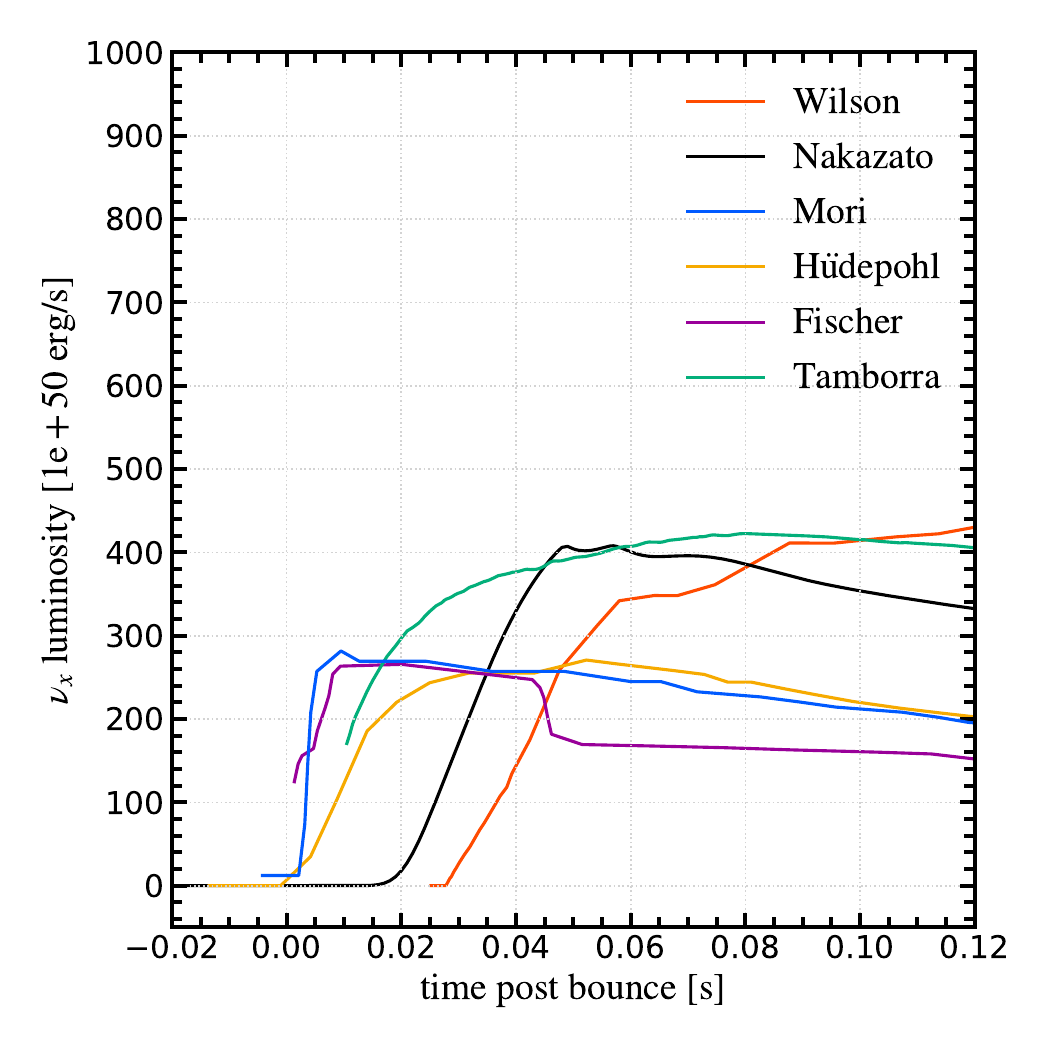}{0.33\textwidth}{(c) $\nu_x$ luminosity}
}
\gridline{
    \fig{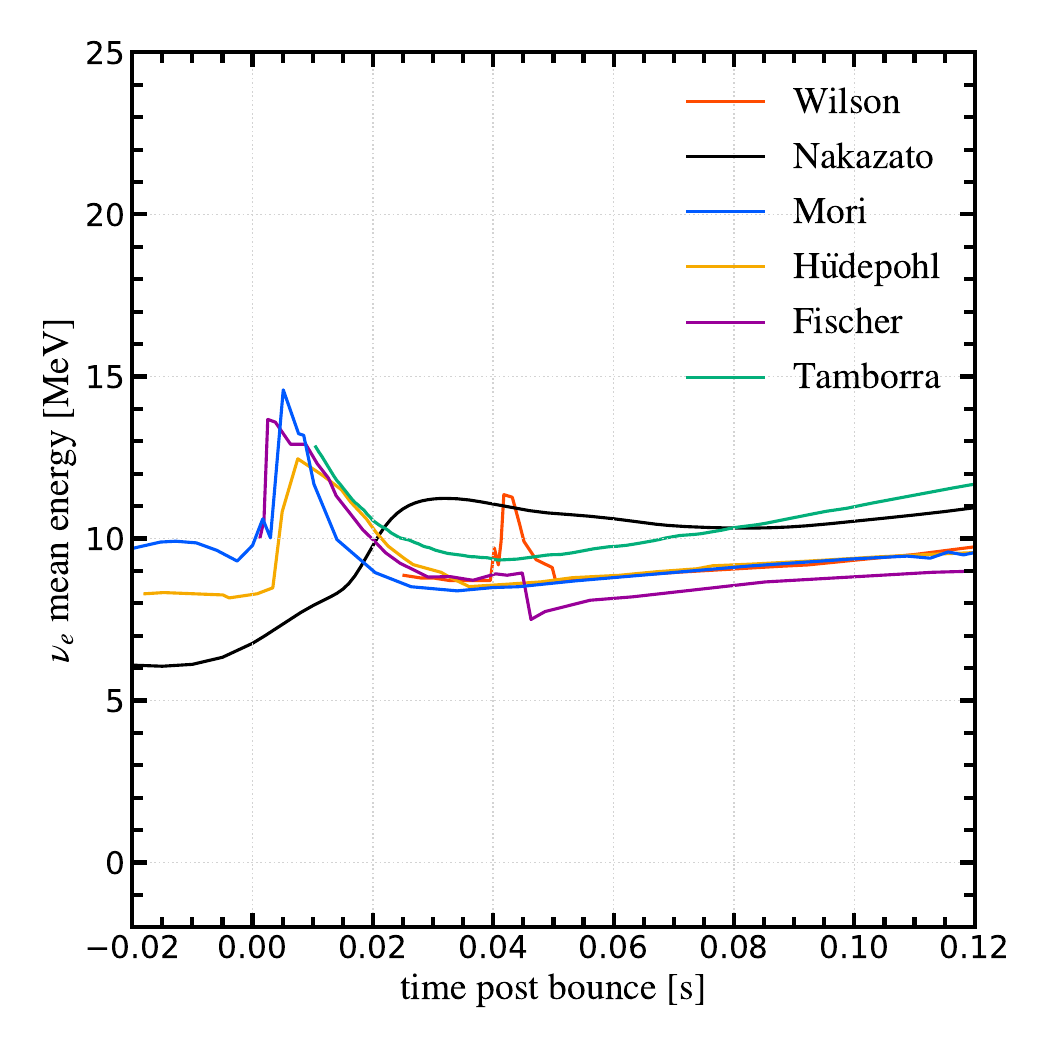}{0.33\textwidth}{(d) $\nu_\mathrm{e}$ mean energy}
    \fig{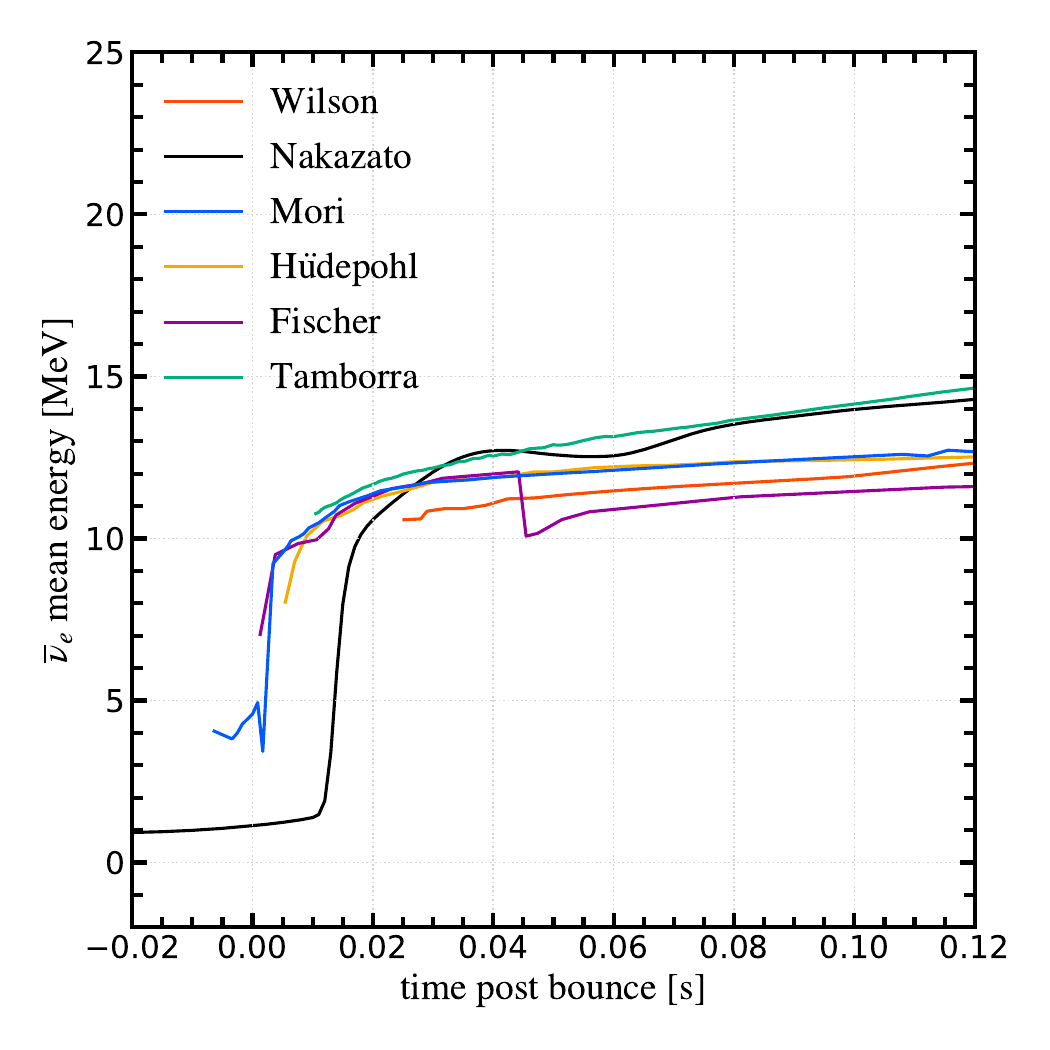}{0.33\textwidth}{(e) $\bar{\nu}_\mathrm{e}$ mean energy}
    \fig{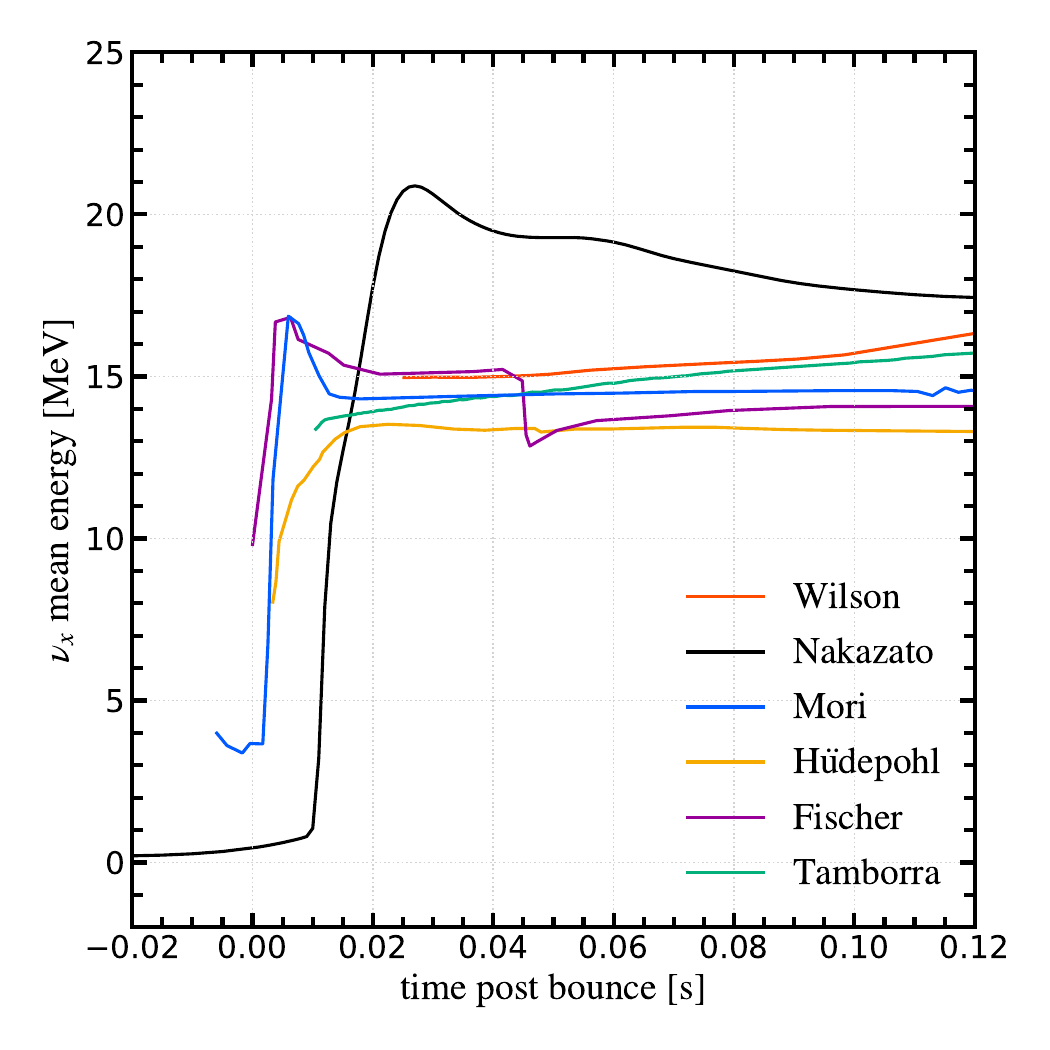}{0.33\textwidth}{(f) $\nu_x$ mean energy}
}
     \caption{Time evolution of luminosity (top) in the unit of 10$^{50}$~erg~s$^{-1}$ and mean energy (bottom) in the unit of MeV for six SN models listed in Table \ref{tab:6modelSummary} for each neutrino flavor, from -0.02~s to 0.12~s (the core bounce occurs at 0~s).  Here, $\nu_x$ means any of $\nu_\mu, \nu_\tau, \bar{\nu}_\mu,$ and $\bar{\nu}_\tau$.  The red, black, blue, orange, purple, and green lines represent the Wilson model, the Nakazato model, the Mori model, the H\"{u}depohl model, the Fischer model, and the Tamborra model, respectively, as indicated by the legend in each panel.  The striking peak in (a) for some models corresponds to the neutronization burst, and a dip before the neutronization burst seen in the Mori model and the H\"{u}depohl model corresponds to neutrino trapping. }
     \label{fig:TimeEvolutionLumiAndEave}
  \end{figure}

\begin{figure}[htb!]
\gridline{
    \fig{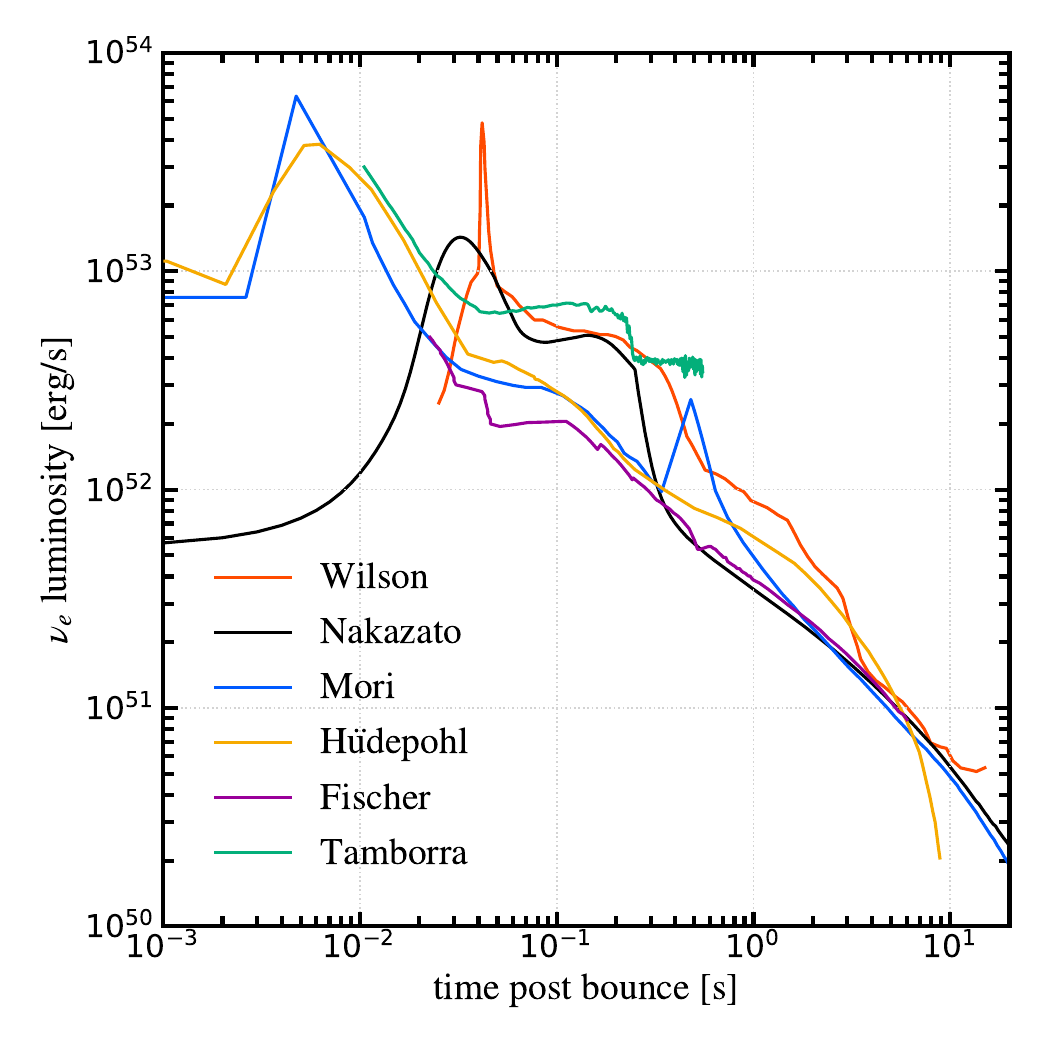}{0.33\textwidth}{(a) $\nu_\mathrm{e}$ luminosity}
    \fig{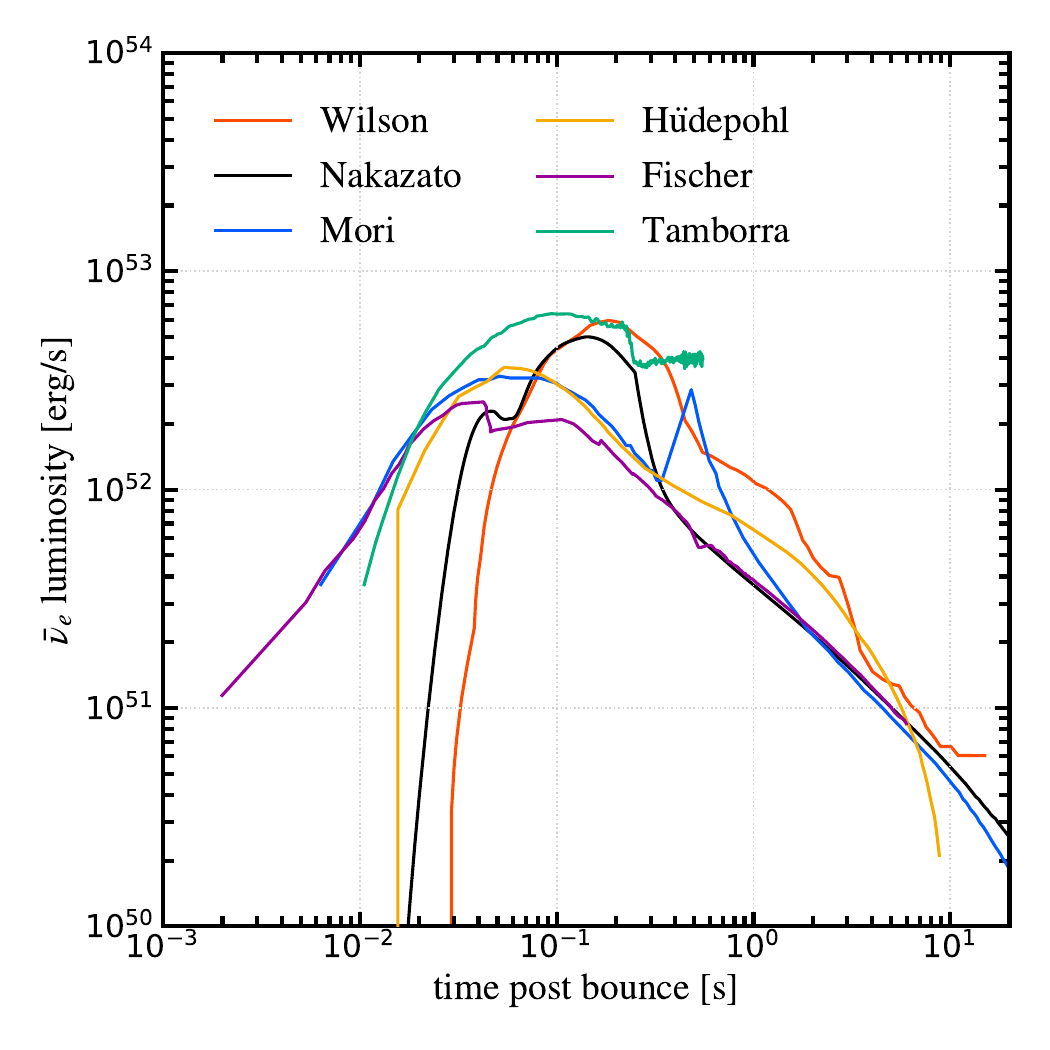}{0.33\textwidth}{(b) $\bar{\nu}_\mathrm{e}$ luminosity}
    \fig{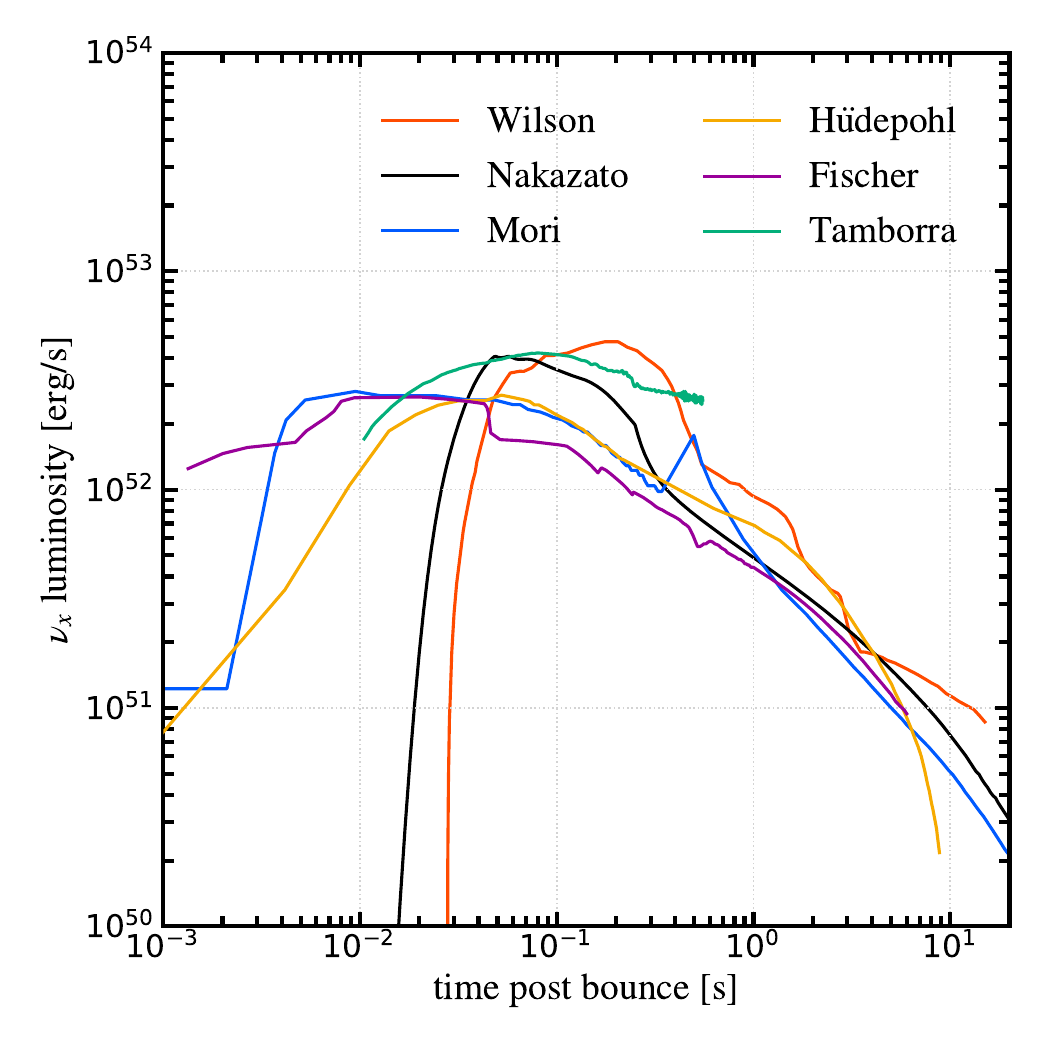}{0.33\textwidth}{(c) $\nu_x$ luminosity}
}
\gridline{
    \fig{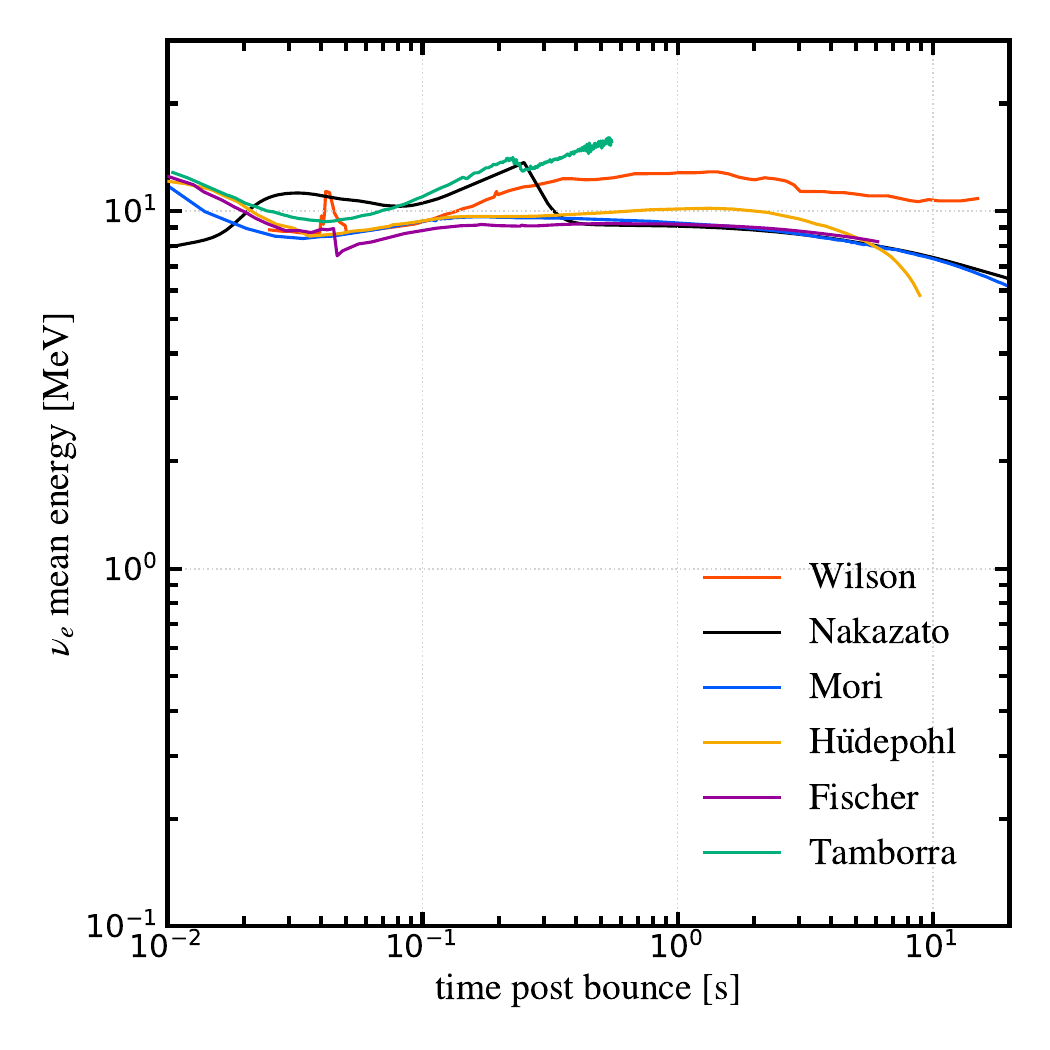}{0.33\textwidth}{(d) $\nu_\mathrm{e}$ mean energy}
    \fig{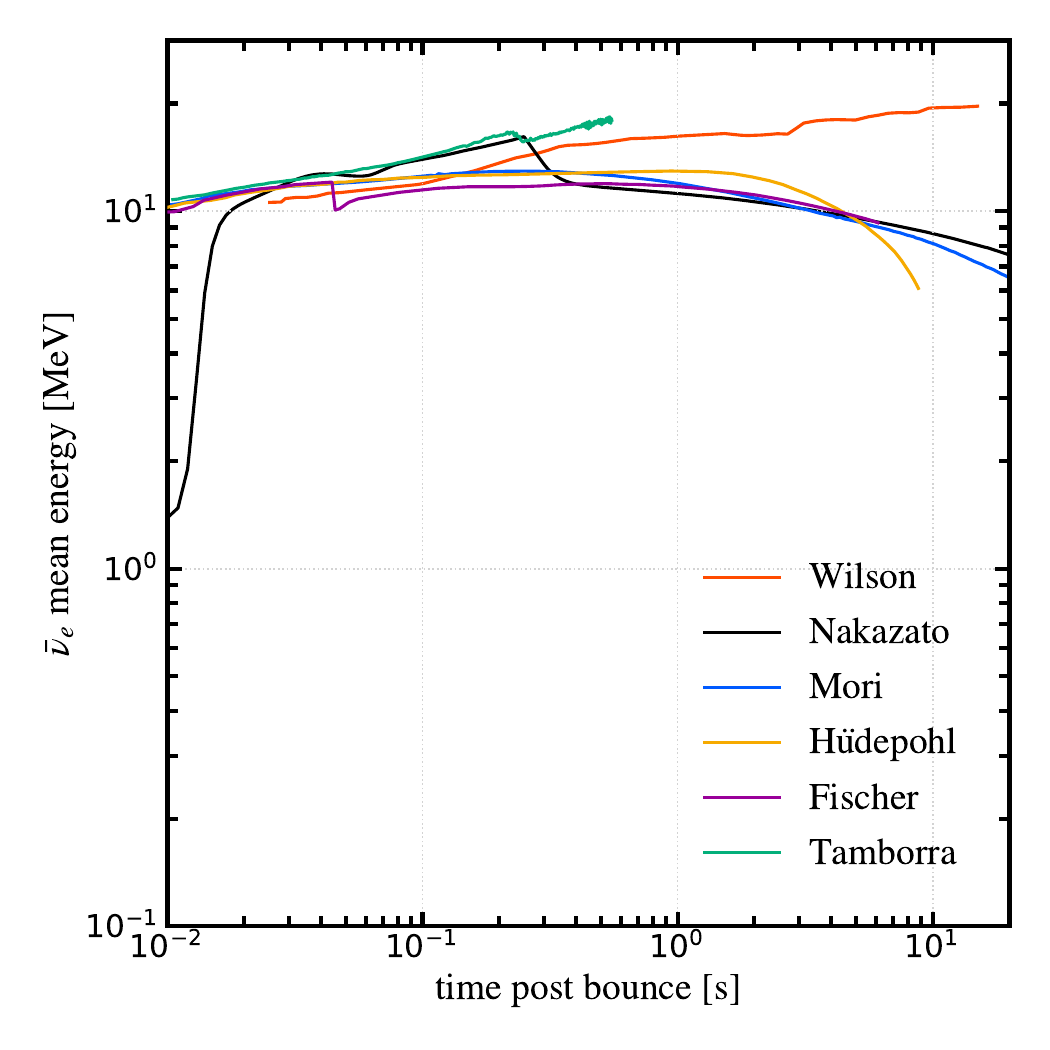}{0.33\textwidth}{(e) $\bar{\nu}_\mathrm{e}$ mean energy}
    \fig{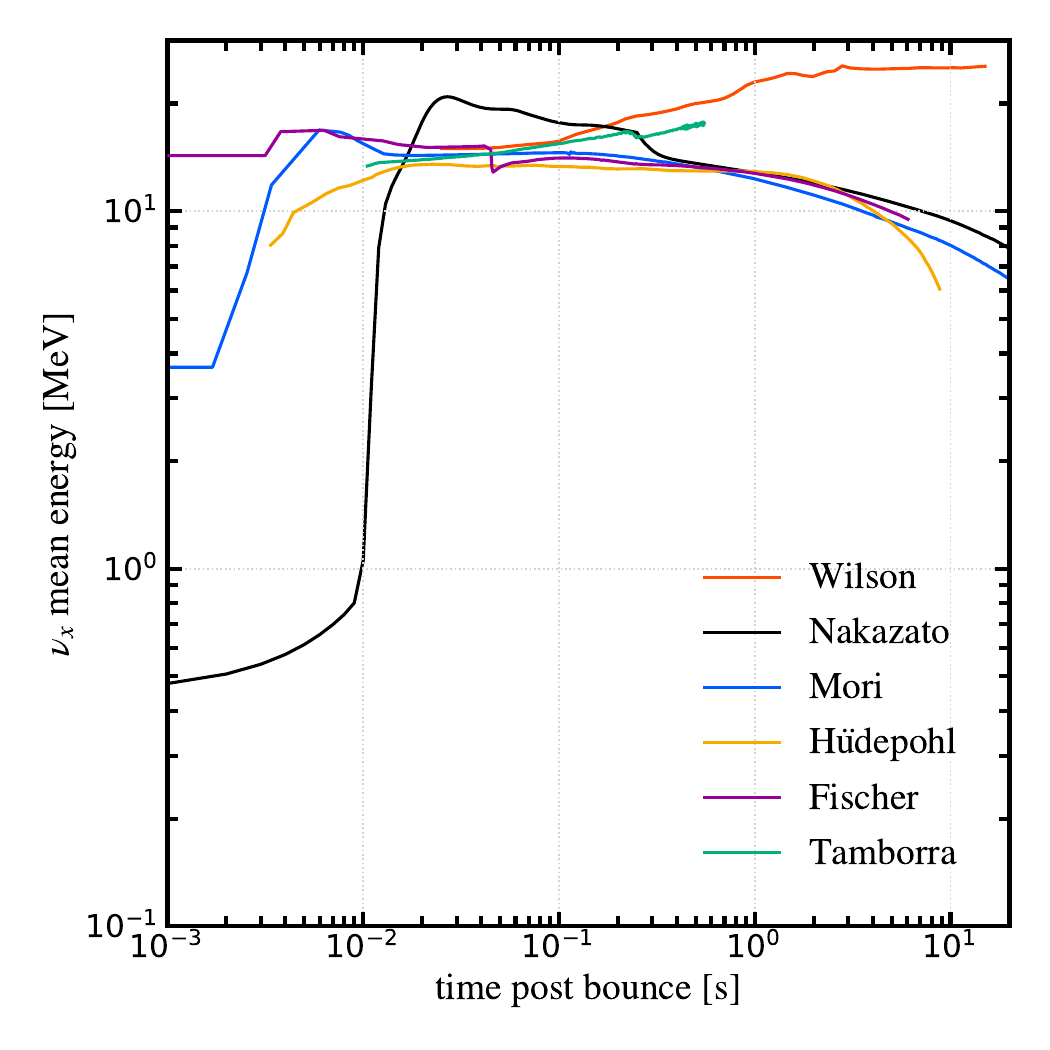}{0.33\textwidth}{(f) $\nu_x$ mean energy}
}
     \caption{Same as Figure~\ref{fig:TimeEvolutionLumiAndEave} but shows longer time range up to 20~s (the core bounce occurs at 0~s).  Note that luminosity is expressed in logarithmic scale in the unit of ~erg~s.$^{-1}$}
     \label{fig:LongTimeEvolutionLumiAndEave}
  \end{figure}

To look at SK's response to different models and to demonstrate our capability to differentiate between them, we selected the following five relatively long-term one-dimensional SN models: the historically significant Wilson model (see Section~\ref{subsubsec:WilsonModel}), two long-term one-dimensional models (the Nakazato model and the Mori model, see Section~\ref{subsubsec:NakazatoModel} and Section~\ref{subsubsec:MoriModel}, respectively), and two electron-capture SN models (the H\"{u}depohl model and the Fischer model, see Section~\ref{subsubsec:HudepohlModel} and Section~\ref{subsubsec:FischerModel}, respectively).  
We also study the Tamborra model, a pioneering three-dimensional model (see Section~\ref{subsubsec:TamborraModel}).  
Table~\ref{tab:6modelSummary} summarizes the main characteristics of the six models above. 
The time evolution of each neutrino flavor's luminosity and mean energy for these models over the first ${\sim}$~0.1~s and over the whole time range (20~s) are shown in Figure~\ref{fig:TimeEvolutionLumiAndEave} and Figure~\ref{fig:LongTimeEvolutionLumiAndEave}, respectively.
The following sub-sections summarize each model.

\begin{table}[htb!]
    \centering
    \caption{Summary of six SN model data employed in this article.  Shen, DD2, and LS mean equation of state (EoS) by \cite{shen1998aEoS, shen1998bEoS}, EoS based on density-dependent relativistic mean-field model~\citep{hempel2010DD2}, and EoS by \cite{lattimer1991generalized}, respectively.  Note that the start times and the durations shown in the table are after the linear extrapolation described in Appendix~\ref{subsec:LinearIntpAndExtrp} and do not necessarily correspond to that of the published models.}
    \label{tab:6modelSummary} 
	\begin{tabular}{ccccccc}\hline
        Model Name & Wilson & Nakazato & Mori & H\"{u}depohl & Fischer & Tamborra \\\hline
        Dimension & 1D & 1D & 1D & 1D & 1D & 3D\\
        progenitor mass [$M_\odot$] & 20 & 20 & 9.6 & 8.8 & 8.8 & 27 \\
        start time [s] & 0.03 & -0.05 & -0.256 & -0.02 & 0.0 & 0.011 \\
        duration [s] & 14.96 & 20.05 & 19.95 & 8.98 & 6.10 & 0.54\\
        EoS & - & Shen & DD2 & Shen & Shen & LS\\
        \hline
    \end{tabular} 
\end{table}

\subsubsection{The Wilson model}\label{subsubsec:WilsonModel}
The Wilson model~\citep{totani1998future} (also called ``the Livermore model'') is a one-dimensional core-collapse SN model with a long duration, stretching from the start of the collapse to ${\sim}$~18~s after the core bounce, which was a pioneering work in the 1990s.  Though it is now dated and no longer preferred in the literature, it remains a baseline reference in SN model papers today~\citep{abe2016snwatch, hk2021discriminate}.  
It is based on a numerical simulation code developed by Wilson and Mayle~\citep{wilson1986stellar, mayle1987neutrinos} 
and with a 20$M_\odot$ progenitor, it can reproduce SN1987A's light curve.
In this article, data scanned from Figure~1 of~\cite{totani1998future} are used in our simulations, covering neutrino emission up to  15~s after the core bounce. 

\subsubsection{The Nakazato model}\label{subsubsec:NakazatoModel}
The Nakazato model~\citep{nakazato2013supernova} is a long-time one-dimensional model that includes eight sets of progenitor masses and metallicities (four masses: 13, 20, 30, 50~$M_\odot$ and two metallicities: $Z=$0.02, 0.004).  
The simulation realizes neutrino emission from the start of the collapse to 20~s after the core bounce by combining a general relativistic neutrino-radiation hydrodynamic simulation ($\nu$RHD) for the early phase and quasi-static evolutionary neutrino diffusion calculations for the PNS cooling phase.  
This model uses the Shen EoS~\cite{shen1998aEoS, shen1998bEoS} and makes use of three times (100, 200, and 300~ms after the bounce) for the possible shock revival time.
We focus on the 20~$M_\odot$, $Z=0.02$ (solar metallicity) progenitor model with a shock revival at 200~ms after the bounce in this article.  Hereafter, ``the Nakazato model'' refers to this case.

\subsubsection{The Mori model}\label{subsubsec:MoriModel}
The Mori model~\citep{mori2021developing} is a one-dimensional core-collapse SN model with a 9.6~$M_\odot$ progenitor that simulates neutrino emission from the onset of core collapse through the PNS cooling phase up to 20~s after the core bounce.  
It uses a general relativistic hydrodynamics code with a spherically symmetric geometry for the accretion phase known as \textsc{GR1D}~\citep{o2015gr1d} to cover the PNS cooling phase.  
This model employs an EoS based on the density-dependent relativistic mean field (DD2)~\citep{mori2021developing}.

\subsubsection{The H\"{u}depohl model}\label{subsubsec:HudepohlModel}
The H\"{u}depohl model~\citep{hudepohl2010neutrino} is a one-dimensional electron-capture SN model with a spherically symmetric geometry.  
An 8.8~$M_\odot$ progenitor with an O-Ne-Mg core modeled by \cite{nomoto1987evolution} is adopted.
The simulation was performed using the neutrino radiation hydrodynamics code \textsc{PROMETHEUS-VERTEX}, which is composed of the hydrodynamics solver \textsc{PROMETHEUS}~\citep{fryxell1991instabilities} and the neutrino transport code \textsc{VERTEX}~\citep{rampp2002radiation}.  
Shen's EoS~\citep{shen1998aEoS, shen1998bEoS} is used.  
For the set of neutrino interactions, the model considers two cases: the Sf case with a ``full'' set of neutrino interactions listed in Appendix A in \cite{buras2006two} and the Sr case with a ``reduced'' set of neutrino interactions omitting pure neutrino interactions listed in \cite{bruenn1985stellar}.  
We use the results of the Sf case in this article, referred to as ``the H\"{u}depohl model'' hereafter.  
This case covers the PNS cooling phase but has a shortened simulation period lasting up until ${\sim}$~9~s because its high-density effects suppress neutrino opacities.

\subsubsection{The Fischer model}\label{subsubsec:FischerModel}
The Fischer model \citep{fischer2010protoneutron} is a one-dimensional SN model including three progenitors: the 8.8~$M_\odot$ O-Ne-Mg core progenitor modeled by \cite{nomoto1987evolution}, which is the same as the one used in the H\"{u}depohl model, and the 10.8 and 18~$M_\odot$ Fe-core progenitors in \cite{woosley2002evolution}.  
This model utilizes \textsc{AGILE-BOLTZTRAN}~\citep{liebendorfer2004agile}, a spherically symmetric, general relativistic hydrodynamics code (\textsc{AGILE}) with a Boltzmann neutrino transport solver (\textsc{BOLTZTRAN}).  
It also uses Shen's EoS~\citep{shen1998aEoS, shen1998bEoS}.
While the 8.8~$M_\odot$ progenitor explodes successfully, the 10.8, 18~$M_\odot$ progenitors do not explode naturally.  
In the latter two cases, neutrino reaction rates have been artificially enhanced to produce a successful explosion.  
In this article we have selected the naturally-exploding 8.8~$M_\odot$ model and use data scanned from Figures 2 and 14 of ~\cite{fischer2010protoneutron} which include emission up to ${\sim}$~6~s after the core bounce. 
Hereafter, ``the Fischer model'' indicates these data.

\subsubsection{The Tamborra model}\label{subsubsec:TamborraModel}
The Tamborra model~\citep{tamborra2014neutrino} is a pioneering three-dimensional simulation with energy-dependent neutrino transport.  It has three progenitors with different masses (11.2, 20, 27~$M_\odot$). 
The simulated time range is up to ${\sim}$~350~ms for the 11.2 and 20~$M_\odot$ progenitor cases and up to ${\sim}$~550~ms for the 27~$M_\odot$ progenitor case, however none of these explodes successfully within these simulation periods.  
The model uses the neutrino radiation hydrodynamics code \textsc{PROMETHEUS-VERTEX} and the Lattimer EoS with compressibility $K=$220~MeV~\cite{lattimer1991generalized}, often referred to as LS EoS.  
For the 20 and 27~$M_\odot$ progenitor cases, quasi-periodic oscillations appear in the neutrino signal due to SASI effects~\cite{2009ApJ...694..664M}. 
Since this signal modulation depends on the angular direction of the observer relative to the progenitor, the model considers three observer directions for each progenitor mass.
We use data for the 27~$M_\odot$ progenitor with the ``black'' observer direction in \cite{tamborra2014neutrino} provided by the SN simulation pipeline SNEWPY~\citep{baxter2022snewpy}.  Hereafter, ``the Tamborra model'' indicates this case.

\subsection{Event generation}\label{subsec:EventGeneration}
We generate events in SK using SKSNSim (Super-Kamiokande Supernova Simulation), an event generator for SN-related neutrino interactions~\citep{nakanishi2024supernova}.
As illustrated in Figure~\ref{fig:SKSimulationOverview}, for an SN at a given distance in an arbitrarily chosen position and a given neutrino oscillation case, SKSNSim computes the expected number of the neutrino interactions listed in Section~\ref{sec:SK-Gd} and generates events from the fluxes of the input SN model and cross sections.
For each interaction $r$, the expected number of events in each true time and energy bin $(E_{\nu},t)$ is calculated as:
\begin{equation}
	\frac{d^2N_r(E_{\nu},t)}{dtdE_{\nu}} = N_{r}\sigma_r(E_{\nu})\frac{d\phi(E_{\nu},t)}{dE_{\nu}}
\label{equ:expectedEventsperbin}
\end{equation}
where $N_r$ represents the number of targets (protons, electrons, or oxygen nuclei, available to interaction $r$) in the full inner volume of SK (32.5k~m$^3$), $\sigma(E_{\nu})$ is the cross section for interaction $r$ as a function of the neutrino energy, $E_\nu$, and  $d\phi(E_{\nu},t)/dE_{\nu}$ stands for the neutrino flux.  
We note that the neutrino flavor in equation~(\ref{equ:expectedEventsperbin}) depends on the interaction $r$.
The cross section calculations are taken from \cite{strumia2003precise} for IBD, from \cite{bahcall1995solar} for ES, from \cite{suzuki2018oxycc} and \cite{nakazato2018charged} for $^{16}$O~CC interactions, and from \cite{langanke1996oxync} and \cite{kolbe2002oxync} for $^{16}$O~NC interactions.  
The backward anisotropy of the angular distribution in the event topology of the $^{16}$O~CC interactions~\citep{tomas2003supernova} and the isotropic gamma-ray emission from the $^{16}$O~NC interaction are considered.
By integrating equation~(\ref{equ:expectedEventsperbin}) over time and energy, the expected number of events for the interaction $r$ for a single SN explosion is obtained.
SKSNSim then generates events using this value as the mean of a Poisson random process for each $r$.

Three-flavor neutrino oscillations including matter effects inside the SN are considered based on the calculation in \cite{dighe2000identifying}.
Assuming the normal mass ordering (NMO), the relationship between the number of neutrinos generated in a collapsing star $N_{\nu_i}^{gen}$ and the number of neutrinos at the stellar surface $N_{\nu_i}^{sur}$ for each flavor $i=e, \mu, \tau$ is given by
\begin{eqnarray}
    N_{\nu_\mathrm{e}}^{sur} &=& N_{\nu_x}^{gen} \label{eq:NMOnue}\\
    N_{\nu_x}^{sur} &=&N_{\nu_\mathrm{e}}^{gen}+N_{\nu_x}^{gen} \label{eq:NMOnux}\\
    N_{\bar{\nu}_\mathrm{e}}^{sur} &=& N_{\bar{\nu}_\mathrm{e}}^{gen}\times \cos^2\theta_{12}+N_{\bar{\nu}_x}^{gen}\times \sin^2\theta_{12} \label{eq:NMOneb}\\
    N_{\bar{\nu}_x}^{sur} &=& N_{\bar{\nu}_\mathrm{e}}^{gen}\times\sin^2\theta_{12}+N_{\bar{\nu}_x}^{gen}\times(1+\cos^2\theta_{12}), \label{eq:NMOnxb}
\end{eqnarray}
where $N_{\nu_x}^{gen}$ represents $N_{\nu_\mu}^{gen}$ or $N_{\nu_\tau}^{gen}$, while $N_{\nu_x}^{sur}=N_{\nu_\mu}^{sur}+N_{\nu_\tau}^{sur}$.
For the inverted mass ordering (IMO) the relations are 
\begin{eqnarray}
    N_{\nu_\mathrm{e}}^{sur} &=& N_{\nu_\mathrm{e}}^{gen}\times\sin^2\theta_{12} +N_{\nu_x}^{gen}\times\cos^2\theta_{12} \label{eq:IMOnue}\\
    N_{\nu_x}^{sur} &=& N_{\nu_\mathrm{e}}^{gen}\times\cos^2\theta_{12}+N_{\nu_x}^{gen}\times(1+\sin^2\theta_{12}) \label{eq:IMOnux}\\
    N_{\bar{\nu}_\mathrm{e}}^{sur} &=& N_{\bar{\nu}_x}^{gen} \label{eq:IMOneb}\\
    N_{\bar{\nu}_x}^{sur} &=& N_{\bar{\nu}_\mathrm{e}}^{gen}+N_{\bar{\nu}_x}^{gen}. \label{eq:IMOnxb}
\end{eqnarray}
Note that matter effects within the earth are not considered in SKSNSim.

SKSNSim was initially developed to read flux data from the Nakazato model's data format (denoted ``Nakazato format'' hereafter) which provides the differential neutrino flux and the differential neutrino luminosity at each time $t$ for each neutrino flavor in bins of energy.
To process other SN models with different data formats, we converted each to ``Nakazato format''(data format unification)
as presented  in Appendix~\ref{appendix:DataFormatUnification}.

\subsection{Detector simulation}\label{subsec:DetectorSimulation}
Events generated with SKSNSim are next processed with SKG4 (Super-Kamiokande \textsc{Geant4}-based simulation) and mccomb\_sn (Monte Carlo Combine for Supernova) as illustrated in Figure~\ref{fig:SKSimulationOverview}.
SKG4 \citep{harada2020geant4} includes Cherenkov radiation, photon scattering, absorption, and reflection in the water (or Gd solution) as well as similar processes on other detector materials.
Further it simulates the response of the SK PMTs to photons and the subsequent response of their readout electronics. 
In this work SKG4 adopts \textsc{Geant4} version 4.10.05.p01 with several modifications.
In particular, the Gd neutron capture model of neutron capture ANNRI-Gd~\citep{10.1093/ptep/ptz002, 10.1093/ptep/ptaa015} has been implemented for this work.
All analyses in this article are performed under the assumption of 0.033\% Gd concentration.

After processing with SKG4 the mccomb\_sn program is responsible for adding realistic dark noise, building events, and applying the software trigger.
Given the complexity of radioactive backgrounds in the SK detector, it is impossible to reliably simulate them with MC models alone.
Instead, hits from randomly triggered data 1~ms in length in the actual detector are added to the simulated events.
Since most of these hits are generated by backgrounds, this creates a realistic simulation of how SN burst events would appear in SK.
In the next step, sequences of simulated events are altered to have the same time structure as expected during a real neutrino burst.
Ideally, each neutrino interaction would represent a single event in the detector. 
However, multiple neutrino interactions can occur within the standard SK DAQ timing block of 17~$\mathrm{\mu}$s~\citep{Nishino:2007ccp, Orii:2015khh, Orii:2016qvf} and further hits from neighboring interactions may overlap in time with one another.
To simulate this, events are assigned random times consistent with their expected event rate evolution, and all hits from all events are then laid out in time. 
These are then separated into 17~$\mathrm{\mu}$s blocks, as is done with the real data, and the software trigger is applied. 
Based on the triggers raised, the hits are repackaged into events using hits before and after the trigger condition was met. 
For this reason, all hits in the final event may not necessarily have a one-to-one correspondence with those from the initial interaction that created them; they may be associated with a neighboring event via this process. 

\section{Simulated Interaction Events in SK-G\lowercase{d} and Their Detector Responses}\label{sec:Results-truth}
In the following, we discuss simulated interactions in SK-Gd and the response of the detector assuming an SN burst located at 10~kpc for each of the six models as described in Section~\ref{sec:Simulations}.
For each model, three types of neutrino oscillation scenarios are studied: without oscillation (No Osc.) and neutrino oscillations under NMO and IMO.  
The calculation assumes $\sin^2\theta_{12}=0.307$~\citep{pdg2022review}.
As mentioned in Section~\ref{subsec:SNWarnWithDirInfo}, though SNWATCH typically uses 3000~MC trials to determine the pointing accuracy to an SN, due to limited computing resources for this study, 1000 MC trials are used here for each combination of SN model and oscillation scenario. 
We note that all simulations use a common SN position, arbitrarily chosen to be the position of the Sun at 0:00 on 23rd March 2011 (i.e., RA~$=1^\mathrm{h}41^\mathrm{m}43.^\mathrm{s}6$, Dec~$=+0^\circ39'12.''5$).

Table~\ref{tab:AverageGenEvents1} shows the average number of events taken from 1000~MC trials of each SN model with events in each trial generated in the  full 32.5k m$^3$ of the ID. 
The number of oxygen-interaction events in Table~\ref{tab:AverageGenEvents1} shows significant differences among models (c.f. Figure~\ref{fig:NMOvarModelEnergyInteractions}). 
These differences originate from the energy spectra of incoming neutrinos; since the inelastic interaction cross section is larger at higher neutrino energies (c.f. Figure~\ref{fig:SNnuCrossSections}), the detected number of oxygen interactions depends on the neutrino energy.  

  \begin{table}[htb!]
	\caption{Average number of events at SK generated by SKSNSim (corresponding to the events after ``SKSNSim (event generation)'' in Figure~\ref{fig:SKSimulationOverview}) in the 32.5k~m$^3$ full volume for 1000 simulated SN bursts located at 10~kpc for the Wilson model, the Nakazato model, the Mori model, the H\"{u}depohl model, the Fischer model, and the Tamborra models.  Note that $\nu_x$ represents $\nu_\mu$ and $\nu_\tau$. No energy cut is applied.}
    \label{tab:AverageGenEvents1}
    \hspace{-1.5cm}
    \begin{tabular}{cr||rr|r||rr|r||rr}\hline
		\textbf{Generated by} & \multicolumn{3}{l|}{Wilson} & \multicolumn{3}{l|}{Nakazato} & \multicolumn{3}{l}{Mori}\\\cline{2-10}
        \textbf{SKSNSim}& No~Osc. & NMO & IMO & No~Osc. & NMO & IMO & No~Osc. & NMO & IMO\\\hline
        IBD ($\bar{\nu}_\mathrm{e}$) &	7431	&	8207	&	9970	&	3542	&	3893	&	4693	&	3275	&	3422	&	3745	\\
        ES ($\nu_\mathrm{e}$) &	223	&	231	&	229	&	173	&	172	&	171	&	177	&	148	&	156	\\
        ES ($\bar{\nu}_\mathrm{e}$) &	97	&	97	&	98	&	63	&	66	&	72	&	60	&	61	&	63	\\
        ES ($\nu_x$) &	80	&	79	&	80	&	60	&	60	&	60	&	52	&	57	&	56	\\
        ES ($\bar{\nu}_x$) &	69	&	69	&	69	&	52	&	51	&	48	&	45	&	45	&	44	\\
        $^{16}$O~CC ($\nu_\mathrm{e}$) &	44	&	1034	&	729	&	48	&	180	&	139	&	8	&	86	&	62	\\
        $^{16}$O~CC ($\bar{\nu}_\mathrm{e}$) &	195	&	329	&	633	&	46	&	68	&	116	&	30	&	42	&	71	\\
        $^{16}$O~NC ($\nu_\mathrm{e}, ^{15}$N) &	4	&	89	&	63	&	4	&	15	&	12	&	1	&	8	&	5	\\
        $^{16}$O~NC ($\bar{\nu}_\mathrm{e}, ^{15}$N) &	22	&	43	&	89	&	5	&	8	&	16	&	3	&	4	&	8	\\
        $^{16}$O~NC ($\nu_x, ^{15}$N) &	177	&	93	&	119	&	31	&	20	&	23	&	15	&	8	&	10	\\
        $^{16}$O~NC ($\bar{\nu}_x, ^{15}$N) &	177	&	156	&	112	&	31	&	28	&	21	&	15	&	14	&	10	\\
        $^{16}$O~NC ($\nu_\mathrm{e}, ^{15}$O) &	1	&	24	&	17	&	1	&	4	&	3	&	0	&	2	&	1	\\
        $^{16}$O~NC ($\bar{\nu}_\mathrm{e}, ^{15}$O) &	6	&	12	&	24	&	1	&	2	&	4	&	1	&	1	&	2	\\
        $^{16}$O~NC ($\nu_x, ^{15}$O) &	48	&	25	&	32	&	9	&	5	&	6	&	4	&	2	&	3	\\
        $^{16}$O~NC ($\bar{\nu}_x, ^{15}$O) &	48	&	42	&	30	&	8	&	8	&	5	&	4	&	4	&	3	\\\hline
        total &	8622	&	10530	&	12294	&	4074	&	4580	&	5389	&	3690	&	3904	&	4239	\\\hline\hline
		\textbf{Generated by}& \multicolumn{3}{l|}{H\"{u}depohl}  &  \multicolumn{3}{l|}{Fischer} & \multicolumn{3}{l}{Tamborra} \\\cline{2-10}
        \textbf{SKSNSim}& No~Osc. & NMO & IMO & No~Osc. & NMO & IMO & No~Osc. & NMO & IMO\\\hline
        IBD ($\bar{\nu}_\mathrm{e}$) &	3048	&	3052	&	3049	&	1884	&	1990	&	2242	&	3830	&	3487	&	2718	\\
        ES ($\nu_\mathrm{e}$) &	146	&	124	&	132	&	90	&	87	&	88	&	135	&	82	&	99	\\
        ES ($\bar{\nu}_\mathrm{e}$) &	53	&	53	&	53	&	35	&	35	&	37	&	50	&	45	&	35	\\
        ES ($\nu_x$) &	43	&	47	&	46	&	31	&	31	&	31	&	28	&	38	&	35	\\
        ES ($\bar{\nu}_x$) &	38	&	38	&	38	&	27	&	26	&	25	&	25	&	26	&	30	\\
        $^{16}$O~CC ($\nu_\mathrm{e}$) &	12	&	32	&	26	&	5	&	27	&	21	&	55	&	90	&	80	\\
        $^{16}$O~CC ($\bar{\nu}_\mathrm{e}$) &	30	&	31	&	33	&	15	&	18	&	27	&	97	&	90	&	77	\\
        $^{16}$O~NC ($\nu_\mathrm{e}, ^{15}$N) &	1	&	3	&	2	&	0	&	2	&	2	&	5	&	8	&	7	\\
        $^{16}$O~NC ($\bar{\nu}_\mathrm{e}, ^{15}$N) &	3	&	3	&	3	&	1	&	2	&	2	&	11	&	10	&	8	\\
        $^{16}$O~NC ($\nu_x, ^{15}$N) &	6	&	4	&	4	&	5	&	3	&	4	&	16	&	13	&	14	\\
        $^{16}$O~NC ($\bar{\nu}_x, ^{15}$N) &	6	&	6	&	6	&	5	&	4	&	4	&	16	&	17	&	19	\\
        $^{16}$O~NC ($\nu_\mathrm{e}, ^{15}$O) &	0	&	1	&	1	&	0	&	1	&	1	&	1	&	2	&	2	\\
        $^{16}$O~NC ($\bar{\nu}_\mathrm{e}, ^{15}$O) &	1	&	1	&	1	&	0	&	0	&	1	&	3	&	3	&	2	\\
        $^{16}$O~NC ($\nu_x, ^{15}$O) &	1	&	1	&	1	&	1	&	1	&	1	&	4	&	3	&	4	\\
        $^{16}$O~NC ($\bar{\nu}_x, ^{15}$O) &	2	&	2	&	1	&	1	&	1	&	1	&	4	&	5	&	5	\\\hline
        total &	3390	&	3398	&	3396	&	2100	&	2228	&	2487	&	4280	&	3919	&	3135	\\\hline
    \end{tabular}
\end{table}

Table~\ref{tab:AverageEvents1} shows the average number of events for each interaction reconstructed in SK-Gd's 22.5k~m$^3$ fiducial volume.
Note that the original interaction information generated by SKSNSim is lost during the mccomb\_sn process as explained in Section~\ref{subsec:DetectorSimulation}.
Therefore, the interaction type that created each reconstructed event is estimated by matching the time of the reconstruction to the time generated by SKSNSim.
The closest MC event within 0.02~$\mathrm{\mu s}$ of the reconstructed time is considered the parent. 
Although the fiducial volume is about 70\% of the full volume of the detector, more than half of ES events are lost during the event selection process. 
Low energy ES events and $^{16}$O~NC events are removed mainly by the 7~MeV cut in the ``prompt'' candidate selection (c.f. Figure~\ref{fig:NMOEnergyInteractionNakazato}).
 
\begin{table}[htb!]
	\caption{Average number of reconstructed events at SK-Gd (with 0.03\% Gd concentration) (corresponding to the events after ``Event Rectonstruction'' in Figure~\ref{fig:SKSimulationOverview}) in the 22.5k~m$^3$ fiducial volume for 1000 simulated SN bursts located at 10~kpc with the Wilson model, the Nakazato model, the Mori model, the H\"{u}depohl model, the Fischer model, and the Tamborra model.  Here, the word ``reconstructed events'' means the events reconstructed in SNWATCH from the PMT signal of the SKSNSim-generated events, and the IBD tagging is not applied to them yet. The event with energy 7~MeV or more are selected as ``prompt'' candidates (see Section~\ref{subsec:EventReconstruction} for detail).  With this 7~MeV energy cut, the number of reconstructed $^{16}$O~NC events are zero.  As the original information generated by SKSNSim is lost during the mccomb\_sn process, to identify the interaction of the reconstructed event, the reconstructed events are associated with the SKSNSim-generated event; the condition of less than 0.02~$\mathrm{\mu s}$ time difference between the SKSNSim-generated event and the closest reconstructed event is applied.  Note that $\nu_x$ represents $\nu_\mu$ and $\nu_\tau$.}
    \label{tab:AverageEvents1}
    \hspace{-1.5cm}
    \begin{tabular}{cr||rr|r||rr|r||rr}\hline
		\textbf{Reconstructed}& \multicolumn{3}{l|}{Wilson} & \multicolumn{3}{l|}{Nakazato} & \multicolumn{3}{l}{Mori}\\\cline{2-10}
        & No~Osc. & NMO & IMO & No~Osc. & NMO & IMO & No~Osc. & NMO & IMO\\\hline
        IBD ($\bar{\nu}_\mathrm{e}$) &	4879	&	5364	&	6465	&	2221	&	2434	&	2921	&	2048	&	2144	&	2355	\\
        ES ($\nu_\mathrm{e}$) &	69	&	106	&	95	&	43	&	57	&	53	&	44	&	46	&	45	\\
        ES ($\bar{\nu}_\mathrm{e}$) &	22	&	25	&	30	&	10	&	11	&	13	&	9	&	9	&	10	\\
        ES ($\nu_x$) &	34	&	28	&	30	&	18	&	16	&	17	&	15	&	14	&	14	\\
        ES ($\bar{\nu}_x$) &	28	&	27	&	26	&	15	&	14	&	13	&	12	&	11	&	11	\\
        $^{16}$O~CC ($\nu_\mathrm{e}$) &	27	&	660	&	465	&	29	&	108	&	83	&	4	&	56	&	40	\\
        $^{16}$O~CC ($\bar{\nu}_\mathrm{e}$) &	126	&	208	&	394	&	28	&	41	&	69	&	19	&	27	&	45	\\
        $^{16}$O~NC ($\nu_\mathrm{e}, ^{15}$N) &	0	&	0	&	0	&	0	&	0	&	0	&	0	&	0	&	0	\\
        $^{16}$O~NC ($\bar{\nu}_\mathrm{e}, ^{15}$N) &	0	&	0	&	0	&	0	&	0	&	0	&	0	&	0	&	0	\\
        $^{16}$O~NC ($\nu_x, ^{15}$N) &	0	&	0	&	0	&	0	&	0	&	0	&	0	&	0	&	0	\\
        $^{16}$O~NC ($\bar{\nu}_x, ^{15}$N) &	0	&	0	&	0	&	0	&	0	&	0	&	0	&	0	&	0	\\
        $^{16}$O~NC ($\nu_\mathrm{e}, ^{15}$O) &	0	&	0	&	0	&	0	&	0	&	0	&	0	&	0	&	0	\\
        $^{16}$O~NC ($\bar{\nu}_\mathrm{e}, ^{15}$O) &	0	&	0	&	0	&	0	&	0	&	0	&	0	&	0	&	0	\\
        $^{16}$O~NC ($\nu_x, ^{15}$O) &	0	&	0	&	0	&	0	&	0	&	0	&	0	&	0	&	0	\\
        $^{16}$O~NC ($\bar{\nu}_x, ^{15}$O) &	0	&	0	&	0	&	0	&	0	&	0	&	0	&	0	&	0	\\\hline
        total &	5185	&	6418	&	7505	&	2364	&	2681	&	3169	&	2151	&	2307	&	2520	\\\hline\hline
		\textbf{Reconstructed}& \multicolumn{3}{l|}{H\"{u}depohl}  &  \multicolumn{3}{l|}{Fischer} & \multicolumn{3}{l}{Tamborra} \\\cline{2-10}
        & No~Osc. & NMO & IMO & No~Osc. & NMO & IMO & No~Osc. & NMO & IMO\\\hline
        IBD ($\bar{\nu}_\mathrm{e}$) &	1936	&	1939	&	1935	&	1186	&	1260	&	1437	&	2505	&	2283	&	1786	\\
        ES ($\nu_\mathrm{e}$) &	38	&	39	&	39	&	22	&	29	&	26	&	46	&	33	&	37	\\
        ES ($\bar{\nu}_\mathrm{e}$) &	9	&	8	&	8	&	5	&	6	&	6	&	12	&	10	&	8	\\
        ES ($\nu_x$) &	12	&	12	&	12	&	9	&	8	&	8	&	10	&	12	&	12	\\
        ES ($\bar{\nu}_x$) &	10	&	10	&	10	&	7	&	7	&	7	&	8	&	9	&	10	\\
        $^{16}$O~CC ($\nu_\mathrm{e}$) &	7	&	19	&	16	&	3	&	17	&	13	&	35	&	58	&	51	\\
        $^{16}$O~CC ($\bar{\nu}_\mathrm{e}$) &	19	&	20	&	21	&	9	&	11	&	17	&	62	&	58	&	49	\\
        $^{16}$O~NC ($\nu_\mathrm{e}, ^{15}$N) &	0	&	0	&	0	&	0	&	0	&	0	&	0	&	0	&	0	\\
        $^{16}$O~NC ($\bar{\nu}_\mathrm{e}, ^{15}$N) &	0	&	0	&	0	&	0	&	0	&	0	&	0	&	0	&	0	\\
        $^{16}$O~NC ($\nu_x, ^{15}$N) &	0	&	0	&	0	&	0	&	0	&	0	&	0	&	0	&	0	\\
        $^{16}$O~NC ($\bar{\nu}_x, ^{15}$N) &	0	&	0	&	0	&	0	&	0	&	0	&	0	&	0	&	0	\\
        $^{16}$O~NC ($\nu_\mathrm{e}, ^{15}$O) &	0	&	0	&	0	&	0	&	0	&	0	&	0	&	0	&	0	\\
        $^{16}$O~NC ($\bar{\nu}_\mathrm{e}, ^{15}$O) &	0	&	0	&	0	&	0	&	0	&	0	&	0	&	0	&	0	\\
        $^{16}$O~NC ($\nu_x, ^{15}$O) &	0	&	0	&	0	&	0	&	0	&	0	&	0	&	0	&	0	\\
        $^{16}$O~NC ($\bar{\nu}_x, ^{15}$O) &	0	&	0	&	0	&	0	&	0	&	0	&	0	&	0	&	0	\\\hline
        total &	2031	&	2037	&	2041	&	1241	&	1338	&	1514	&	2678	&	2463	&	1953	\\\hline
    \end{tabular}
\end{table}

\subsection{SK-Gd's Response with the Nakazato Model}\label{subsubsec:SK-Gd'sResponseWithNakazatoNMO}
This section presents SK-Gd's response to an SN burst located at 10~kpc simulated with the Nakazato model (20$M_\odot$, $Z=0.02$) assuming the NMO scenario as a representative example.
Similar studies and figures for other models under the IMO scenario are presented in Appendix~\ref{app:differenceAmongModels}.
Reconstructed events in the following figures correspond to events after reconstruction in SNWATCH in Figure~\ref{fig:SNWATCH_system} and after the time matching between the reconstructed time and the generated time by SKSNSim described in Section~\ref{sec:Results-truth}.

\begin{figure}[htb!]
\gridline{\fig{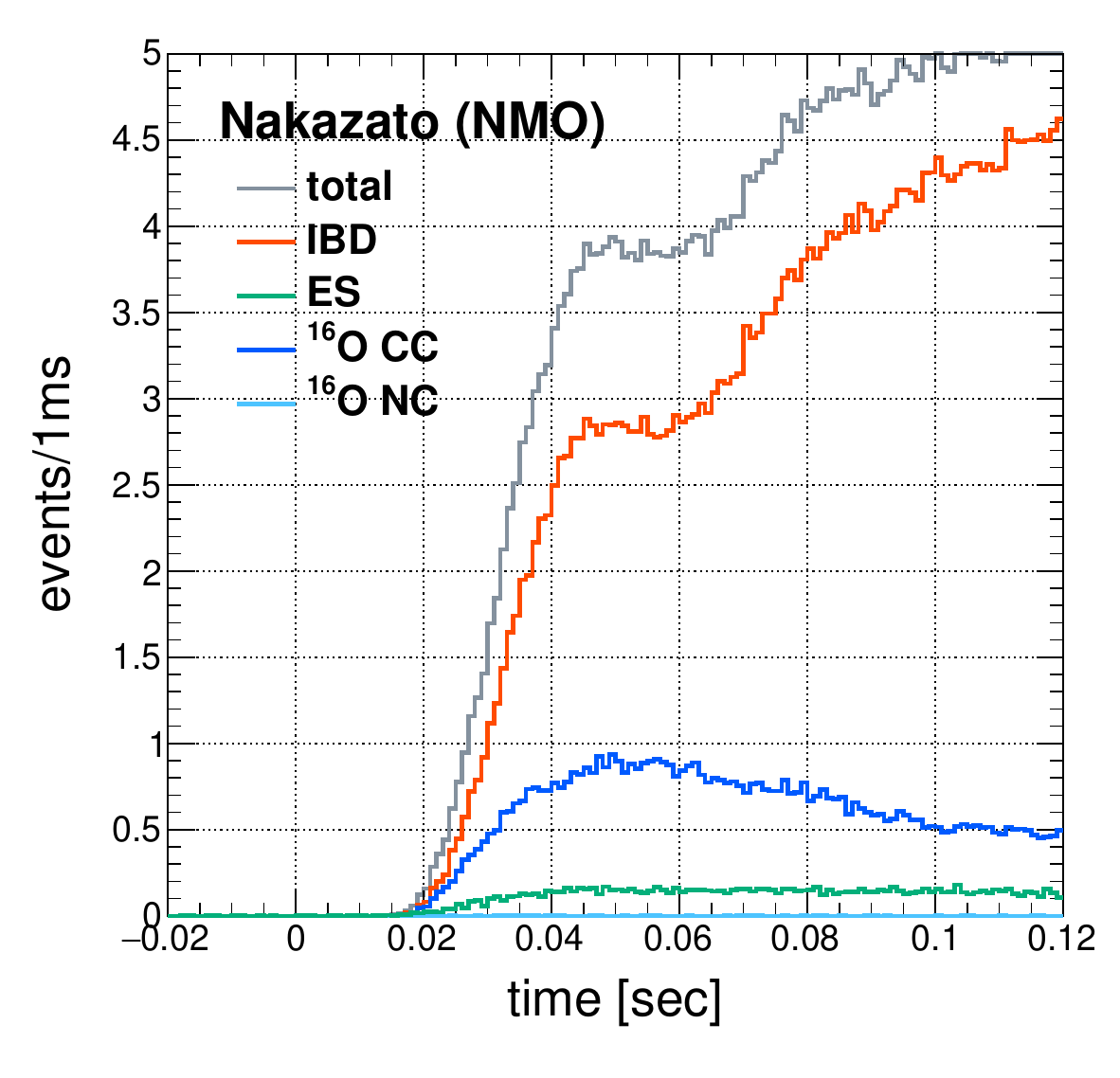}{0.3\textwidth}{(a) up to 0.12~s}
\fig{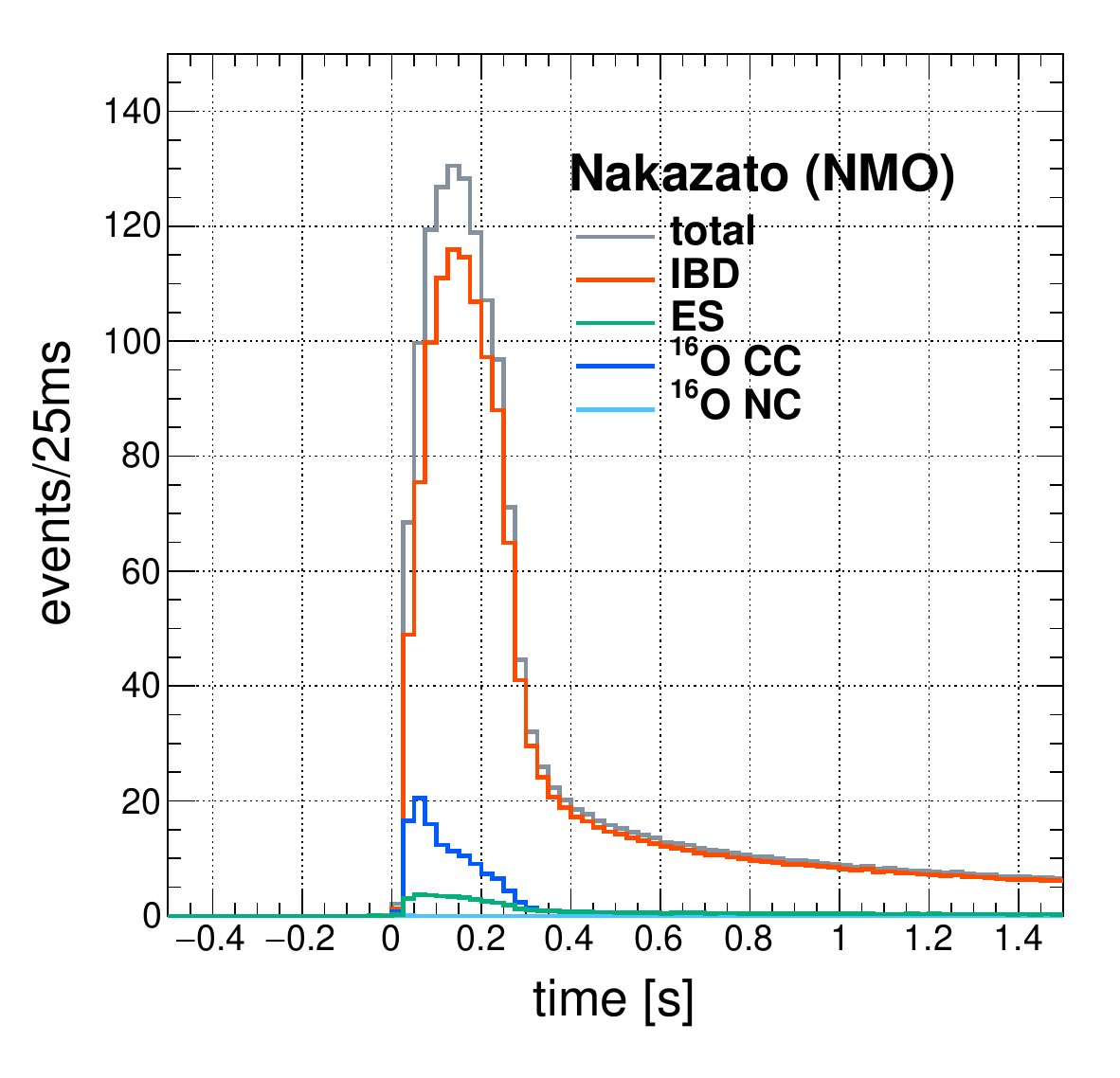}{0.3\textwidth}{(b) up to 1.5~s}
\fig{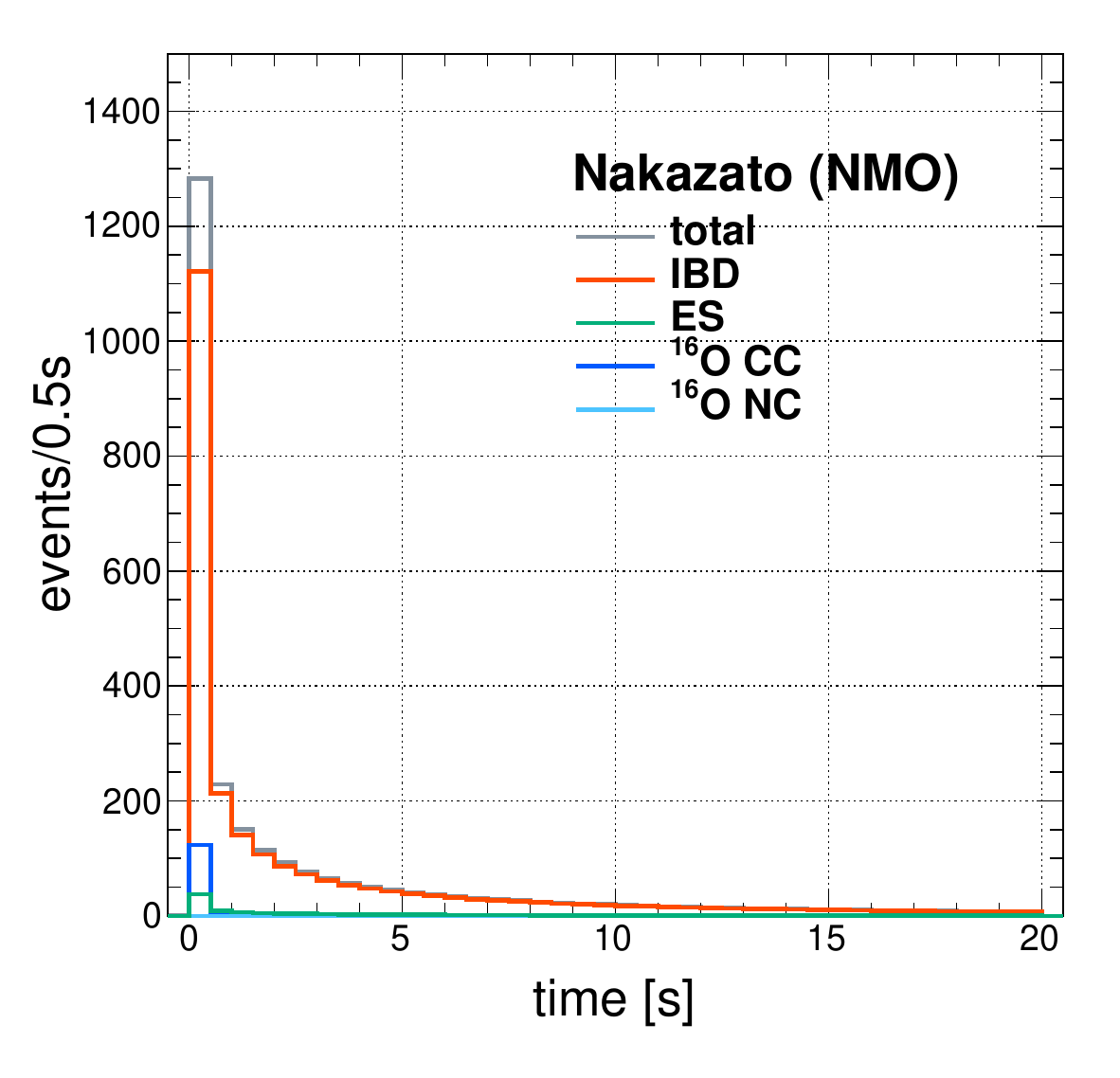}{0.3\textwidth}{(c) up to 20~s}
}
\caption{Time evolution of each interaction for the Nakazato model for an SN burst located at 10~kpc with neutrino oscillation in NMO.  (a) up to 0.12~s, (b) up to 1.5~s, and (c) up to 20~s.  The red, green, blue, and light blue histograms represent IBD events, ES events, $^{16}$O~CC events, and $^{16}$O~NC events, respectively.  The grey histogram includes all the interactions.}
\label{fig:NMOTimeInteractionsNakazato}
\end{figure}

\begin{figure}
\gridline{
\fig{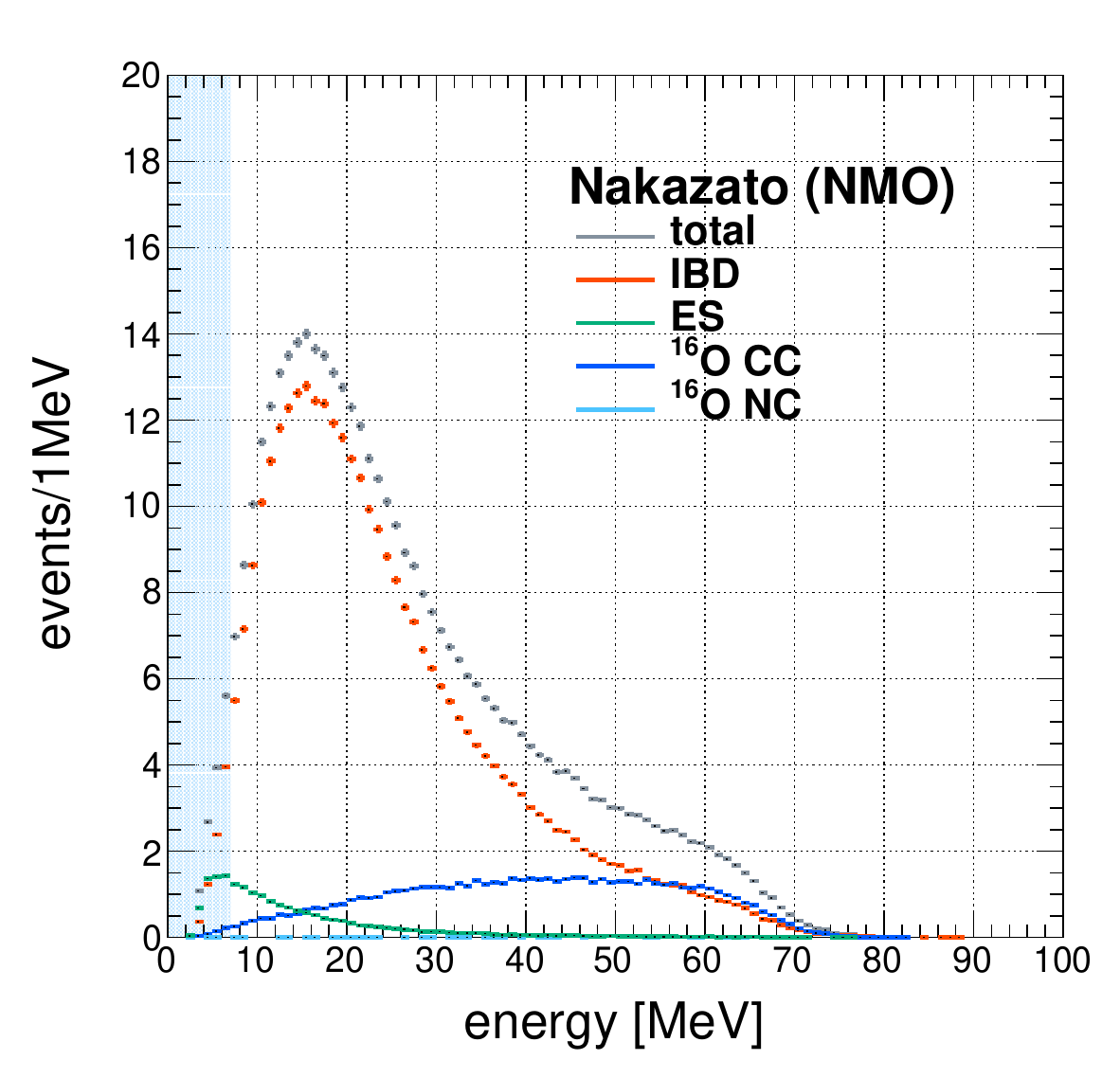}{0.3\textwidth}{(a) up to 0.12~s}\fig{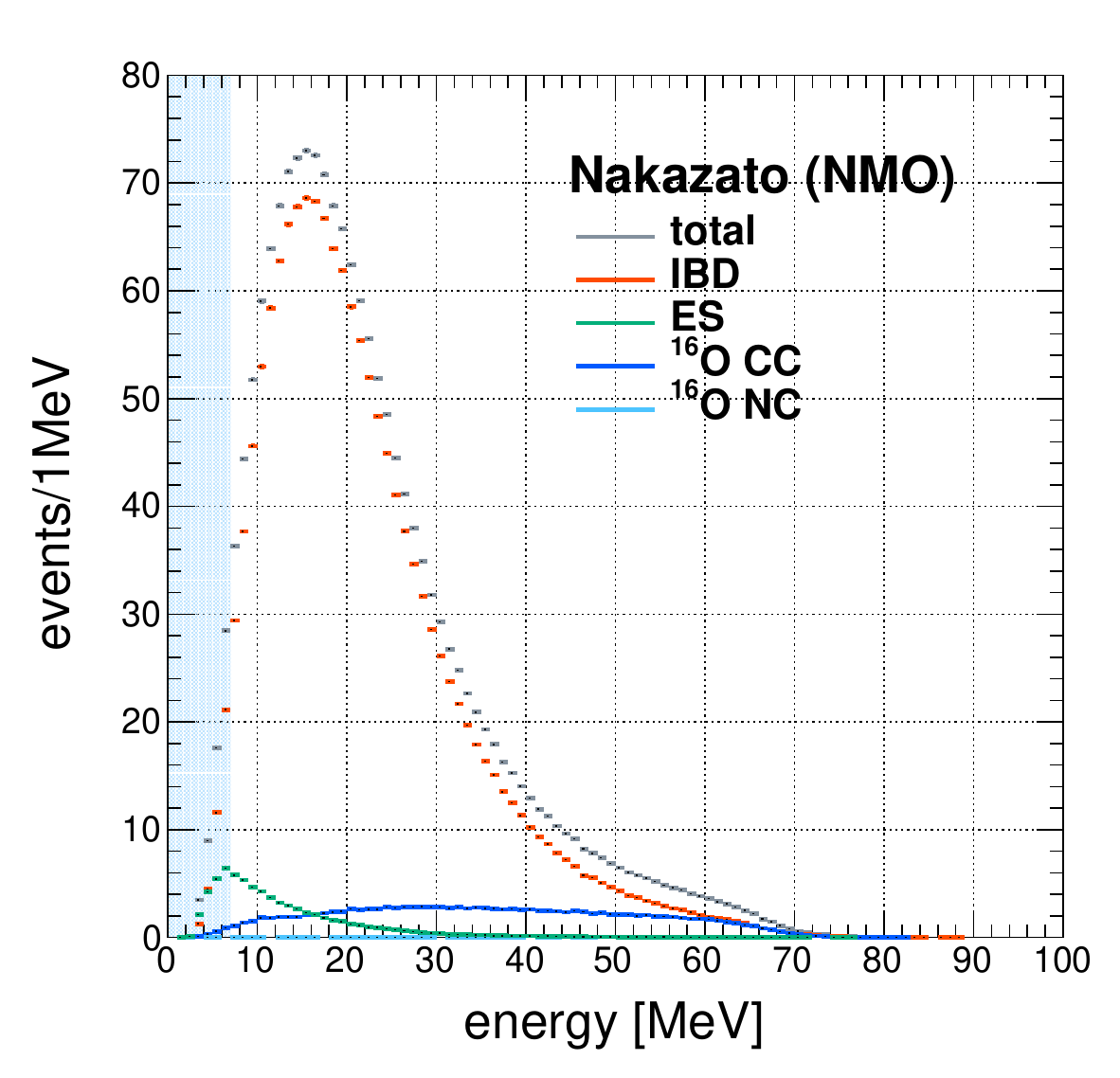}{0.3\textwidth}{(b) up to 1.5~s}\fig{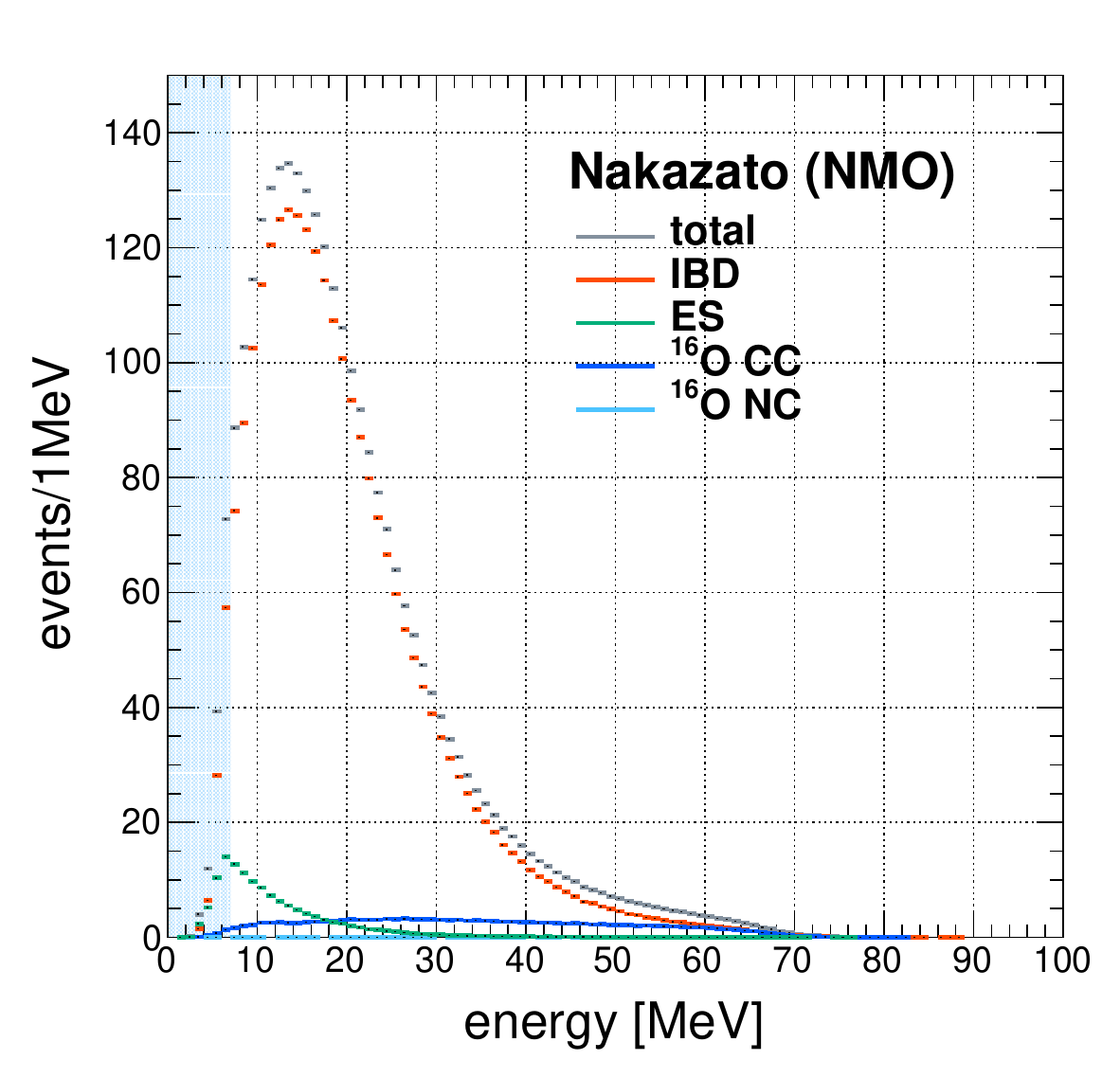}{0.3\textwidth}{(c) up to 20~s}
}
    \caption{The time-integrated energy spectra of each interaction for the Nakazato model for an SN burst located at 10~kpc with neutrino oscillation in NMO, in the different time range: (a) up to 0.12~s, (b) up to 1.5~s, and (c) up to 20~s.  Here, the energy indicates the reconstructed energy of e$^+$ for IBD, e$^-$, for ES, e$^+$ and e$^-$ for $^{16}$O~CC, and the gamma rays for $^{16}$O~NC events.  The energy region below the 7~MeV threshold for selecting ``prompt'' candidates is shaded in light blue.}
    \label{fig:NMOEnergyInteractionNakazato}
\end{figure}

\begin{figure}
\gridline{
\fig{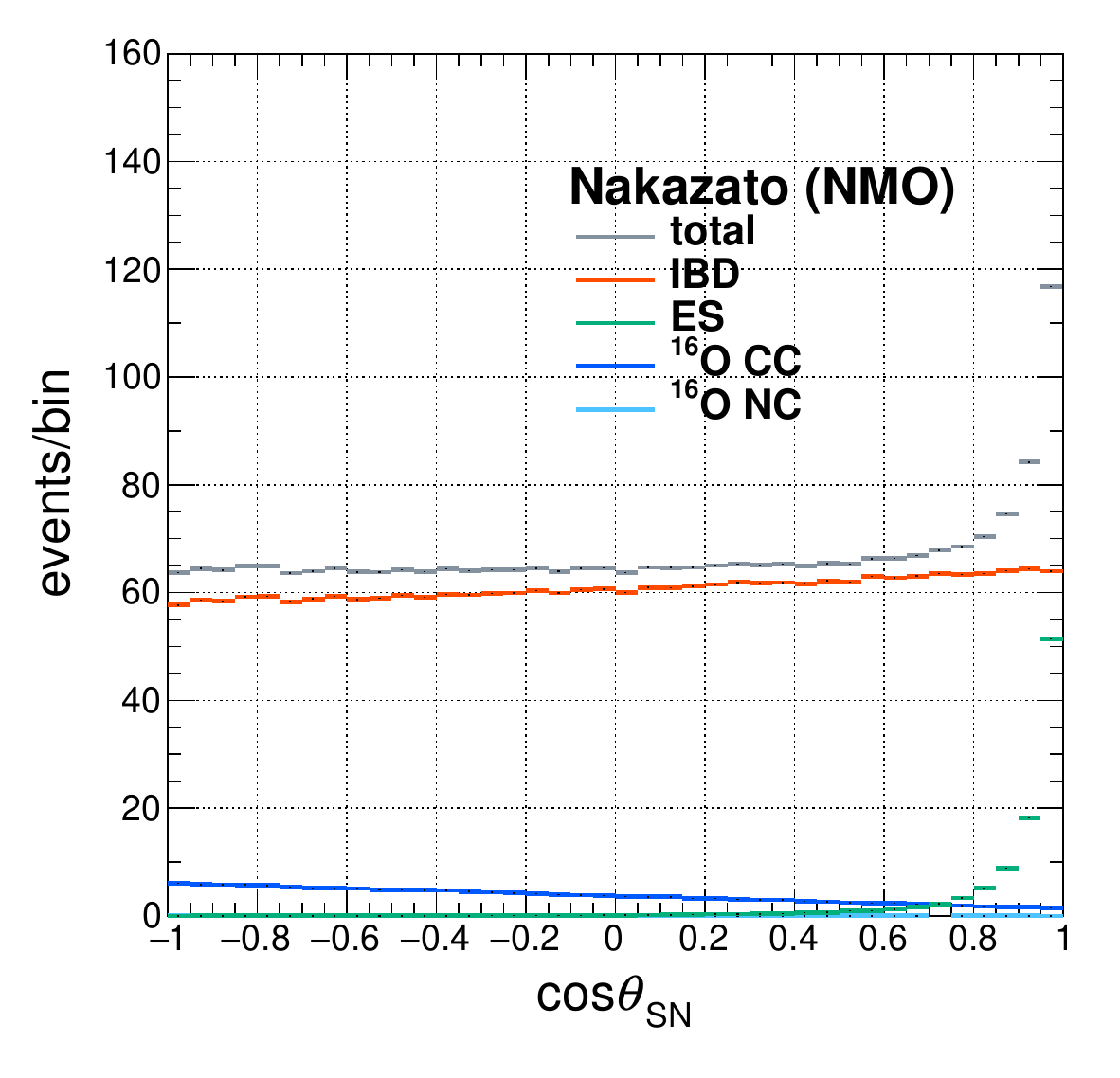}{0.33\textwidth}{(a) 1D $\cos\theta_{\mathrm{SN}}$ distribution}
\fig{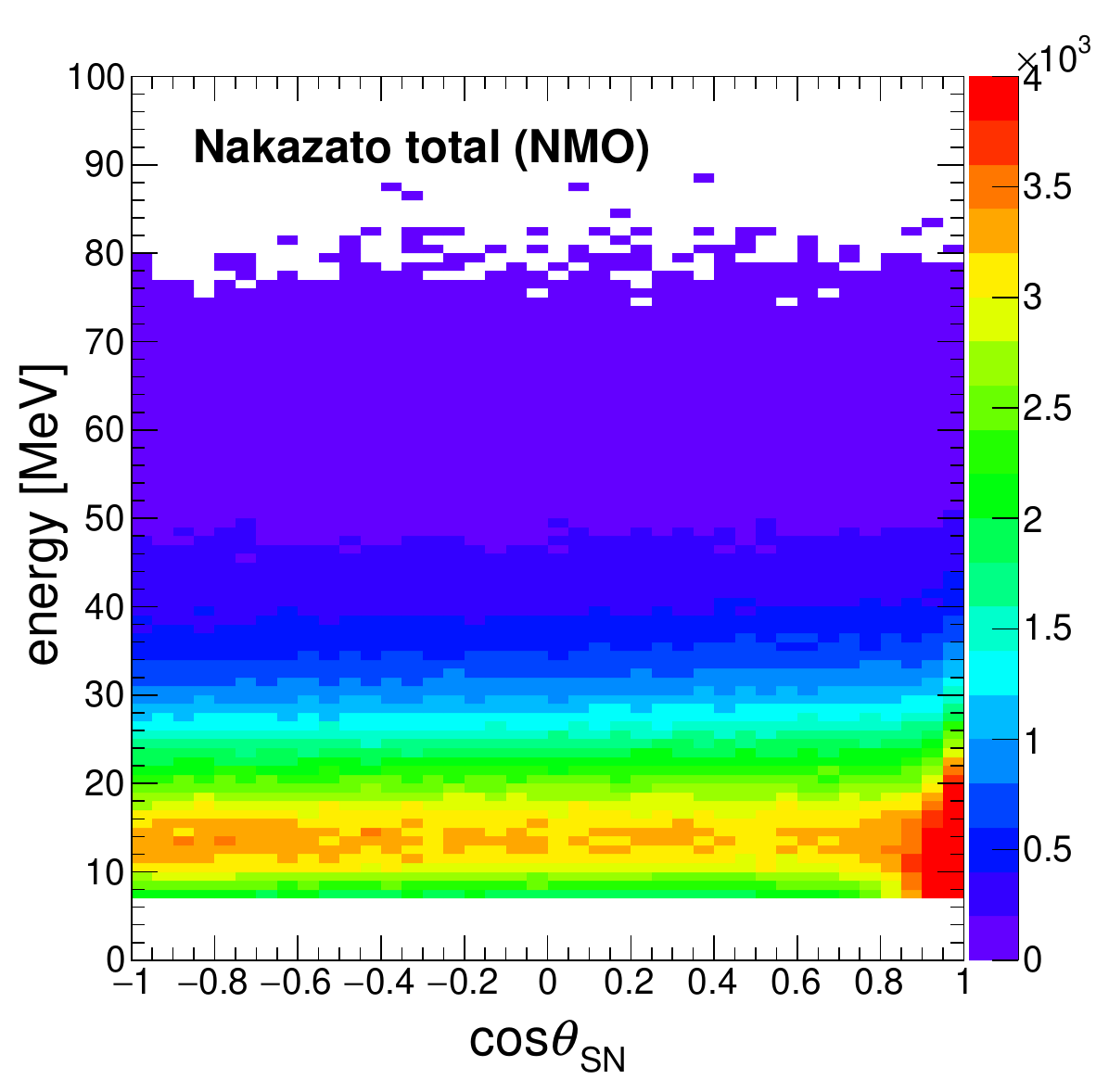}{0.33\textwidth}{(b) Energy~vs.~$\cos\theta_{\mathrm{SN}}$ for total events}
\fig{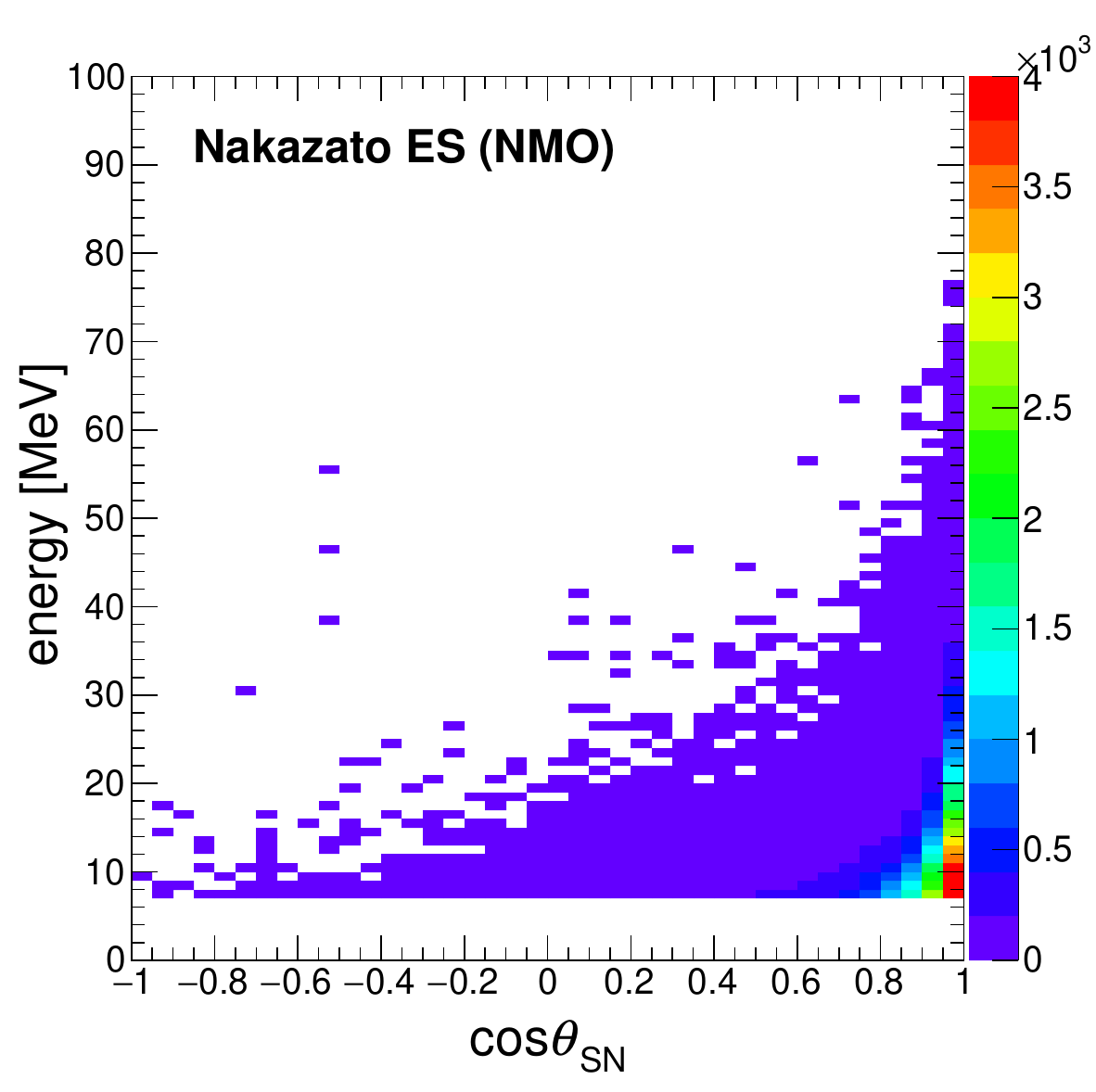}{0.33\textwidth}{(c) Energy~vs.~$\cos\theta_{\mathrm{SN}}$ for ES events}
}
\caption{Angular distribution of events for the Nakazato model for an SN burst located at 10~kpc with neutrino oscillation in NMO. (a) 1D $\cos\theta_{\mathrm{SN}}$ distribution.  The red, green, blue, and light blue histograms represent IBD events, ES events, $^{16}$O~CC events, and $^{16}$O~NC events, respectively.  The grey histogram includes all the interactions. 
(b) Energy~vs.~$\cos\theta_{\mathrm{SN}}$ for total events. (c) Energy~vs.~$\cos\theta_{\mathrm{SN}}$ for ES events.}
\label{fig:NMOCosInteractionsNakazato}
\end{figure}

Figure~\ref{fig:NMOTimeInteractionsNakazato} shows the time evolution of each interaction across different time ranges 
and Figure~\ref{fig:NMOEnergyInteractionNakazato} shows the corresponding time-integrated energy spectra.
Most events are concentrated in the first second, as is expected from the luminosity evolution (Figure~\ref{fig:LongTimeEvolutionLumiAndEave}), and IBD events dominate.
In Figure~\ref{fig:NMOEnergyInteractionNakazato} , energy indicates the reconstructed energy of each mode's visible particles: e$^+$ for IBD events,  e$^-$ for ES, e$^+$ or e$^-$ for $^{16}$O~CC, and gamma rays from $^{16}$O~NC events.  
These energy spectra are similar in shape over the considered time ranges. 
The three panels in Figure~\ref{fig:NMOCosInteractionsNakazato}
describe the time-integrated angular distribution of the reconstructed events that satisfy the same criteria described in Section~\ref{subsec:EventReconstruction}.
Figure~\ref{fig:NMOCosInteractionsNakazato} (a) shows the angular distribution for each interaction.
Here, $\theta_\mathrm{SN}$ represents the angle between the true SN direction and the reconstructed direction of the visible particle in each interaction. 
The shape of the $\cos\theta_\mathrm{SN}$ distribution reflects the event topology assumed in SKSNSim as mentioned in Section~\ref{subsec:EventGeneration}.  
The distribution of IBD events (red) has an almost flat but slightly forward inclination due positrons being slightly more likely to be emitted in the forward direction at higher neutrino energies.
 ES (green)  events are the result of a forward scattering process, so the distribution peaks at $\cos\theta_\mathrm{SN}{\sim}1$.
On the other hand, $^{16}$O~CC events have a backward bias and while the distribution for $^{16}$O~NC  should be flat, the number of events is too small for it to be observed here.

Figure~\ref{fig:NMOCosInteractionsNakazato} (b) and (c)  show $\cos{\theta}_\mathrm{SN}$~vs.~energy distribution for all events and for only ES events, respectively.
One might think that using only the high-energy component of ES events would be effective in determining the direction to the SN  because those scattered electrons populate the region $\cos\theta_\mathrm{SN}{\sim}1$.  
However, as Figure~\ref{fig:NMOCosInteractionsNakazato} (c) suggests, the number of high-energy ES events is insufficient to make a significant contribution to the direction estimation.

\subsection{Comparison among Models}\label{subsubsec:DifferenceAmongModels}
This section reports the differences in SK-Gd's response to an SN burst located at 10~kpc, comparing the six SN models described in Section~\ref{subsec:SNmodelsSummary} and focusing on the NMO oscillation scenario. 
Appendix~\ref{app:differenceAmongModels} presents corresponding figures for the IMO scenario.

\begin{figure}[htb!]
\gridline{
\fig{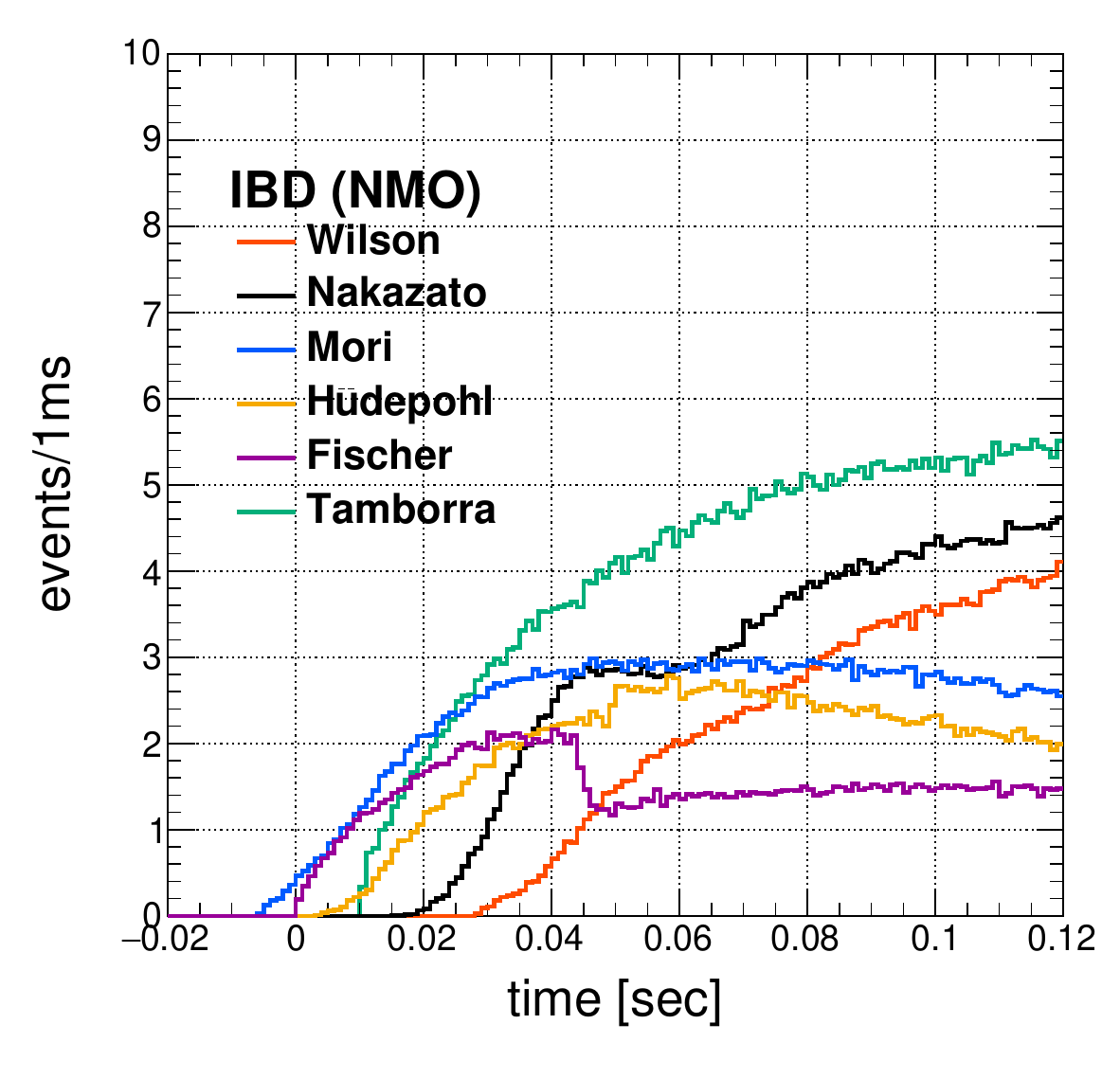}{0.3\textwidth}{(a) IBD, up to 0.12~s}
\fig{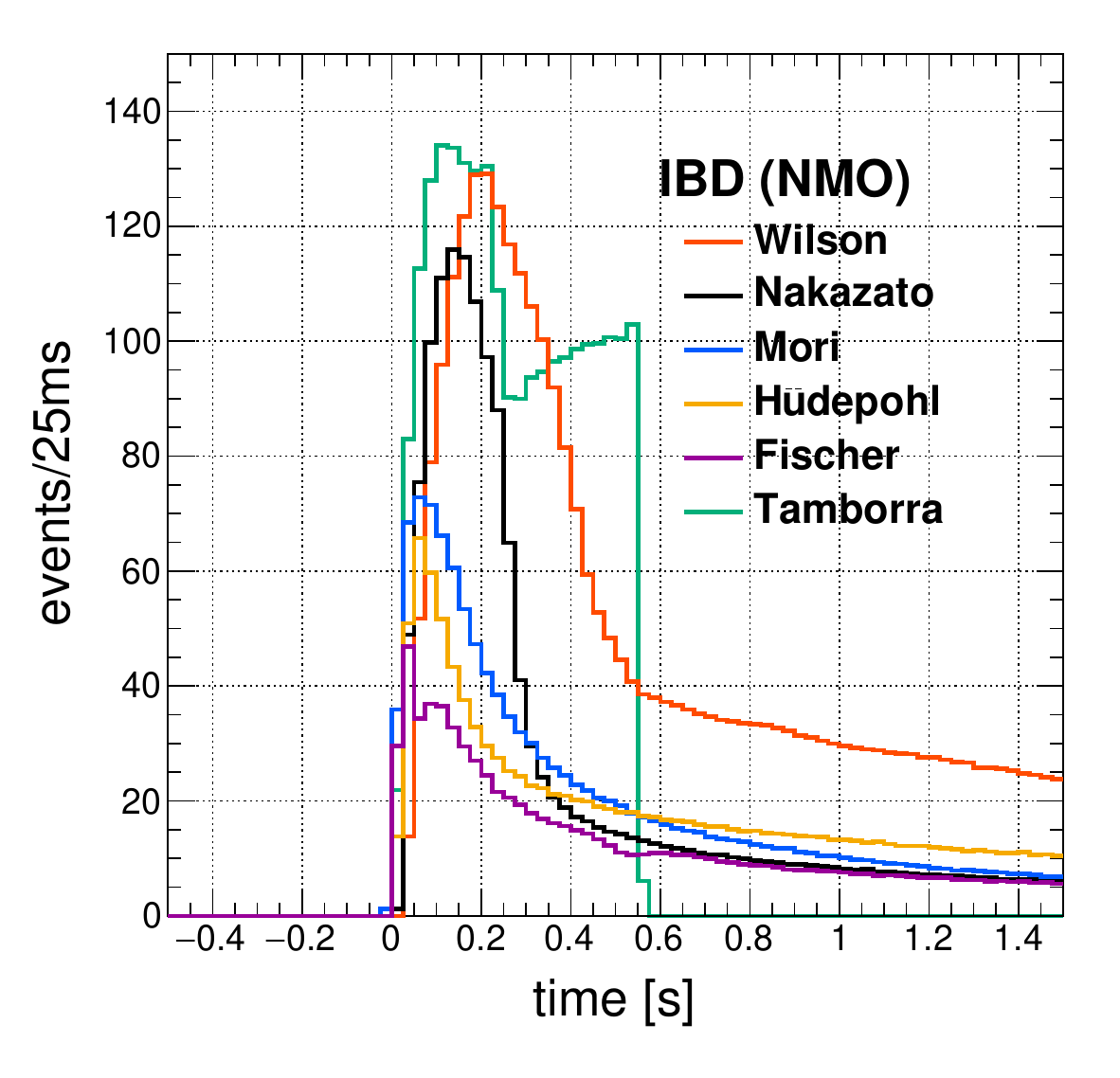}{0.3\textwidth}{(b) IBD, up to 1.5~s}
\fig{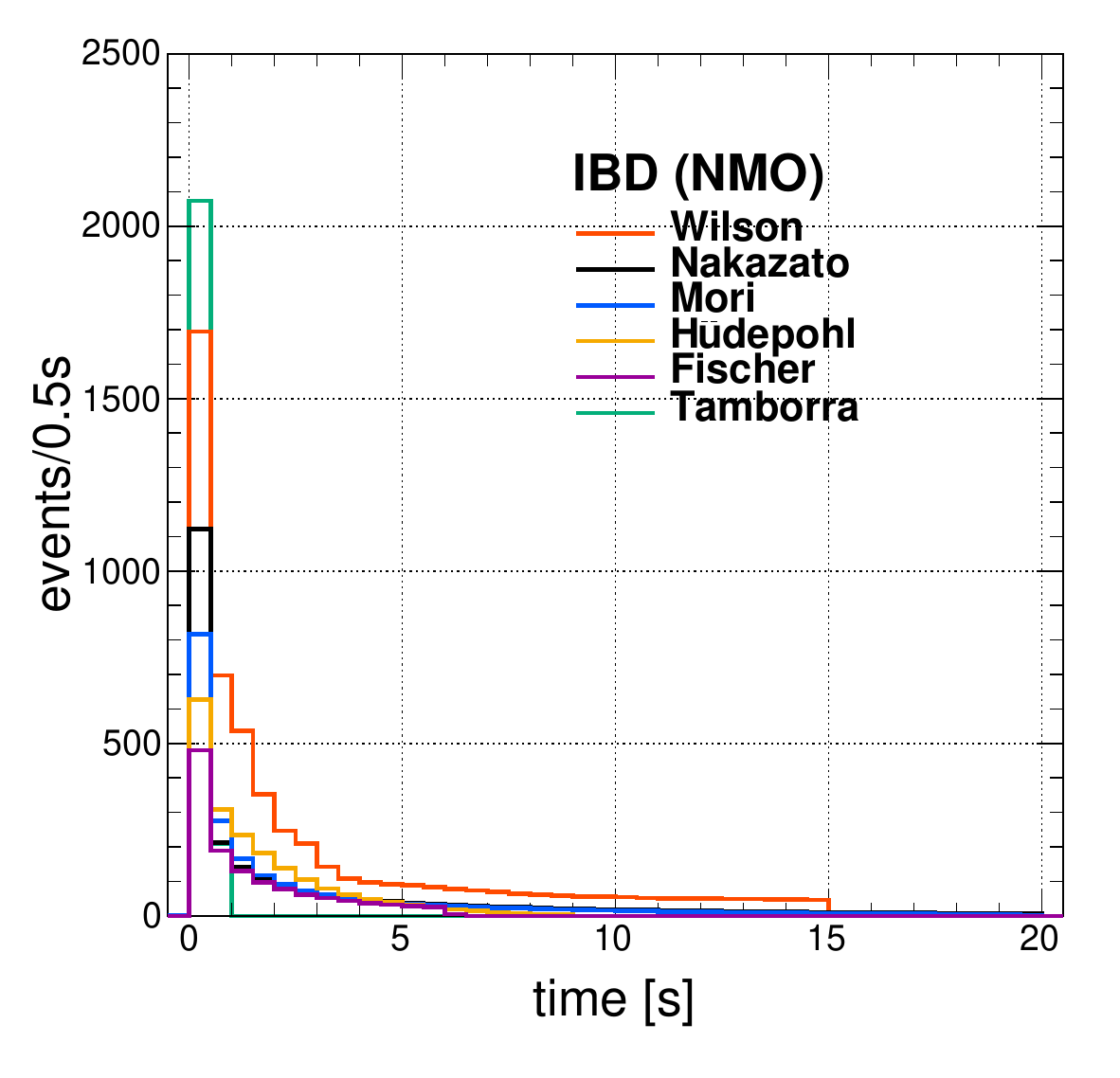}{0.3\textwidth}{(c) IBD, up to 20~s}
}
\vspace{-1.2cm}
\gridline{
\fig{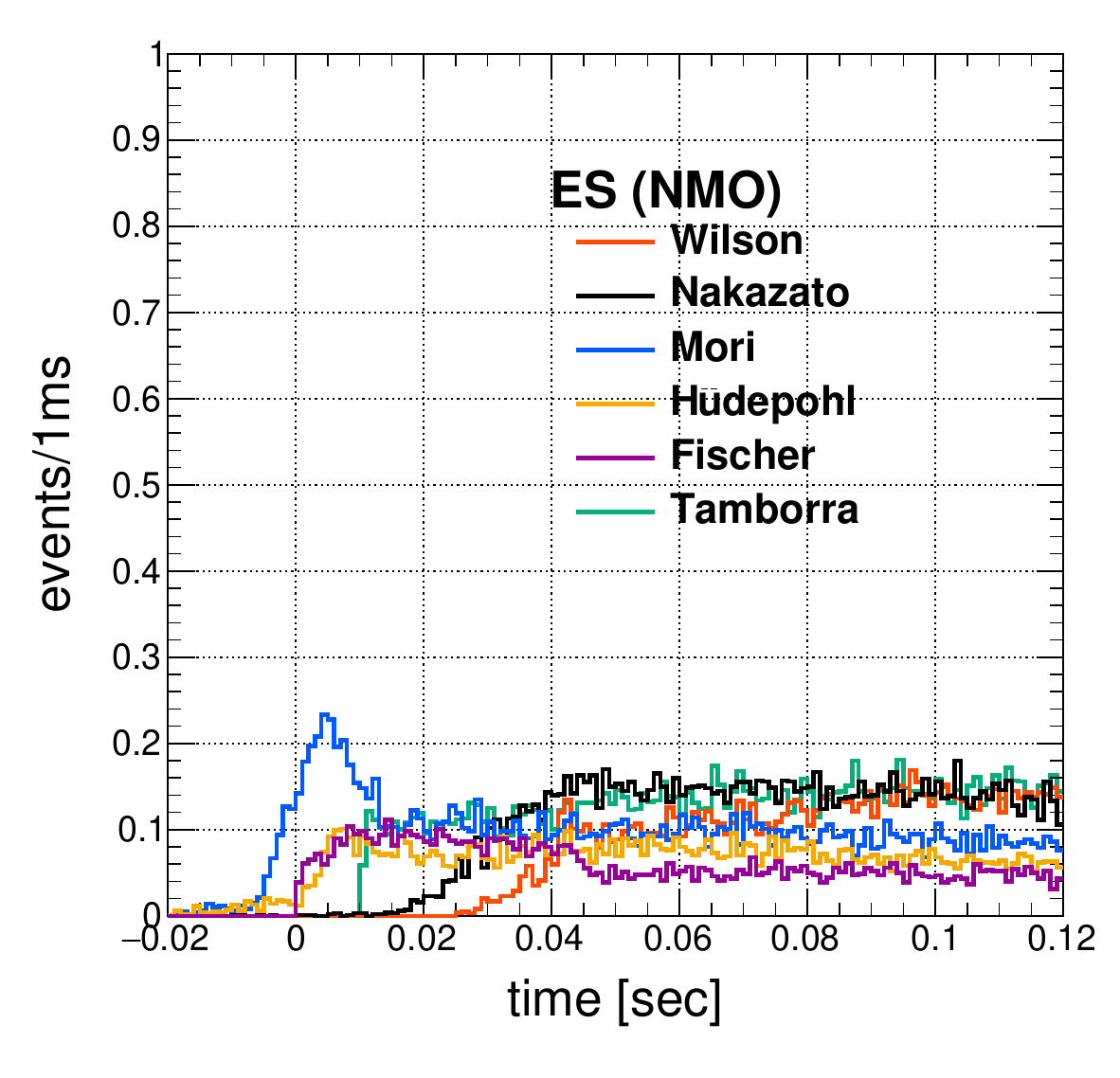}{0.3\textwidth}{(f) ES, up to 0.12~s}
\fig{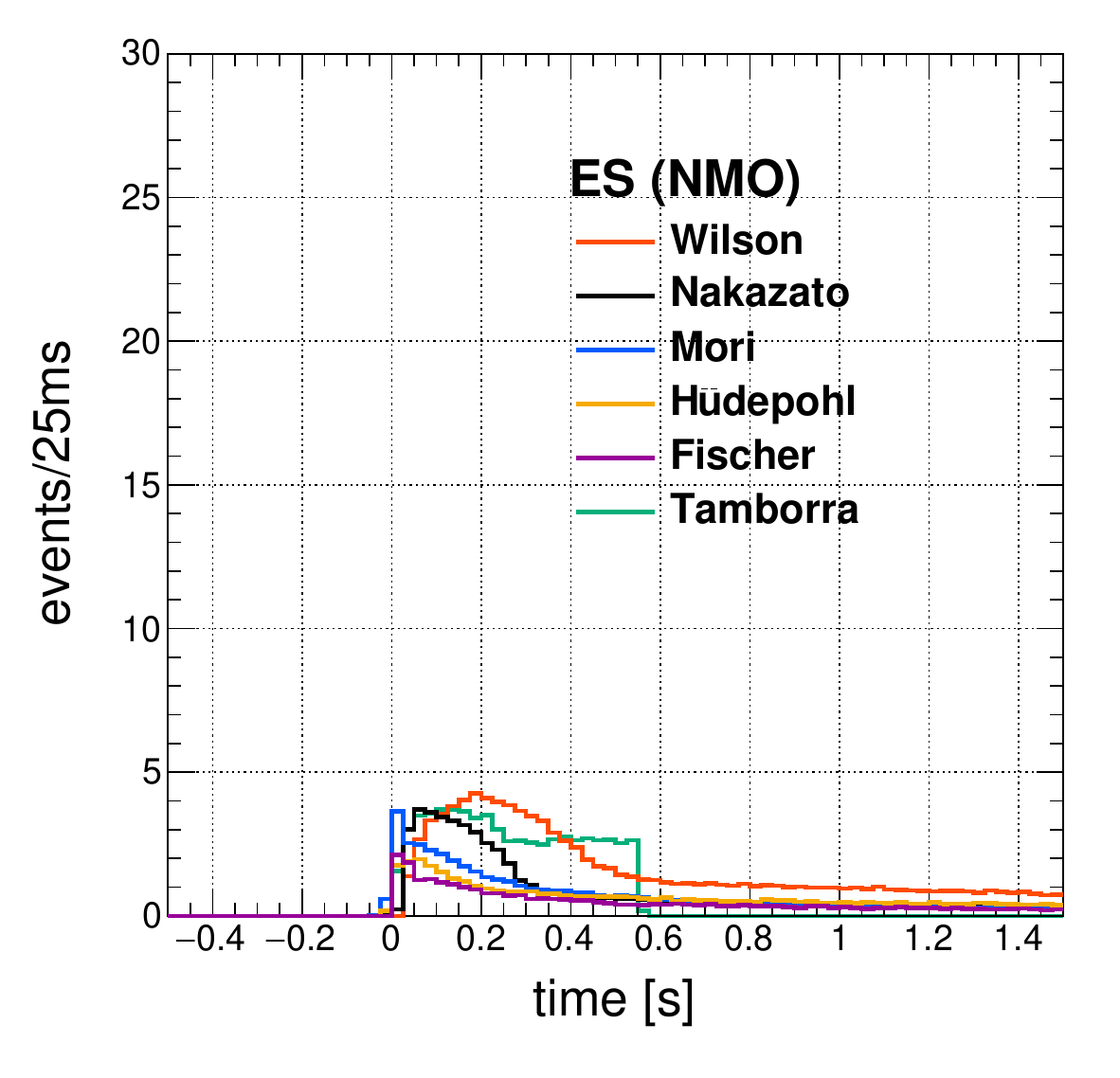}{0.3\textwidth}{(e) ES, up to 1.5~s}
\fig{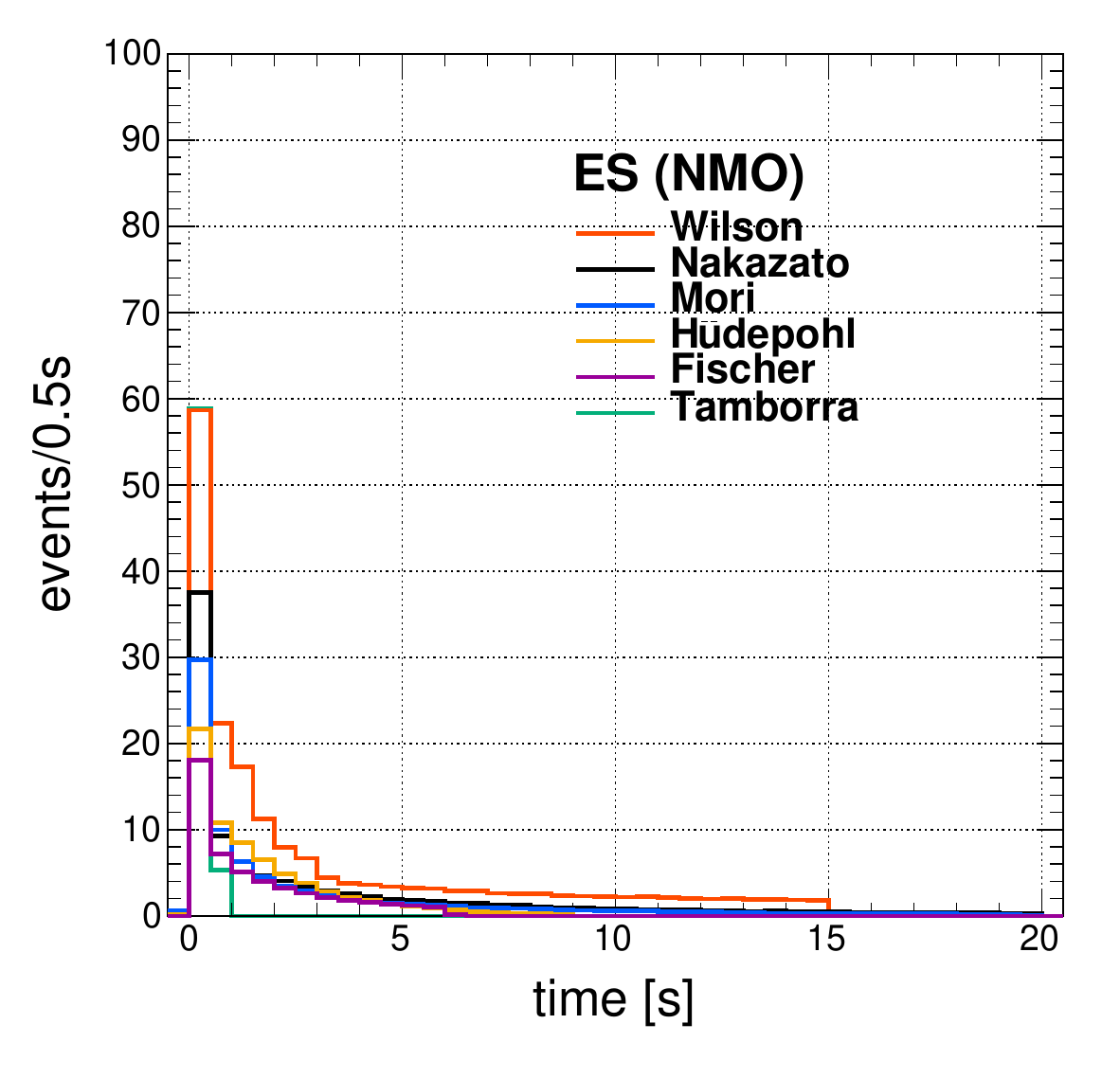}{0.3\textwidth}{(d) ES, up to 20~s}
}
\vspace{-1.2cm}
\gridline{
\fig{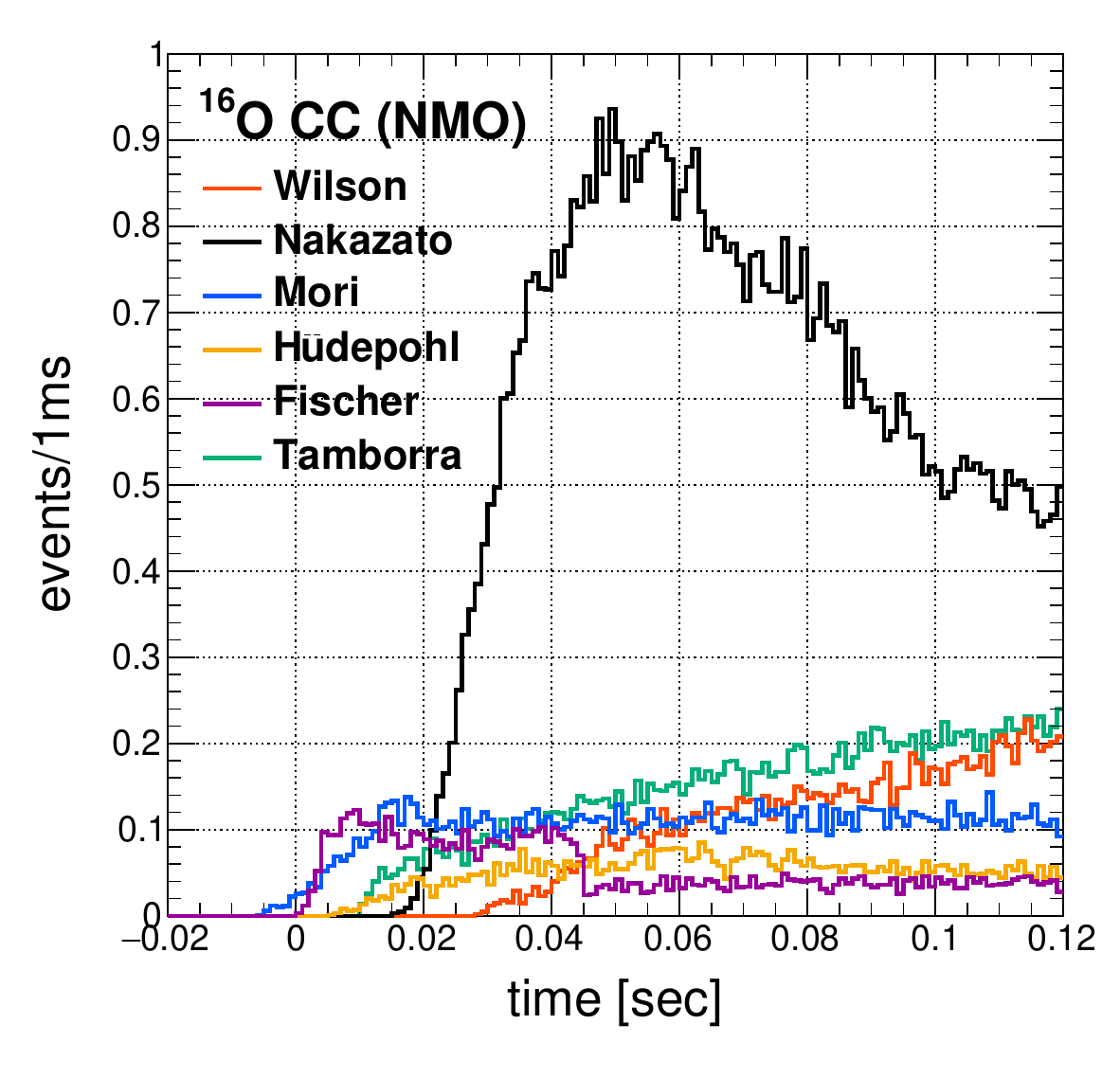}{0.3\textwidth}{(i) $^{16}$O~CC, up to 0.12~s}
\fig{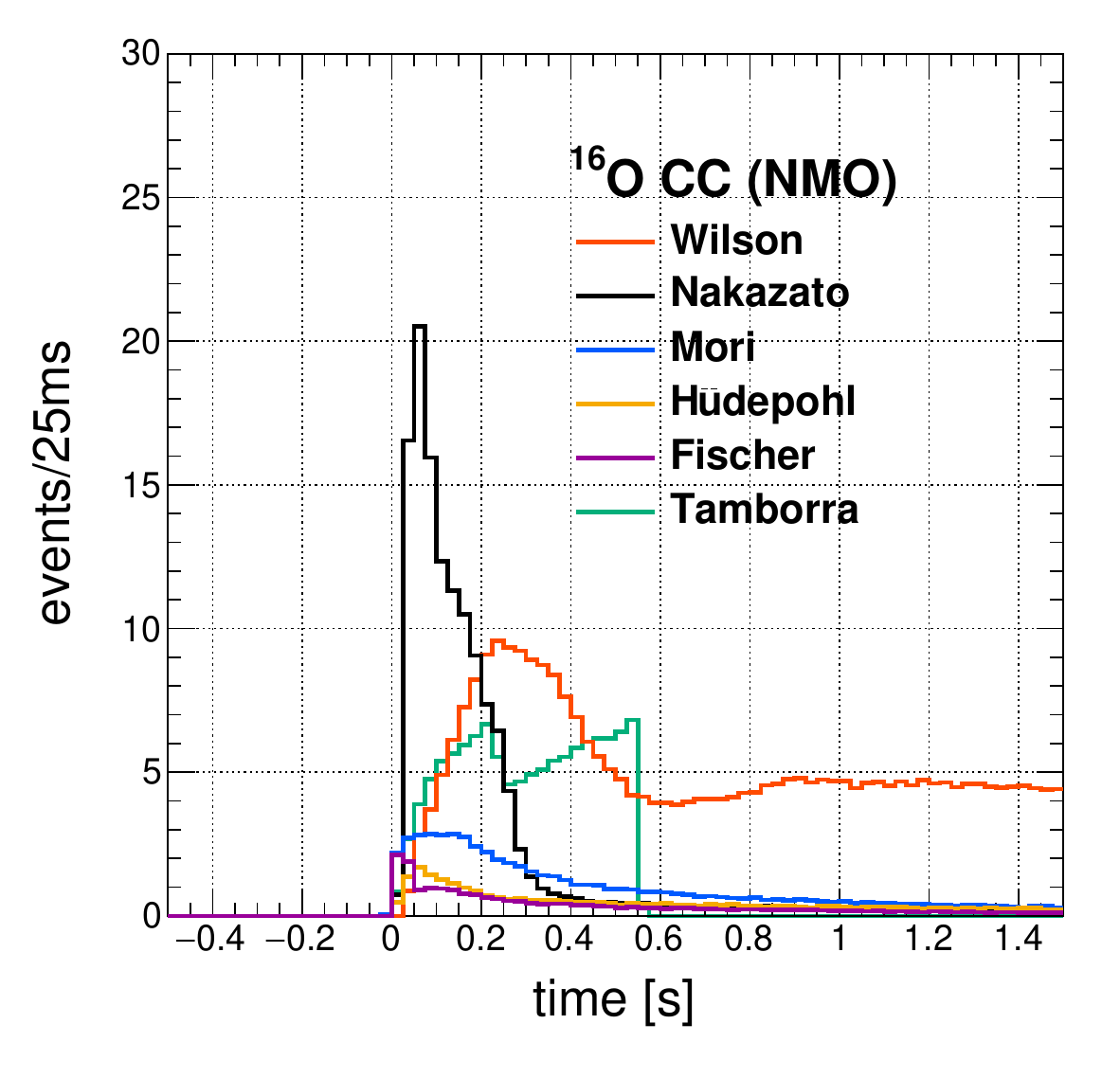}{0.3\textwidth}{(h) $^{16}$O~CC, up to 1.5~s}
\fig{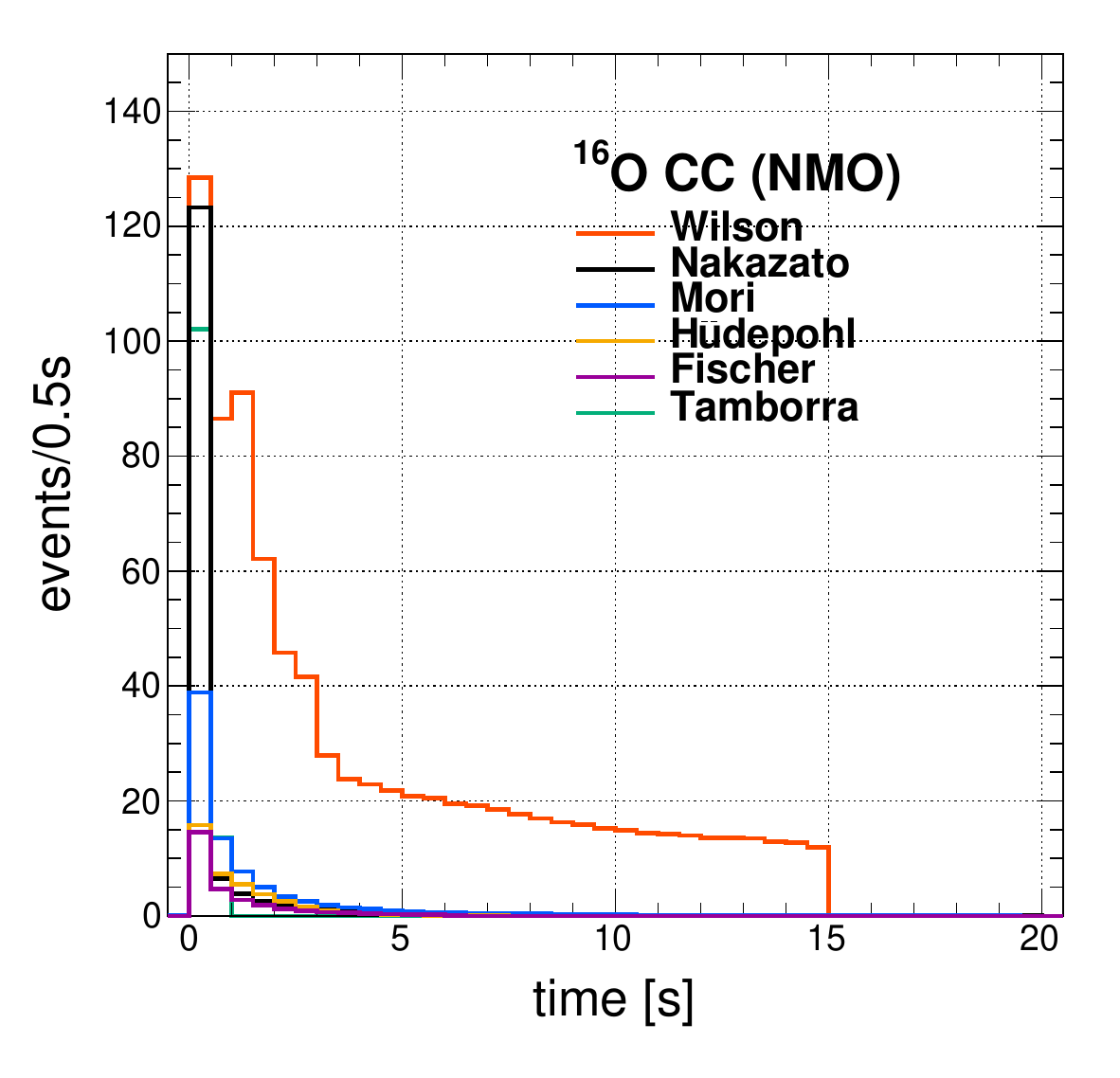}{0.3\textwidth}{(g) $^{16}$O~CC, up to 20~s}
}
\vspace{-1.2cm}
\gridline{
\fig{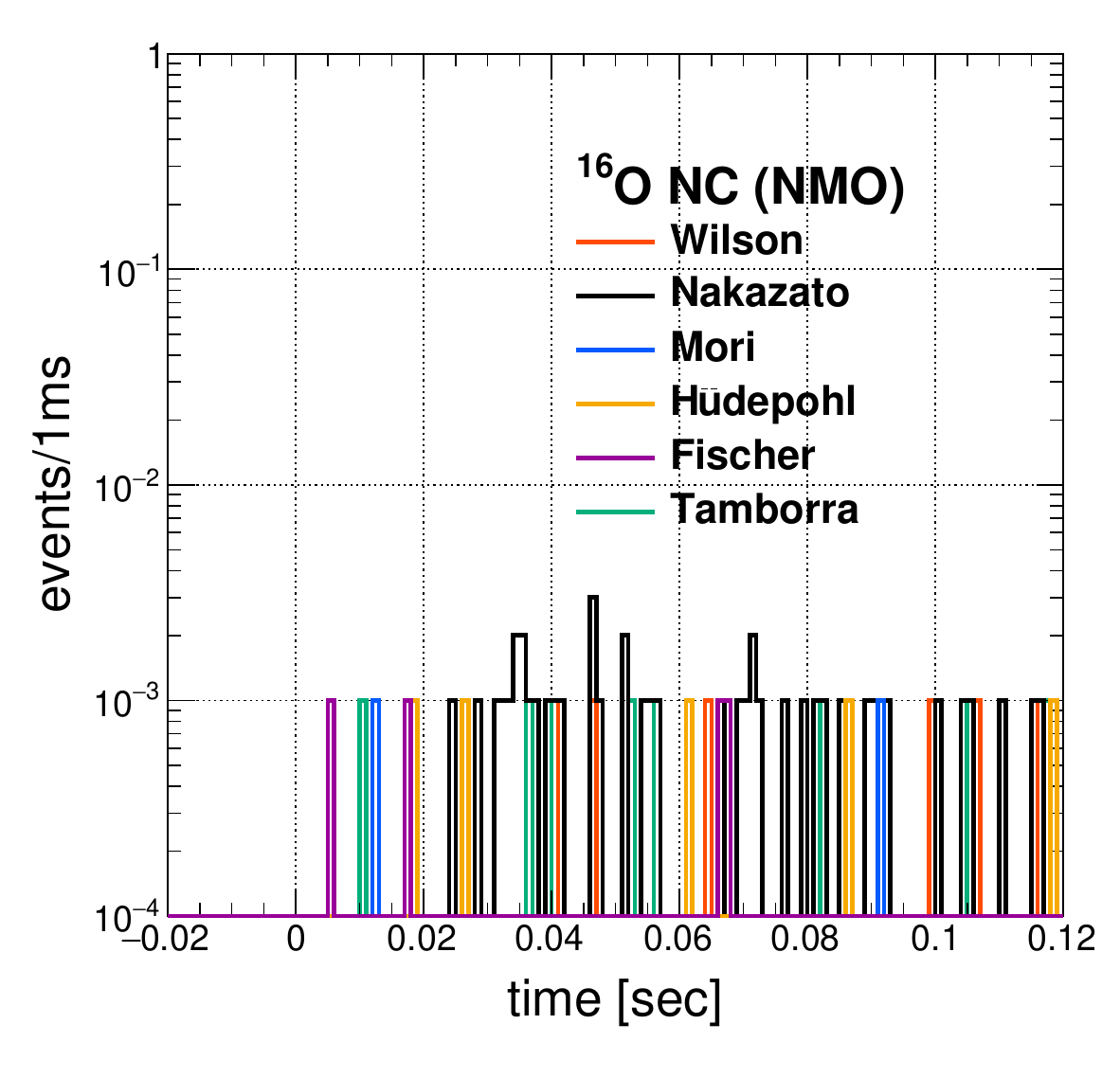}{0.3\textwidth}{}
\fig{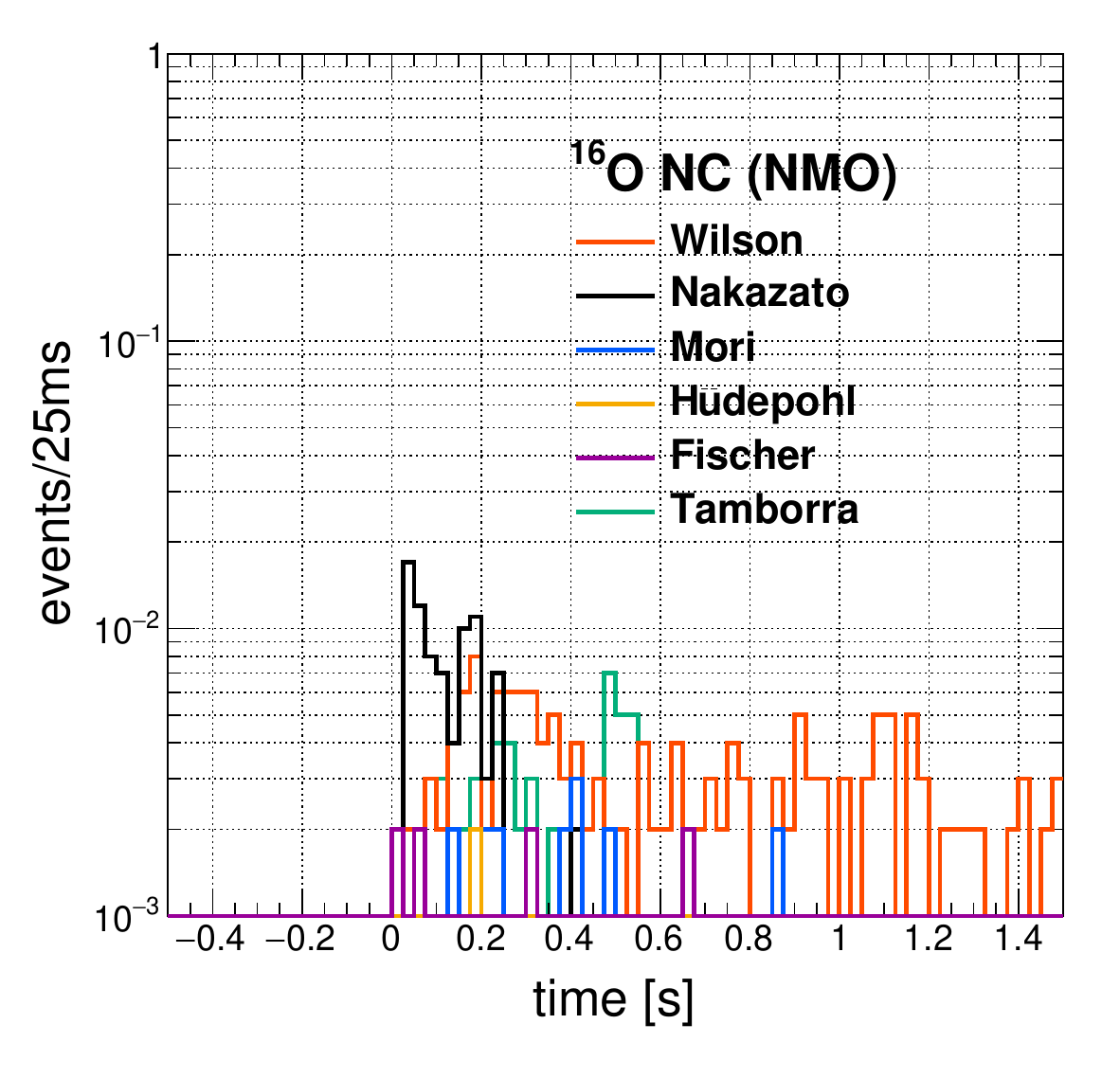}{0.3\textwidth}{}
\fig{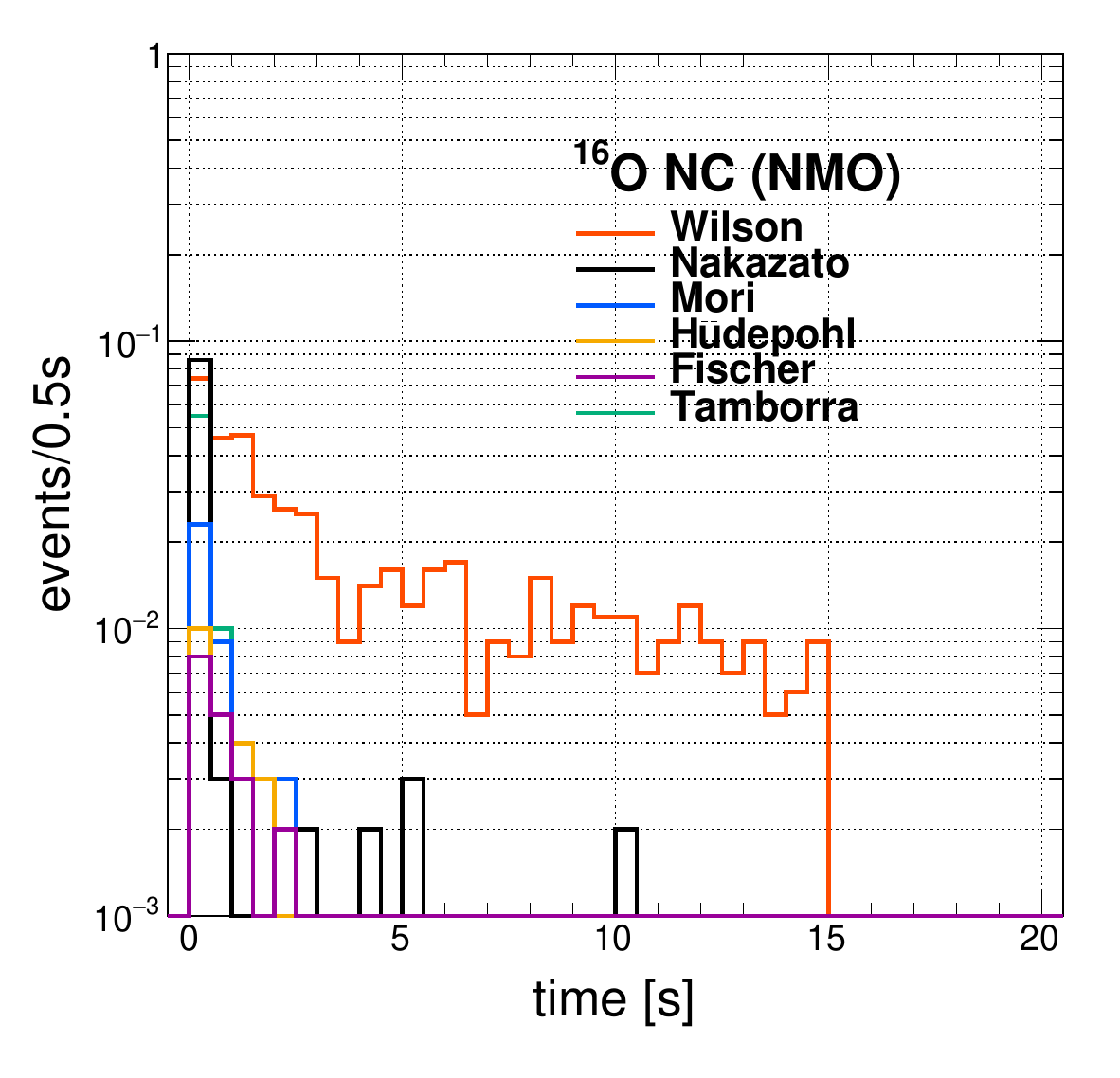}{0.3\textwidth}{}
}
\vspace{-0.5cm}
\caption{Comparison of time evolution among models for each interaction for an SN burst located at 10~kpc with neutrino oscillation in NMO.  The left, middle, and right columns show the time evolution up to 0.12~s, 1.5~s, and 20~s, respectively.  Each row represents IBD, ES, $^{16}$O~CC, and $^{16}$O~NC from top to bottom.}
\label{fig:NMOvarModelTimeInteractions}
\end{figure}

Figure~\ref{fig:NMOvarModelTimeInteractions} compares the time evolution of reconstructed events among models for each interaction type over three different time ranges.  
IBD events in the Tamborra model have an increasing feature after the first peak in the event rate, as shown in the top middle of Figure~\ref{fig:NMOvarModelTimeInteractions}, corresponding to Figure~\ref{fig:LongTimeEvolutionLumiAndEave}~(b).
In the right panel of the second row, the ES peak from the neutronization burst appears in the first 0.1~s for the Mori model (blue).
The same feature is seen at around ${\sim}$0.04~s in the Wilson model (red) but is unclear in the other models.
To determine whether or not the timing of the neutronization burst can be seen in SK-Gd, we consider the optimistic scenario given by the Mori model. 
Note that the excess of ES events over the flat component corresponds to $N_{\nu_e}^{sur}$ in equations~(\ref{eq:NMOnue})--(\ref{eq:NMOnux}).
Integrating from 0.04~s to 0.12~s yields the flat component, 93.8~events per second, or 1.4~events expected from $-0.002$~s~to~$0.017$~s.
This can be subtracted from the peak component, 4.31~events, obtained by integrating over the same time region, to yield 2.9~events.
From the low number of events, it is clear that it would be difficult to observe this difference with any significance for an SN burst located at 10~kpc. 

\begin{figure}[htb!]
\gridline{\fig{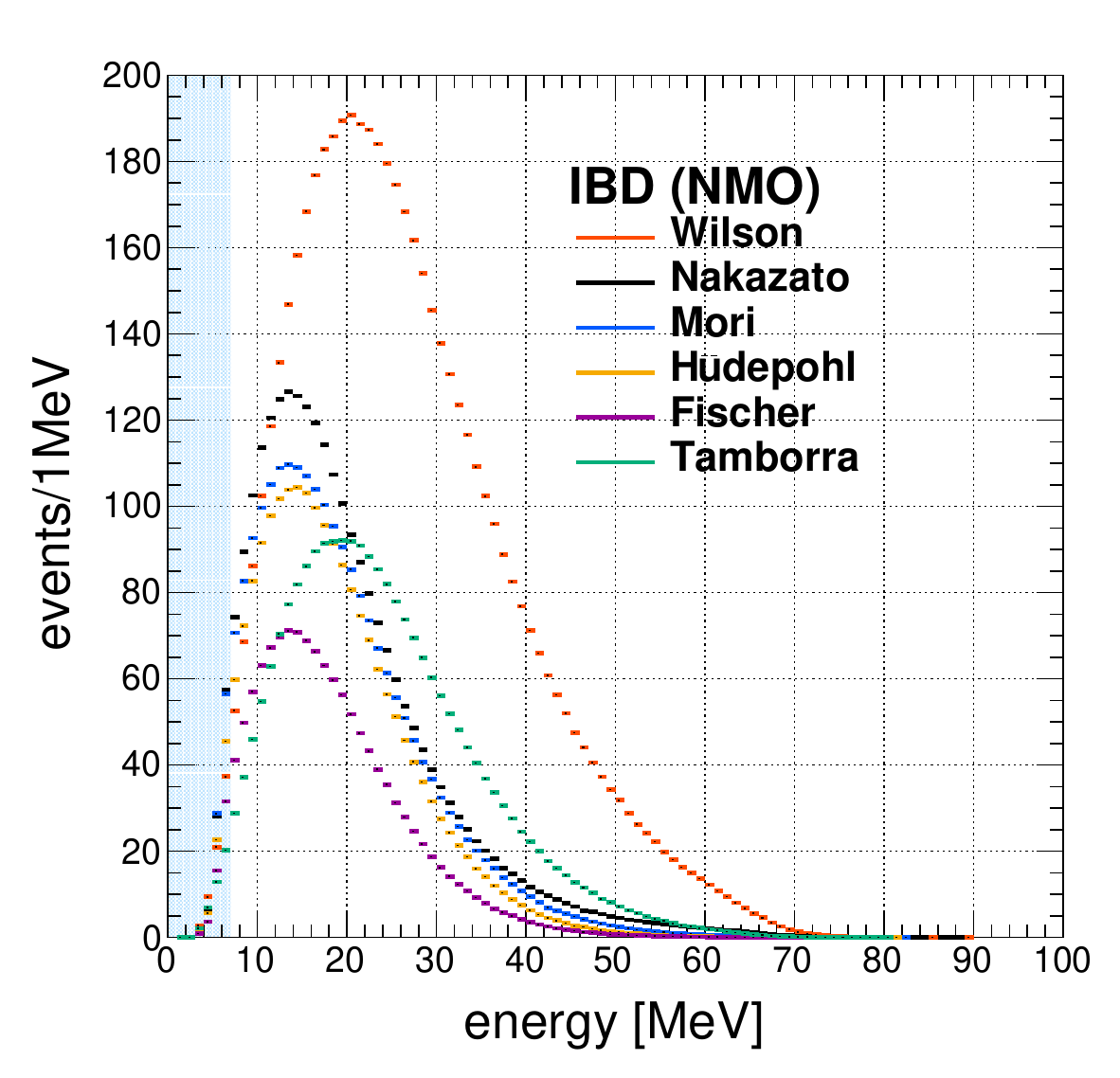}{0.35\textwidth}{(a) IBD}
          \fig{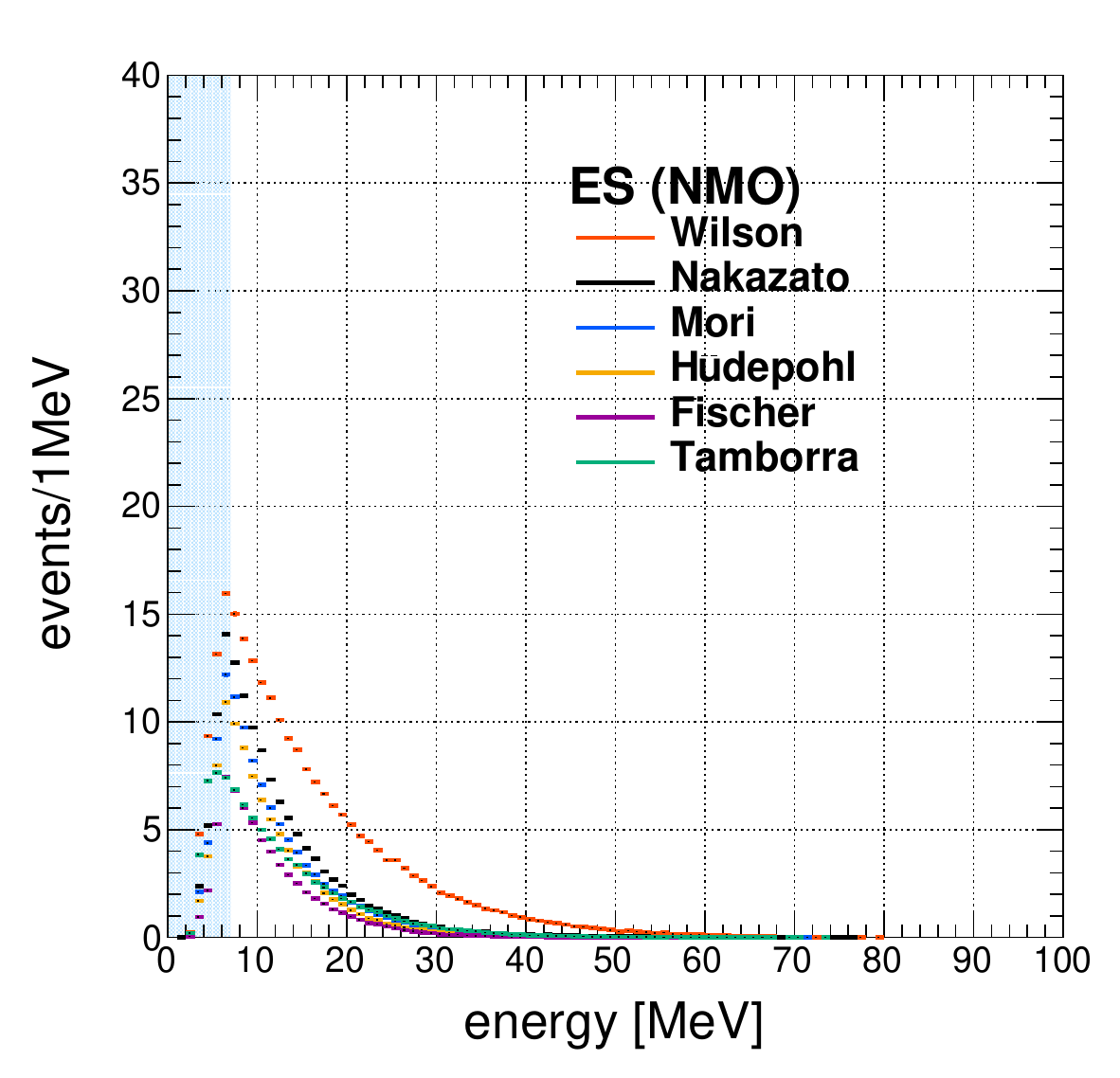}{0.35\textwidth}{(b) ES}}
\gridline{\fig{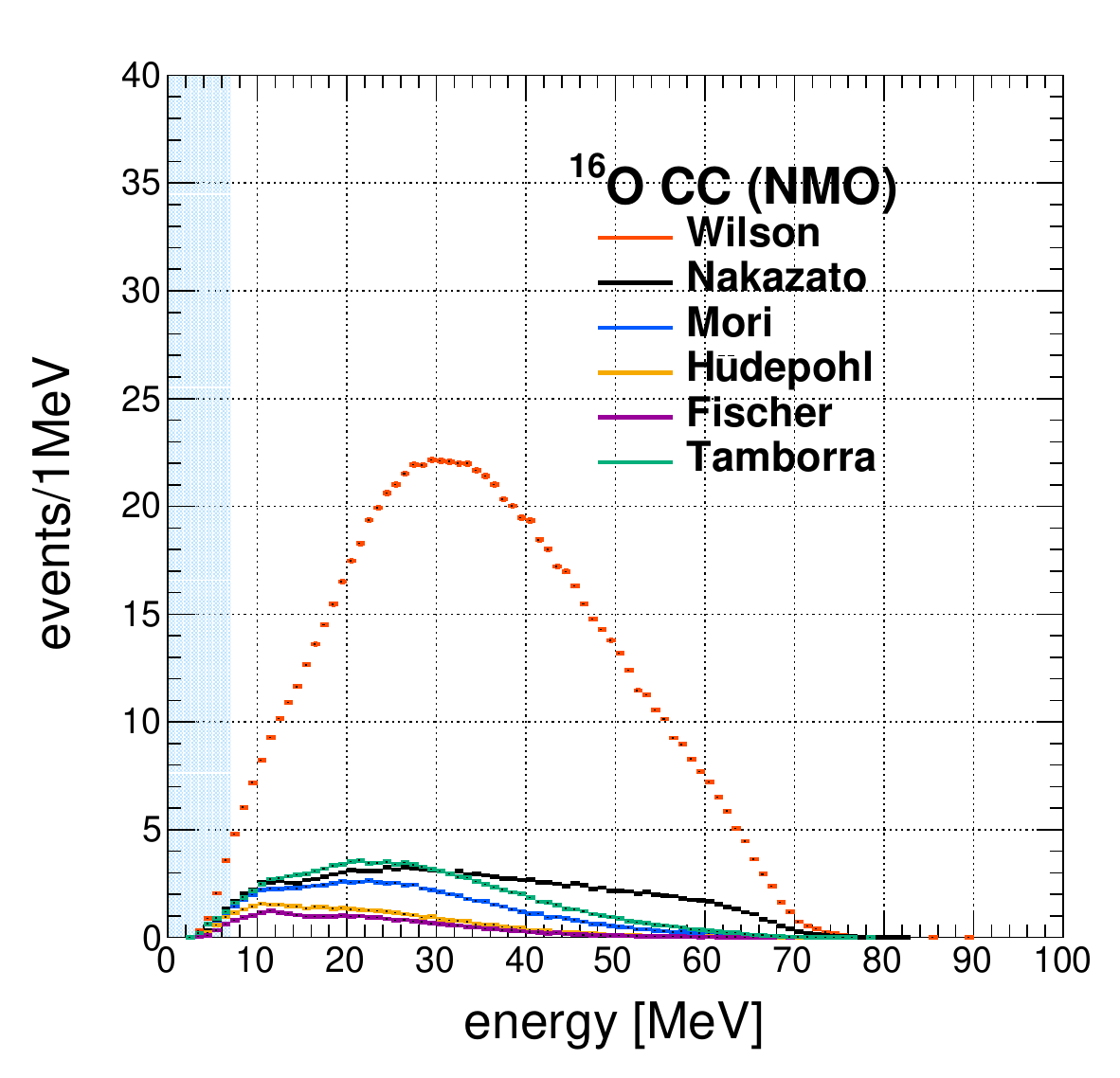}{0.35\textwidth}{(c) $^{16}$O~CC}
          \fig{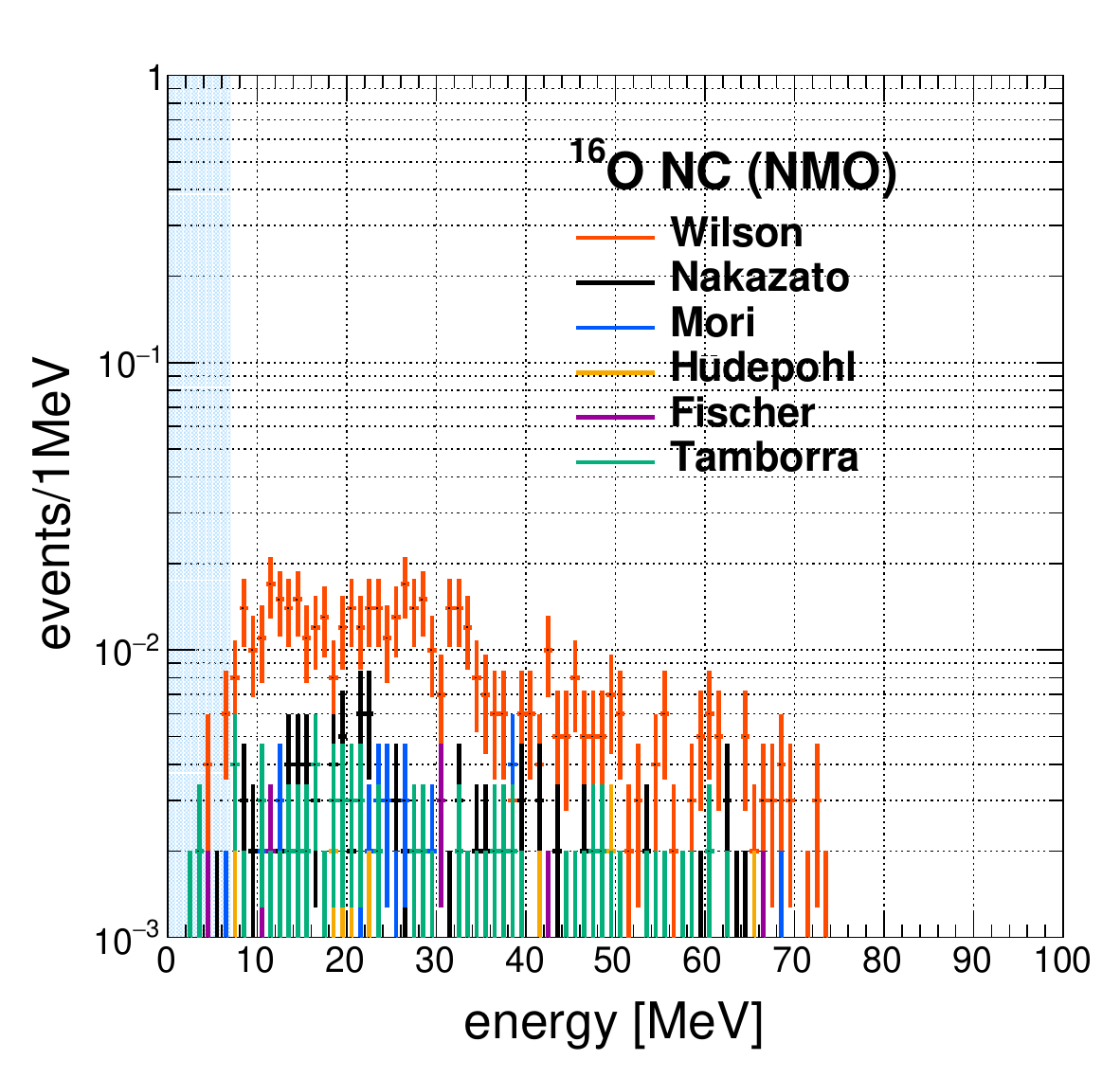}{0.35\textwidth}{(d) $^{16}$O~NC}}
\caption{Comparison of energy spectra among models for each interaction: (a)~IBD, (b)~ES, (c)~$^{16}$O~CC, and (d)~$^{16}$O~NC.  The energy region below the 7~MeV threshold for selecting ``prompt'' candidates is shaded in light blue.}
\label{fig:NMOvarModelEnergyInteractions}
\end{figure}

Figure~\ref{fig:NMOvarModelEnergyInteractions} shows comparison of energy spectra among models considering neutrino oscillation in NMO.
The Wilson model and the Tamborra model have higher energy neutrinos than the other models.
As seen in panel (c), the energy spectra of $^{16}$O~CC interaction have significant differences among models, resulting in a large variation in the number of interactions on oxygen in Table~\ref{tab:AverageEvents1}.
Note that $^{16}$O~NC interaction events shown in panel (d) are low in energy and small in number as presented in Table~\ref{tab:AverageEvents1}.

\begin{figure}[htb!]
\gridline{\fig{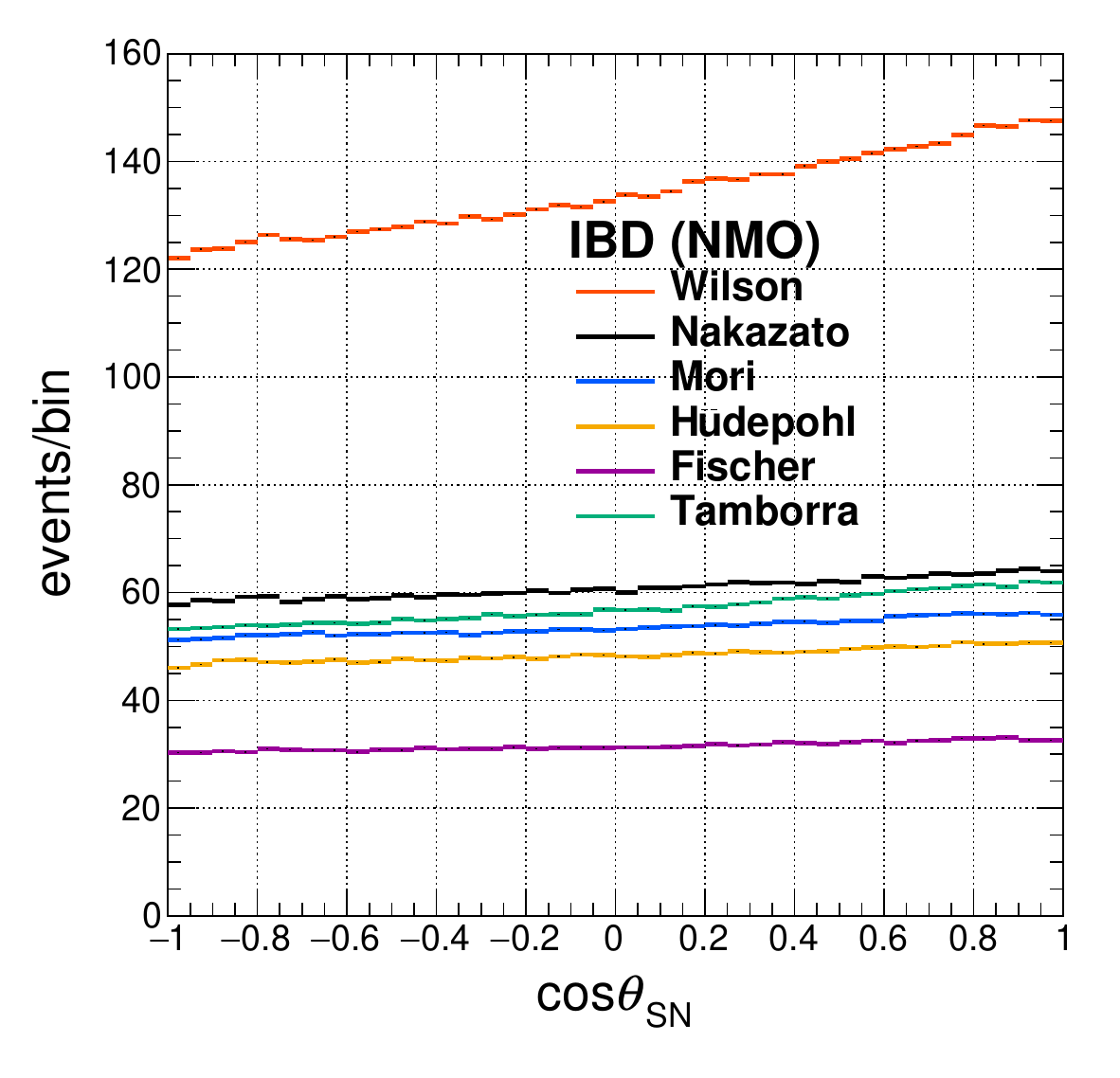}{0.35\textwidth}{(a) IBD}
          \fig{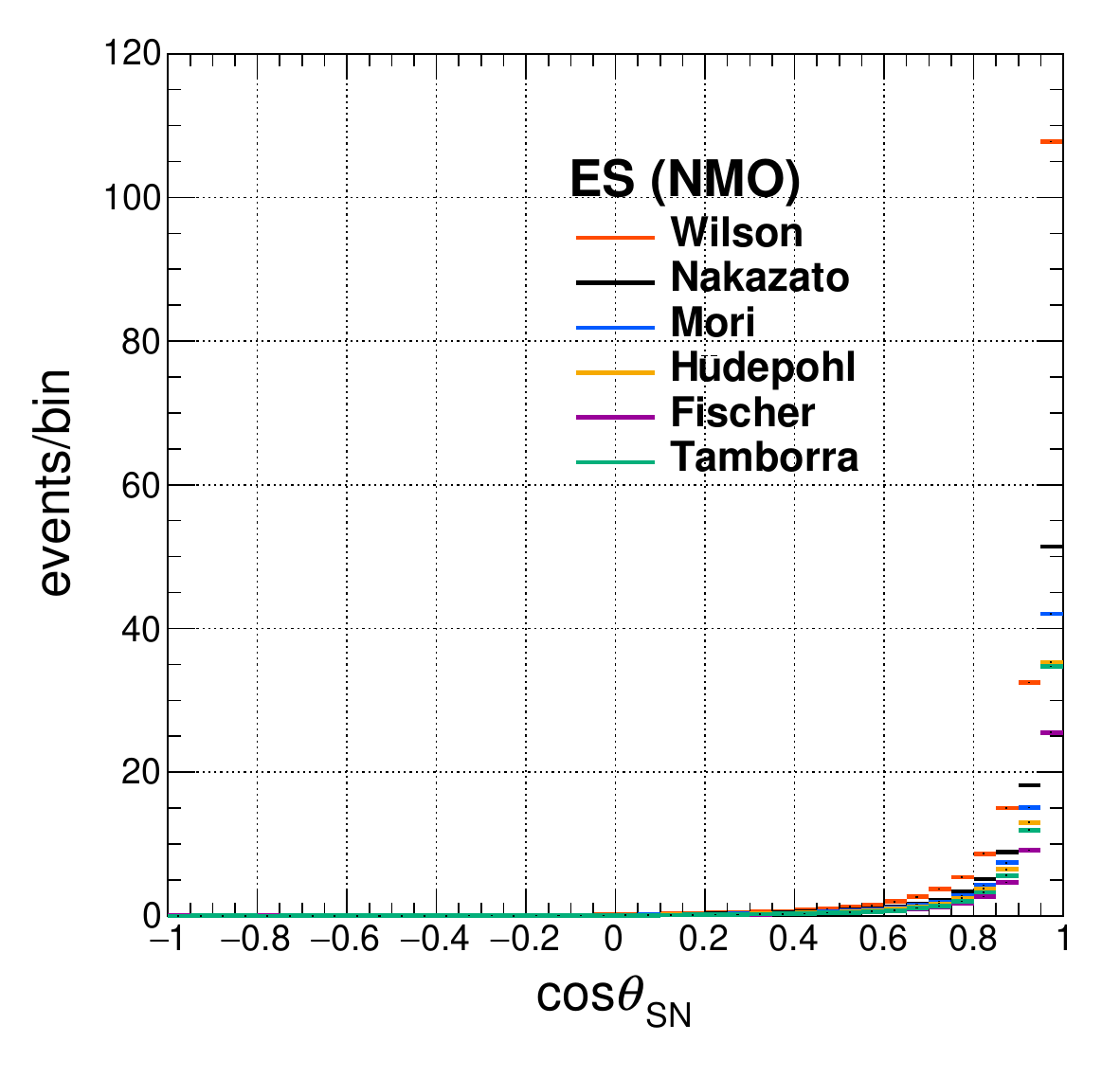}{0.35\textwidth}{(b) ES}}
\gridline{\fig{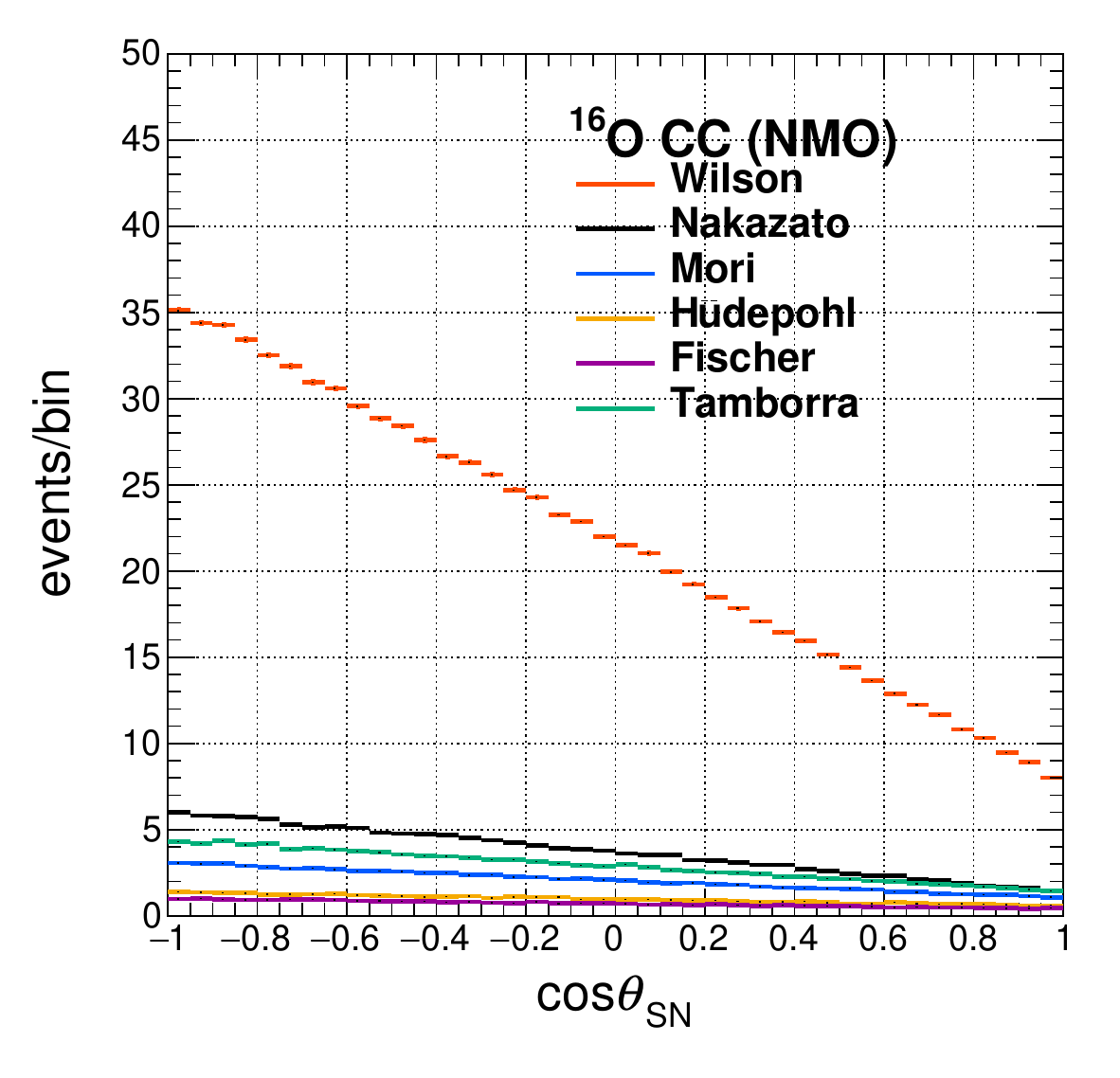}{0.35\textwidth}{(c) $^{16}$O~CC}
          \fig{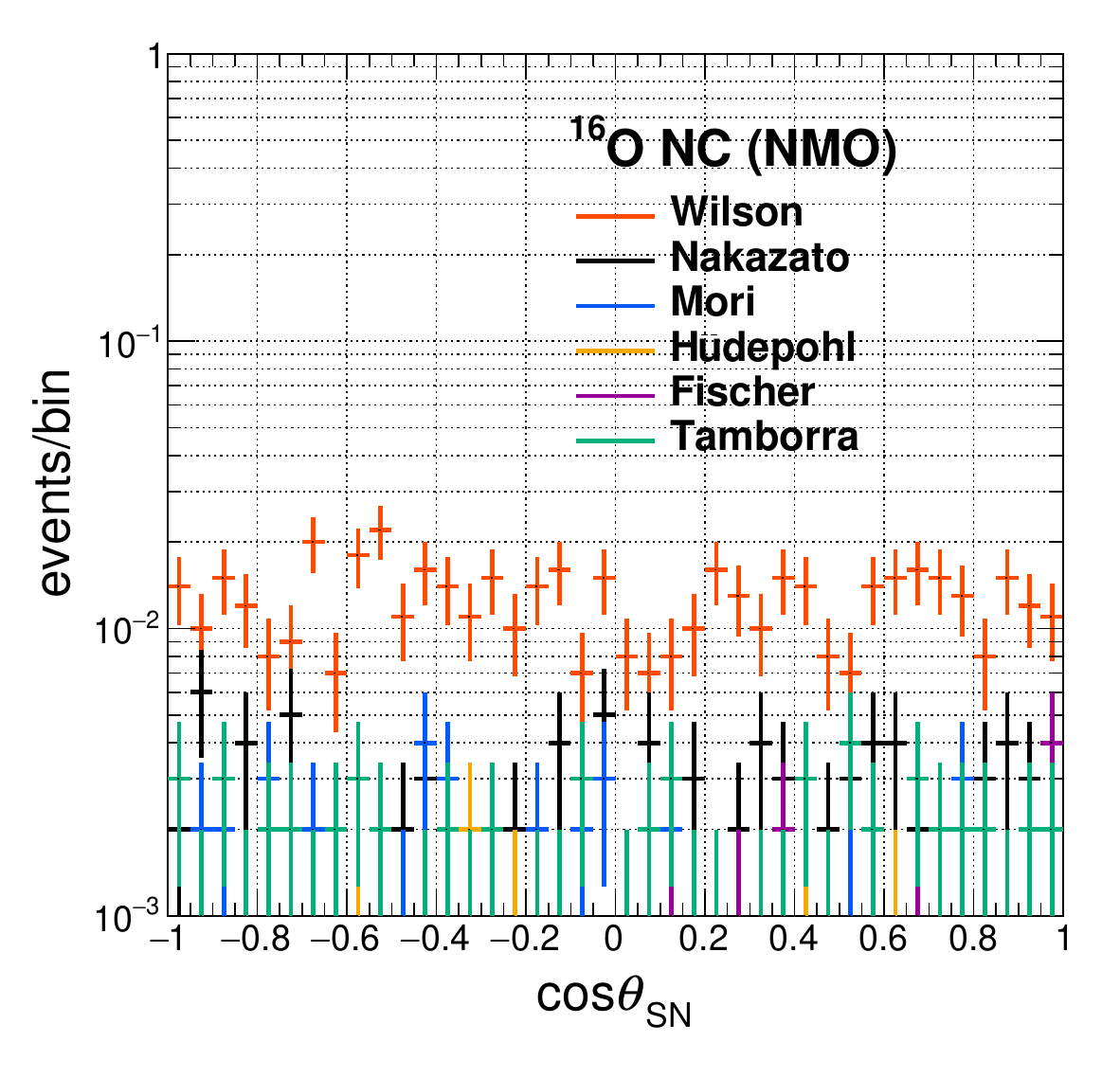}{0.35\textwidth}{(d) $^{16}$O~NC}}
\caption{Comparison of $\cos\theta_\mathrm{SN}$ distribution among models for each interaction: (a)~IBD, (b)~ES, (c)~$^{16}$O~CC, and (d)~$^{16}$O~NC.}
\label{fig:NMOvarModelCosineInteractions}
\end{figure}

Figure~\ref{fig:NMOvarModelCosineInteractions} shows comparison of angular distribution of events among models.
We note that the general shape of the $\cos\theta_\mathrm{SN}$ distribution does not depend on the model choice. 
However, the slope of the $\cos\theta_\mathrm{SN}$ distribution for IBD and $^{16}$O~CC events is steeper in the models with higher mean energy.

\section{Performance of IBD Tagging and SN Direction Fit}\label{sec:Results-SNWATCH}
This section discusses the performance of SNWATCH.
In Section~\ref{subsec:TaggingPerformance}, we report the IBD-tagging performance, focusing on an SN burst located at 10~kpc assuming the NMO scenario.
The SK-Gd's response to events tagged as either IBD-like or otherwise represents the two main event categories that can be observed in SK-Gd. 
In Section~\ref{subsec:PointingAccuracyForSN10kpc}, we present the pointing accuracy for each SN model assuming an SN burst located at 10~kpc for both NMO and IMO assumptions.

\subsection{IBD tagging Performance}\label{subsec:TaggingPerformance}
For each interaction X (IBD, ES, $^{16}$O~CC, $^{16}$O~NC, neutron capture by Gd (Gd-n), or other), the IBD-like selection efficiency, IBD-like purity, and ES-like purity are defined as follows:
\begin{equation}
    \mathrm{IBD\mathchar`-like~selection~efficiency}\equiv\frac{\mathrm{true~interaction~X~tagged~as~IBD}}{\mathrm{total~number~of~true~interaction~X}},
    \label{eq:EfficiencyOfIBDtagging}
\end{equation}

\begin{equation}
    \mathrm{IBD\mathchar`-like~purity}\equiv\frac{\mathrm{true~interaction~X~tagged~as~IBD}}{\mathrm{total~number~of~events~tagged~as~IBD}},
    \label{eq:PurityOfIBDtagging}
\end{equation}

\begin{equation}
    \mathrm{ES\mathchar`-like~purity}\equiv\frac{\mathrm{true~interaction~X~tagged~as~not~IBD}}{\mathrm{total~number~of~events~tagged~as~not~IBD}}.
    \label{eq:ESPurityOfIBDtagging}
\end{equation}

\begin{table}[htb!]
    \centering
    \caption{The IBD-tagging performance for the six SN models for an SN burst located at 10~kpc in the NMO scenario. IBD-like selection efficiency (equation~(\ref{eq:EfficiencyOfIBDtagging})), IBD-like purity (equation~(\ref{eq:PurityOfIBDtagging})), and ES-like purity (equation~(\ref{eq:ESPurityOfIBDtagging})) are shown in units of percent. The errors are the statistical uncertainty of Monte-Carlo samples.}
    \label{tab:IBDtaggingPerformance}
    \hspace{-1.5cm}
    \centering
    
    \begin{tabular}{ccccccc}\hline\hline
         \textbf{Wilson}&	true IBD			&	true ES		&	true $^{16}$O~CC&	true $^{16}$O~NC&	true Gd-n&	Other			\\\hline

efficiency (\%)& 48.98 $\pm$ 0.03             & 0.35 $\pm$ 0.02             & 8.12 $\pm$ 0.03             & 0.13 $\pm$ 0.01             & 0.01 $\pm$ 0.01            & - \\
IBD-like purity (\%)& 97.2 $\pm$ 0.01             & 0.04 $\pm$ 0.01             & 2.68 $\pm$ 0.01             & 0.01 $\pm$ 0.01             & 0.03 $\pm$ 0.01   & 0.03 $\pm$ 0.01\\
ES-like purity (\%)& 69.09 $\pm$ 0.03             & 4.94 $\pm$ 0.02             & 21.16 $\pm$ 0.03             & 0.0 $\pm$ 0.01             & 4.69 $\pm$ 0.02             & 0.12 $\pm$ 0.01\\\hline
& & & & & &\\
    \end{tabular}

    \centering
    \begin{tabular}{ccccccc}\hline\hline
         \textbf{Nakazato}&	true IBD			&	true ES		&	true $^{16}$O~CC&	true $^{16}$O~NC&	true Gd-n&	Other			\\\hline
efficiency (\%)& 46.86 $\pm$ 0.04             & 0.23 $\pm$ 0.02             & 8.0 $\pm$ 0.08             & 0.14 $\pm$ 0.02             & 0.01 $\pm$ 0.01& - \\
IBD-like purity (\%)& 98.82 $\pm$ 0.01             & 0.04 $\pm$ 0.01             & 1.07 $\pm$ 0.01             & 0.01 $\pm$ 0.01             & 0.04 $\pm$ 0.01              & 0.02 $\pm$ 0.01\\
ES-like purity (\%)& 78.7 $\pm$ 0.04             & 6.52 $\pm$ 0.03             & 9.06 $\pm$ 0.03             & 0.0 $\pm$ 0.01             & 5.49 $\pm$ 0.02             & 0.23 $\pm$ 0.01\\\hline
& & & & & &\\
    \end{tabular}

    \centering
    \begin{tabular}{ccccccc}\hline\hline
        \textbf{Mori}&	true IBD			&	true ES		&	true $^{16}$O~CC&	true $^{16}$O~NC&	true Gd-n&	Other			\\\hline
efficiency (\%)& 47.04 $\pm$ 0.04             & 0.21 $\pm$ 0.02             & 8.85 $\pm$ 0.1             & 0.13 $\pm$ 0.03             & 0.01 $\pm$ 0.01& - \\
IBD-like purity (\%)& 99.21 $\pm$ 0.01             & 0.04 $\pm$ 0.01             & 0.7 $\pm$ 0.01             & 0.0 $\pm$ 0.01             & 0.03 $\pm$ 0.01             & 0.02 $\pm$ 0.01\\
ES-like purity (\%)& 81.81 $\pm$ 0.04             & 6.38 $\pm$ 0.03             & 5.93 $\pm$ 0.03             & 0.0 $\pm$ 0.01             & 5.64 $\pm$ 0.03             & 0.24 $\pm$ 0.01\\\hline
& & & & & &\\
    \end{tabular}

    \centering
    \begin{tabular}{ccccccc}\hline\hline
        \textbf{H\"{u}depohl}&	true IBD			&	true ES		&	true $^{16}$O~CC&	true $^{16}$O~NC&	true Gd-n&	Other			\\\hline
efficiency (\%)& 47.78 $\pm$ 0.04             & 0.19 $\pm$ 0.02             & 11.75 $\pm$ 0.16             & 0.12 $\pm$ 0.04             & 0.01 $\pm$ 0.01& - \\
IBD-like purity (\%)& 99.42 $\pm$ 0.01             & 0.03 $\pm$ 0.01             & 0.5 $\pm$ 0.01             & 0.0 $\pm$ 0.01             & 0.03 $\pm$ 0.01             & 0.01 $\pm$ 0.01\\
ES-like purity (\%)& 84.64 $\pm$ 0.04             & 6.26 $\pm$ 0.03             & 3.09 $\pm$ 0.02             & 0.0 $\pm$ 0.01             & 5.77 $\pm$ 0.03             & 0.24 $\pm$ 0.01\\\hline
& & & & & &\\
    \end{tabular}

    \centering
    \begin{tabular}{ccccccc}\hline\hline
        \textbf{Fischer}&	true IBD			&	true ES		&	true $^{16}$O~CC&	true $^{16}$O~NC&	true Gd-n&	Other			\\\hline
efficiency (\%)& 47.64 $\pm$ 0.05             & 0.18 $\pm$ 0.02             & 9.29 $\pm$ 0.17             & 0.17 $\pm$ 0.06             & 0.01 $\pm$ 0.01& - \\
IBD-like purity (\%)& 99.48 $\pm$ 0.01             & 0.03 $\pm$ 0.01             & 0.44 $\pm$ 0.01             & 0.0 $\pm$ 0.01             & 0.03 $\pm$ 0.01             & 0.01 $\pm$ 0.01\\
ES-like purity (\%)& 83.74 $\pm$ 0.05             & 6.81 $\pm$ 0.03             & 3.52 $\pm$ 0.03             & 0.0 $\pm$ 0.01             & 5.68 $\pm$ 0.03             & 0.25 $\pm$ 0.01\\\hline
& & & & & &\\
    \end{tabular}

    \centering
    \begin{tabular}{ccccccc}\hline\hline
        \textbf{Tamborra}&	true IBD			&	true ES		&	true $^{16}$O~CC&	true $^{16}$O~NC&	true Gd-n&	Other			\\\hline
efficiency (\%)& 48.58 $\pm$ 0.04             & 0.51 $\pm$ 0.03             & 14.49 $\pm$ 0.1             & 0.12 $\pm$ 0.02             & 0.01 $\pm$ 0.01& - \\
IBD-like purity (\%)& 98.37 $\pm$ 0.02             & 0.06 $\pm$ 0.01             & 1.52 $\pm$ 0.02             & 0.0 $\pm$ 0.01             & 0.04 $\pm$ 0.01             & 0.01 $\pm$ 0.01\\
ES-like purity (\%)& 82.47 $\pm$ 0.04             & 4.71 $\pm$ 0.02             & 7.25 $\pm$ 0.03             & 0.0 $\pm$ 0.01             & 5.42 $\pm$ 0.02             & 0.16 $\pm$ 0.01\\\hline
& & & & & &\\
    \end{tabular}
\end{table}


\noindent
We note that the  IBD-like selection efficiency for IBD events is about 46--48\% for all six models, but this could potentially be improved by tightening the tagging criterion described in Section~\ref{subsec:EventReconstruction}.
Further, many true $^{16}$O~CC events are mistakenly tagged as IBD-like.
We achieved about 97--99\% IBD-like purity for IBD events from all six models.
The purity of true IBD events tagged as ES-like varies ${\sim}$~69-84\% depending on the model and is consistently the largest contributor to that sample. 
However, the true ES component of the ES-like sample still provides pointing to the SN as demonstrated below. 

\begin{figure}[htb!]
\gridline{
\fig{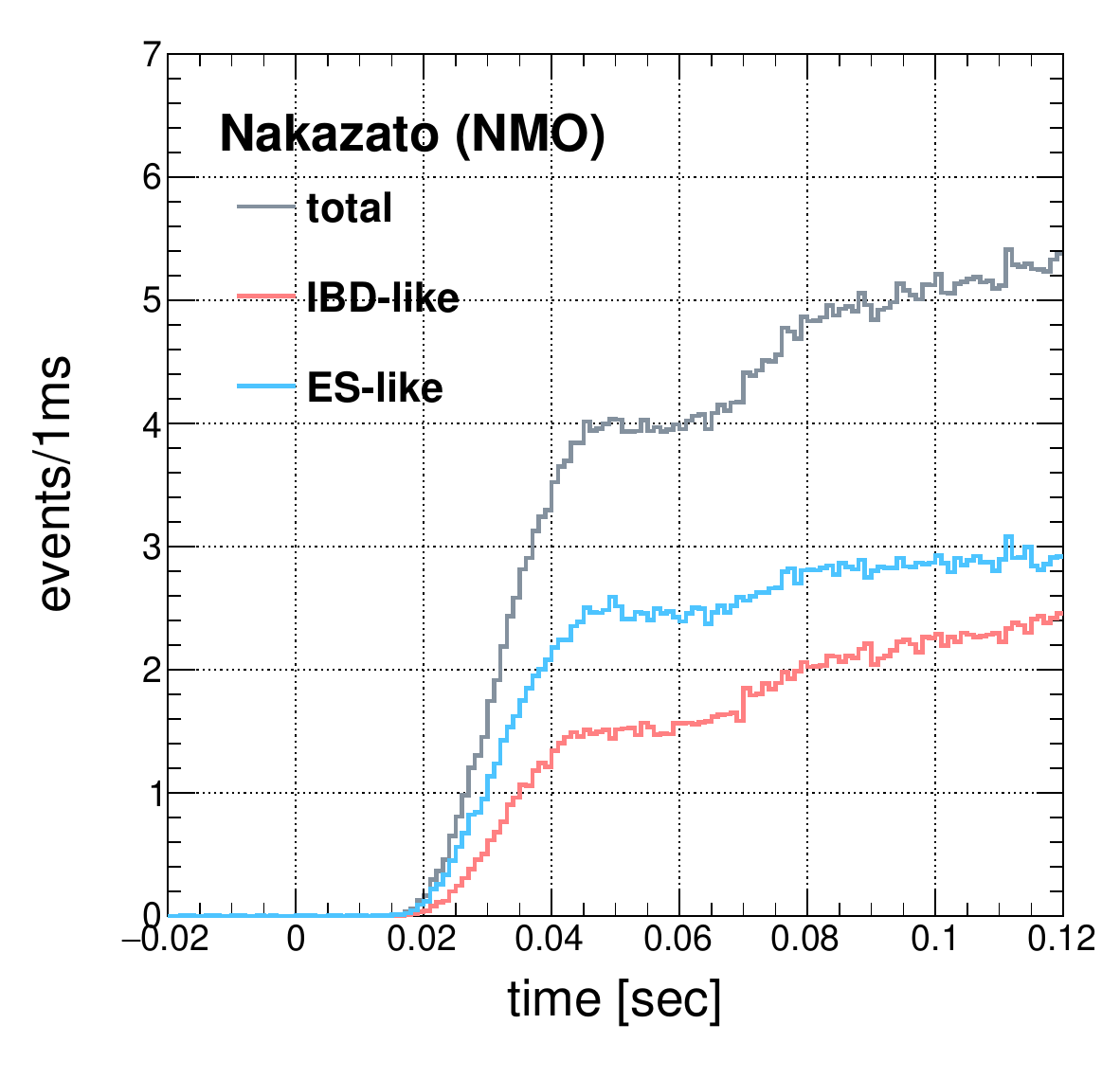}{0.3\textwidth}{(a) up to 0.12~s}
\fig{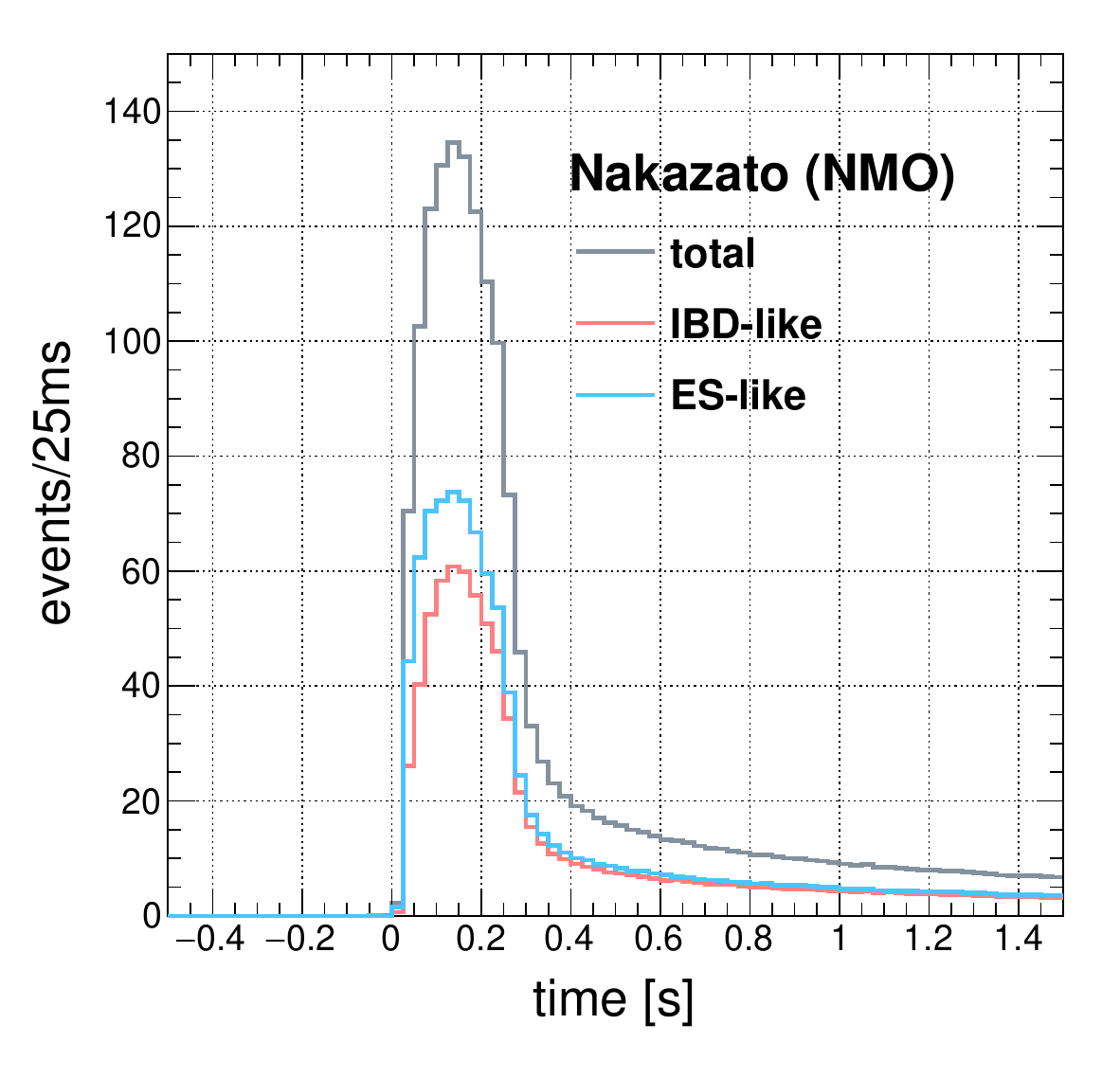}{0.3\textwidth}{(b) up to 1.5~s}
\fig{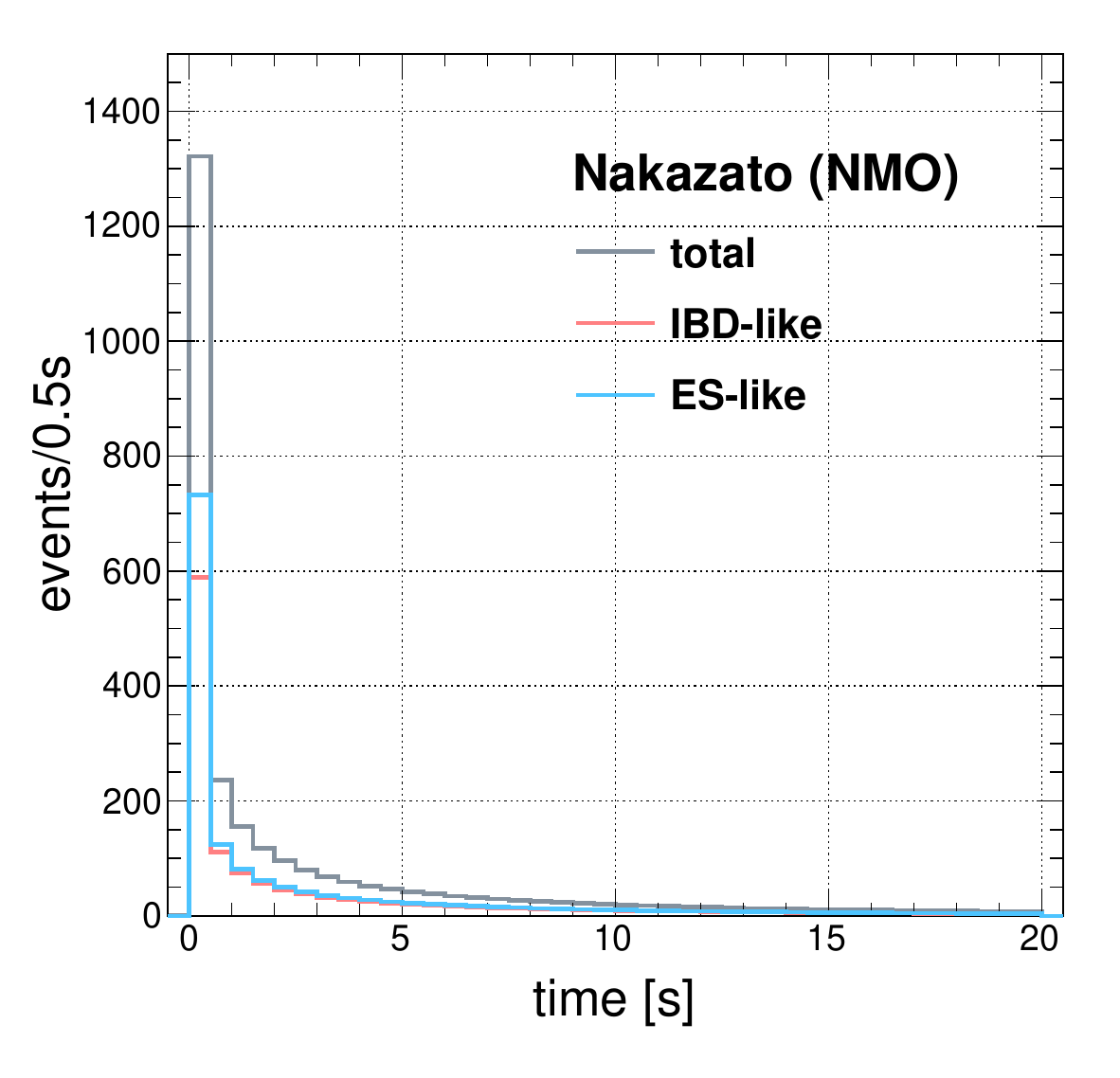}{0.3\textwidth}{(c) up to 20~s}
}
\caption{Time evolution of the IBD-like (pink) and ES-like (light blue) events for the Nakazato model for an SN burst located at 10~kpc with neutrino oscillation in NMO. 
(Same as Figure~\ref{fig:NMOTimeInteractionsNakazato} but for the IBD-like (pink) and ES-like (light blue) events.)
(a)~up to ~0.12~s, (b)~up to 1.5~s, and (c)~up to 20~s.  The grey histogram is the sum of the pink and light-blue histograms.}
\label{fig:NMOTimeTaggedNakazato}
\end{figure}

\begin{figure}[htb!]
\gridline{
\fig{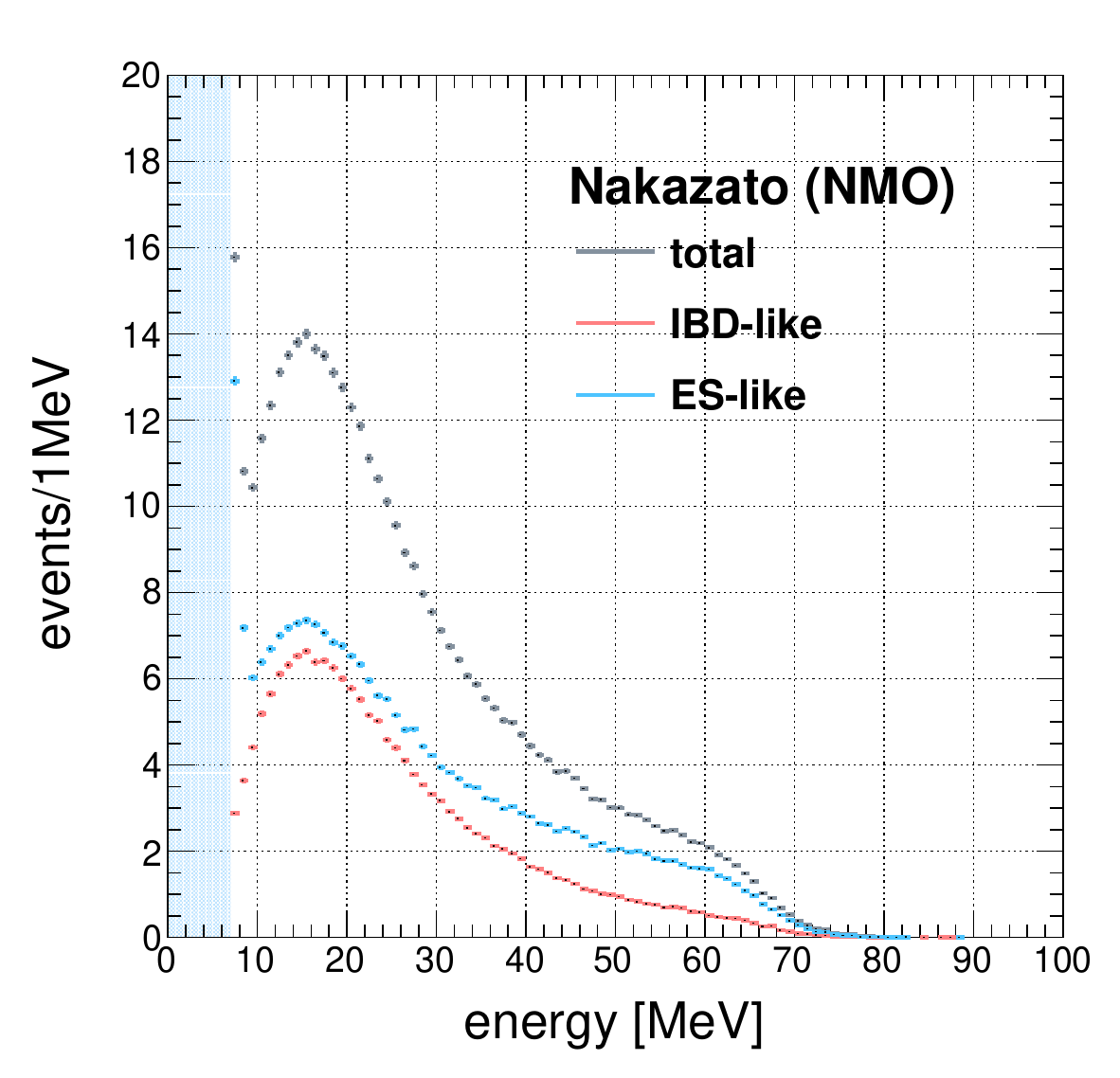}{0.3\textwidth}{(a) up to 0.12~s}
\fig{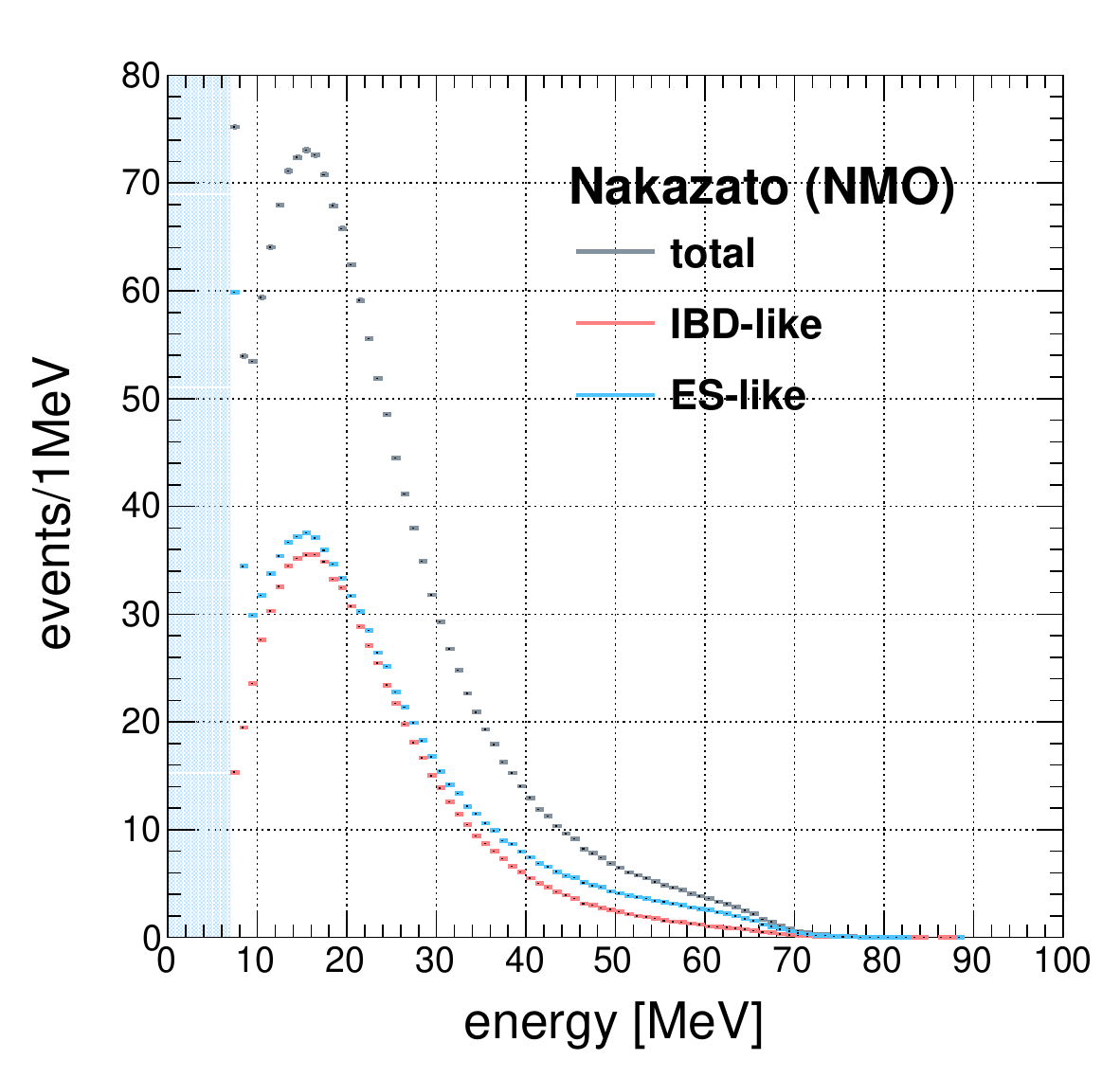}{0.3\textwidth}{(b) up to 1.5~s}
\fig{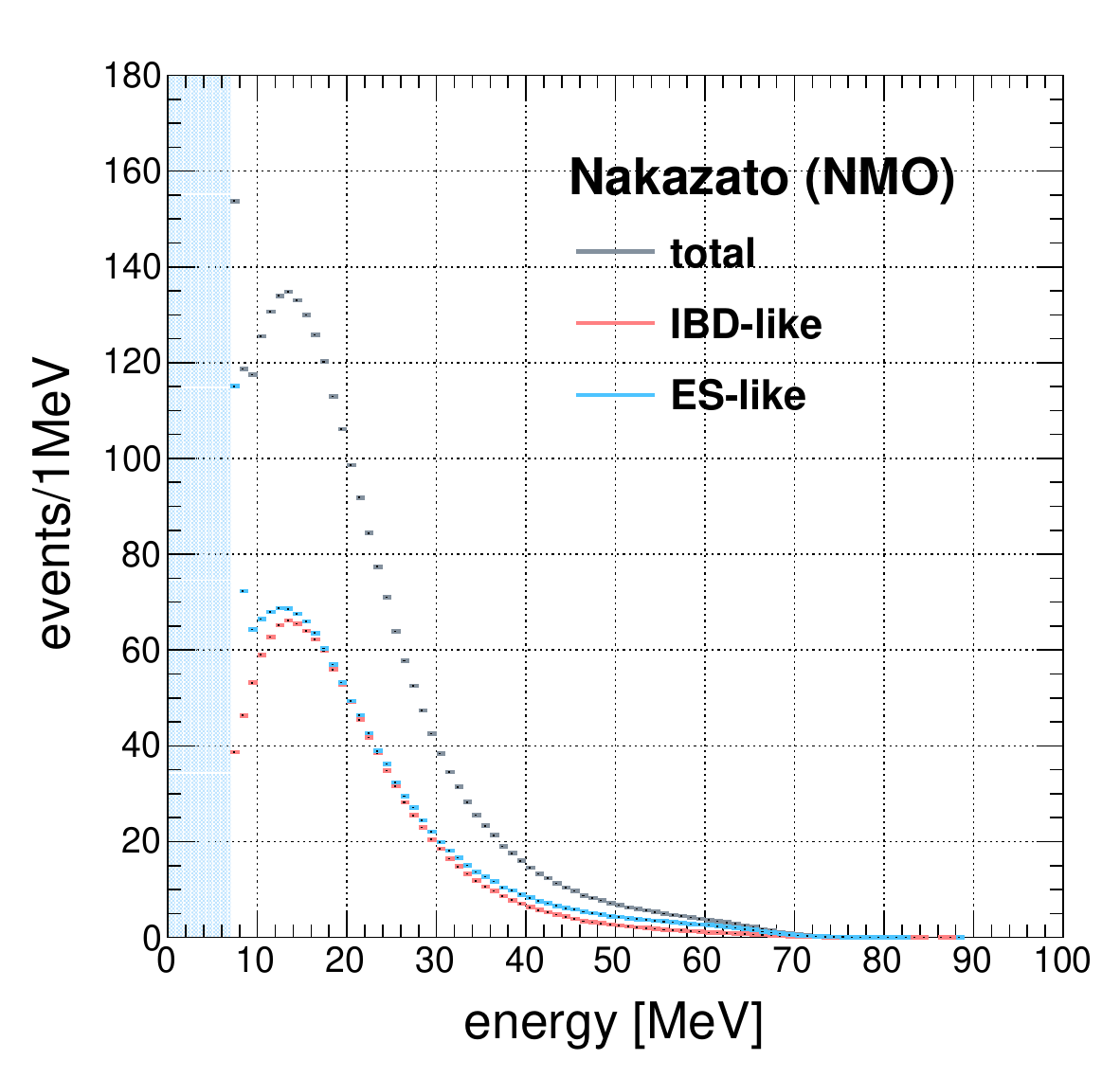}{0.3\textwidth}{(c) up to 20~s}
}
\caption{Time-integrated energy spectra of the IBD-like (pink) and ES-like (light blue) events for the Nakazato model for an SN burst located at 10~kpc with neutrino oscillation in NMO in the different time range: (a)~up to ~0.12~s, (b)~up to 1.5~s, and (c)~up to 20~s.
(Same as Figure~\ref{fig:NMOEnergyInteractionNakazato} but for the IBD-like (pink) and ES-like (light blue) events.)
Here, the energy indicates the reconstructed energy of e$^+$ for IBD, e$^-$, for ES, e$^+$ and e$^-$ for $^{16}$O~CC, the gamma rays for $^{16}$O~NC events, the gamma rays from neutron capture events, and the other events.  
The energy region below the 7~MeV threshold for selecting ``prompt'' candidates is shaded in light blue.
The grey histogram is the sum of the pink and light-blue histograms.}
\label{fig:NMOEnergyTaggedNakazato}
\end{figure}

\begin{figure}[htb!]
\gridline{
\fig{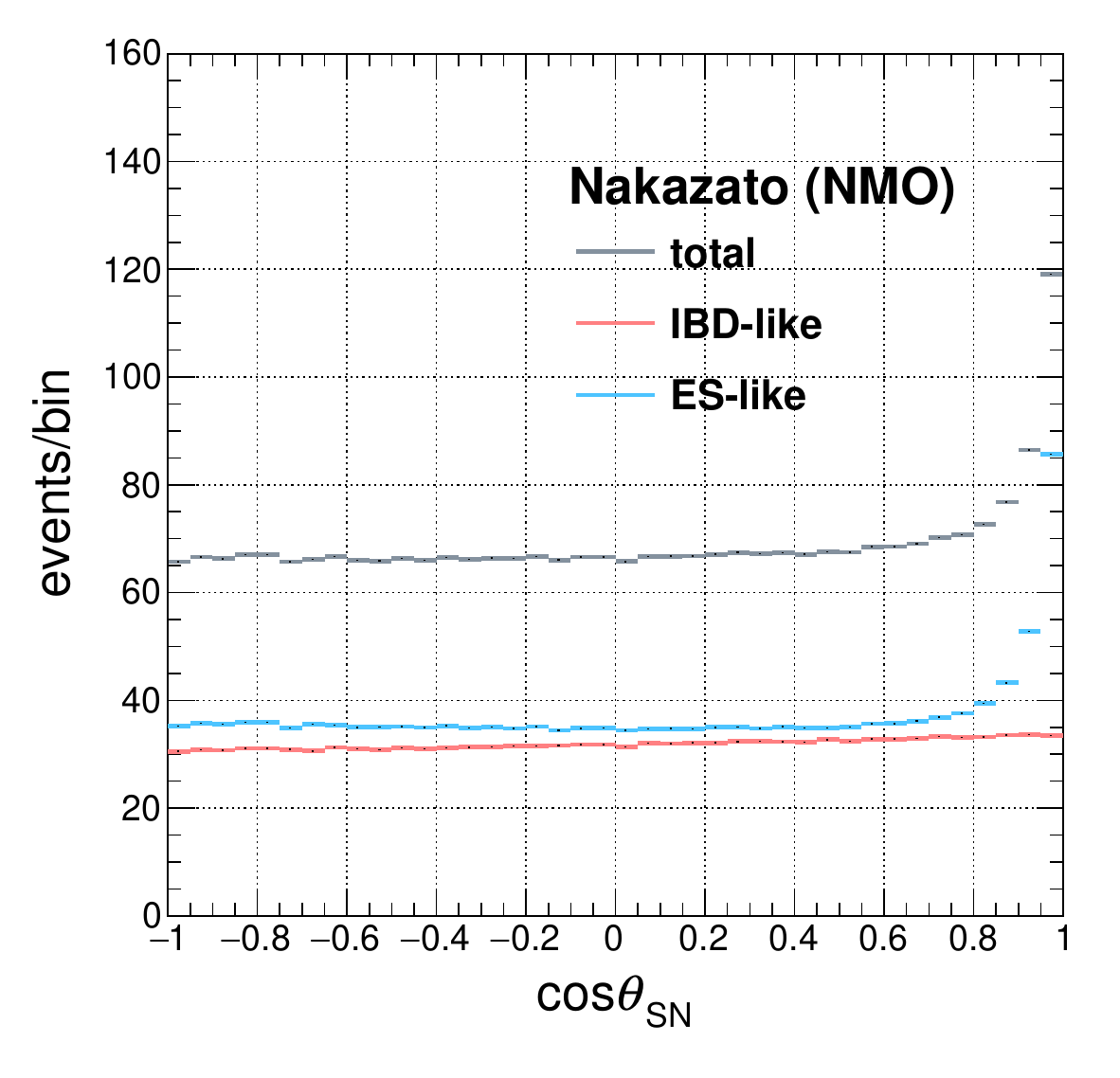}{0.33\textwidth}{(a) 1D $\cos\theta_{\mathrm{SN}}$ distribution}
\fig{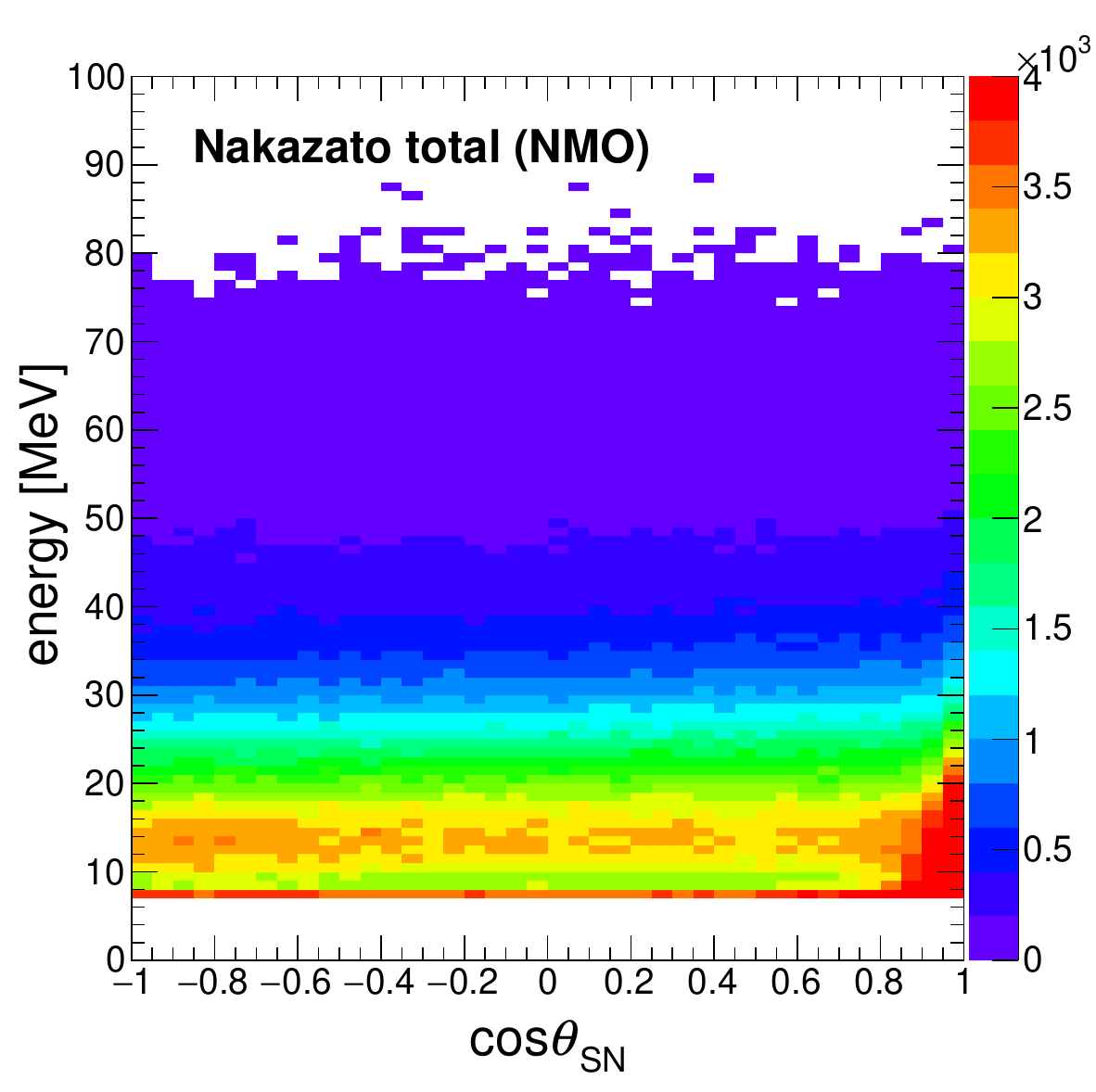}{0.33\textwidth}{(b) Energy~vs.~$\cos\theta_{\mathrm{SN}}$ for total events}
\fig{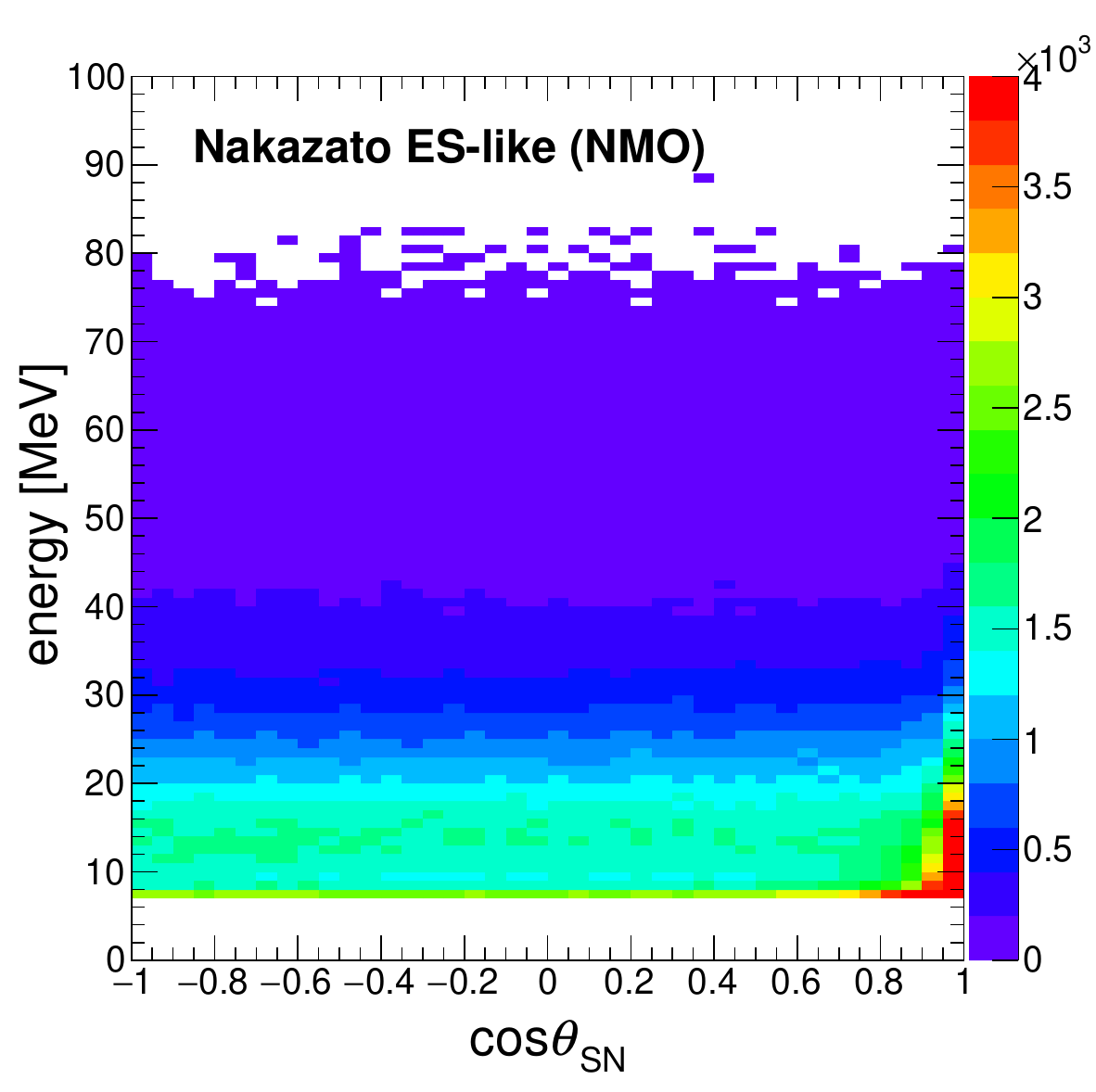}{0.33\textwidth}{(c) Energy~vs.~$\cos\theta_{\mathrm{SN}}$ for ES events}
}
\caption{Angular distribution of events for the Nakazato model for an SN burst located at 10~kpc with neutrino oscillation in NMO. (a) 1D $\cos\theta_{\mathrm{SN}}$ distribution of the IBD-like (pink) and ES-like (light blue) events. The grey histogram is the sum of the red and the green histograms. 
(b) Energy~vs.~$\cos\theta_{\mathrm{SN}}$ for all the reconstructed events. (c) Energy~vs.~$\cos\theta_{\mathrm{SN}}$ for the ES-like events.}
\label{fig:NMOAngleTaggedNakazato}
\end{figure}

SK-Gd's response after IBD-tagging is shown in Figures~\ref{fig:NMOTimeTaggedNakazato}--\ref{fig:NMOAngleTaggedNakazato} for the Nakazato model assuming the NMO as a representative example.
Figure~\ref{fig:NMOTimeTaggedNakazato} shows the time evolution of the IBD-like (pink) and ES-like (light blue) samples.  
After IBD tagging the time evolution of true ES events and oxygen events is washed out as is shown in Figure~\ref{fig:NMOTimeInteractionsNakazato}.
The shape of the time evolution of the ES-like events is similar to that of the IBD-like events in Figure~\ref{fig:NMOTimeInteractionsNakazato} for all time ranges due to the large impurity of true IBD events in the sample.
Figure~\ref{fig:NMOEnergyTaggedNakazato} shows the time-integrated energy spectra of the IBD-like (pink) and ES-like (light blue) samples. 
The peak near 7~MeV in the ES-like sample  corresponds to gamma-ray emission from Gd-n capture events.
Figure~\ref{fig:NMOAngleTaggedNakazato}~(a) shows the angular distribution of the IBD-like (pink) and ES-like (light blue) events.
Panels (b) and (c) show two-dimensional plots of the reconstructed energy and $\cos\theta_\mathrm{SN}$ distributions, respectively. 
The effect of IBD tagging can be seen by comparing these two panels.
Since IBD-like events are removed from (c), there is a stronger peak near $\cos\theta_\mathrm{SN}{\sim}1$ than in (b), which improves the pointing accuracy to the SN.

\begin{figure}[htb!]
\gridline{
\fig{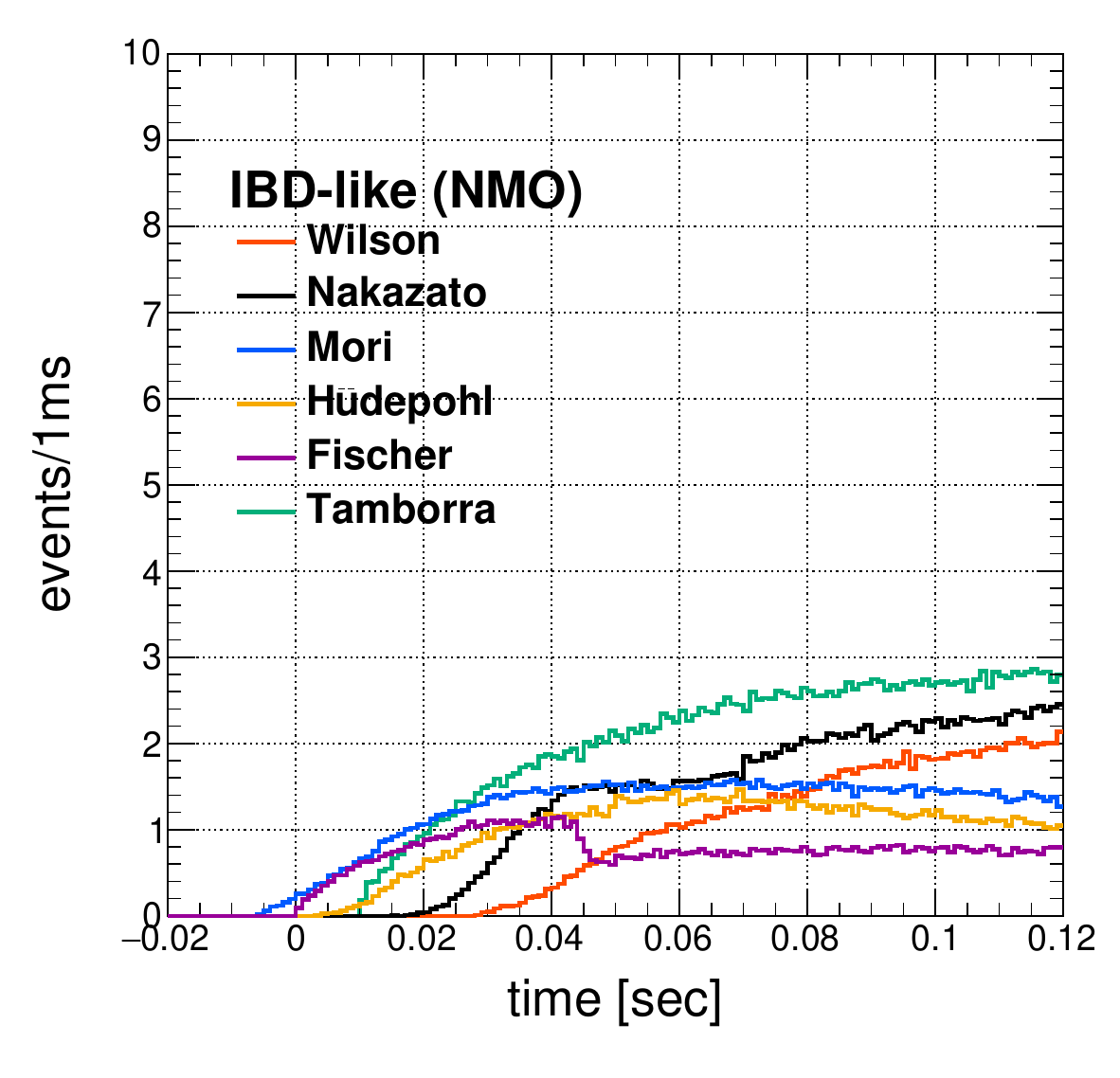}{0.35\textwidth}{(a) IBD-like events}
\fig{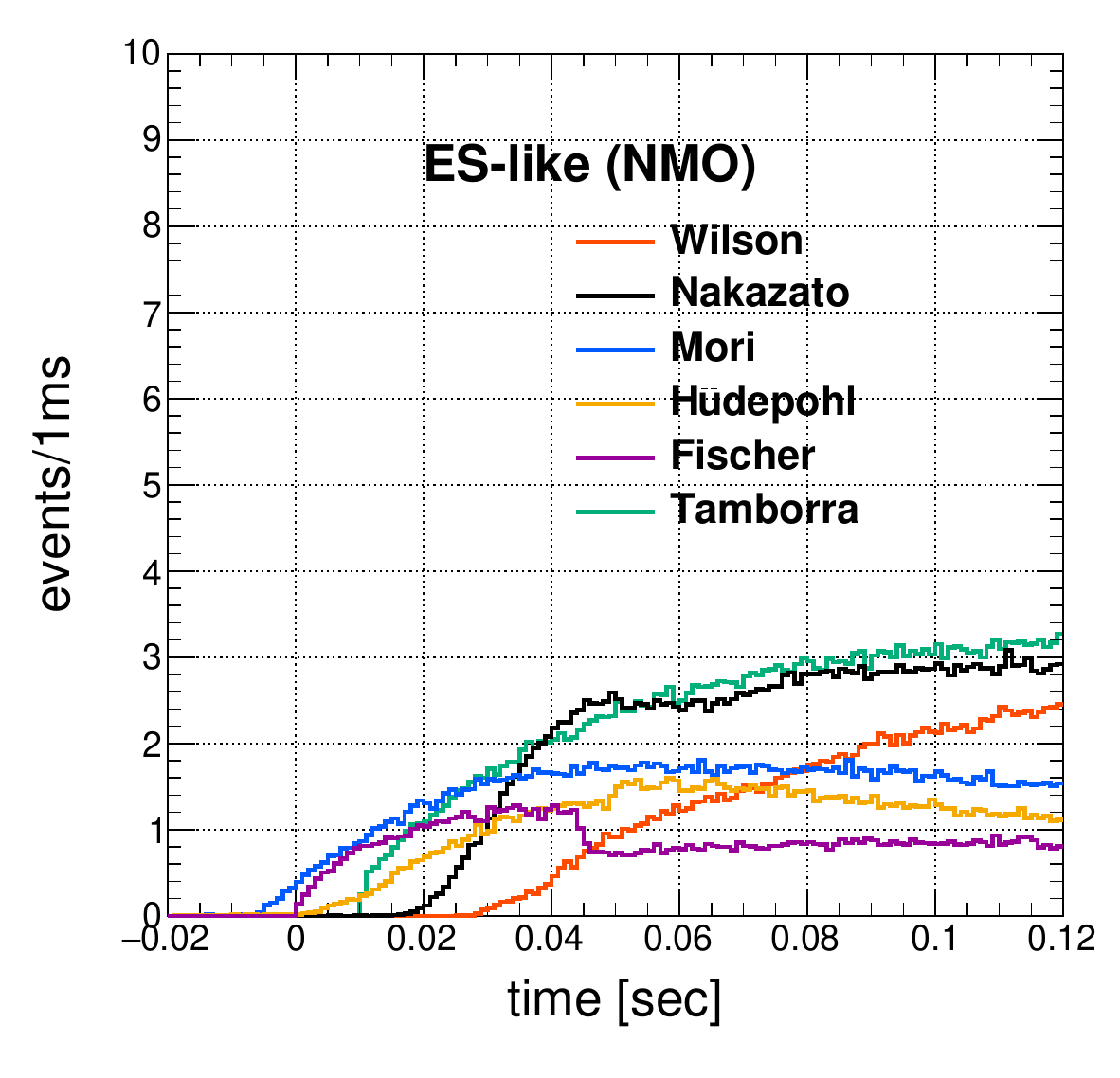}{0.35\textwidth}{(b) ES-like events}
}
\gridline{
\fig{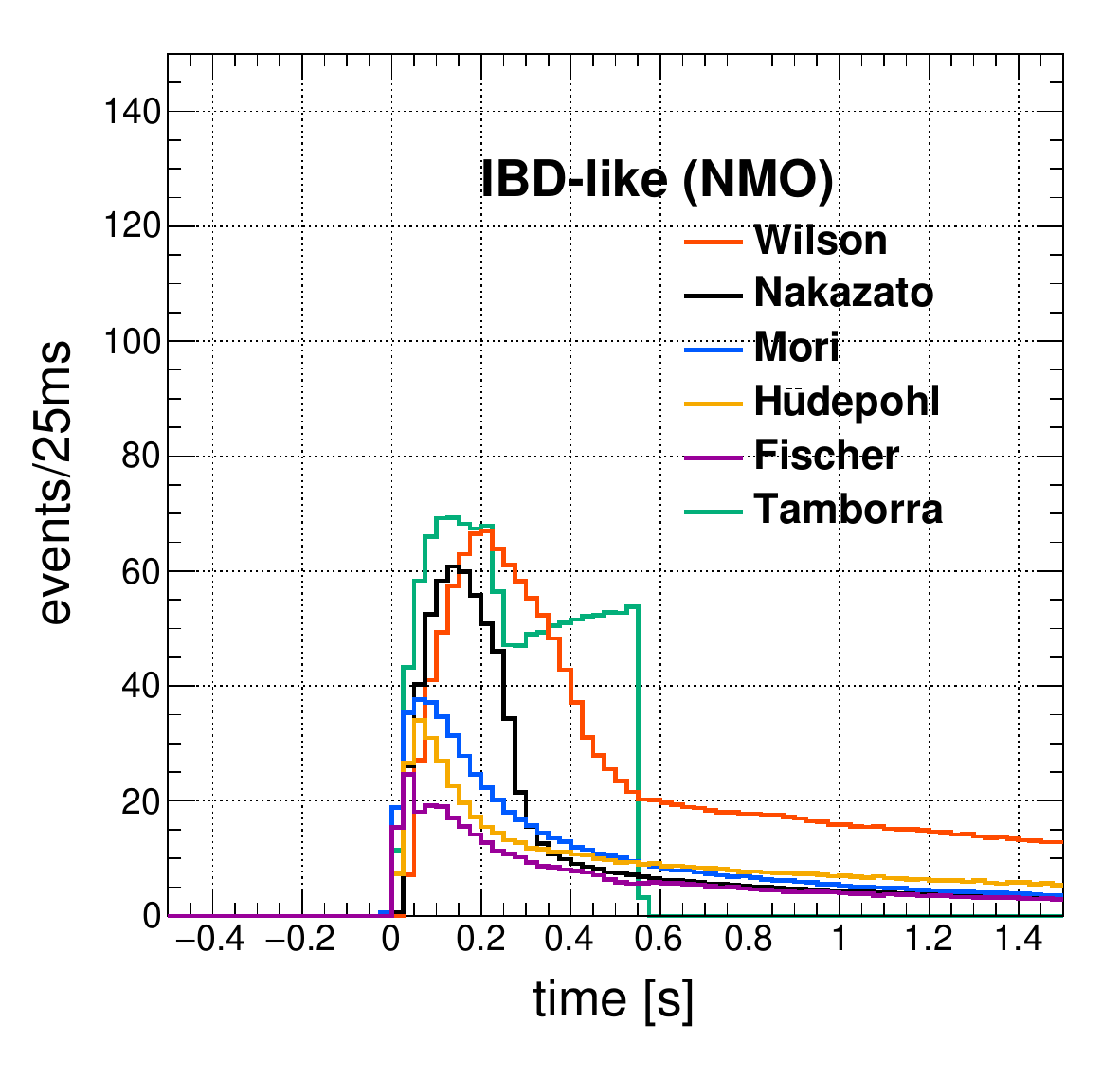}{0.35\textwidth}{(c) IBD-like events}
\fig{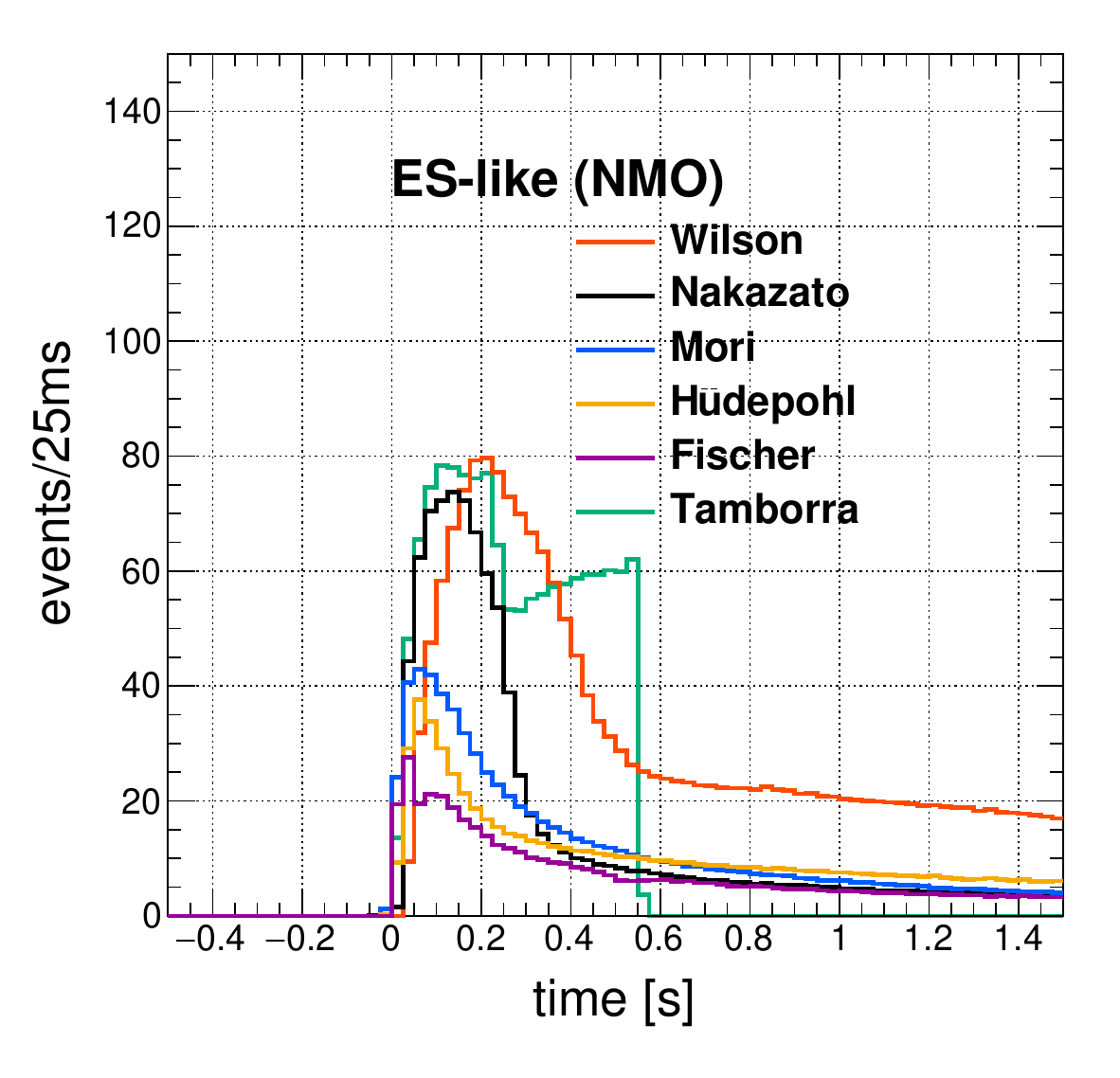}{0.35\textwidth}{(d) ES-like events}
}
\gridline{
\fig{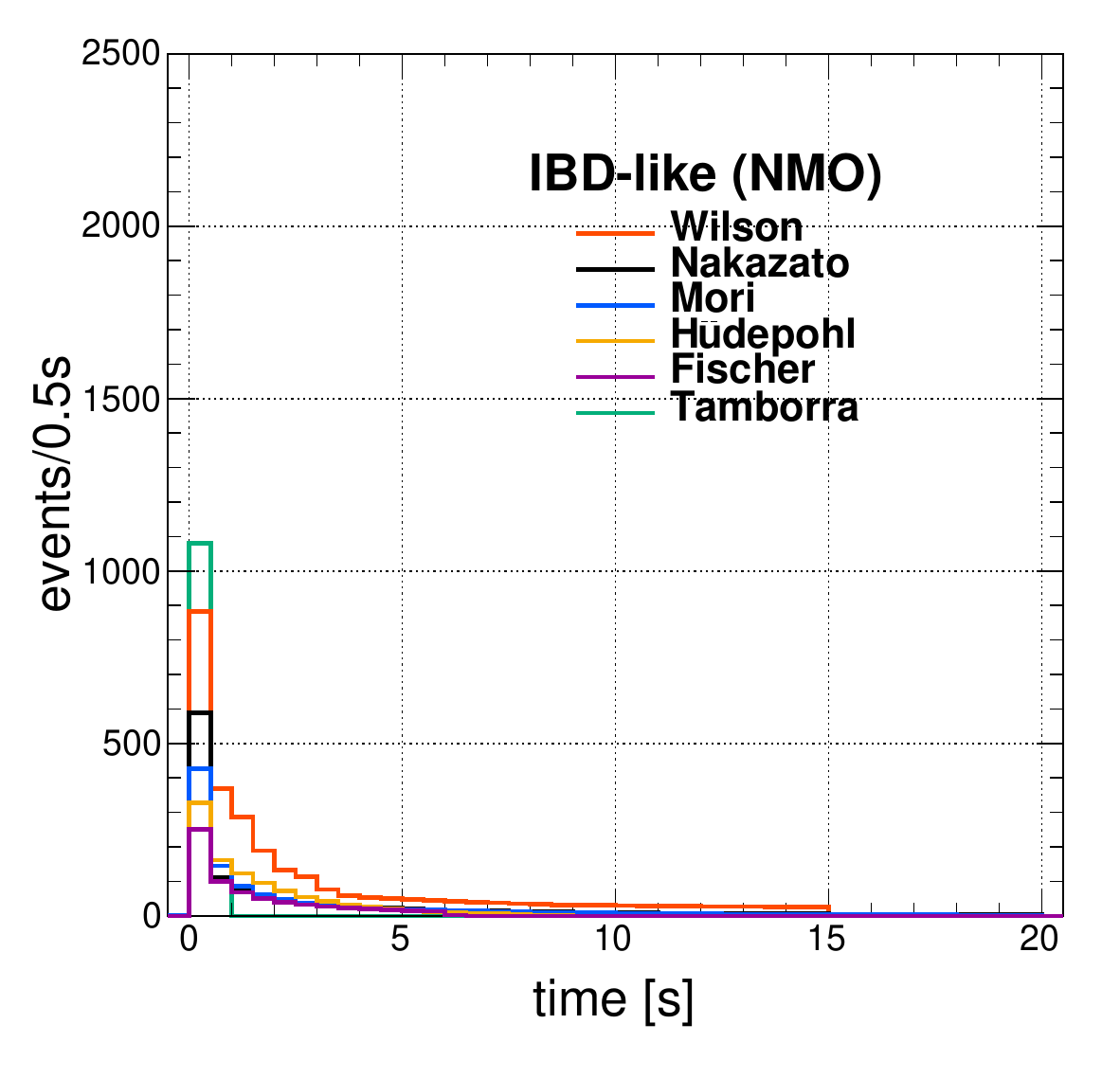}{0.35\textwidth}{(e) IBD-like events}
\fig{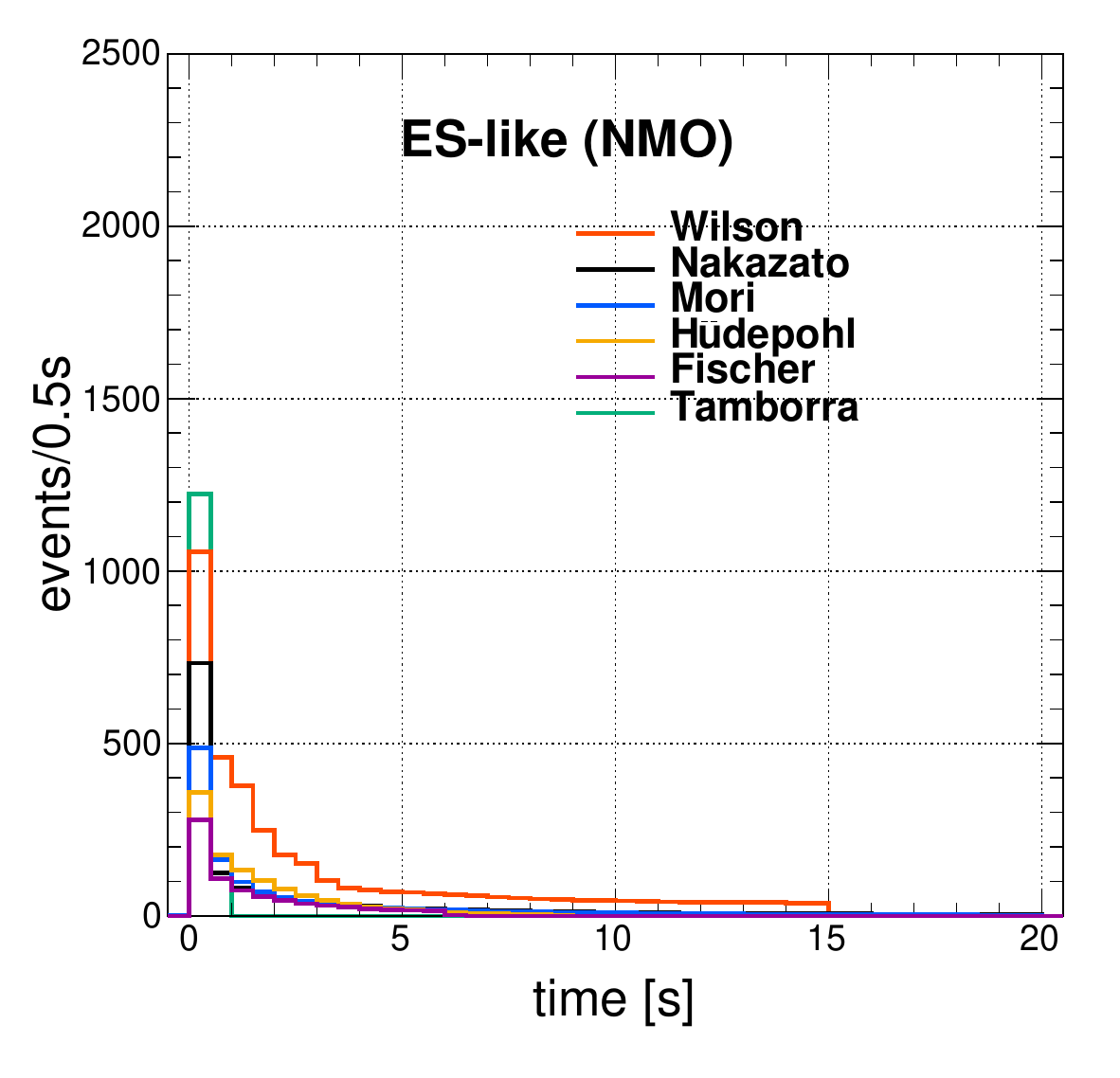}{0.35\textwidth}{(f) ES-like events}
}
\caption{Comparison of time evolution of the IBD-like (left column) and ES-like (right column) events among models for an SN burst located at 10~kpc with neutrino oscillation in the NMO scenario.  The top, middle, and bottom rows show the time evolution up to 0.12~s, 1.5~s, and 20~s, respectively.}
\label{fig:NMOTimeTagged6mddels}
\end{figure}

\begin{figure}[htb!]
\gridline{
\fig{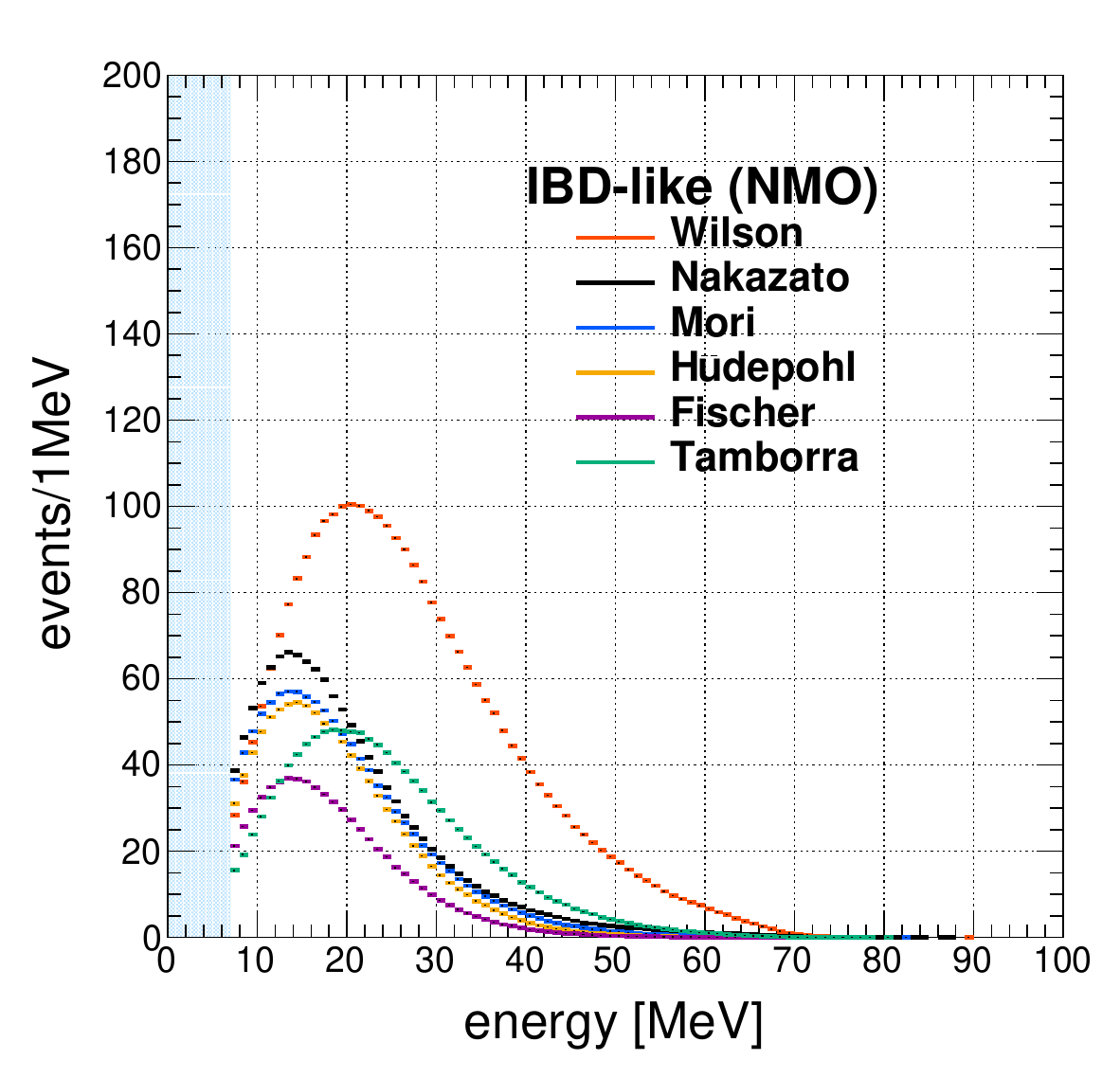}{0.35\textwidth}{(a) IBD-like events}
\fig{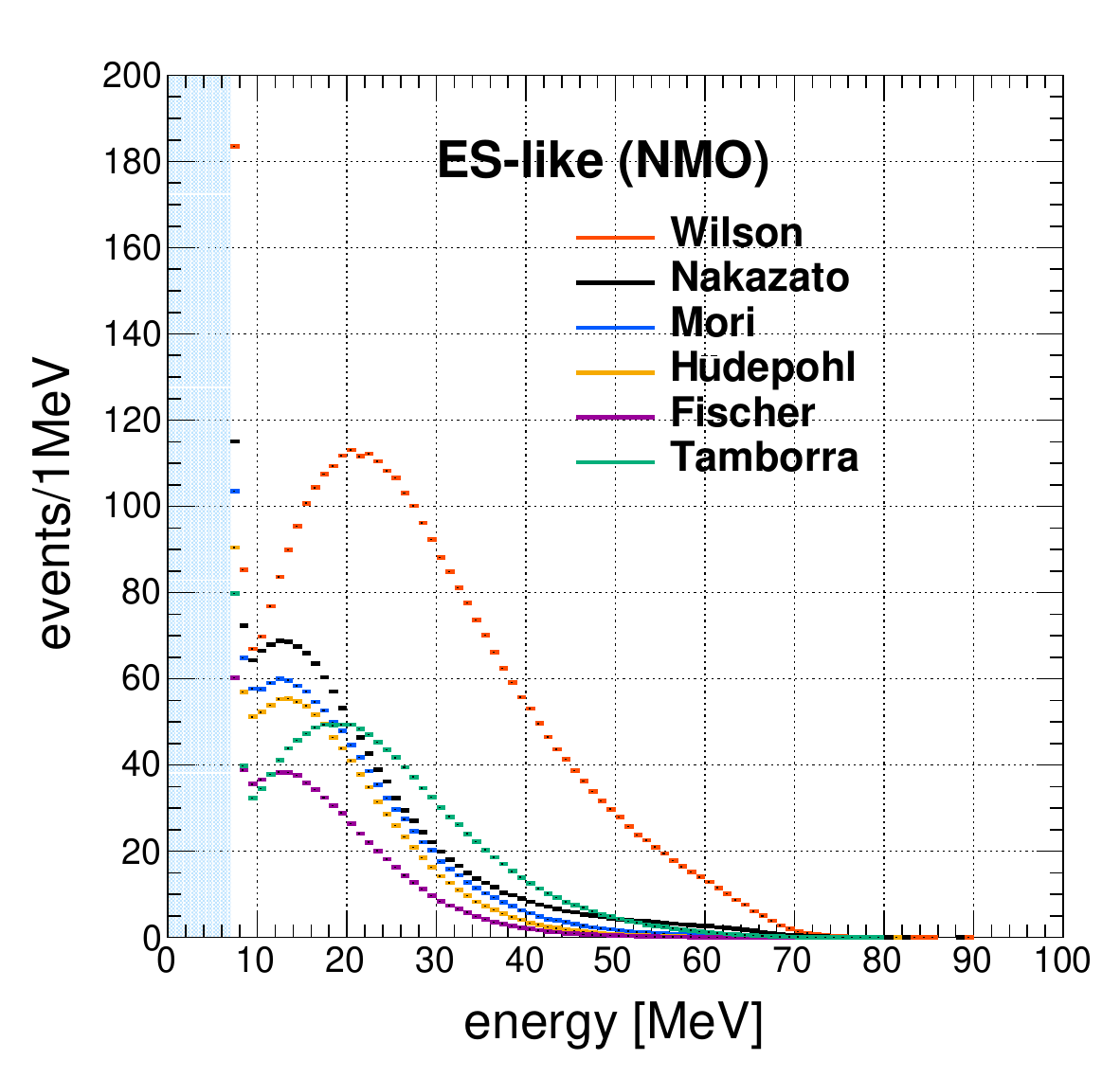}{0.35\textwidth}{(b) ES-like events}
}
\gridline{
\fig{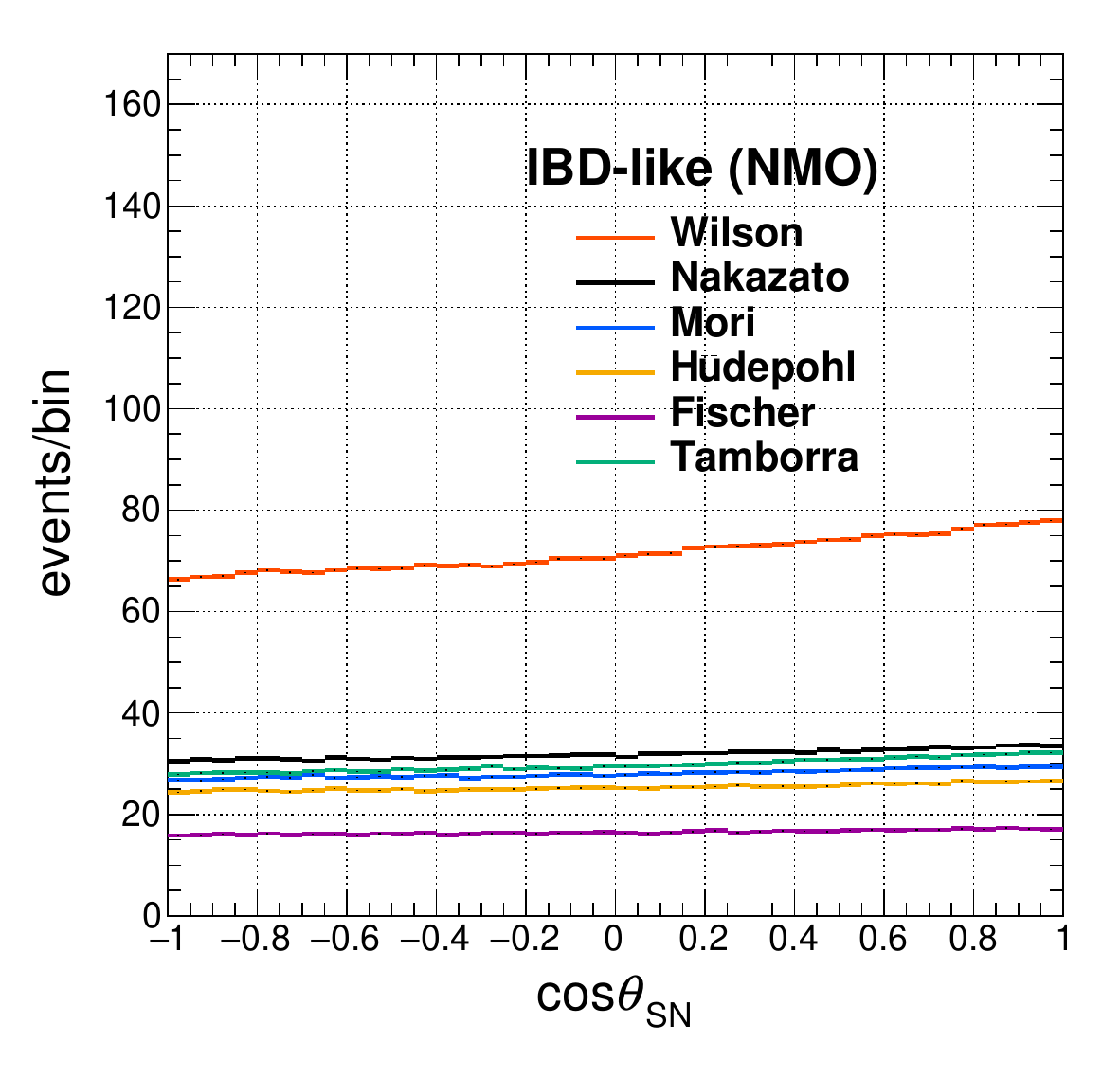}{0.35\textwidth}{(c) IBD-like events}
\fig{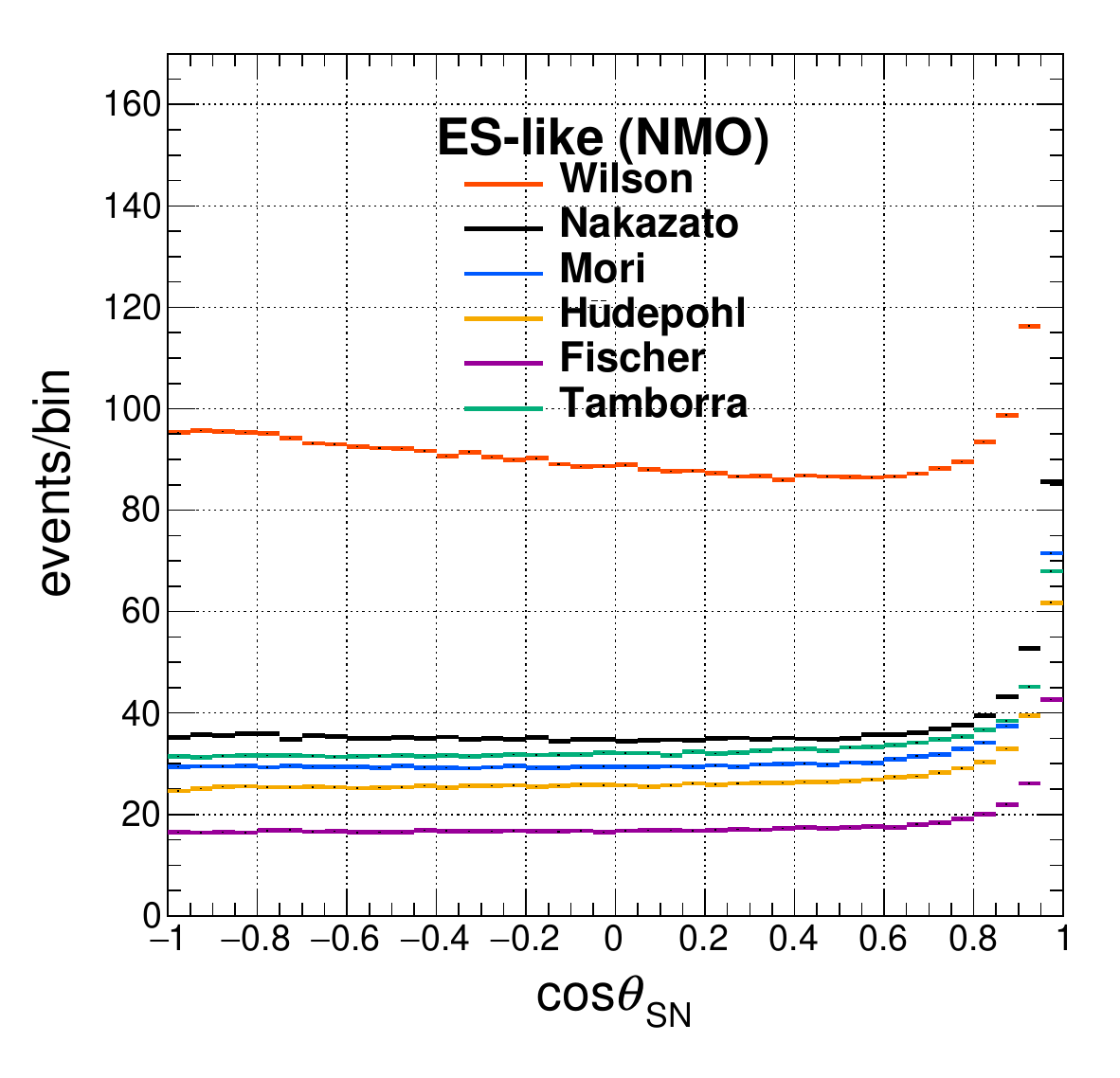}{0.35\textwidth}{(d) ES-like events}
}
\caption{(top) Comparison of energy spectra among models for (a) the IBD-like events and (b) the ES-like events.  The energy region below the 7~MeV threshold for selecting ``prompt'' candidates is shaded in light blue. (bottom) Comparison of $\cos\theta_\mathrm{SN}$ distribution among models for (c) the IBD-like events and (d) the ES-like events.}
\label{fig:NMOEneAndCosTagged6models}
\end{figure}

Similarly the time evolution of the models after sample selection, time-integrated energy, and angular distributions are shown in Figure~\ref{fig:NMOTimeTagged6mddels} and Figure~\ref{fig:NMOEneAndCosTagged6models}. 
The time and energy structures of the events after IBD tagging differ among models. 

\subsection{Pointing Accuracy for Supernova at 10 kpc}\label{subsec:PointingAccuracyForSN10kpc}
\begin{figure}[htb!]
\gridline{
\fig{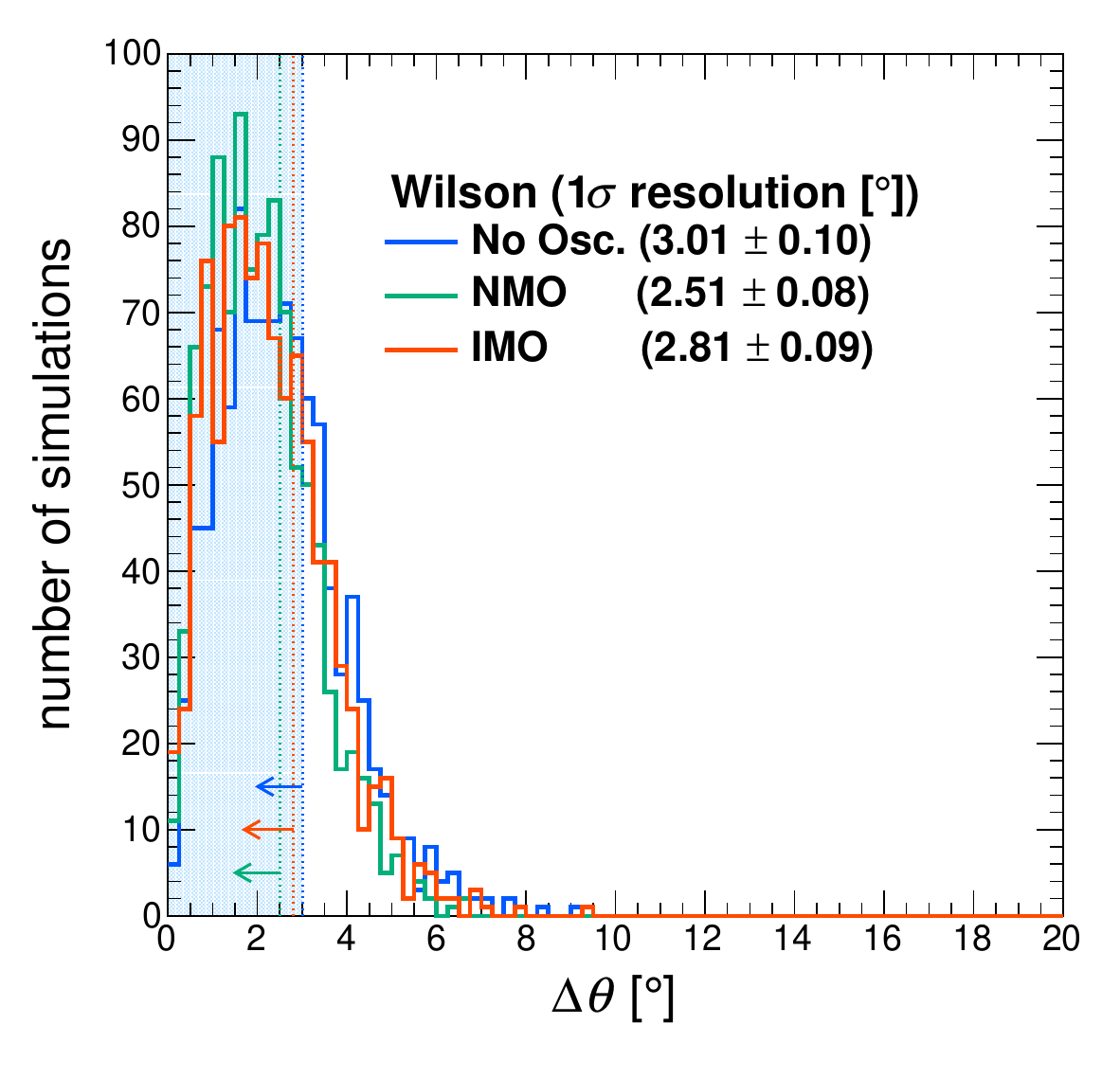}{0.3\textwidth}{(a) the Wilson model}
\fig{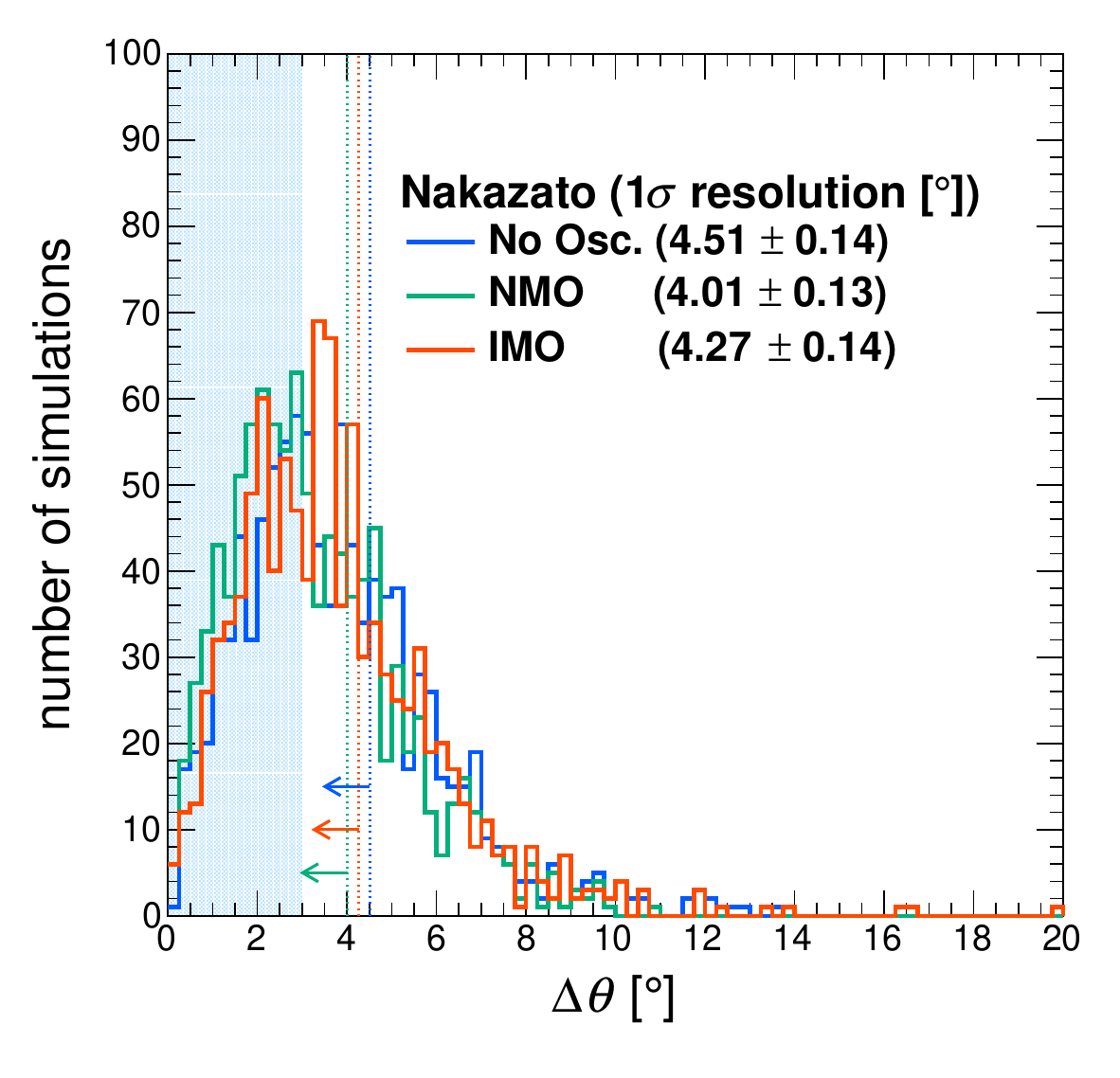}{0.3\textwidth}{(b) the Nakazato model}
\fig{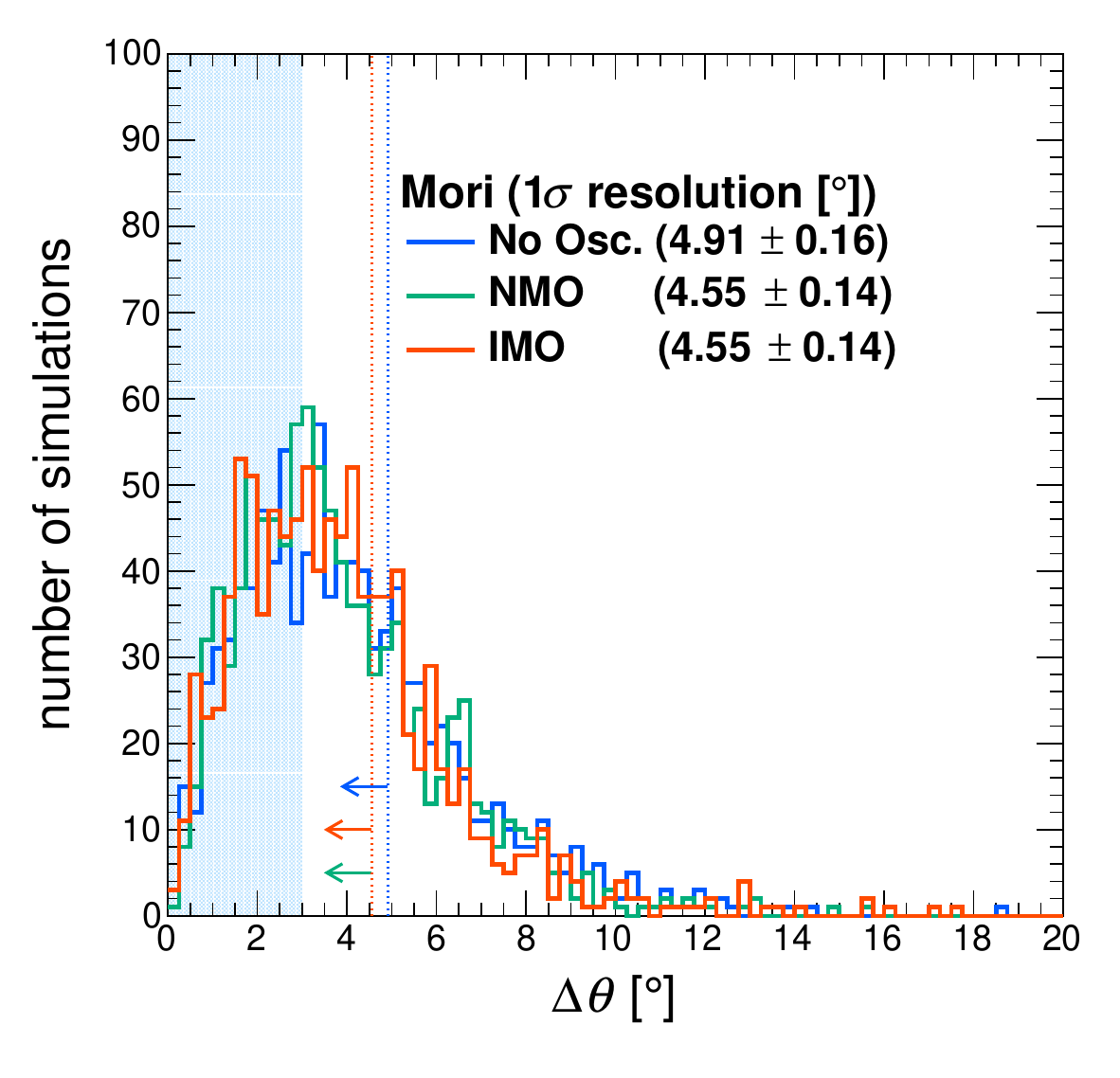}{0.3\textwidth}{(c) the Mori model}
}
\gridline{
\fig{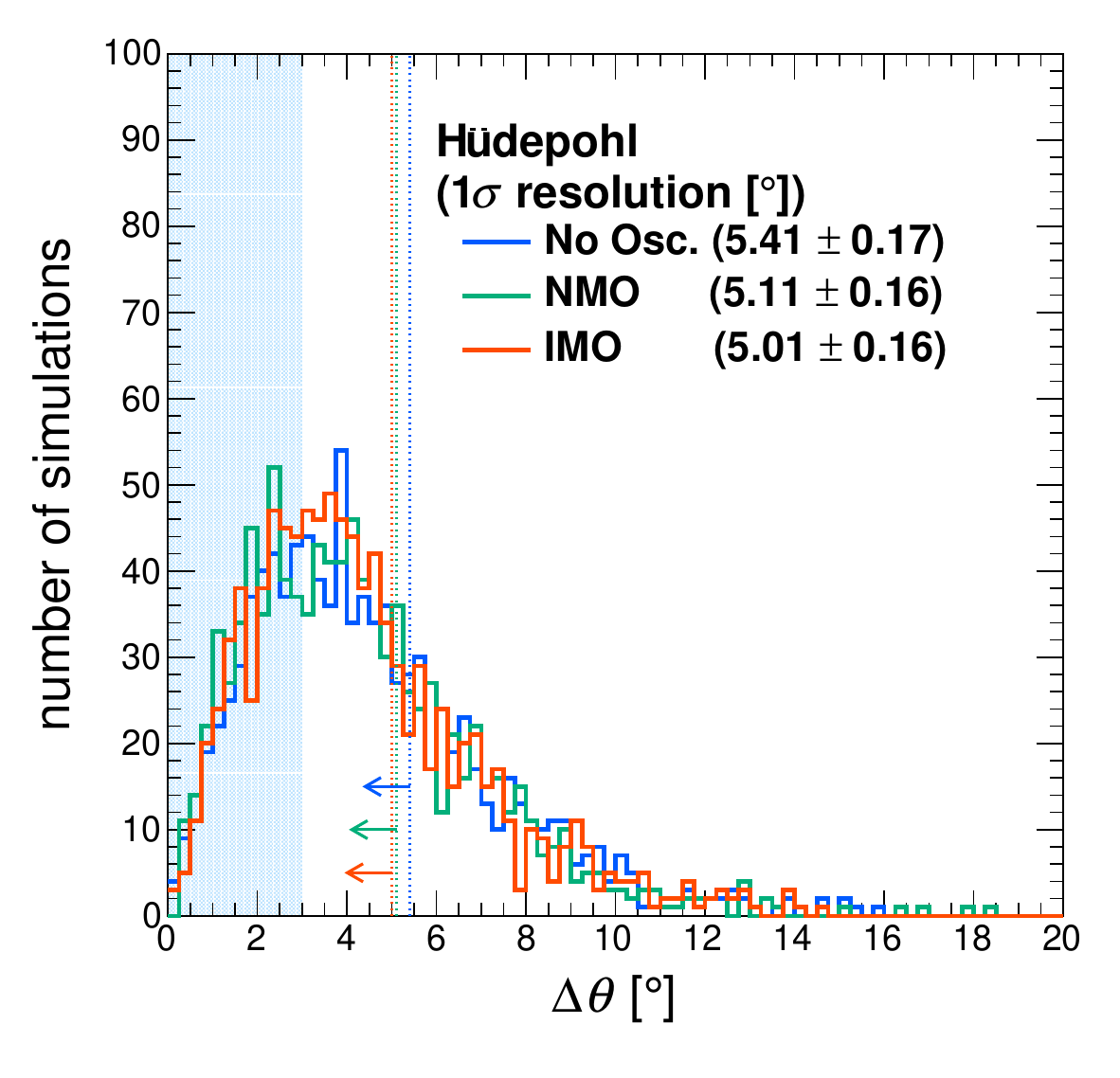}{0.3\textwidth}{(d) the H\"{u}depohl model}
\fig{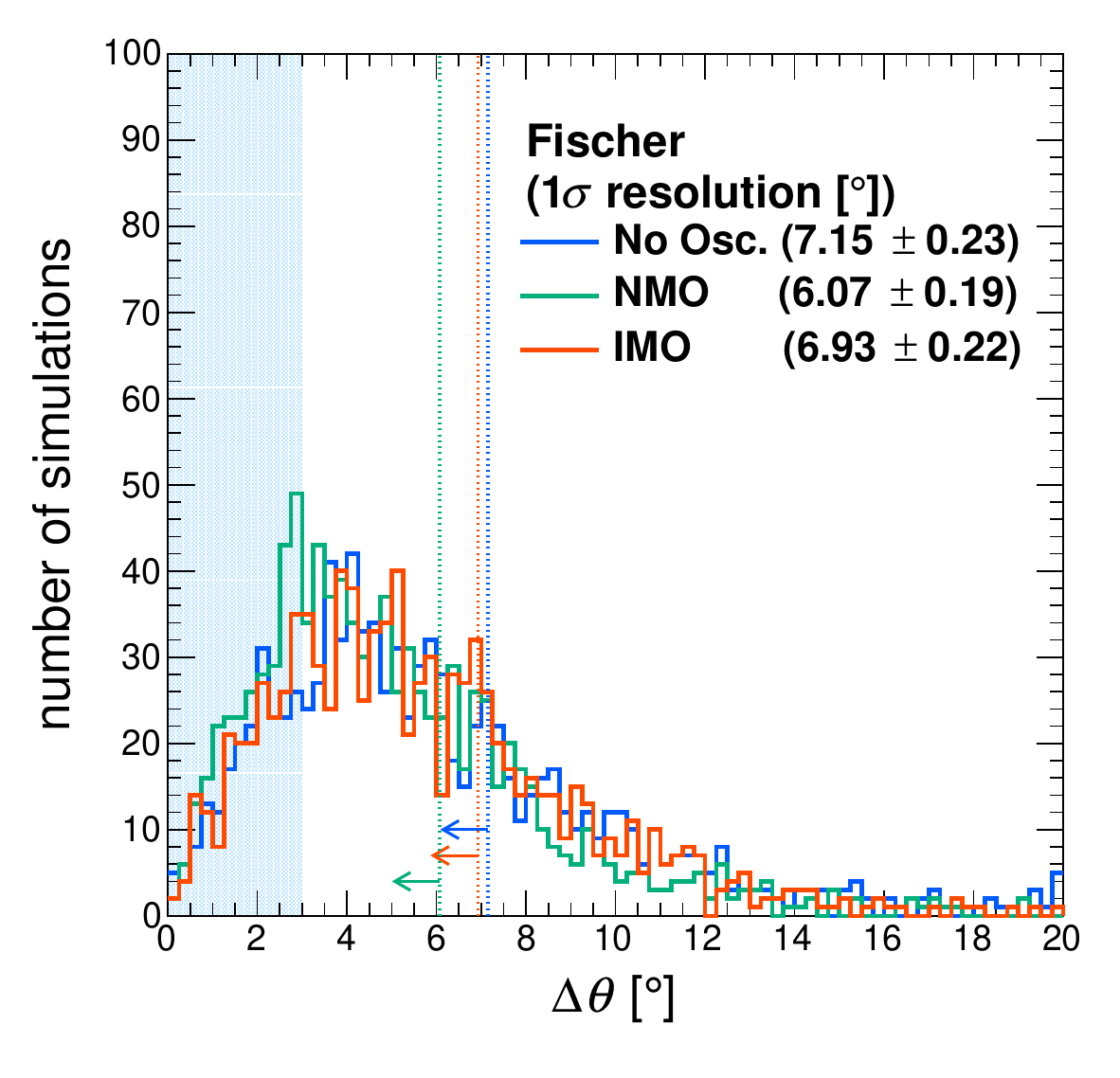}{0.3\textwidth}{(e) the Fischer model}
\fig{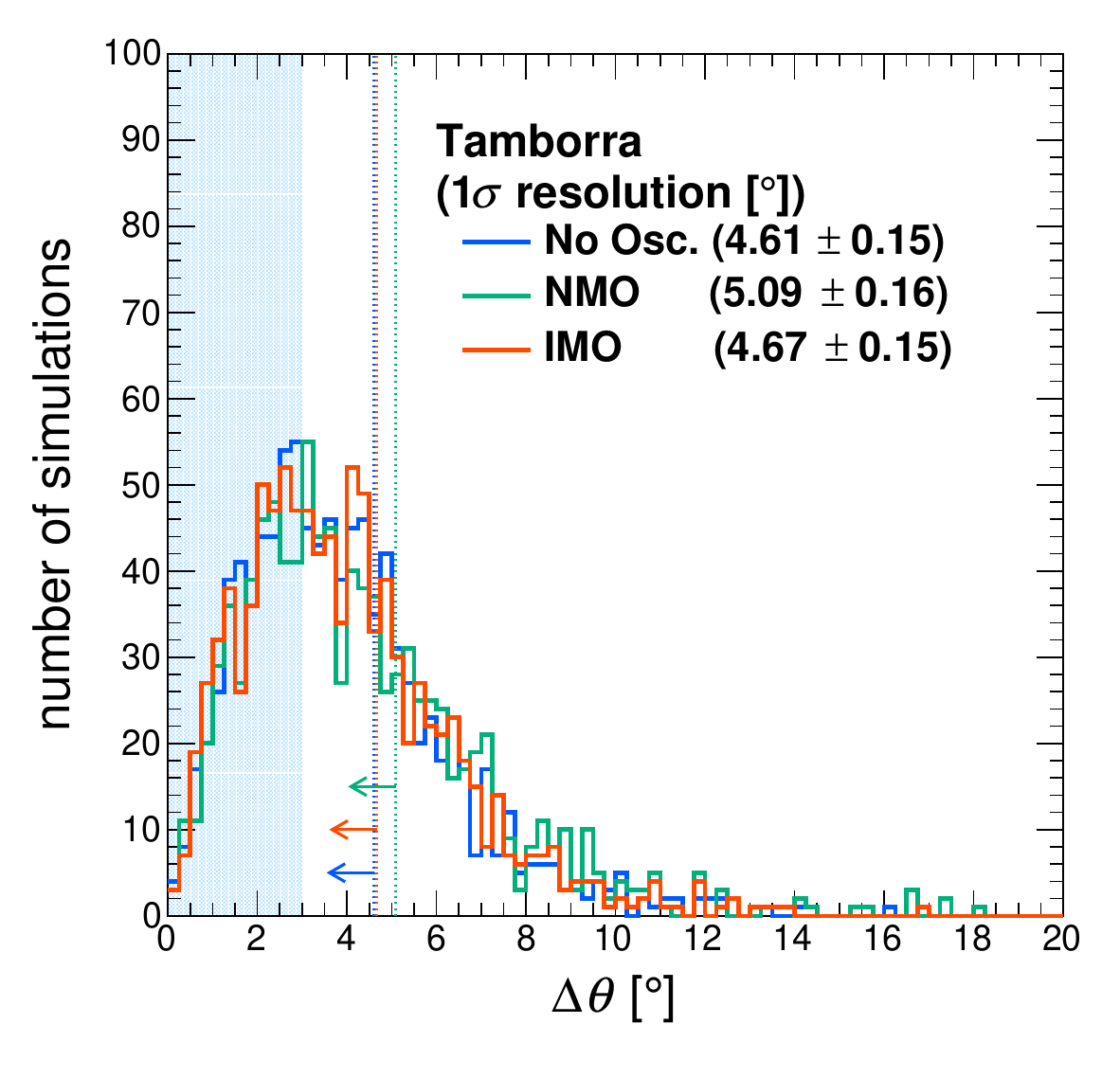}{0.3\textwidth}{(f) the Tamborra model}
}
\caption{The $\Delta\theta$ distribution of each oscillation scenario for each model for an SN burst located at 10~kpc. The blue, green, and red histograms represent No~Osc., NMO, and IMO, respectively.  The dashed line and the arrow in each histogram of the corresponding color indicate the pointing accuracy at 1$\sigma$, i.e., the value of $\Delta\theta$ up to which the integral of the histogram covers 68\% of the 1000 simulations (see Section~\ref{subsec:SNWarnWithDirInfo}).  The pointing accuracy for each oscillation assumption in units of degree is also shown in the legend.  The shaded region (light blue) indicates the target pointing accuracy ($\le3^\circ$).}
\label{fig:varOscDeltaTheta}
 \end{figure}

Finally, we derived pointing accuracy for each SN model using the distribution of $\Delta\theta$, where $\cos(\Delta\theta)=\hat{d}_\mathrm{SN}^{\mathrm{true}}\cdot\hat{d}_{\mathrm{SN}}^\mathrm{reco}$ and where $\hat{d}_\mathrm{SN}^{\mathrm{true}}$ and $\hat{d}_{\mathrm{SN}}^\mathrm{reco}$ are described in Section~\ref{subsec:SNWarnWithDirInfo}.
The pointing accuracy at 1$\sigma$ is then defined as the value of $\Delta\theta$ at which the integral of this distribution 
contains 68\% of the 1000~MC samples when integrating above from zero.
Figure~\ref{fig:varOscDeltaTheta} shows the distribution of $\Delta\theta$ from 1000 simulations of each model and oscillation assumption. 
Table~\ref{tab:PointingAccuracy6models} shows the derived pointing accuracy for the six models, which is seen to vary 
 from ${\sim} 3^\circ$ to ${\sim} 7^\circ$.  
Although the statistical error is large in some models, it can be seen that the best resolution depends on the oscillation assumption.
When neutrino oscillations are considered, better pointing accuracy is achieved in the NMO scenario in the Wilson model, the Nakazato model, and the Fischer model, while in the Mori model, the H\"{u}depohl model, and the Tamborra models, better pointing is achieved in the IMO scenario.  

\begin{table}[htb!]
	\centering
	\caption{Pointing accuracy at $1\sigma$ in the unit of degree for six models with three oscillation scenarios.  
    }
    \label{tab:PointingAccuracy6models}
    \hspace{-1.5cm}
    \begin{tabular}{crrrrrr}\hline
		& \multicolumn{3}{|l|}{Wilson} & \multicolumn{3}{l|}{Nakazato}\\\cline{2-7}
        & \multicolumn{1}{|c||}{No~Osc.} 
        & \multicolumn{1}{c}{NMO} 
        & \multicolumn{1}{|c|}{IMO} 
        & \multicolumn{1}{c||}{No~Osc.} 
        & \multicolumn{1}{c}{NMO} 
        & \multicolumn{1}{|c|}{IMO}\\\hline
        Pointing accuracy [$^\circ$] 
        & \multicolumn{1}{|r||}{3.0$\pm$0.1} 
        & \multicolumn{1}{r}{2.5$\pm$0.1} 
        & \multicolumn{1}{|r|}{2.8$\pm$0.1} 
        &\multicolumn{1}{r||}{4.5$\pm$0.1} 
        & \multicolumn{1}{r}{4.0$\pm$0.1}	
        & \multicolumn{1}{|r|}{4.3$\pm$0.1} \\\hline
        & & & & & & \\\hline
        & \multicolumn{3}{|l|}{Mori} 
        & \multicolumn{3}{l|}{H\"{u}depohl} \\\cline{2-7}
        & \multicolumn{1}{|c||}{No~Osc.} 
        & \multicolumn{1}{c|}{NMO} 
        & \multicolumn{1}{c|}{IMO} 
        & \multicolumn{1}{c||}{No~Osc.} 
        & \multicolumn{1}{c|}{NMO}
        & \multicolumn{1}{c|}{IMO}\\\hline
        Pointing accuracy [$^\circ$] 
        & \multicolumn{1}{|r||}{4.9$\pm$0.2}
        & \multicolumn{1}{r|}{4.6$\pm$0.1} 
        & \multicolumn{1}{r|}{4.6$\pm$0.1} 
        & \multicolumn{1}{r||}{5.4$\pm$0.2}
        & \multicolumn{1}{r|}{5.1$\pm$0.2} 
        & \multicolumn{1}{r|}{5.0$\pm$0.2}\\\hline
        & & & & & & \\\hline
		& \multicolumn{3}{|l|}{Fischer} 
        & \multicolumn{3}{l|}{Tamborra}\\\cline{2-7}
        & \multicolumn{1}{|c||}{No~Osc.}  
        & \multicolumn{1}{c}{NMO} 
        & \multicolumn{1}{|c|}{IMO}  
        & \multicolumn{1}{c||}{No~Osc.}
        & \multicolumn{1}{c|}{NMO}
        & \multicolumn{1}{c|}{IMO}\\\hline
        Pointing accuracy [$^\circ$] 
        & \multicolumn{1}{|r||}{7.2$\pm$0.2} 
        & \multicolumn{1}{r}{6.1$\pm$0.2} 
        & \multicolumn{1}{|r|}{6.9$\pm$0.2}
        & \multicolumn{1}{r||}{4.6$\pm$0.2}
        & \multicolumn{1}{r|}{5.1$\pm$0.2} 
        & \multicolumn{1}{r|}{4.7$\pm$0.2} \\\hline
    \end{tabular}
\end{table}

\begin{figure}[htb!]
\gridline{
    \fig{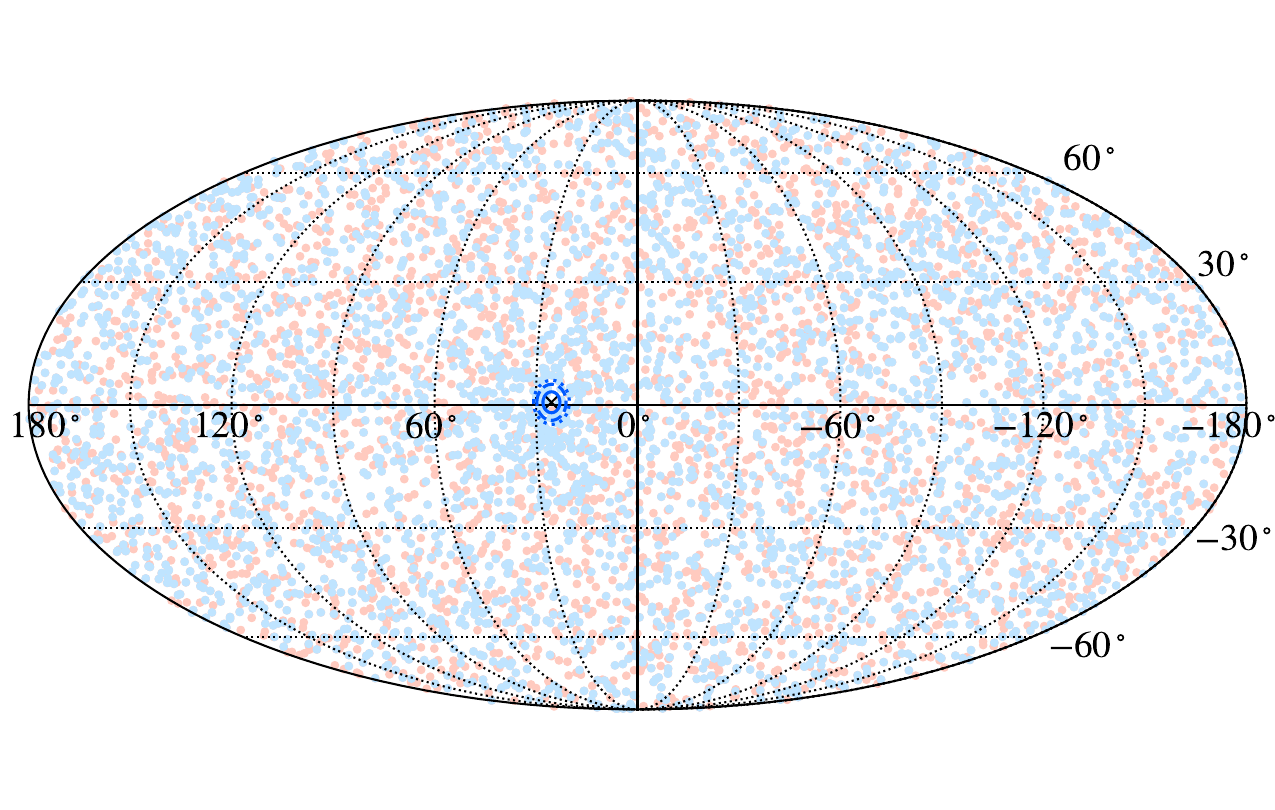}{0.3\textwidth}{(a) the Wilson model}
    \fig{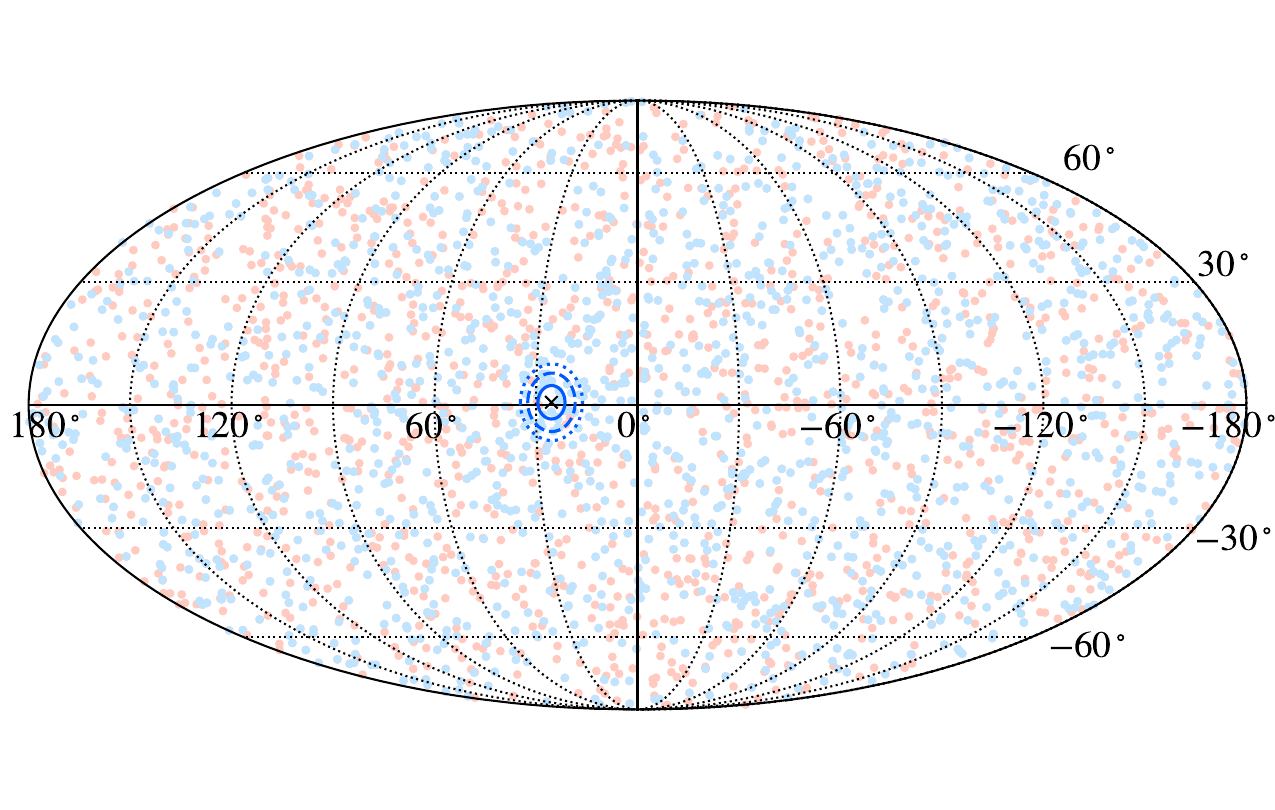}{0.3\textwidth}{(b) the Nakazato model}
    \fig{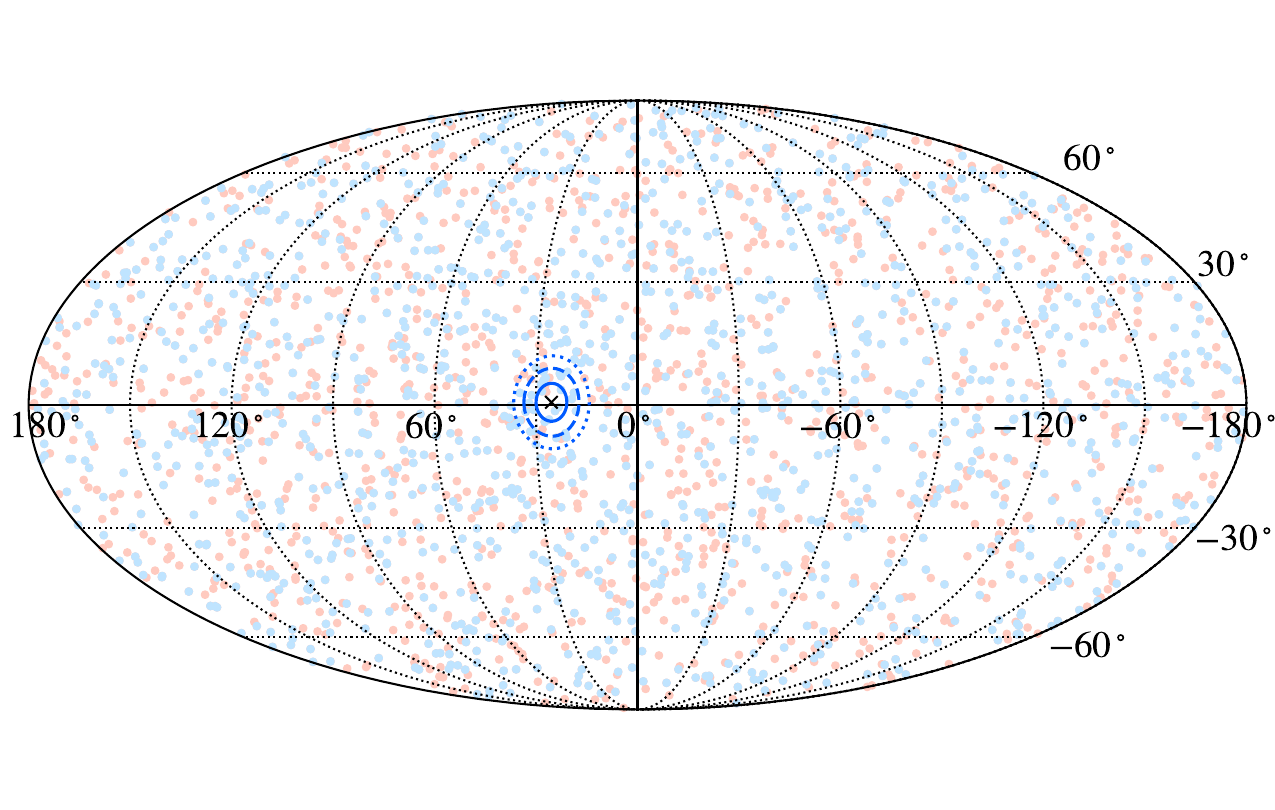}{0.3\textwidth}{(c) the Mori model}
}
\gridline{
    \fig{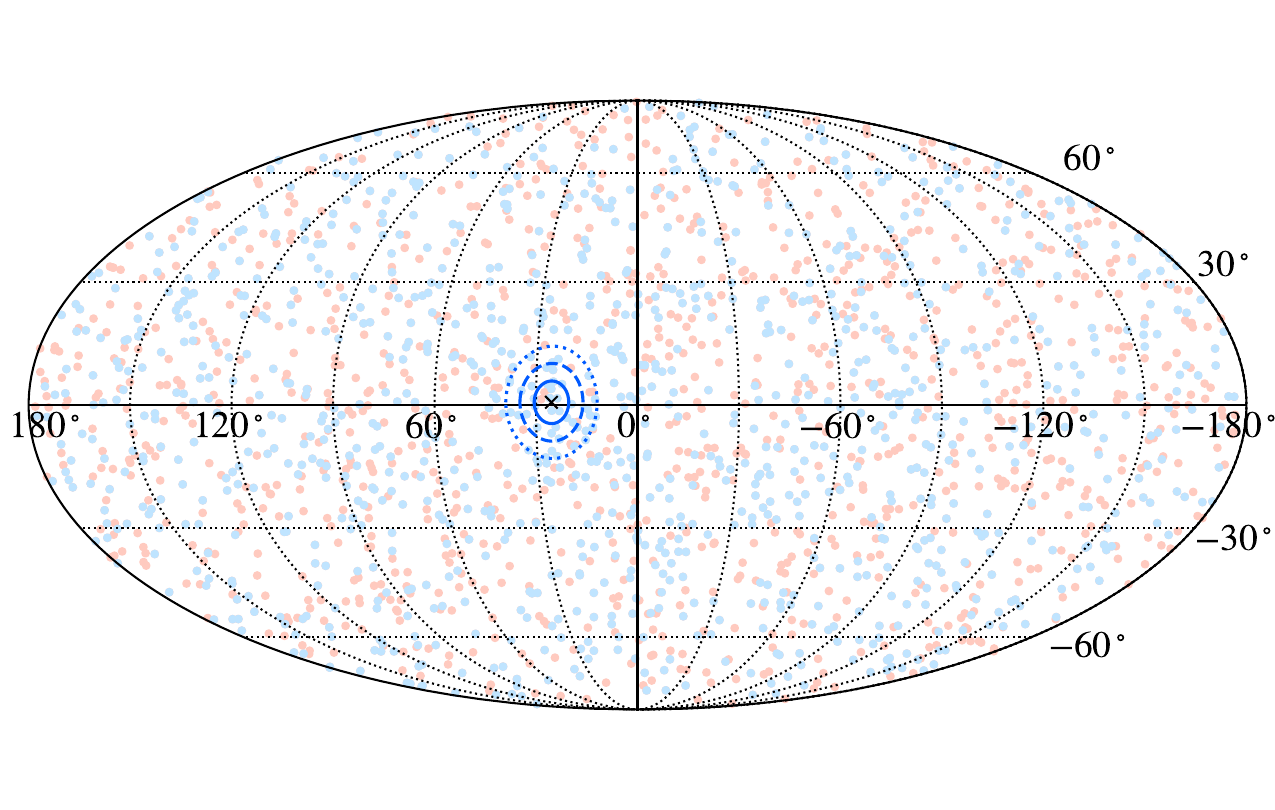}{0.3\textwidth}{(d) the H\"{u}depohl model}
    \fig{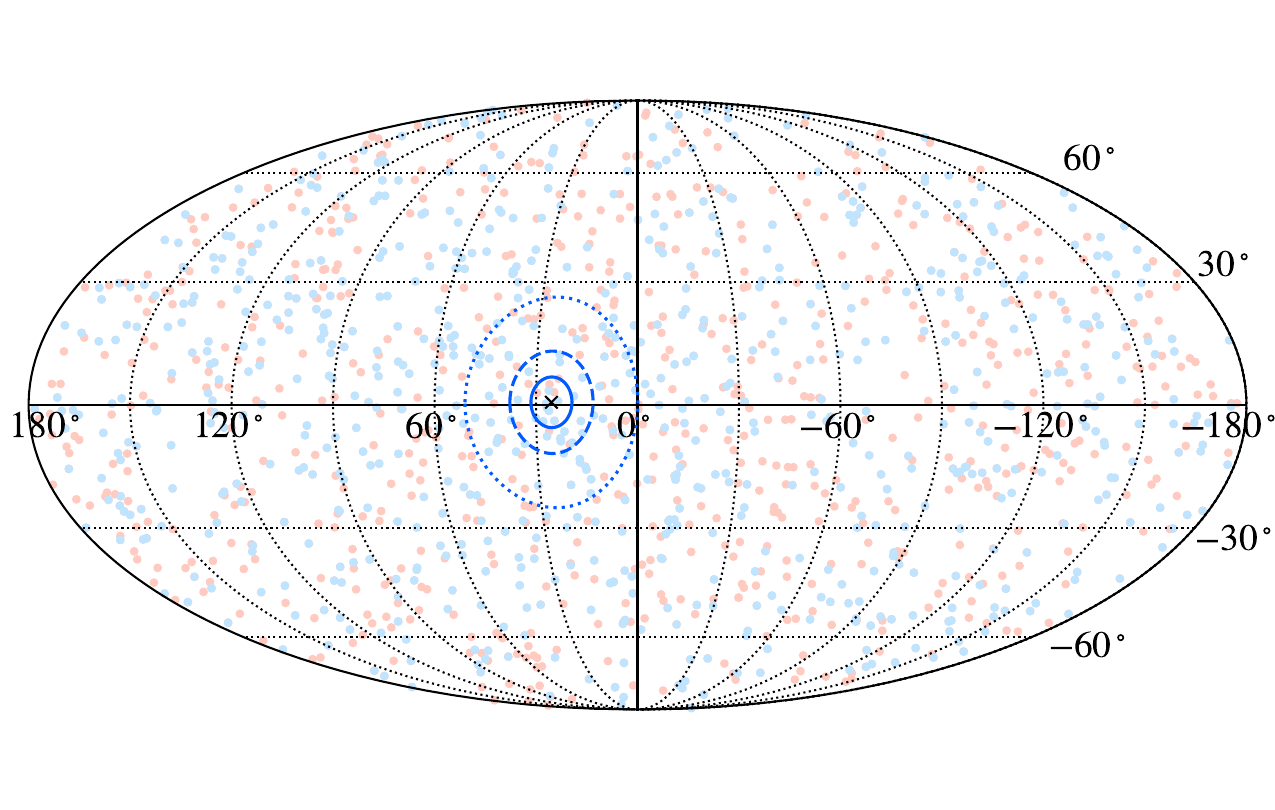}{0.3\textwidth}{(e) the Fischer model}
    \fig{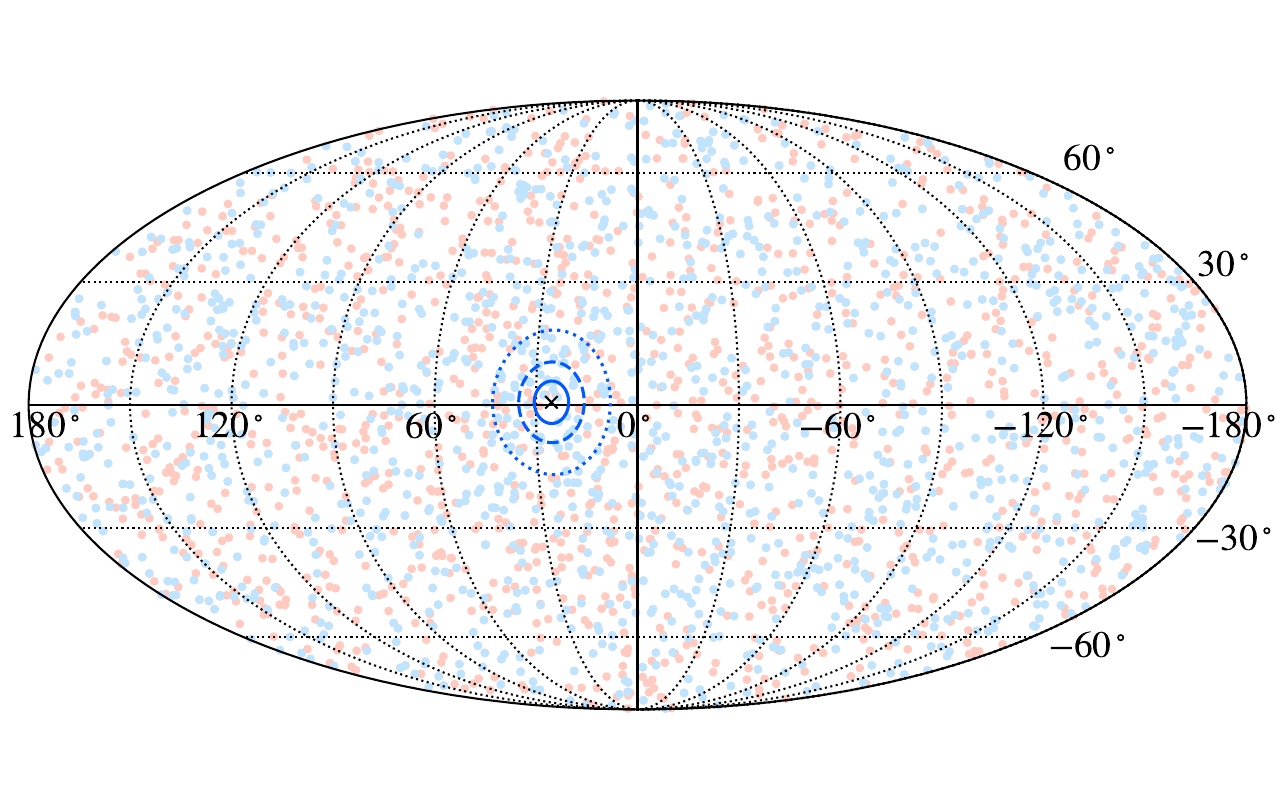}{0.3\textwidth}{(f) the Tamborra model}
}
 \caption{Distribution of the reconstructed SN position from each event using IBD tag information for one MC simulating an SN burst located at 10~kpc with neutrino oscillation in NMO. The light red points represent the reconstructed SN positions from the IBD-like events, and the light blue points stand for the reconstructed SN positions from the ES-like events.  For each skymap, pointing accuracies at 1, 2, 3$\sigma$ are shown as solid, dashed, and dotted contours in blue, respectively.  The black cross at the center of the contour circles represents the true SN position that is arbitrarily selected as 
RA~$=1^\mathrm{h}41^\mathrm{m}43^\mathrm{s}.6$, Dec.~$=0^\circ39'12~5''$).}
\label{fig:skymap1000MC6models}
\end{figure}

Figure~\ref{fig:skymap1000MC6models} shows the distribution of the reconstructed SN position from each event using IBD tag information for each model for one MC simulating an SN burst located at 10~kpc with neutrino oscillation in NMO.  
For each skymap, pointing accuracies at 1, 2, 3$\sigma$ are overlaid as contours in blue.
While the reconstructed SN positions from the IBD-like events (pink points) are uniformly distributed, the ES-like events (light blue points) are concentrated around the true SN position represented as a black cross.
The density of each skymap is in proportion to the number of events in the model, and the contour size corresponds to the pointing accuracy of the model.

\section{Conclusion and Prospects} \label{sec:Summary}
Super-Kamiokande plays a crucial role in multi-messenger observations of upcoming galactic SNe by serving as an SN neutrino detector capable of determining the SN direction.
With enhanced ability to determine the SN direction following the detector upgrade in 2020 (SK-Gd), 
we have improved SK's real-time SN monitoring system, SNWATCH~\citep{abe2016snwatch}, and launched the SN warning system, SK\_SN Notice.
To evaluate the performance of the current SNWATCH,
we have investigated SK's response to an SN burst located at a distance of 10~kpc simulated with neutrino fluxes from six SN models.

The studies described above indicate that the response of SK accurately reflects time and energy differences among the SN models that can be used to discriminate between them in the event of an SN burst.
Further, using the Gd-loaded detector we have demonstrated the capabilities of IBD tagging on this discrimination and on 
determining the direction to an SN burst. 
For example, for the simulation using the Nakazato model with a 20$M_\odot$, $Z=0.02$ progenitor, SNWATCH identifies IBD events 
with an efficiency of 46.86$\pm$0.04\% and results in a sample with 98.82$\pm$0.01\% purity after IBD tagging. 
Separating into IBD-like and ES-like samples, SNWATCH achieves a pointing accuracy ranging from 3$^\circ$ to 7$^\circ$,
depending upon the SN model.

In order to facilitate follow-up observations using optical telescopes with large fields of view, SK-Gd was designed to achieve a 3$^\circ$ precision using the Wilson model. 
This has been demonstrated successfully above and marks an improvement of about 20\% compared to our previous study.
Given that this 3$^\circ$ accuracy has not been attained in other models due to their lower neutrino fluxes, further improvements to SNWATCH are planned. 
Going forward it will be important to further quantify SK-Gd's ability to distinguish among models and to determine its ability to extract features of the progenitor from the neutrino data. 

\begin{acknowledgments}
We gratefully acknowledge the cooperation of the Kamioka Mining and Smelting Company. The Super-Kamiokande experiment has been built and operated from funding by the Japanese Ministry of Education, Culture, Sports, Science and Technology, the U.S. Department of Energy, and the U.S. National Science Foundation. Some of us have been supported by funds from the National Research Foundation of Korea (NRF-2009-0083526 and NRF 2022R1A5A1030700) funded by the Ministry of Science, ICT, the Institute for Basic Science (IBS-R016-Y2), and the Ministry of Education (2018R1D1A1B07049158, 2021R1I1A1A01042256, 2021R1I1A1A01059559), the Japan Society for the Promotion of Science, the National Natural Science Foundation of China under Grants No. 11620101004, the Spanish Ministry of Science, Universities and Innovation (grant PGC2018-099388-B-I00), the Natural Sciences and Engineering Research Council (NSERC) of Canada, the Scinet and Westgrid consortia of Compute Canada, the National Science Centre (UMO-2018/30/E/ST2/00441) and the Ministry of Education and Science (DIR/WK/2017/05), Poland, the Science and Technology Facilities Council (STFC) and GridPPP, UK, the European Union's Horizon 2020 Research and Innovation Programme under the Marie Sklodowska-Curie grant agreement no. 754496, H2020-MSCA-RISE-2018 JENNIFER2 grant agreement no.822070, H2020-MSCA-RISE-2019 SK2HK grant agreement no. 872549.
\end{acknowledgments}

%

\vspace{5mm}
\facilities{Super-Kamioaknde}


\software{ROOT~\citep{brun1997root},
    matplotlib~\citep{matplotlib_Hunter2007},
    NumPy~\citep{Numpy_vanderWalt+11},
    SciPy~\citep{Scipy_Virtanen+20},
    astropy~\citep{2013A&A...558A..33A,2018AJ....156..123A},  
    Cloudy~\citep{2013RMxAA..49..137F}, 
    Source Extractor~\citep{1996A&AS..117..393B}
          }



\appendix

\section{Data Format Unification}\label{appendix:DataFormatUnification}
SKSNSim has a class to read ``Nakazato format'' in which the differential neutrino flux $\Delta N_{k, \nu_i}(t_n)/\Delta E_k$ and differential neutrino luminosity $\Delta L_{k, \nu_i}(t_n)/\Delta E_k$ at the time $t_n$ are provided for each neutrino flavor $\nu_i$ $(i=e, \mu, \tau)$ and energy bin $E_k$ as described in Appendix of \cite{nakazato2013supernova}\footnote{The ``Nakazato format'' data of~\cite{nakazato2013supernova} and their guide are available on the Web site~\url{http://asphwww.ph.noda.tus.ac.jp/snn/}}.
We prepared the lists of time-luminosity pairs ($t, L_{\nu_i}(t)$) and time-mean-energy pairs ($t, \langle{E_{\nu_i}}\rangle(t)$) by scanning the plots of the time evolution of luminosity and mean energy in the published papers.  How this ``data format unification'' is conducted is described below.

\subsection{Linear Interpolation and Extrapolation}\label{subsec:LinearIntpAndExtrp}
Besides the Nakazato model, the Mori model is provided in ``Nakazato format'' with slightly different energy binning $E_k$.  Other non-``Nakazato format'' models (the Wilson model, the H\"{u}depohl model, the Fischer model, and the Tamborra model) need data format unification.   For these four models, we used the same energy binning as that of the Nakazato model.
For the H\"{u}depohl model, the lists of ($t, L_{\nu_i}(t)$)-pairs and ($t, \langle{E_{\nu_i}}\rangle(t)$)-pairs are already provided in the repository of SK database\footnote{Prepared for developing SNWATCH.  The data are confirmed to reproduce the published plots.}.  Regarding the Wilson model and the Fischer model, we scanned the figures of luminosity~vs.~time and mean energy~vs.~time using a web-based plot digitizing tool WebPlotDigitizer~\citep{Rohatgi2022} for each neutrino flavor to make the lists of ($t, L_{\nu_i}(t)$) and ($t, \langle{E_{\nu_i}}\rangle(t)$).  Since the time data points $t$ of these three models are not identical among neutrino flavors $\nu_i$, we performed linear interpolations and extrapolations to each list before making them into ``Nakazato format'' (described in the following Sections \ref{subsec:DiffNfluxCalc}--\ref{subsec:NormalizationToReproducePlot}).

Note that the Fischer model's $\nu_\mathrm{e}$ luminosity curve (the upper left panel of Figure~2 in \cite{fischer2010protoneutron}) starts later than that of other flavors, which requires extrapolation of $\nu_\mathrm{e}$ luminosity so that it has the same time range as the other flavor's luminosity data points.
Since the maximum luminosity and the rising edge of $\nu_\mathrm{e}$ luminosity is out of the plotted region, linear extrapolation is the only possible way to realize this.
For the Tamborra model, we took advantage of SNEWPY~\citep{baxter2022snewpy}\footnote{Available on the GitHub \url{https://github.com/SNEWS2/snewpy}} that has the lists of ($t, L_{\nu_i}(t)$) and ($t, \langle{E_{\nu_i}}\rangle(t)$) with the identical time data points among flavors.  

\subsection{Differential Neutrino Flux Calculation}\label{subsec:DiffNfluxCalc}
The goal of this ``data format unification'' is to obtain the differential neutrino flux $\Delta N_{k, \nu_i}(t_n)/\Delta E_k$~[s$^{-1}$MeV$^{-1}$] and differential neutrino luminosity $\Delta L_{k, \nu_i}(t_n)/\Delta E_k$ [erg~s$^{-1}$MeV$^{-1}$] at the time $t_n$~[s] for each neutrino flavor $\nu_i$ $(i=e, \mu, \tau)$ and energy bin $E_k$ from the available pairs of ($t, L_{\nu_i}(t)$) and ($t, \langle{E_{\nu_i}}\rangle(t)$).
The time-integrated flux of SN neutrino [MeV$^{-1}$~kpc$^{-2}$] is represented as
\begin{equation}
    \frac{\mathrm{d}F(E_\nu)}{\mathrm{d}E_\nu}=\frac{1}{4\pi d_{\mathrm{SN}}^2}\frac{E_{\nu_i, \mathrm{total}}}{\langle{E_{{\nu}_i}}\rangle}f(E_{\nu}).
    \label{eq:SNdFluxdE}
\end{equation}
where $d_{\mathrm{SN}}$~[kpc] is the distance to the SN, $E_{\nu_i, \mathrm{total}}$~[MeV] represents the total energy emitted by $\nu_i$, $\langle{E_{{\nu}_i}}\rangle$~[MeV] stands for the average energy of $\nu_i$, and $f(E_{\nu})$ is a normalized distribution function~\citep{nakazato2018charged}.
Since SKSNSim calculates the distance term $1/4\pi d_{\mathrm{SN}}^2$, it does not need to be considered here.
As $f(E_{\nu})$, we assume a Fermi-Dirac distribution 
\begin{equation}
    f_{\mathrm{FD}}(E_{\nu})=\frac{2}{3\zeta(3)T_{\nu_i}^3}\frac{{E_\nu}^2}{e^{E_\nu/T_{{\nu}_i}}+1}
    \label{eq:FDdistribution}
\end{equation}
where $\zeta(3)\approx1.202$ is the zeta function and 
\begin{equation}
    T_{\nu_i}=\frac{180}{7\pi^4}\zeta(3)\langle{E_{{\nu}_i}}\rangle\approx\frac{\langle{E_{{\nu}_i}}\rangle}{3.151}
    \label{eq:NeutrinoTemperature}
\end{equation}
is the neutrino temperature in the units of MeV.
By replacing the total energy $E_{\nu_i, \mathrm{total}}$~[MeV] with the luminosity $L_{\nu_i}(t)$~[erg~s$^{-1}$] and substituting $\langle{E_{{\nu}_i}}\rangle=\langle{E_{\nu_i}}\rangle(t)$ in equations~(\ref{eq:SNdFluxdE})--(\ref{eq:NeutrinoTemperature}), the expression of the differential neutrino flux~[s$^{-1}$MeV$^{-1}$] is obtained as
\begin{equation}
    \frac{\mathrm{d}^2N_\nu}{\mathrm{d}E_\nu\mathrm{d}t}=\frac{L_{{\nu}_i}(t)}{\langle{E_{{\nu}_i}}\rangle(t)}\frac{2}{3\zeta(3)T_{\nu_i}^3}\frac{{E_\nu}^2}{e^{E_\nu/T_{{\nu}_i}}+1}
    \label{eq:diffNfluxFromScan}
\end{equation}
where $T_{\nu_i}\approx\langle{E_{{\nu}_i}}\rangle(t)/3.151$~[MeV]. 

The differential neutrino luminosity~[ergs$^{-1}$MeV$^{-1}$] can be computed from the differential neutrino flux~[s$^{-1}$MeV$^{-1}$] as
\begin{equation}
    \frac{\Delta L_{k, \nu_i}(t_n)}{\Delta E_k} = \frac{\Delta N_{k, \nu_i}(t_n)}{\Delta E_k} \times E_k \times \frac{1.6022\times 10^{-6}\mathrm{erg}}{\mathrm{MeV}}
    \label{eq:erg2mevConversion}
\end{equation}

\subsection{Normalization for Reproducing Original Luminosity}\label{subsec:NormalizationToReproducePlot}
In order to reproduce the original luminosity~vs.~time plot and mean energy~vs.~time plot from the ``Nakazato format'' table, a normalization factor is necessary.  Such normalization factor is written as
\begin{equation}
    \mathrm{Normalizing~Factor~at}~t = \frac{\mathrm{original~luminosity~at}~t}{\mathrm{reproduced~luminosity~at}~t},
    \label{eq:NormalizingFactor}
\end{equation}
where the original luminosity is $L_{\nu_i}(t)$ provided in the list and the reconstructed luminosity $L_{\nu_i, \mathrm{reco}}(t)$ is obtained by the energy-integration of the differential neutrino luminosity as
\begin{equation}
    L_{\nu_i, \mathrm{reco}}(t) = \sum_k \frac{\Delta L_{k, \nu_i}(t)}{\Delta E_k} \times (E_{k+1}-E_k).
    \label{eq:diffLumiEnergyInteg}
\end{equation}
By multiplying equation~(\ref{eq:NormalizingFactor}) by the results of equations~(\ref{eq:diffNfluxFromScan})--(\ref{eq:erg2mevConversion}) and arranging them, the normalized ``Nakazato format'' tables of other models are obtained.

\section{Comparison of SK-Gd's Response among Models for Supernova at 10 kpc}\label{app:differenceAmongModels}
\begin{figure}[htb!]
\gridline{
    \fig{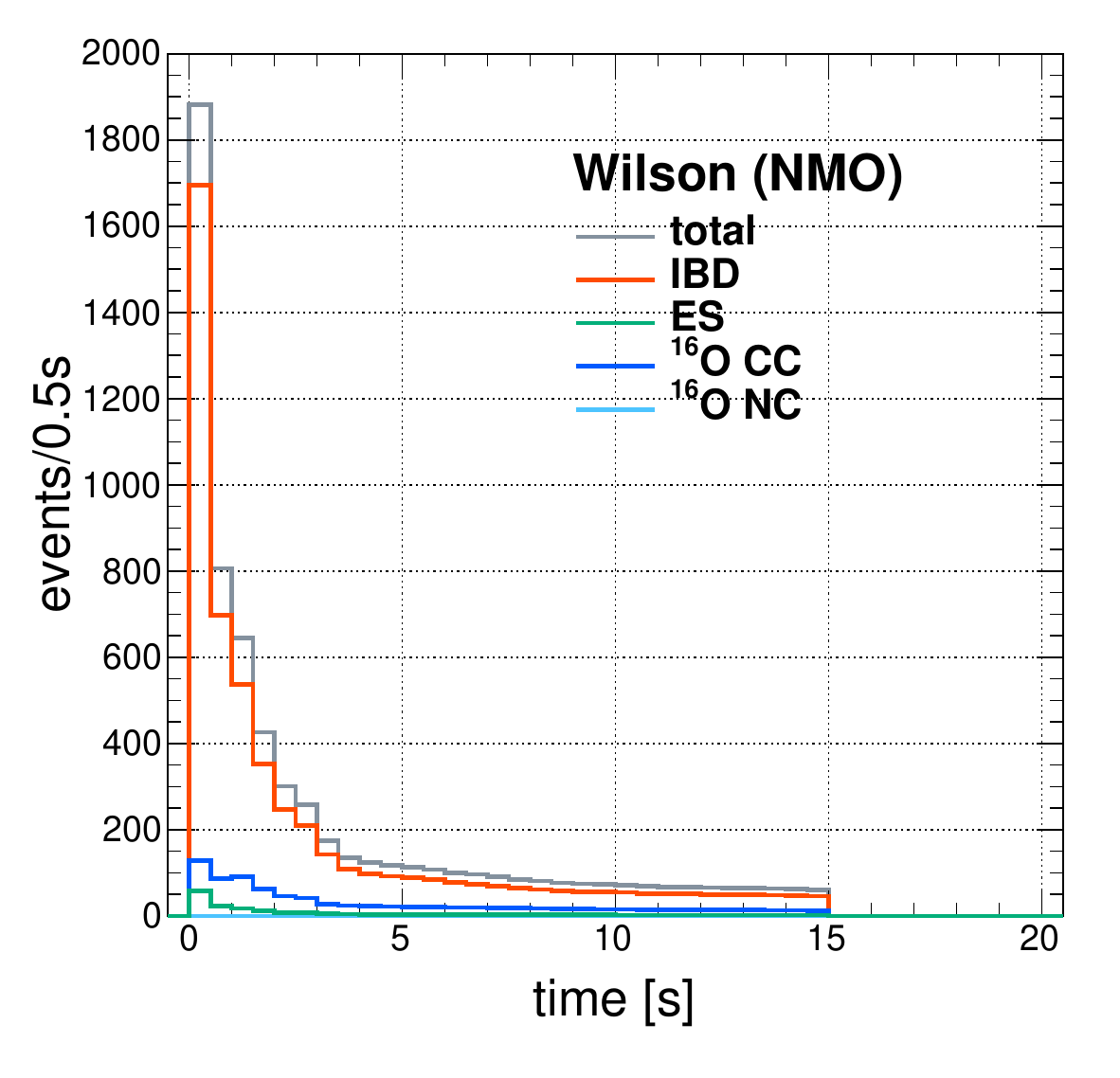}{0.33\textwidth}{(a) the Wilson model}
    \fig{Modification_23Dec_WholeTime_mtimePrompt_10kpc_NMO_recoTime_reactions_Nakazato.pdf}{0.33\textwidth}{(b) the Nakazato model}
    \fig{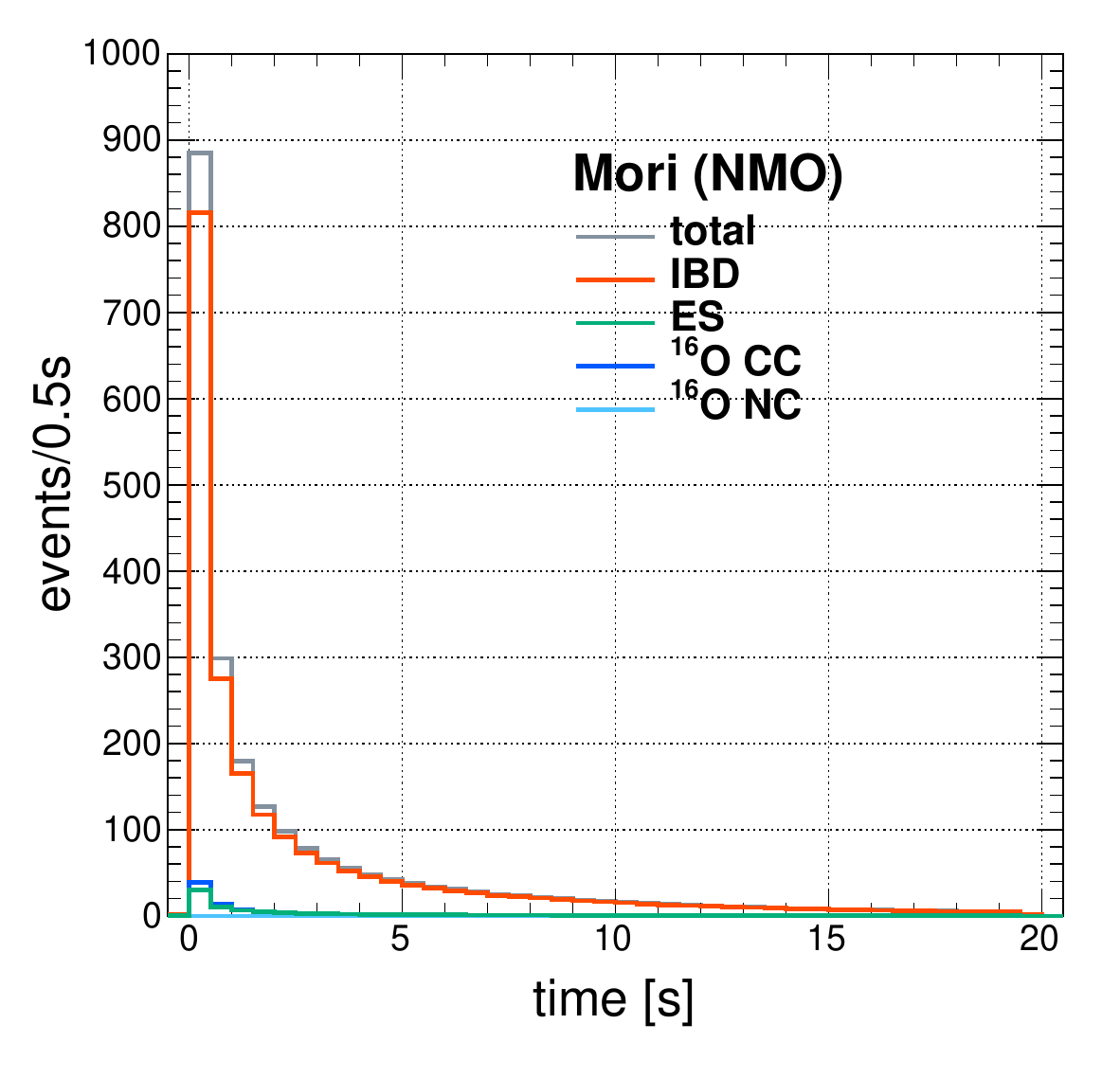}{0.33\textwidth}{(c) the Mori model}
}
\vspace{-1.2cm}
\gridline{
    \fig{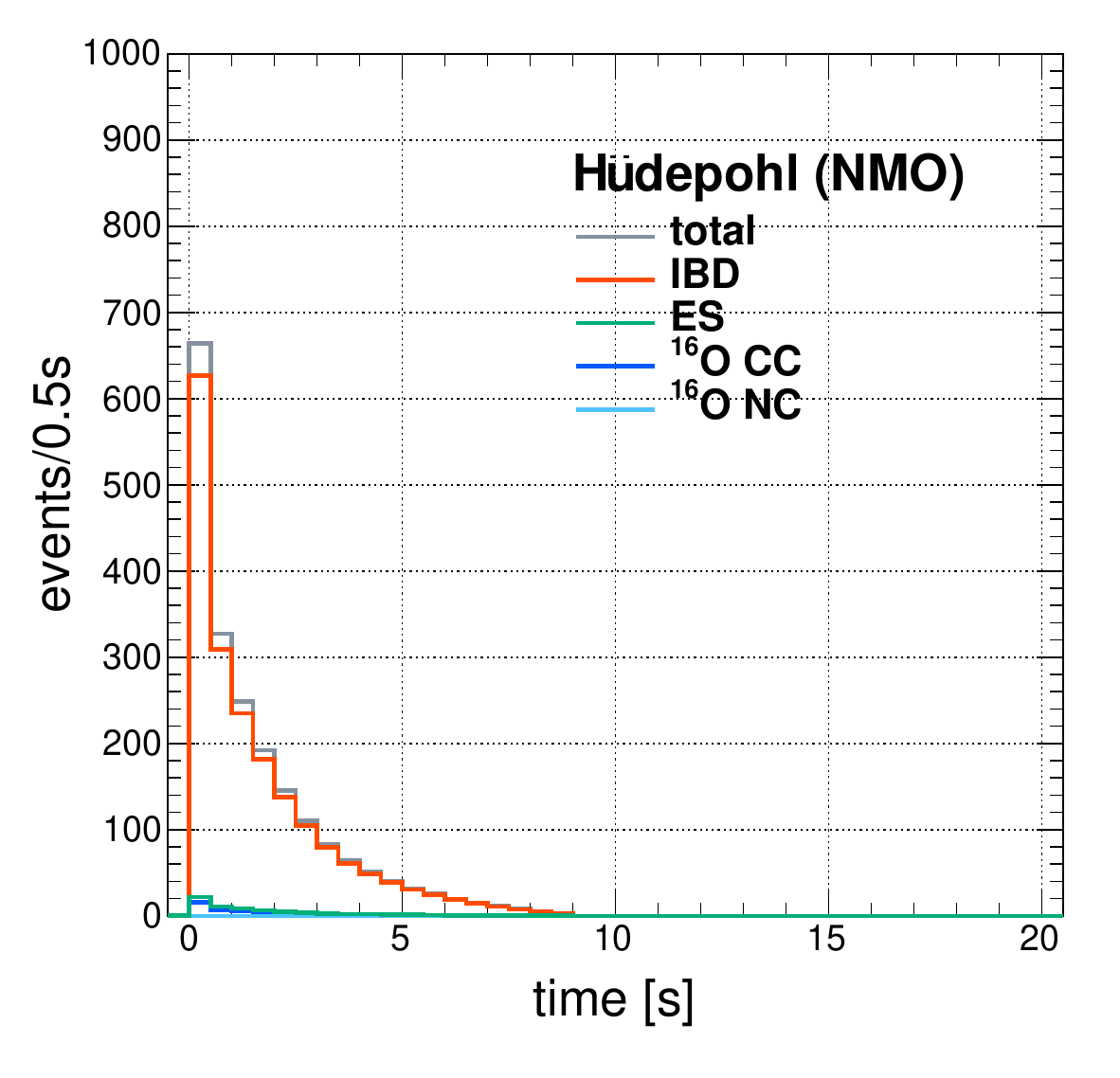}{0.33\textwidth}{(d) the H\"{u}depohl model}
    \fig{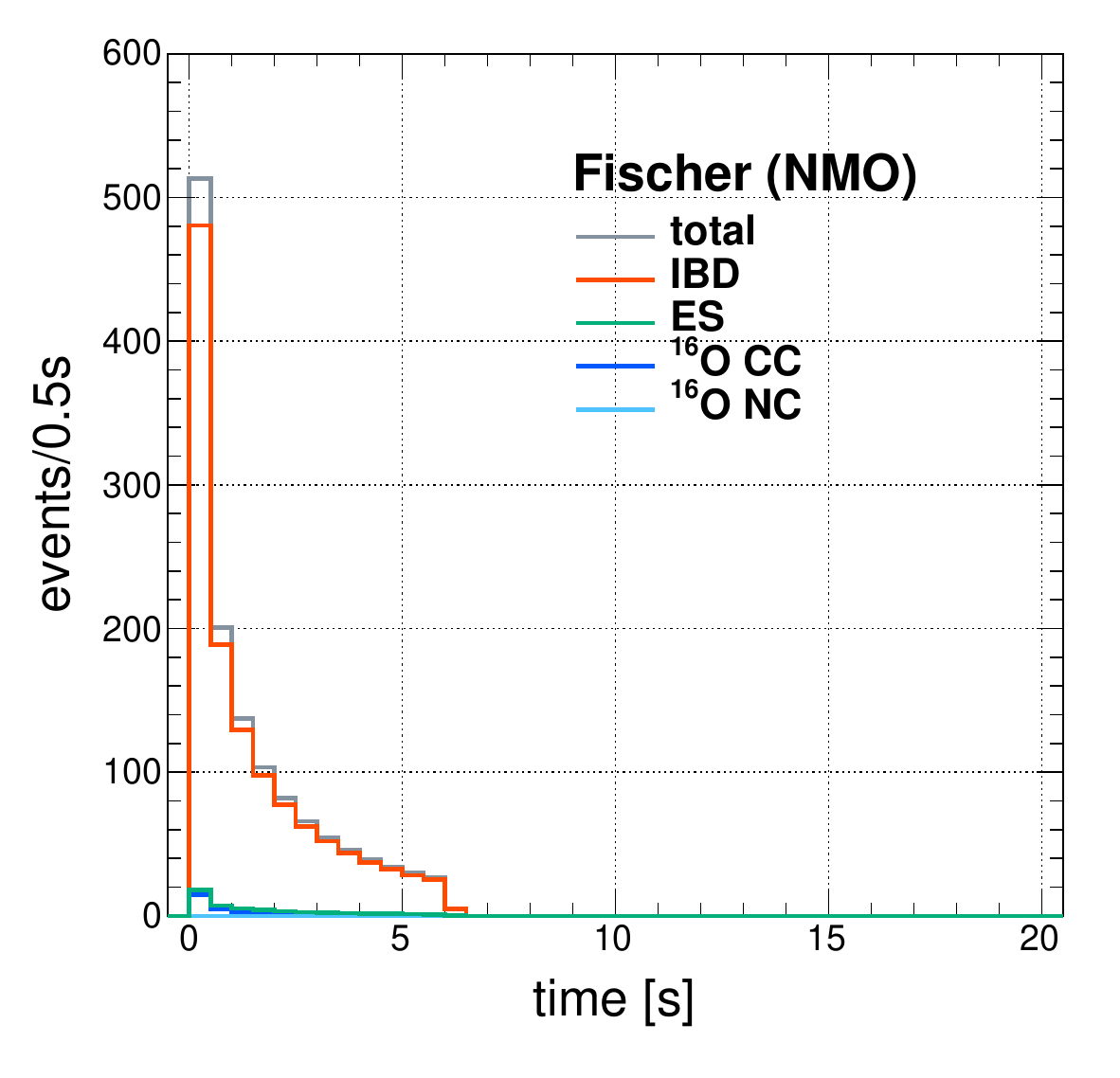}{0.33\textwidth}{(e) the Fischer model}
    \fig{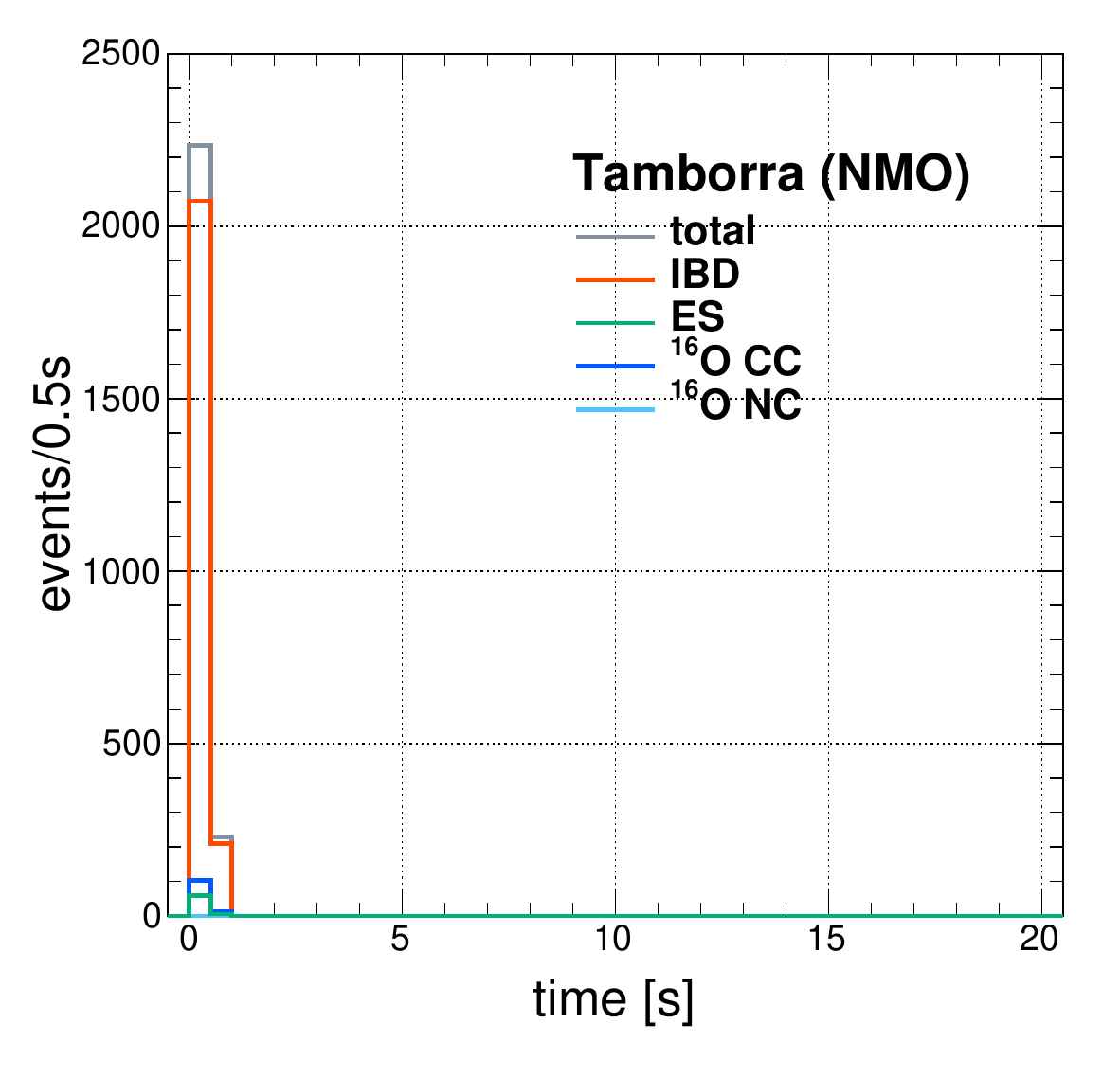}{0.33\textwidth}{(f) the Tamborra model}
    }
\vspace{-1.2cm}
\gridline{
    \fig{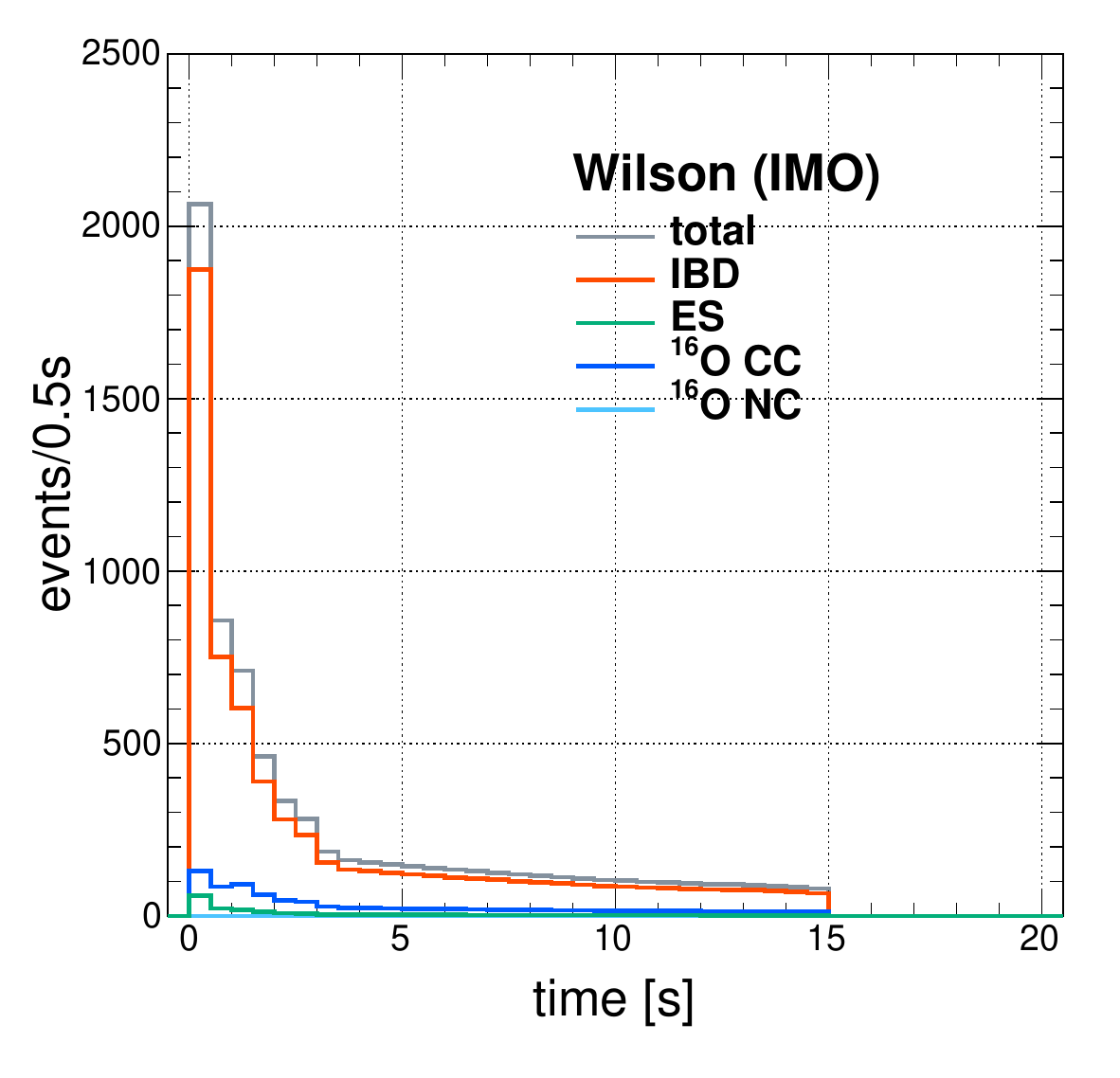}{0.33\textwidth}{}
    \fig{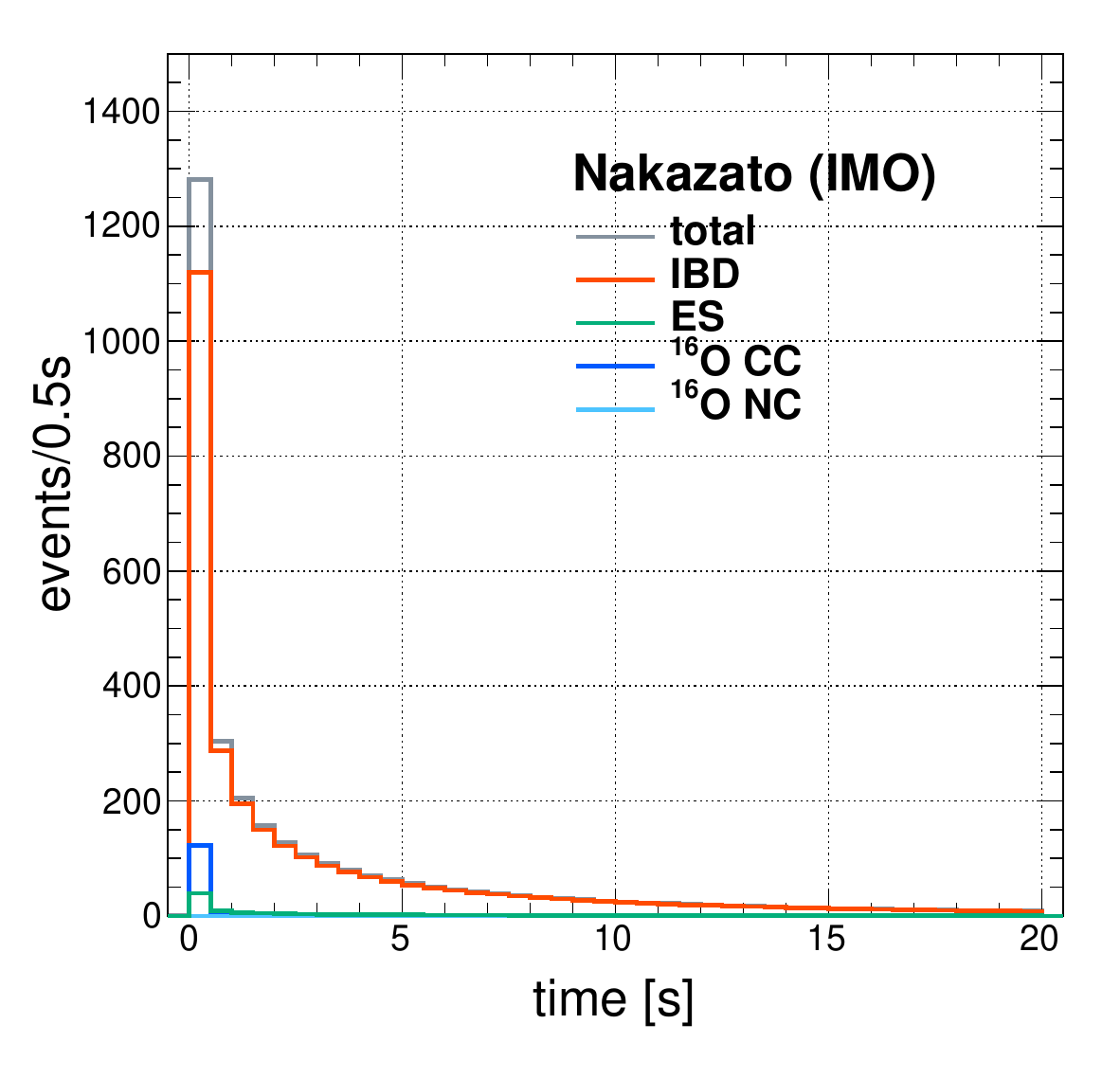}{0.33\textwidth}{}
    \fig{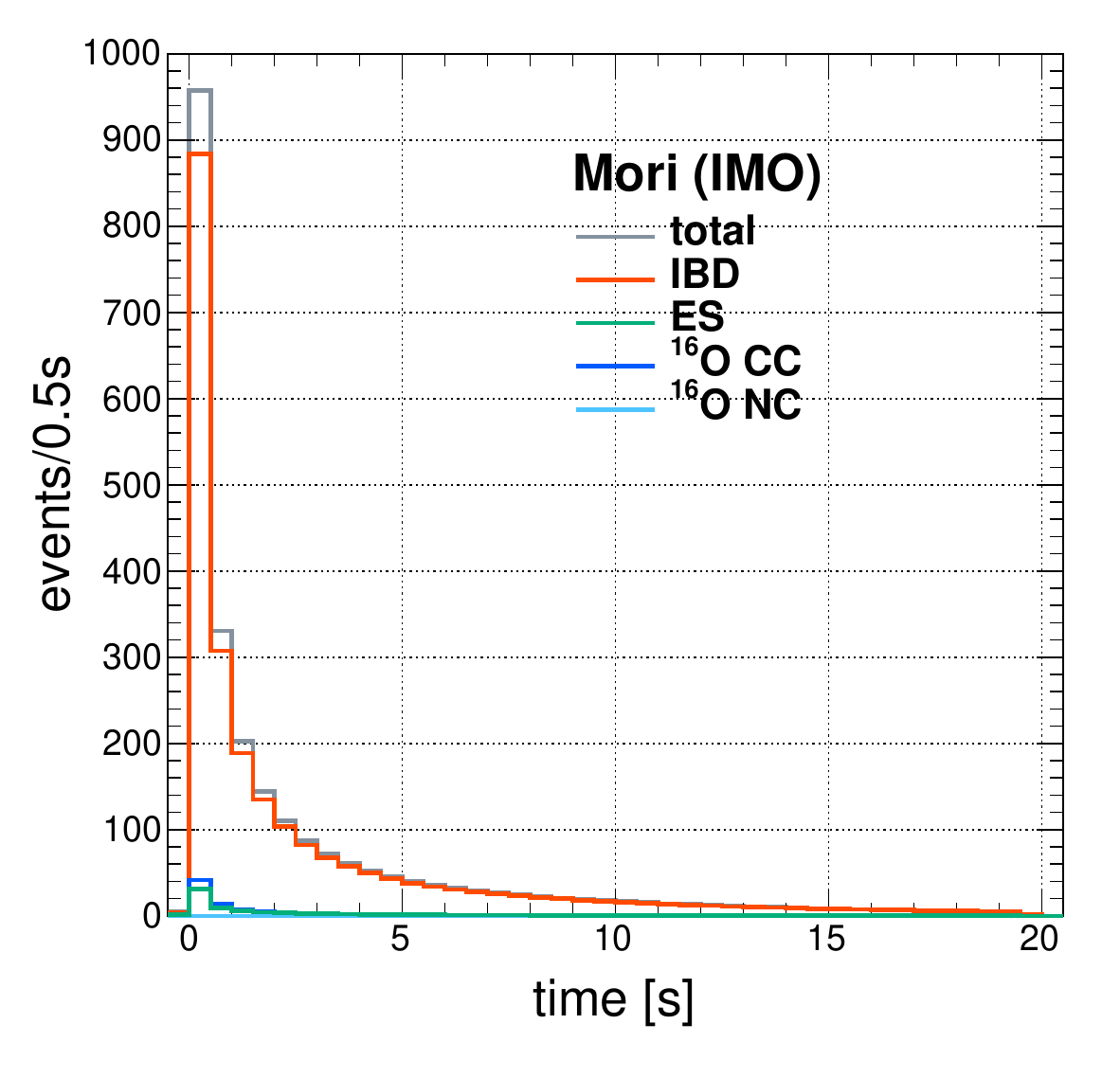}{0.33\textwidth}{}
}
\vspace{-1.2cm}
\gridline{
    \fig{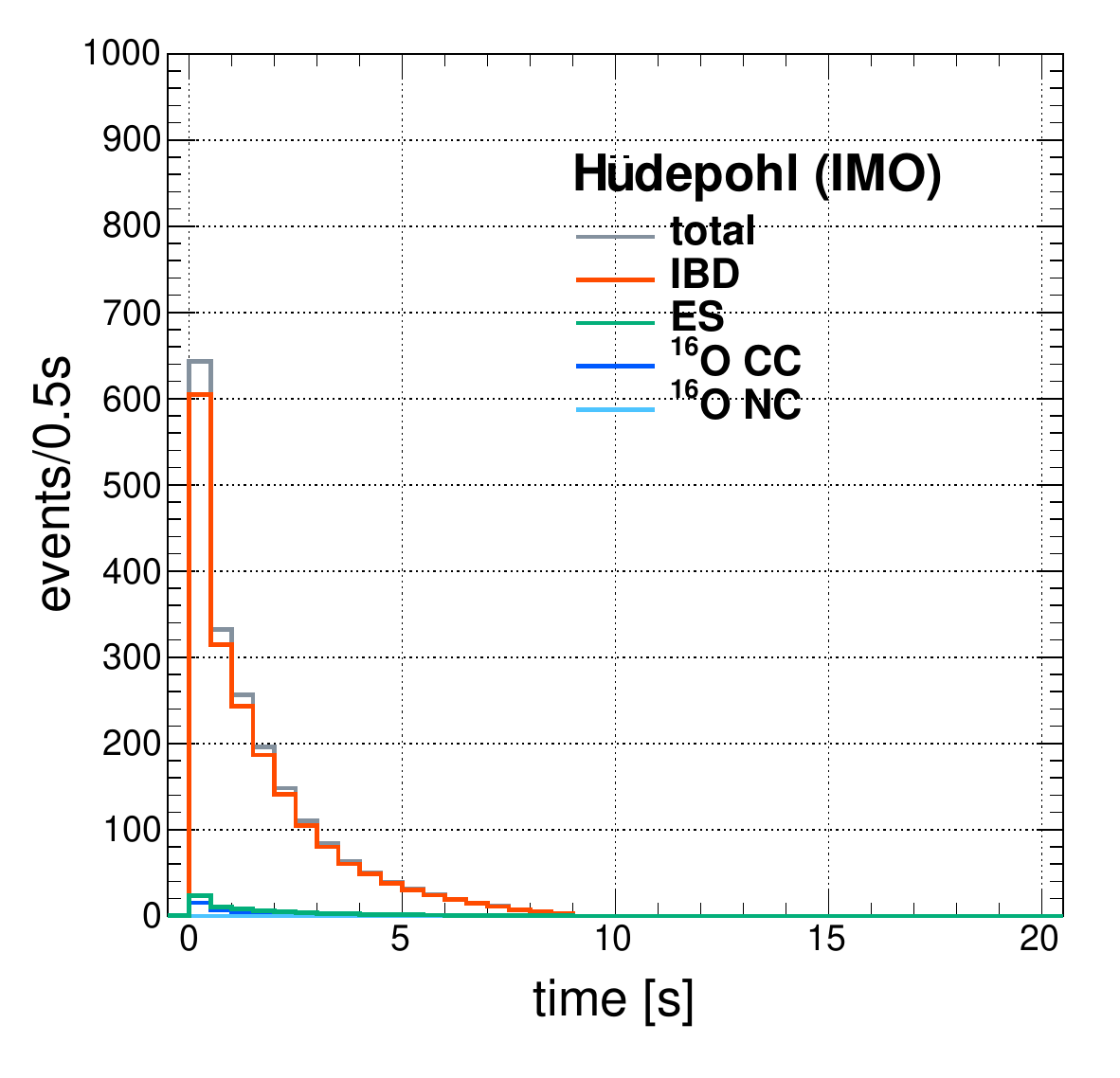}{0.33\textwidth}{}
    \fig{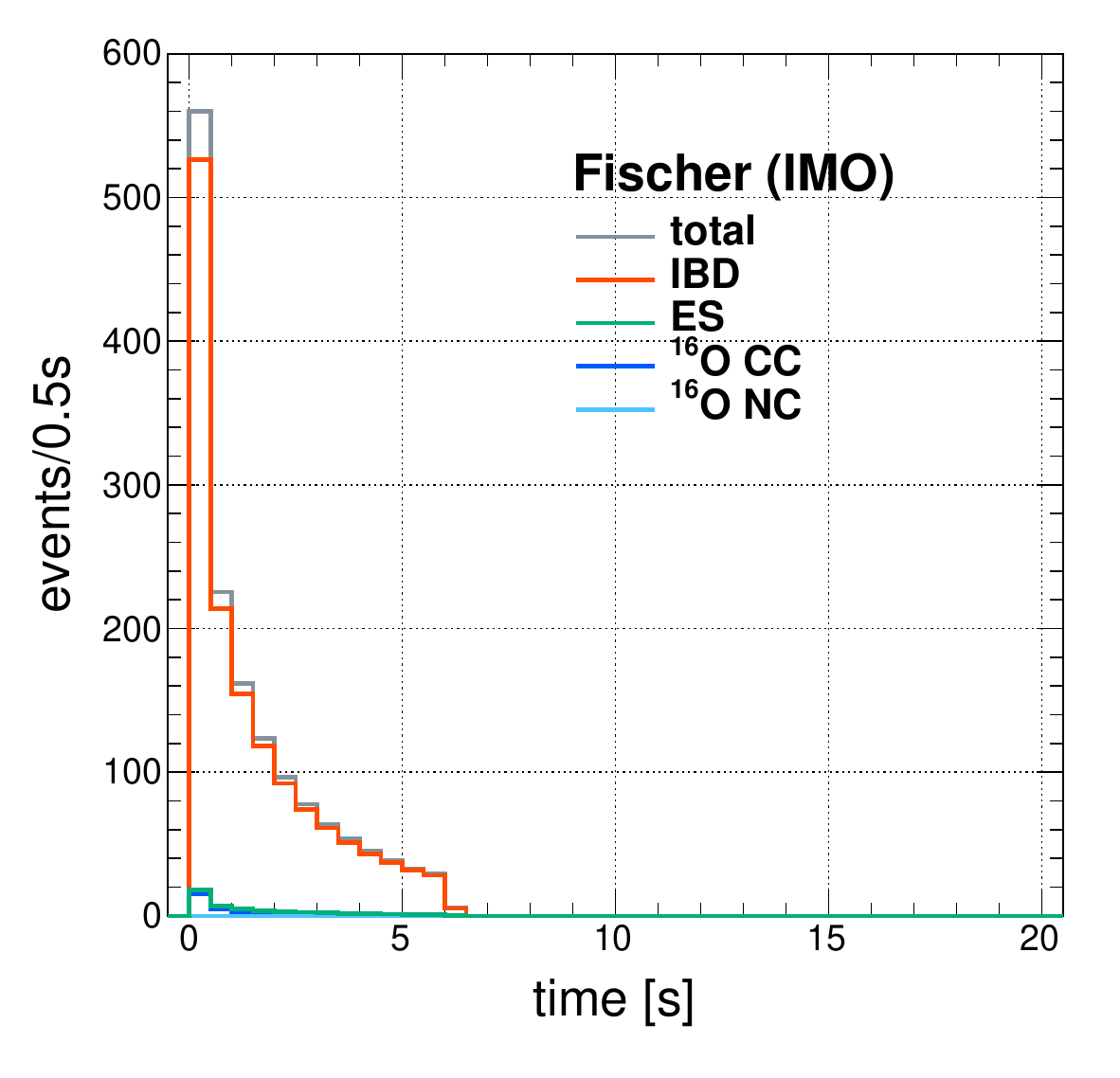}{0.33\textwidth}{}
    \fig{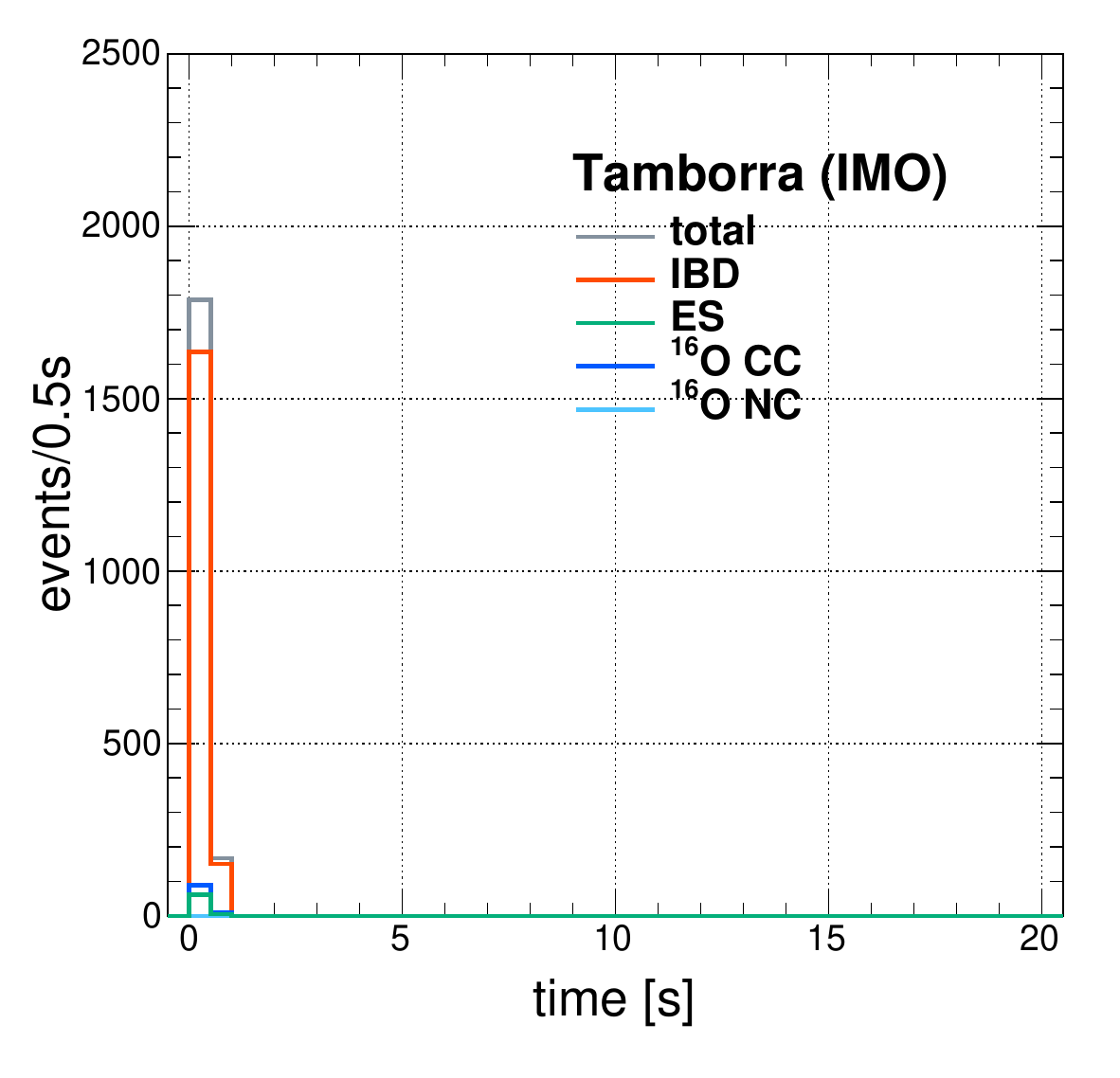}{0.33\textwidth}{}
    }
\vspace{-0.5cm}
\caption{Comparison of time evolution up to 20~s among interactions for each model for an SN burst located at 10~kpc in the NMO scenario (top six panels) and the IMO scenario (bottom six panels).}
\label{fig:NMOandIMOTimeWholeInteractionsEachModel}
\end{figure}

\begin{figure}[htb!]
\gridline{
    \fig{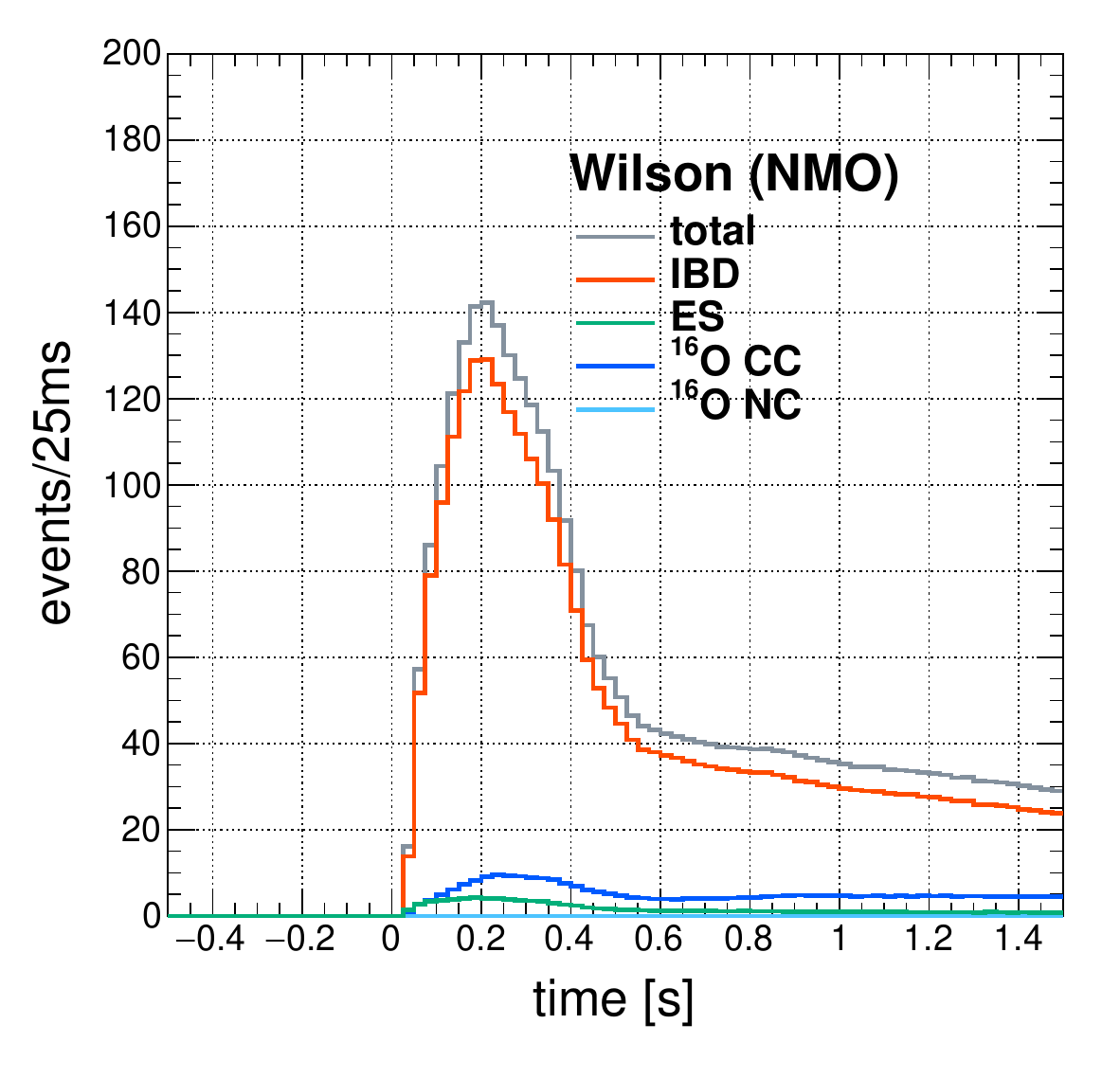}{0.33\textwidth}{(a) the Wilson model}
    \fig{Modification_23Dec_First1s_mtimePrompt_10kpc_NMO_recoTime1s_reactions_Nakazato.pdf}{0.33\textwidth}{(b) the Nakazato model}
    \fig{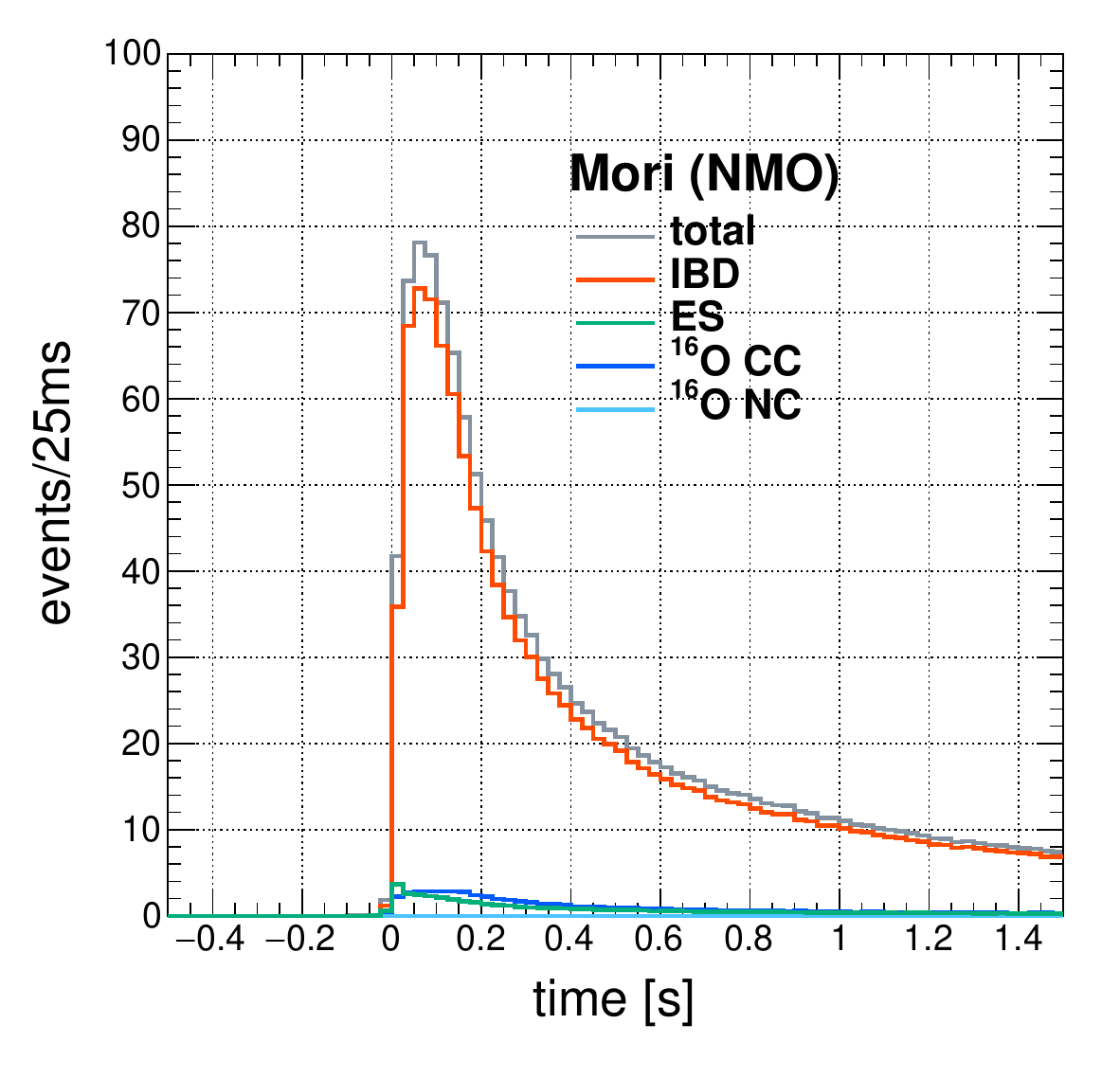}{0.33\textwidth}{(c) the Mori model}
}
\vspace{-1.2cm}
\gridline{
    \fig{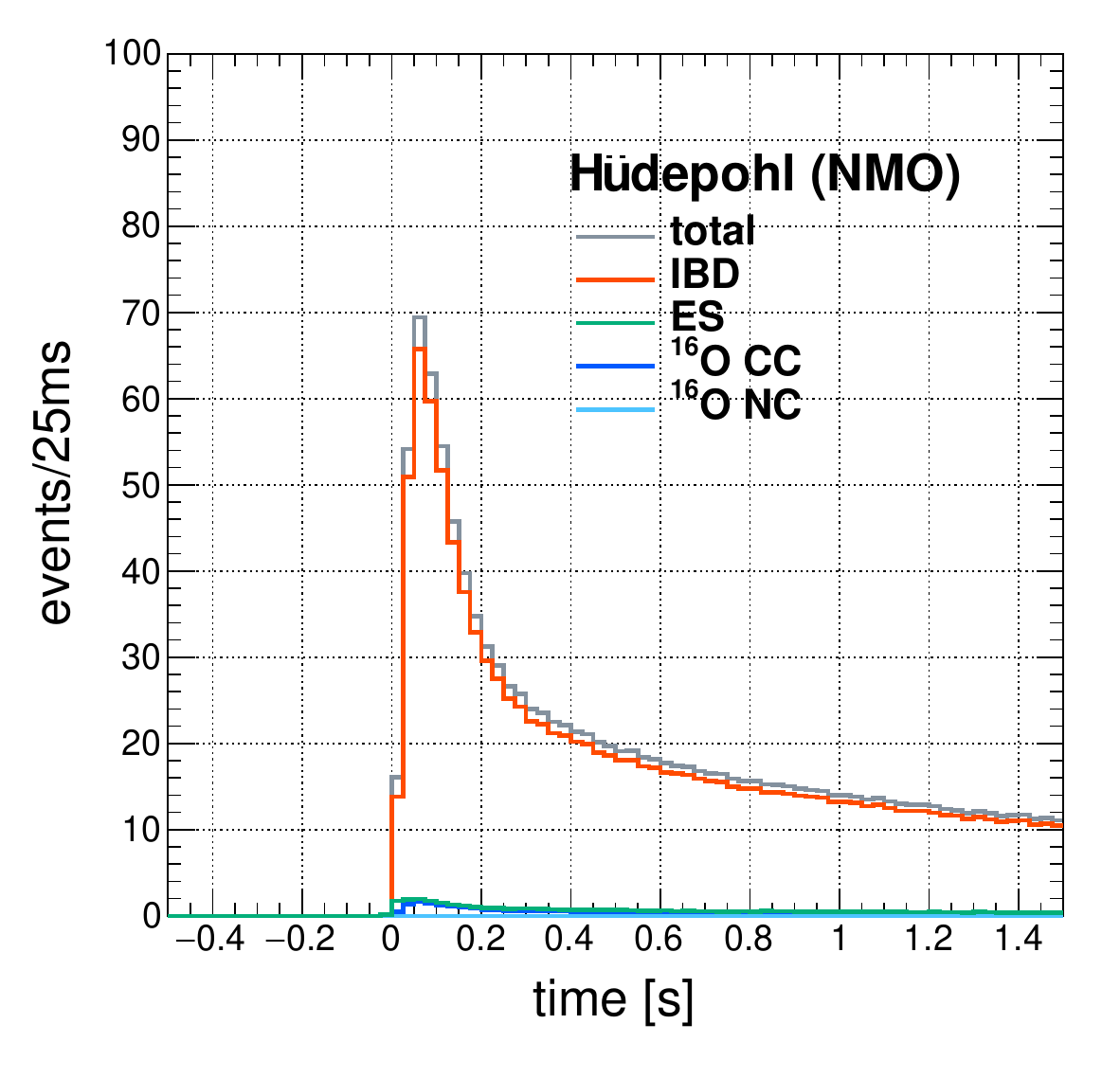}{0.33\textwidth}{(d) the H\"{u}depohl model}
    \fig{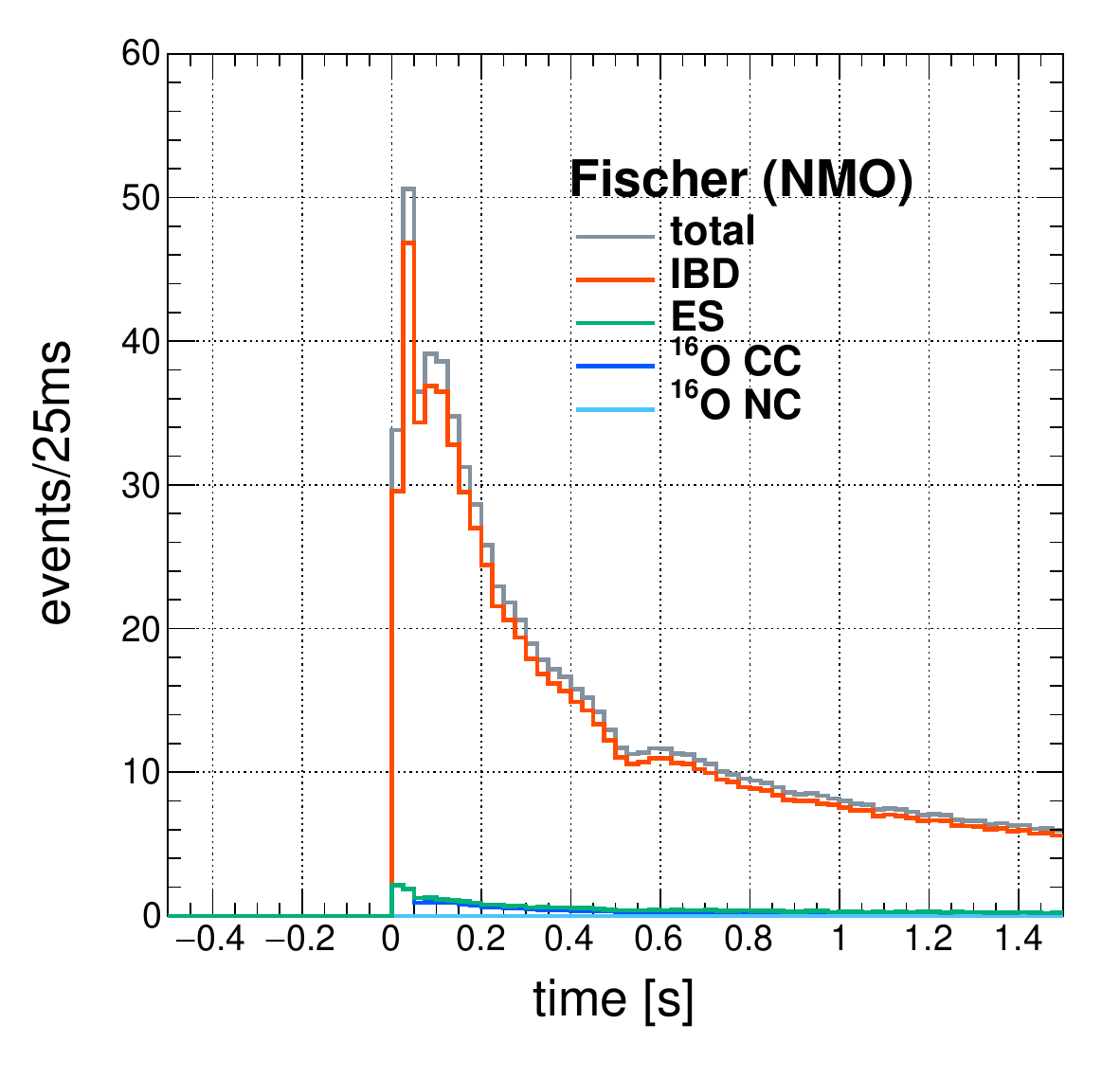}{0.33\textwidth}{(e) the Fischer model}
    \fig{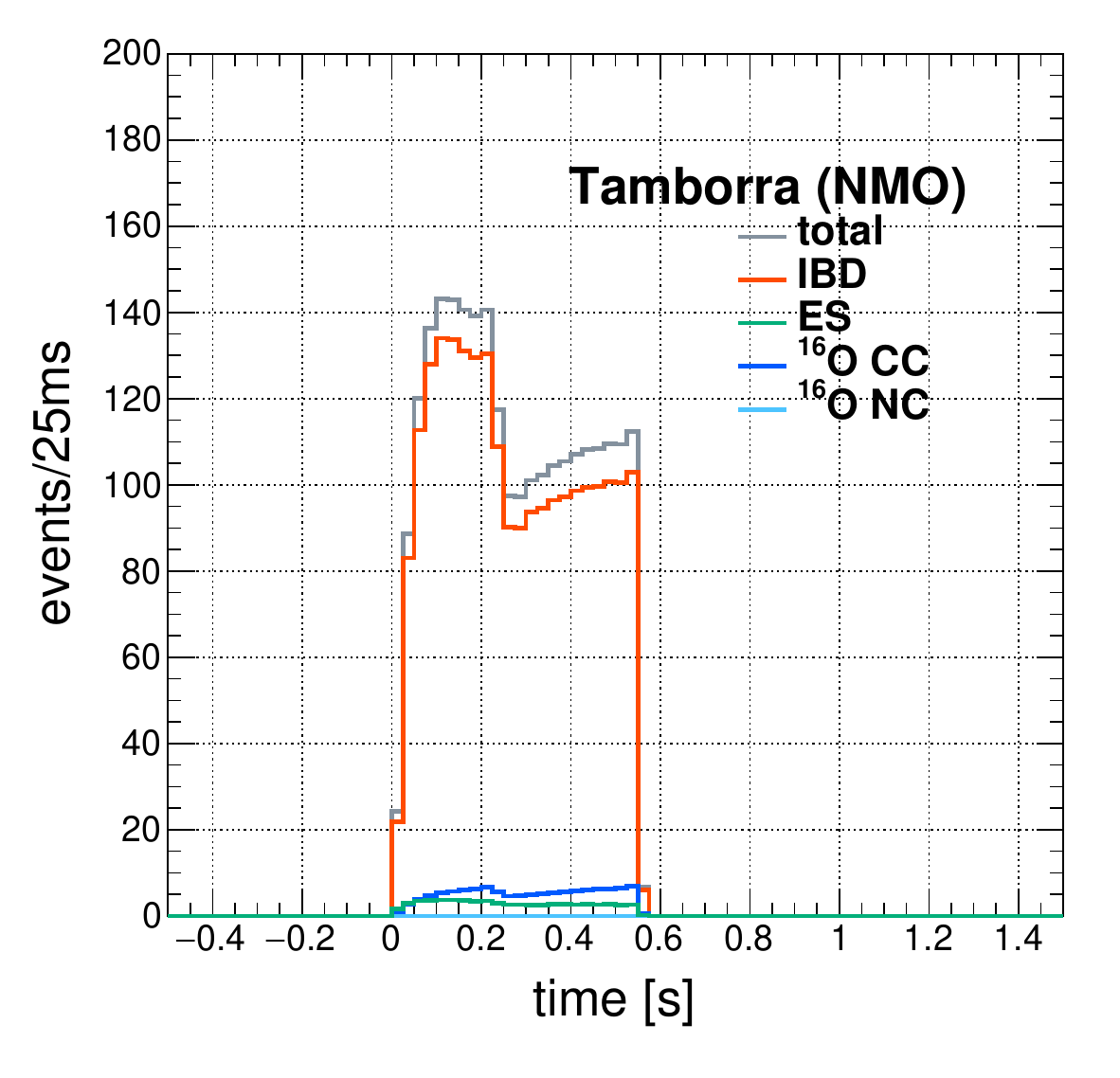}{0.33\textwidth}{(f) the Tamborra model}
}
\vspace{-1.2cm}
\gridline{
    \fig{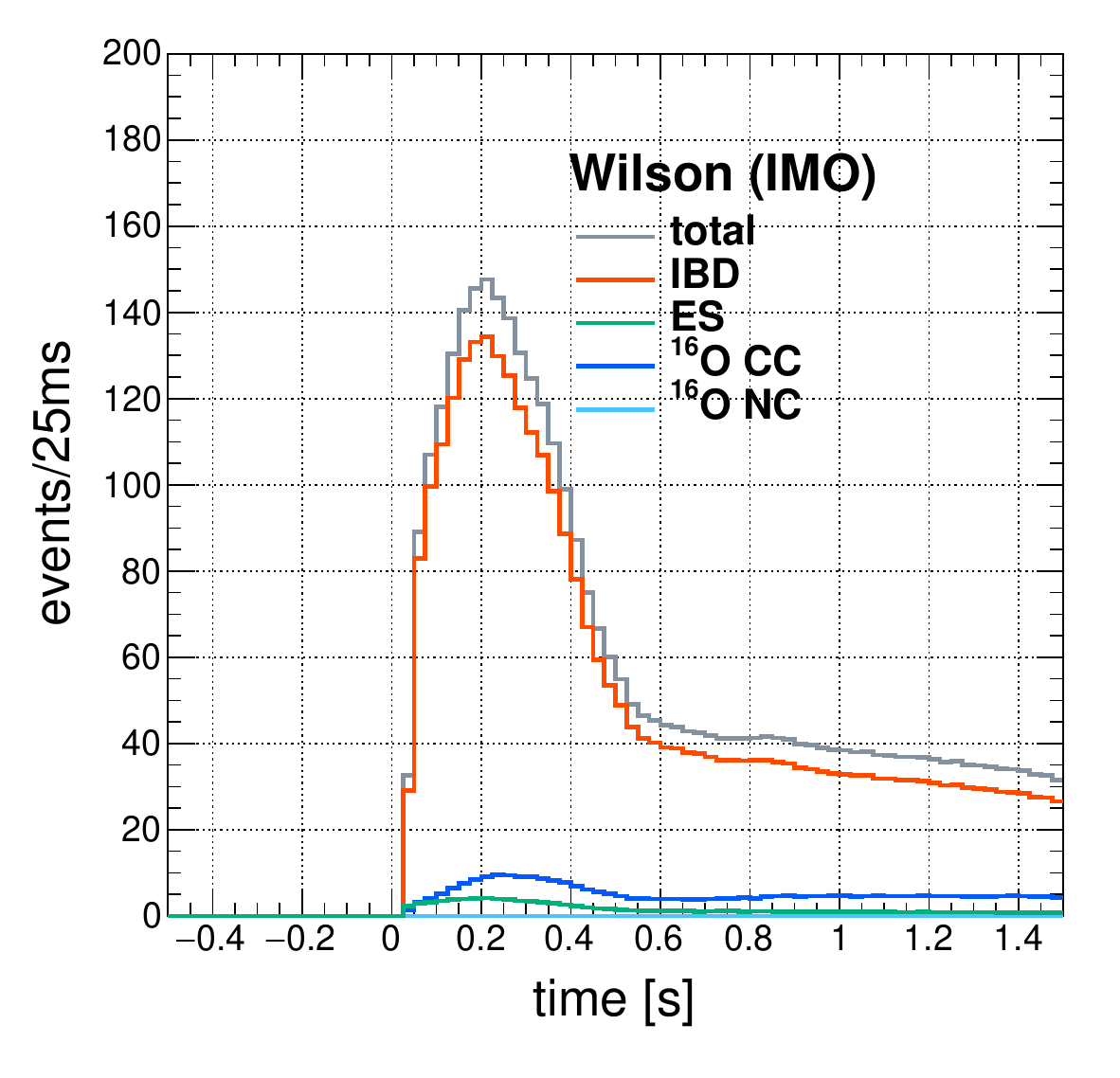}{0.33\textwidth}{(a) the Wilson model}
    \fig{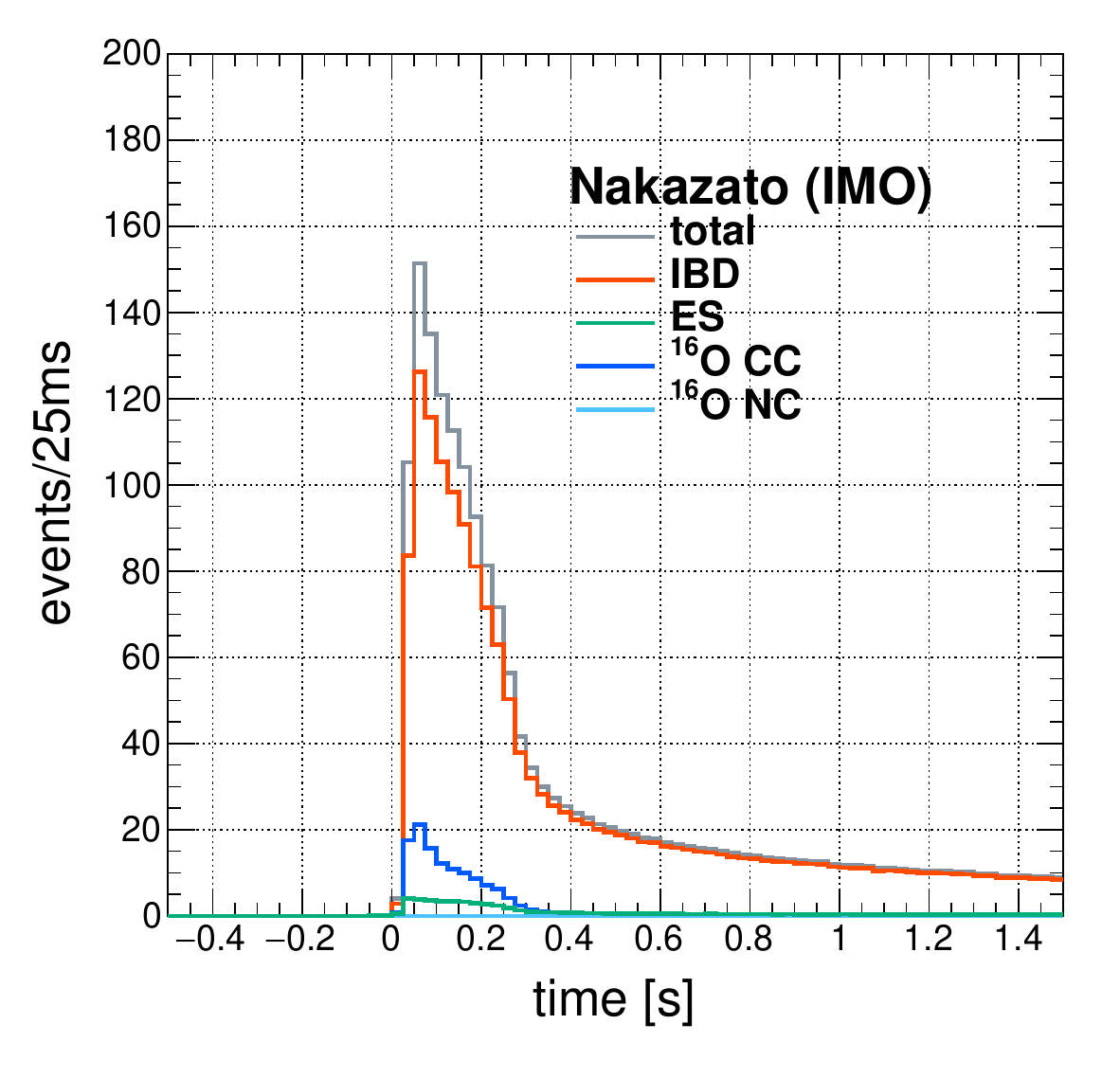}{0.33\textwidth}{(b) the Nakazato model}
    \fig{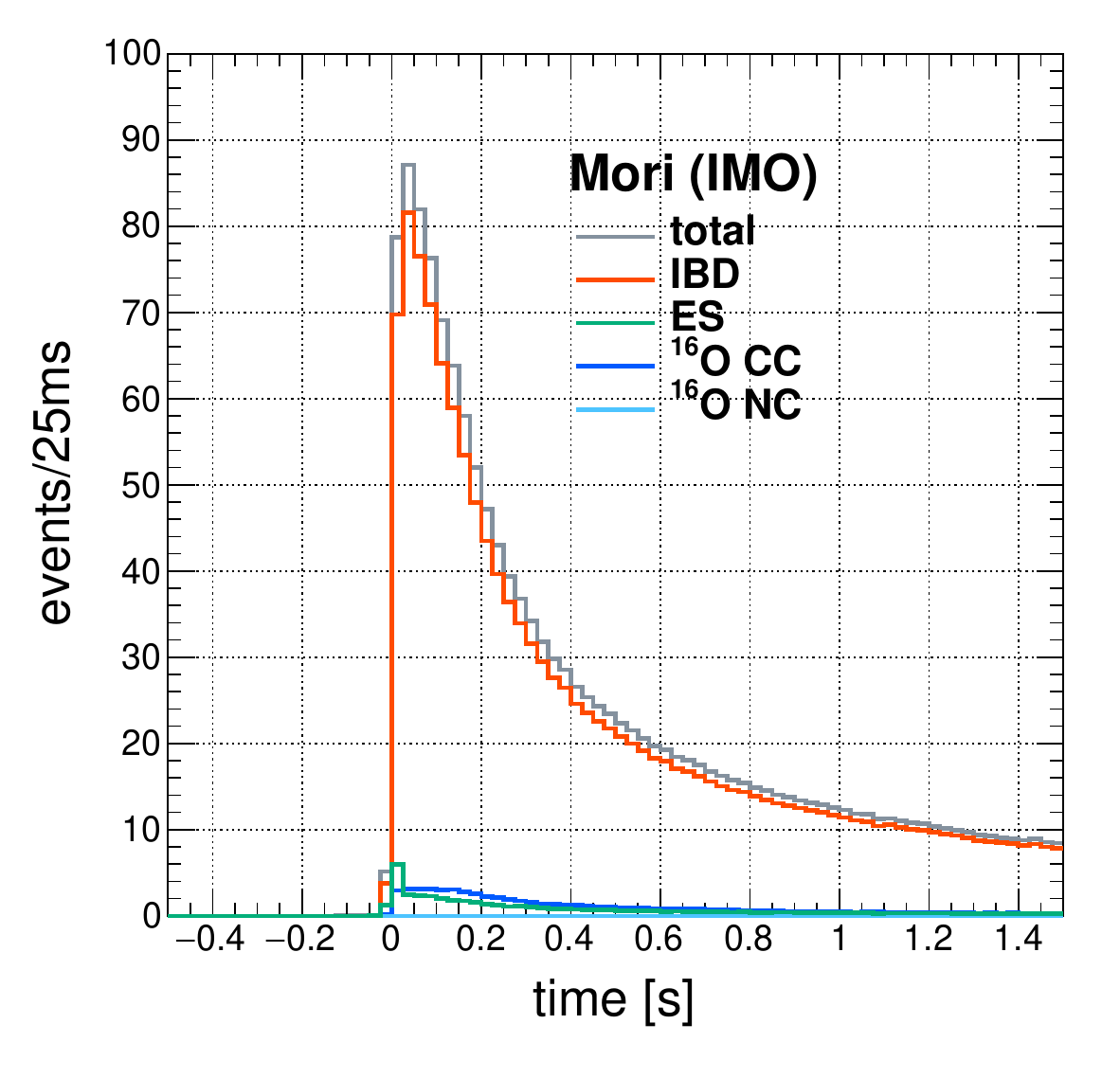}{0.33\textwidth}{(c) the Mori model}
}
\vspace{-1.2cm}
\gridline{
    \fig{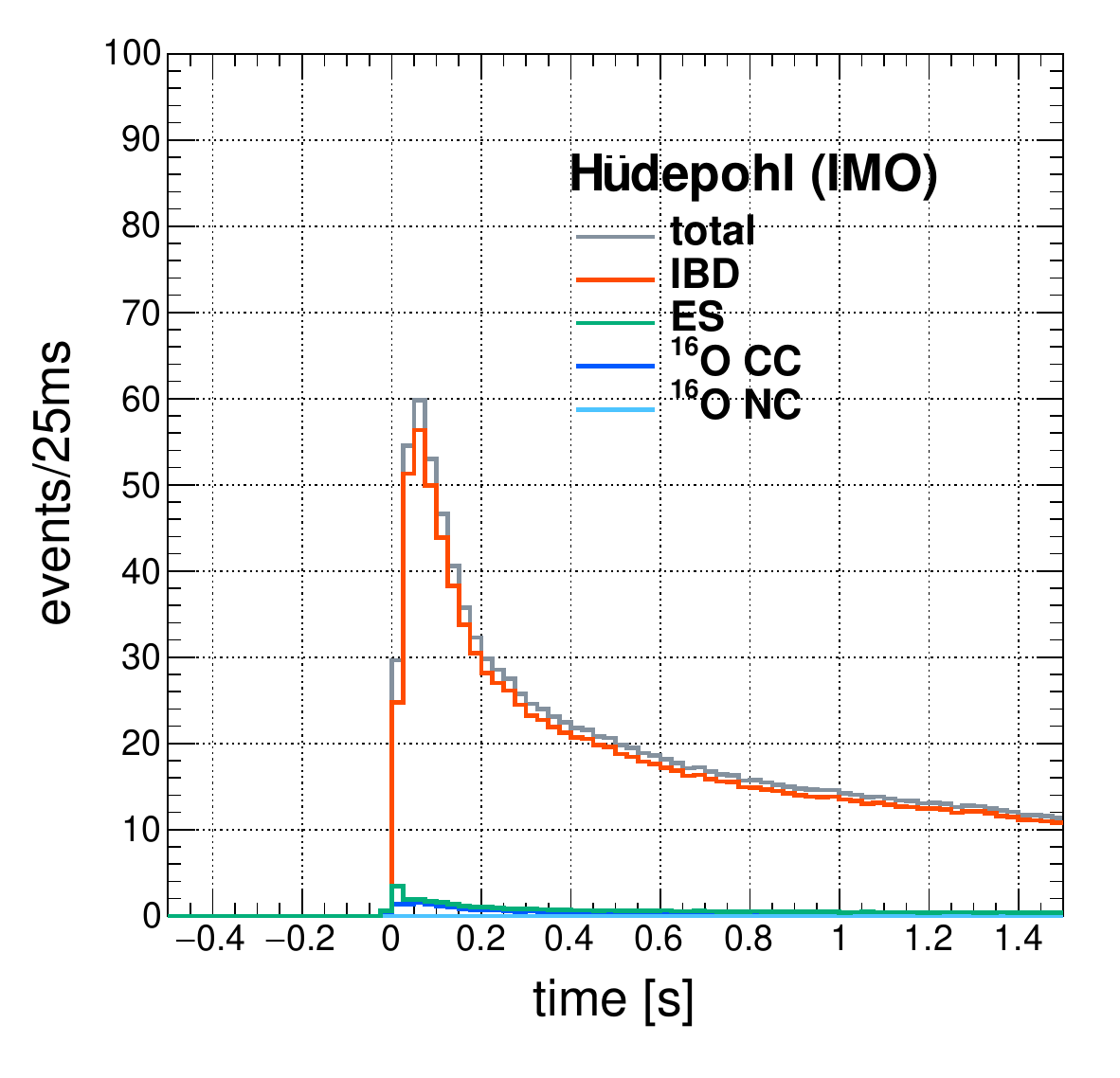}{0.33\textwidth}{}
    \fig{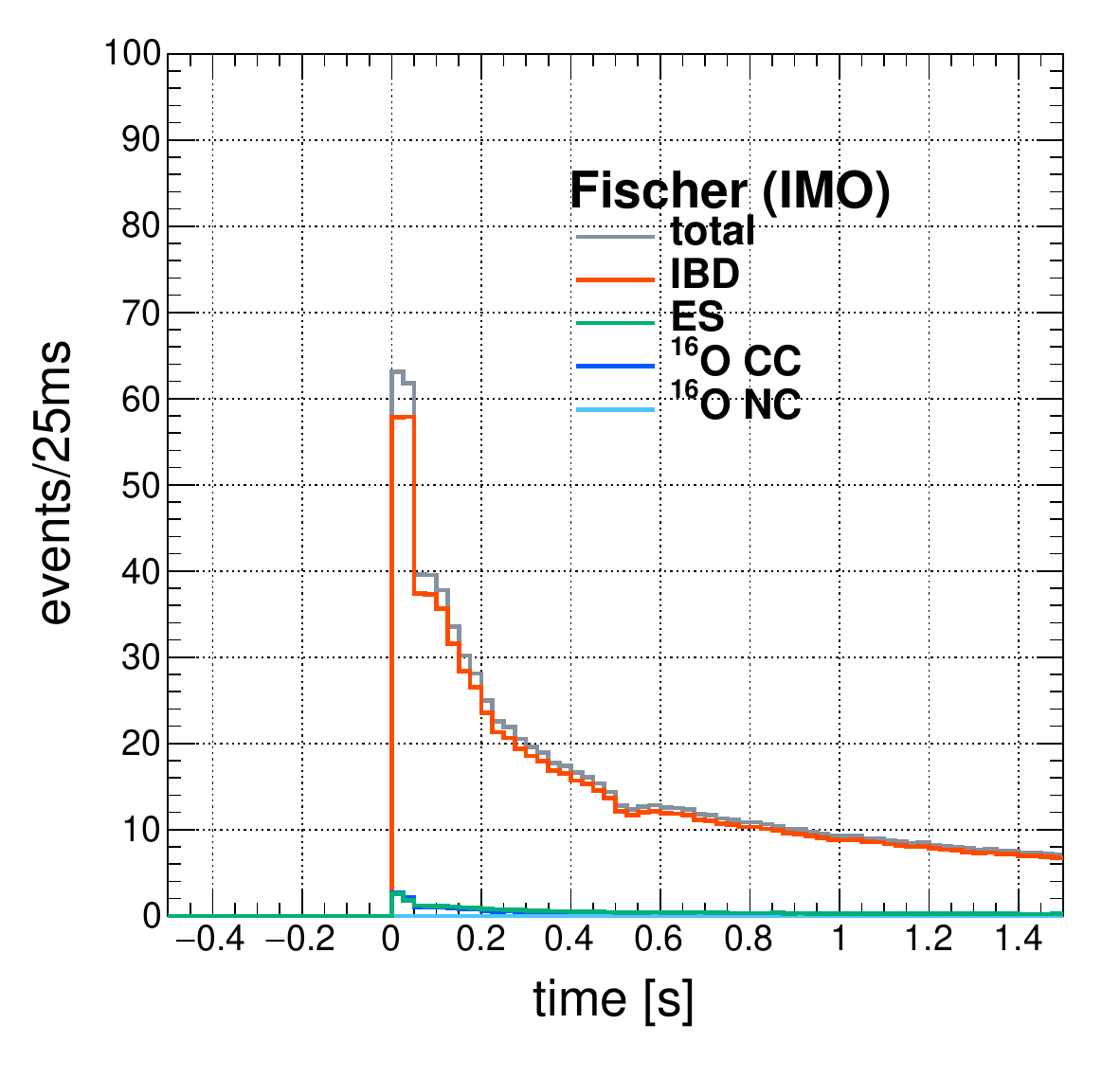}{0.33\textwidth}{}
    \fig{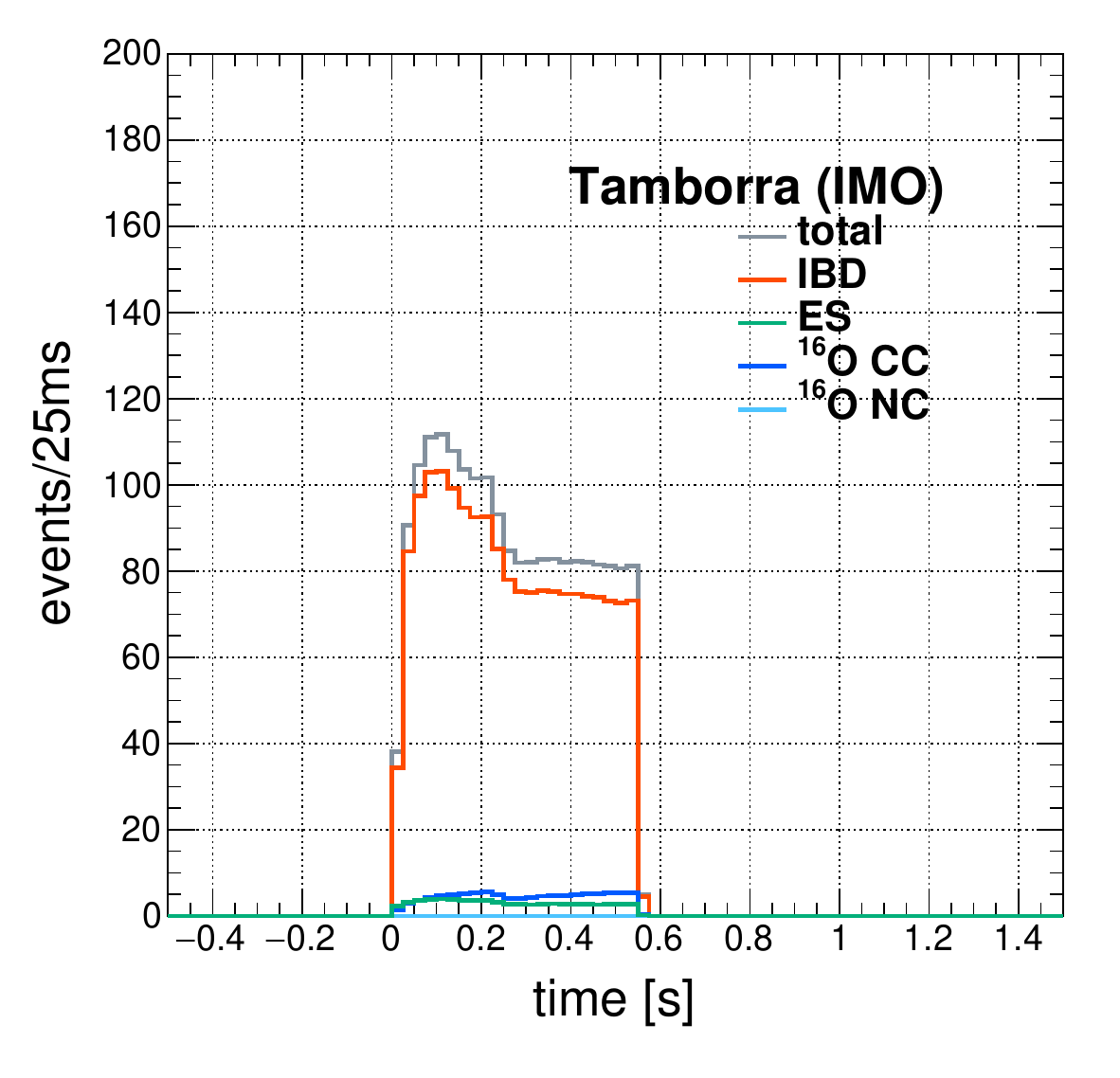}{0.33\textwidth}{}
}
\vspace{-0.5cm}
\caption{Comparison of time evolution up to 1.5~s among interactions for each model for an SN burst located at 10~kpc in the NMO scenario (top six panels) and the IMO scenario (bottom six panels).}
\label{fig:NMOandIMOTime1sInteractionsEachModel}
\end{figure}

\begin{figure}[htb!]
\gridline{
    \fig{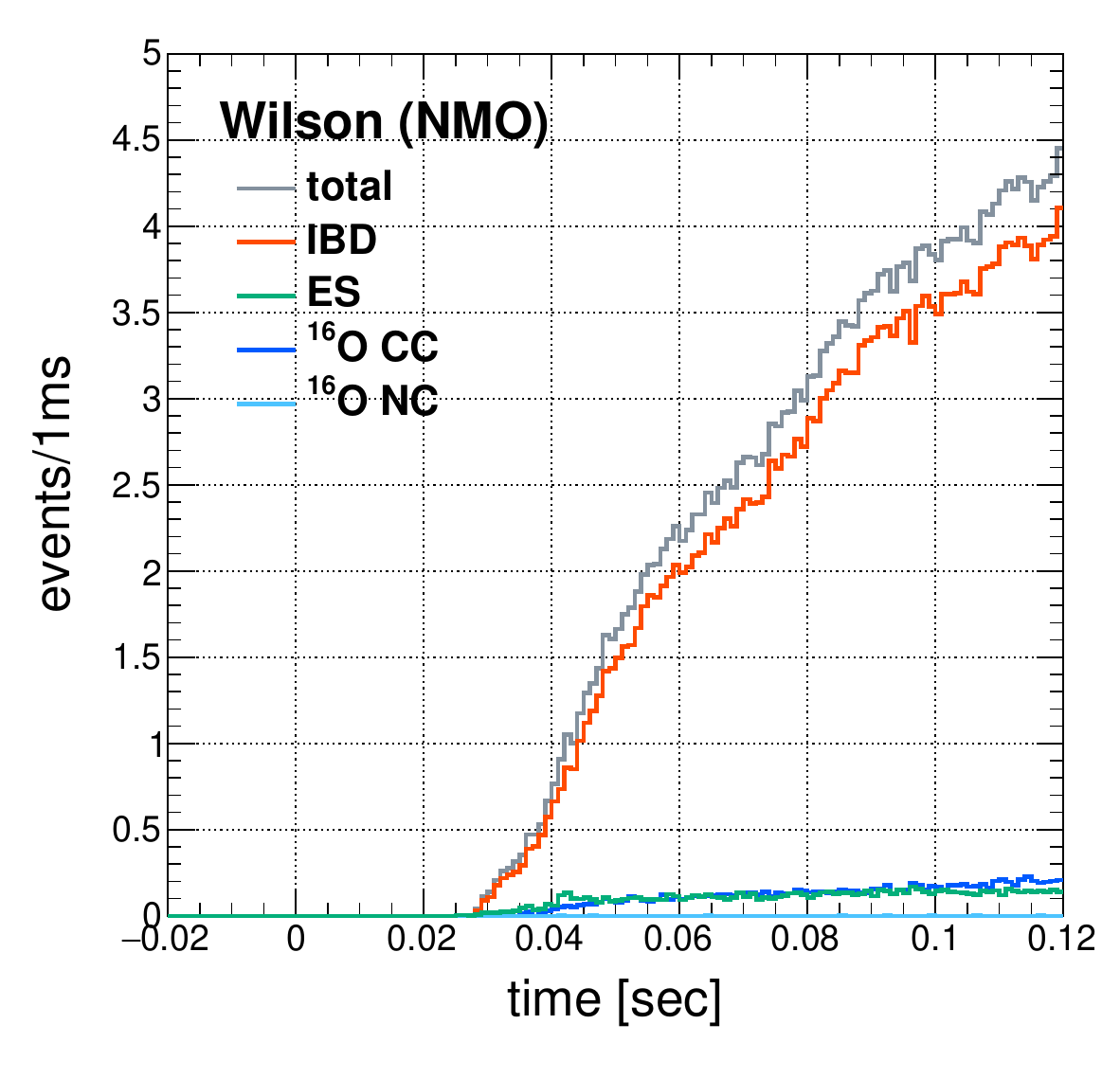}{0.33\textwidth}{(a) the Wilson model}
    \fig{Modification_23Dec_First100ms_mtimePrompt_10kpc_NMO_recoTime100ms_reactions_Nakazato.pdf}{0.33\textwidth}{(b) the Nakazato model}
    \fig{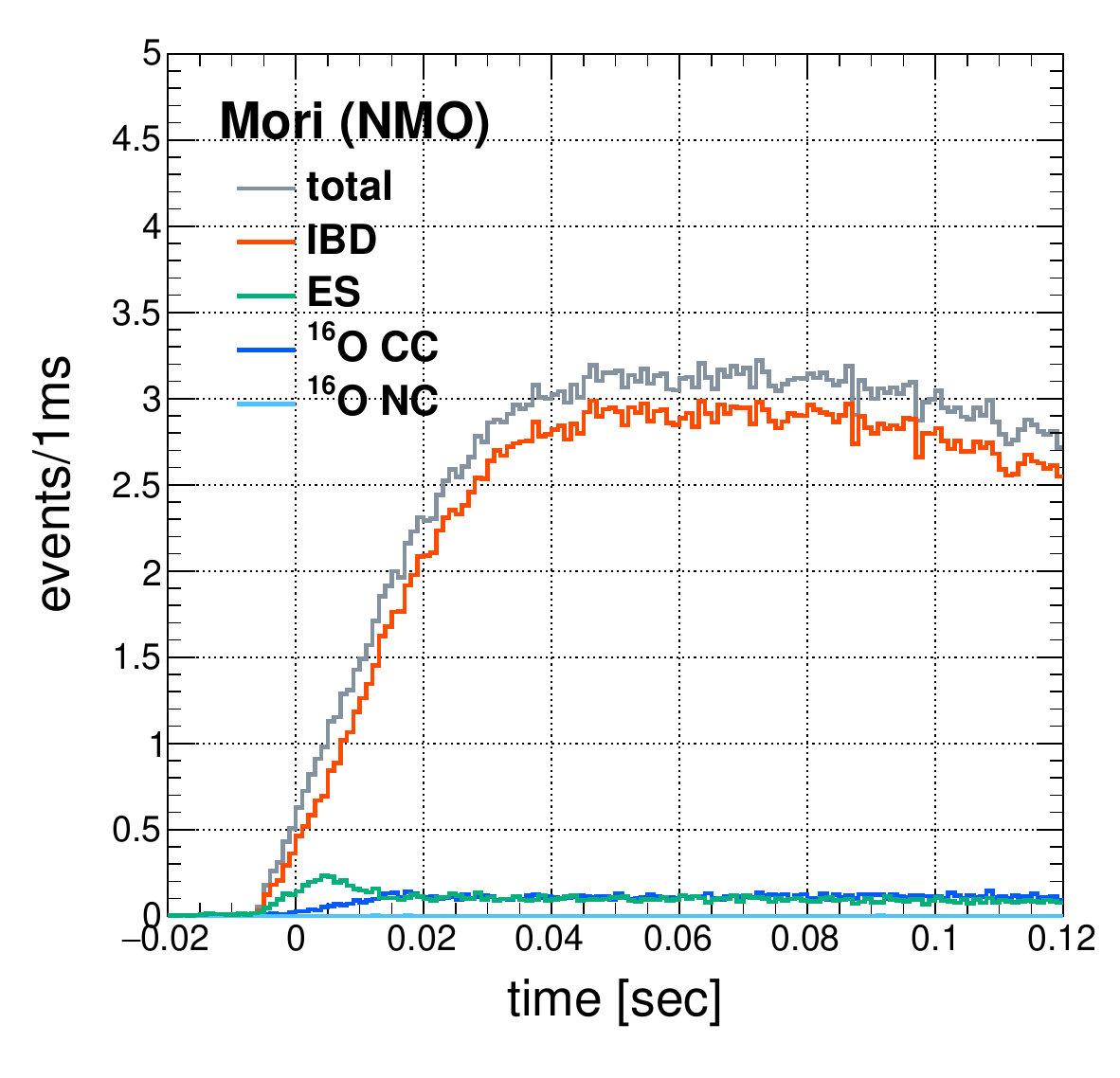}{0.33\textwidth}{(c) the Mori model}
}
\vspace{-1.2cm}
\gridline{
    \fig{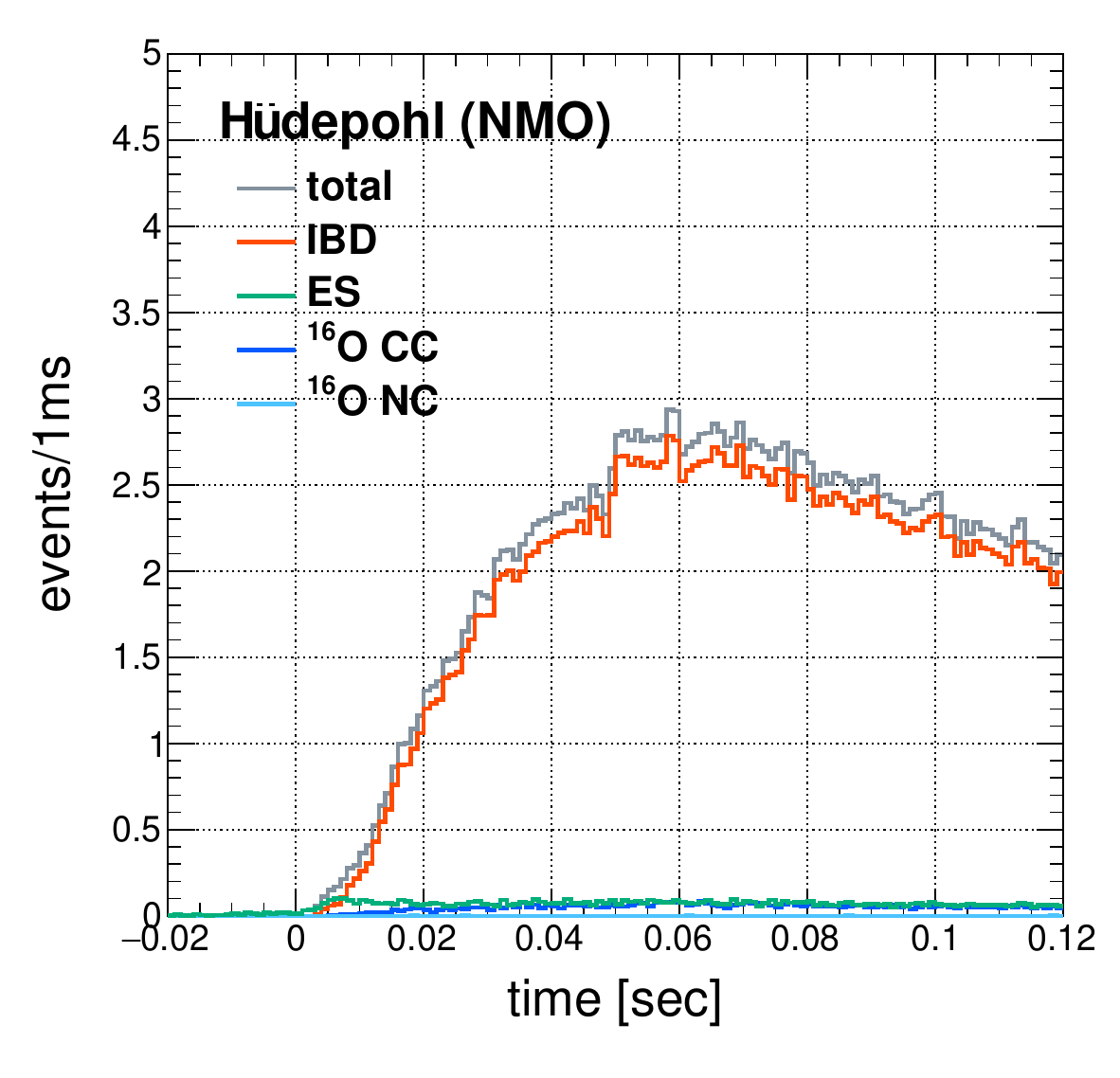}{0.33\textwidth}{(d) the H\"{u}depohl model}
    \fig{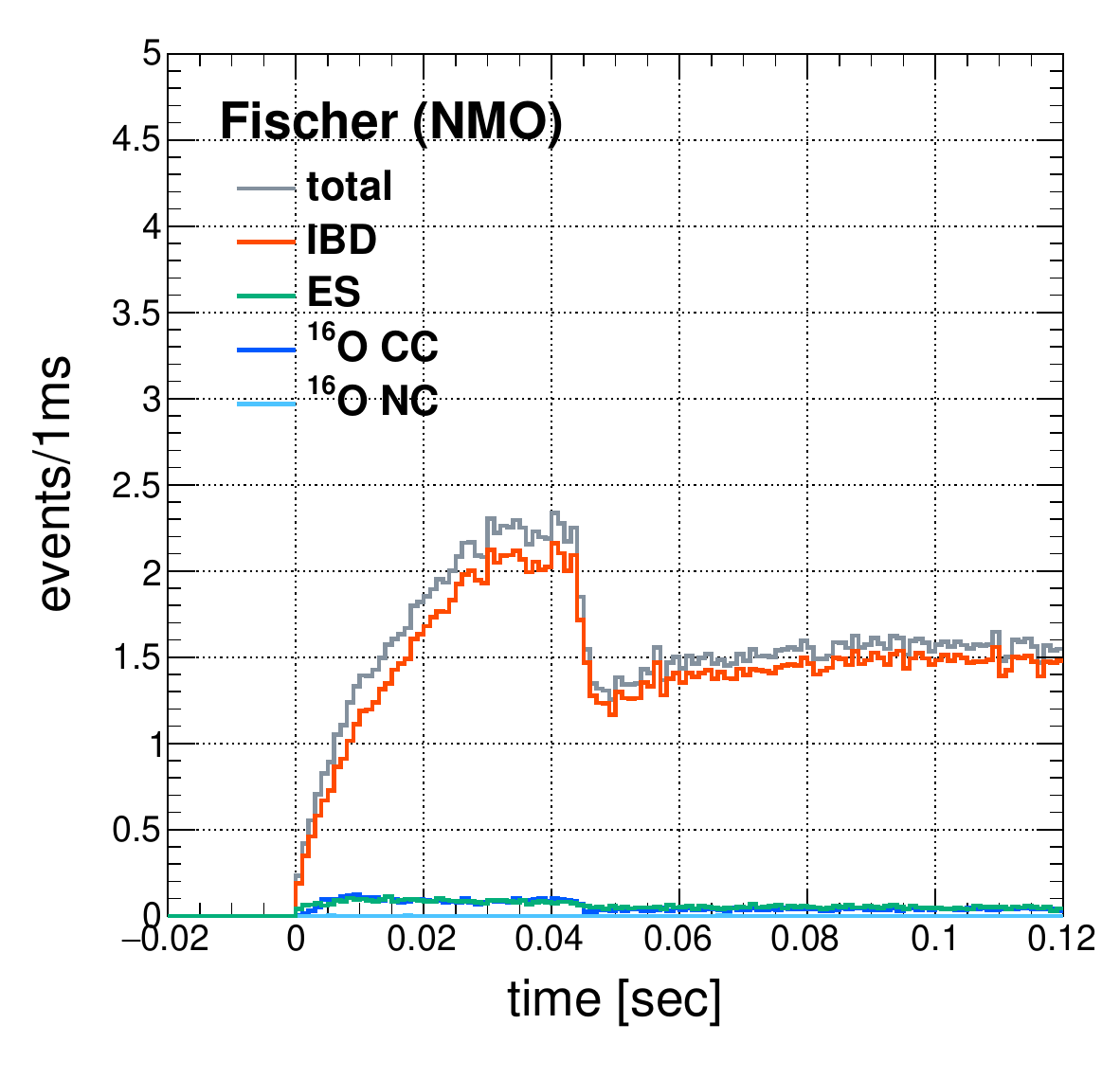}{0.33\textwidth}{(e) the Fischer model}
    \fig{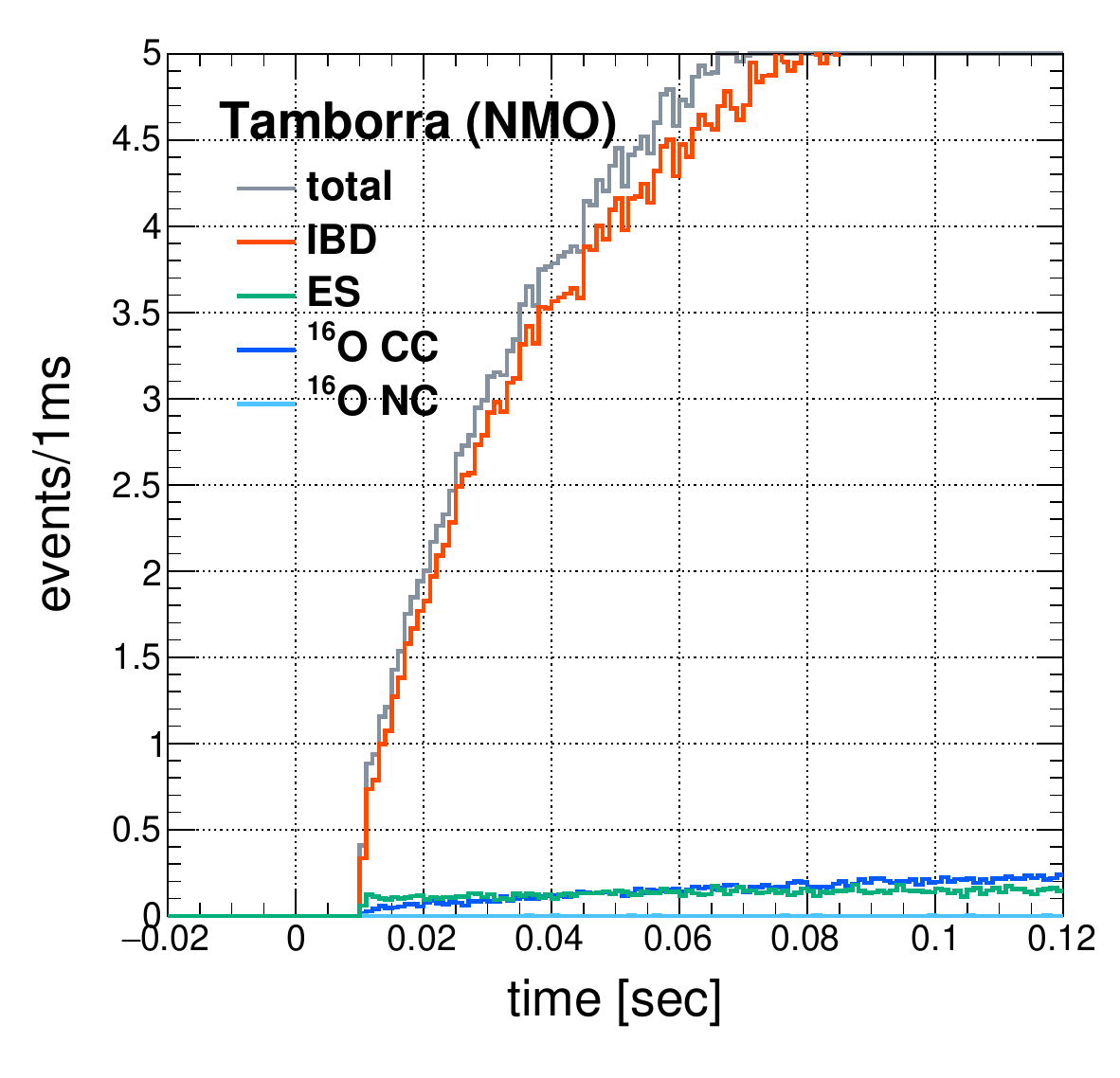}{0.33\textwidth}{(f) the Tamborra model}
}
\vspace{-1.2cm}
\gridline{
    \fig{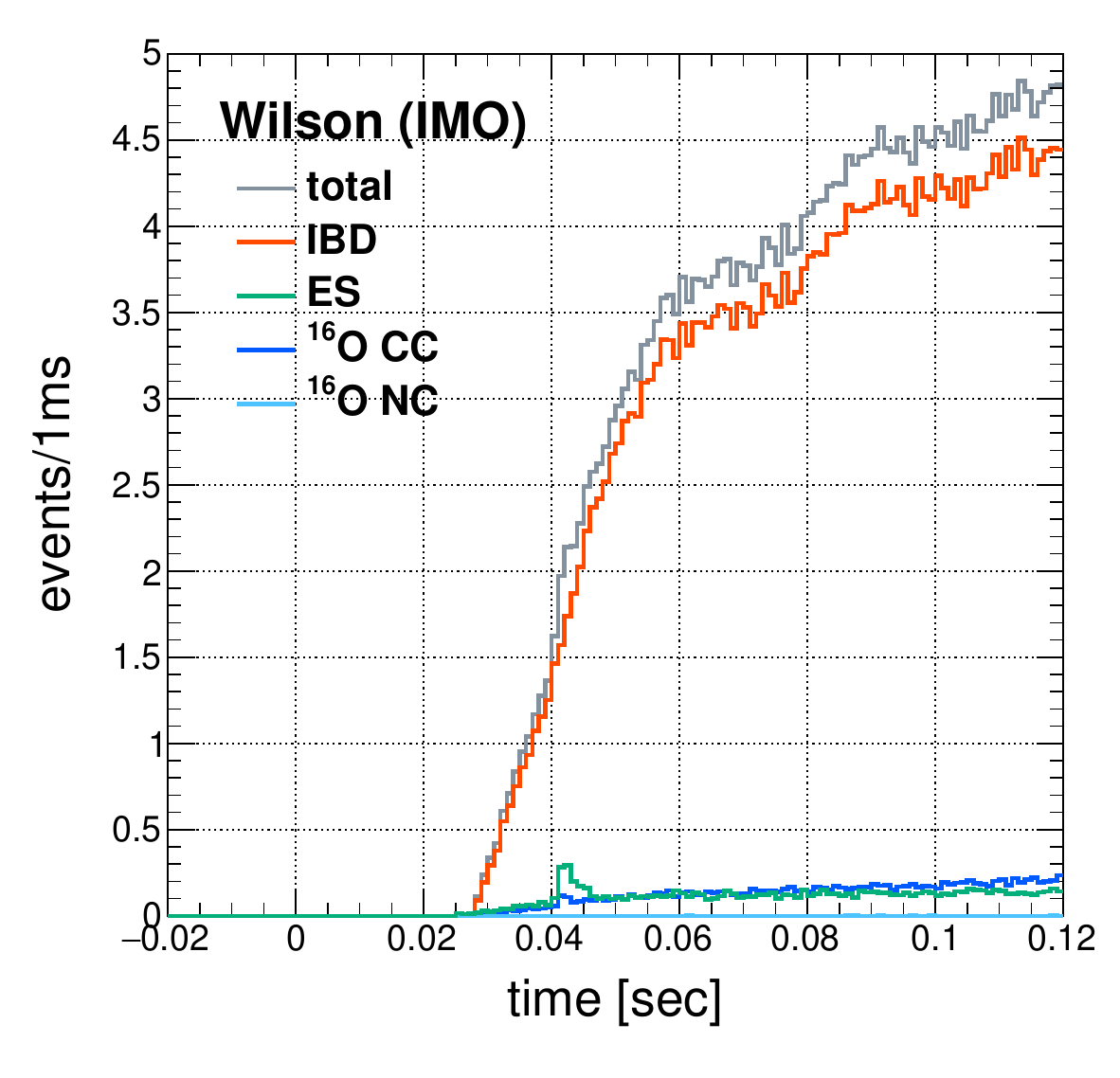}{0.33\textwidth}{(a) the Wilson model}
    \fig{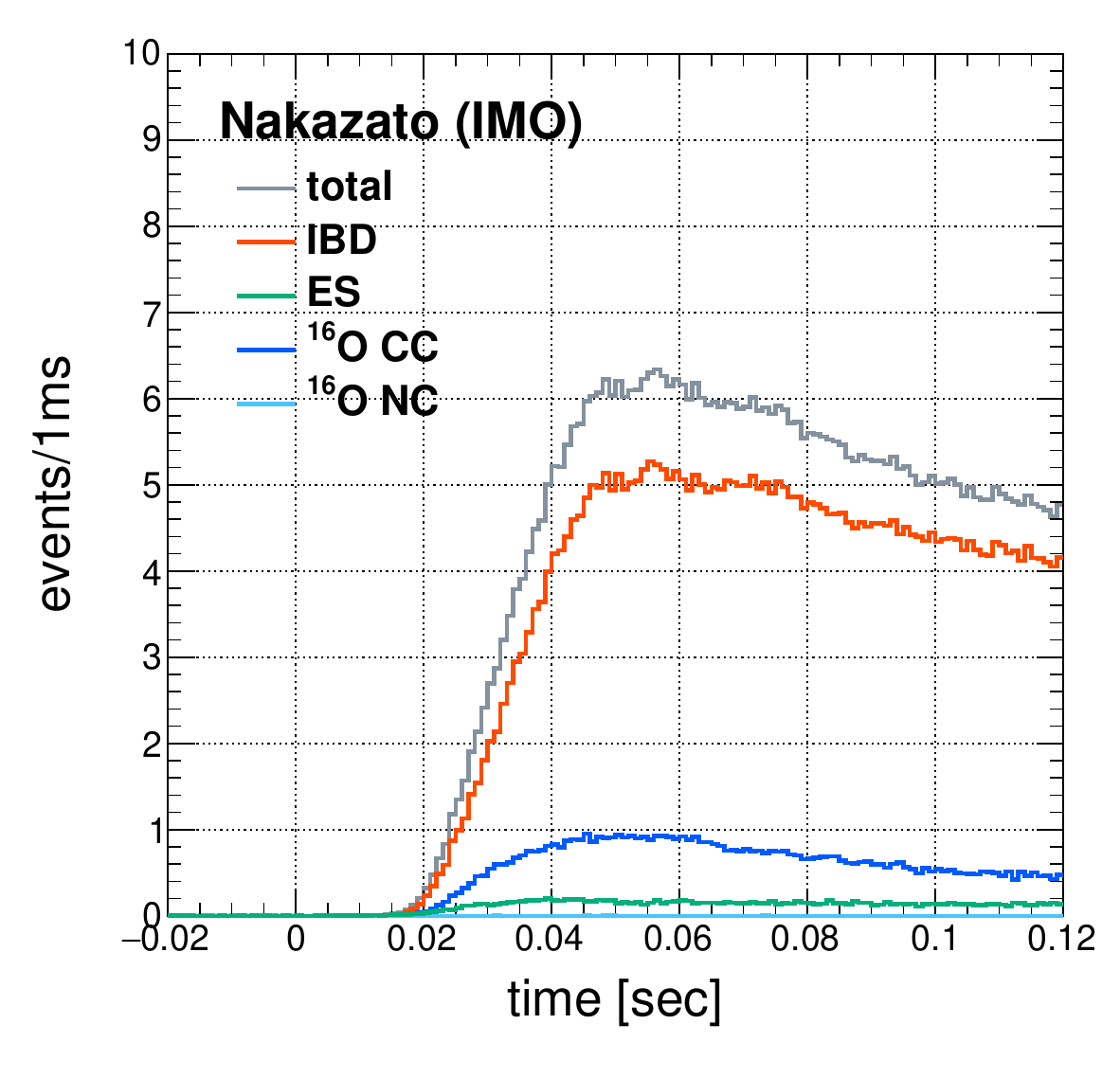}{0.33\textwidth}{(b) the Nakazato model}
    \fig{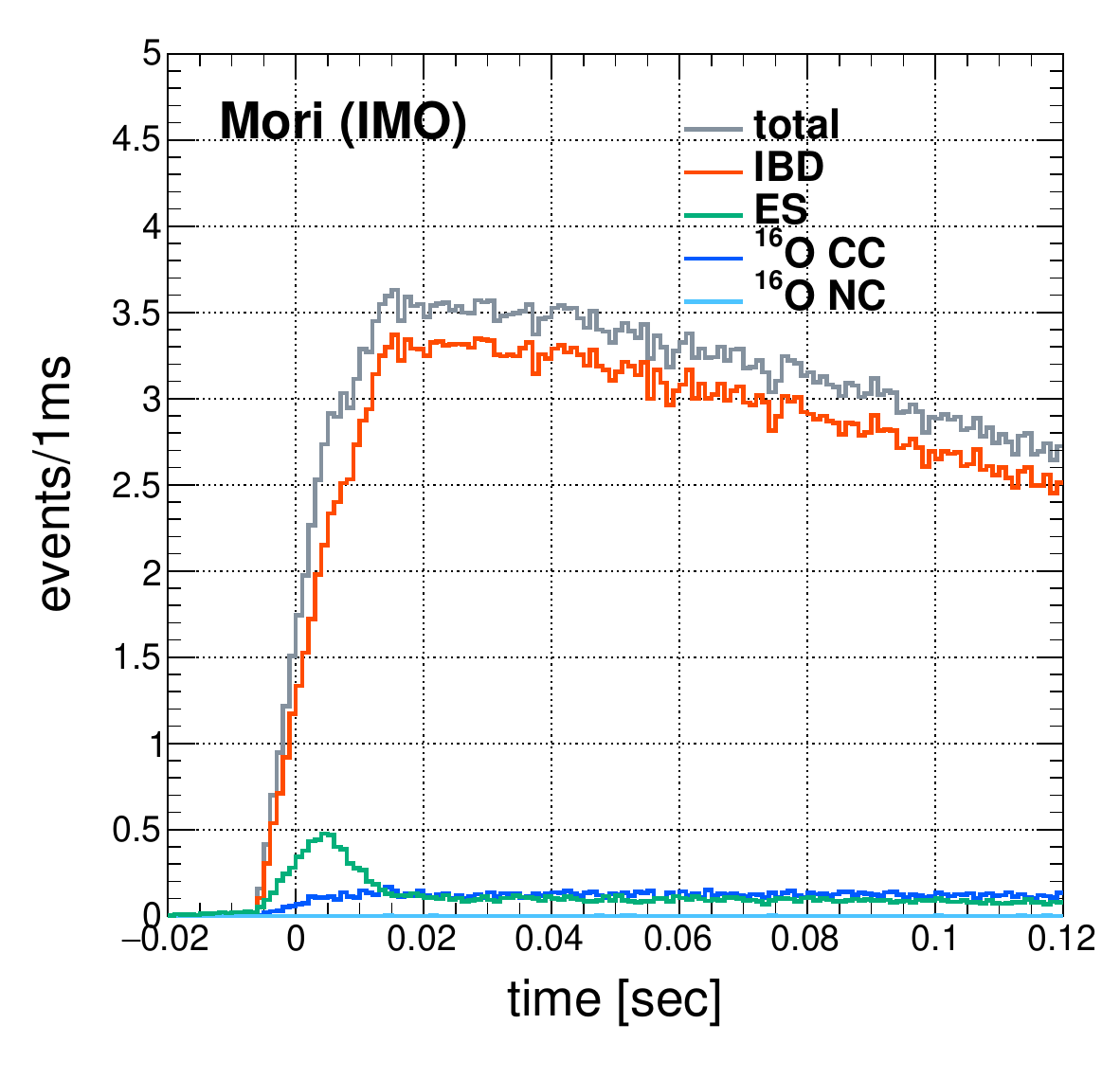}{0.33\textwidth}{(c) the Mori model}
}
\vspace{-1.2cm}
\gridline{
    \fig{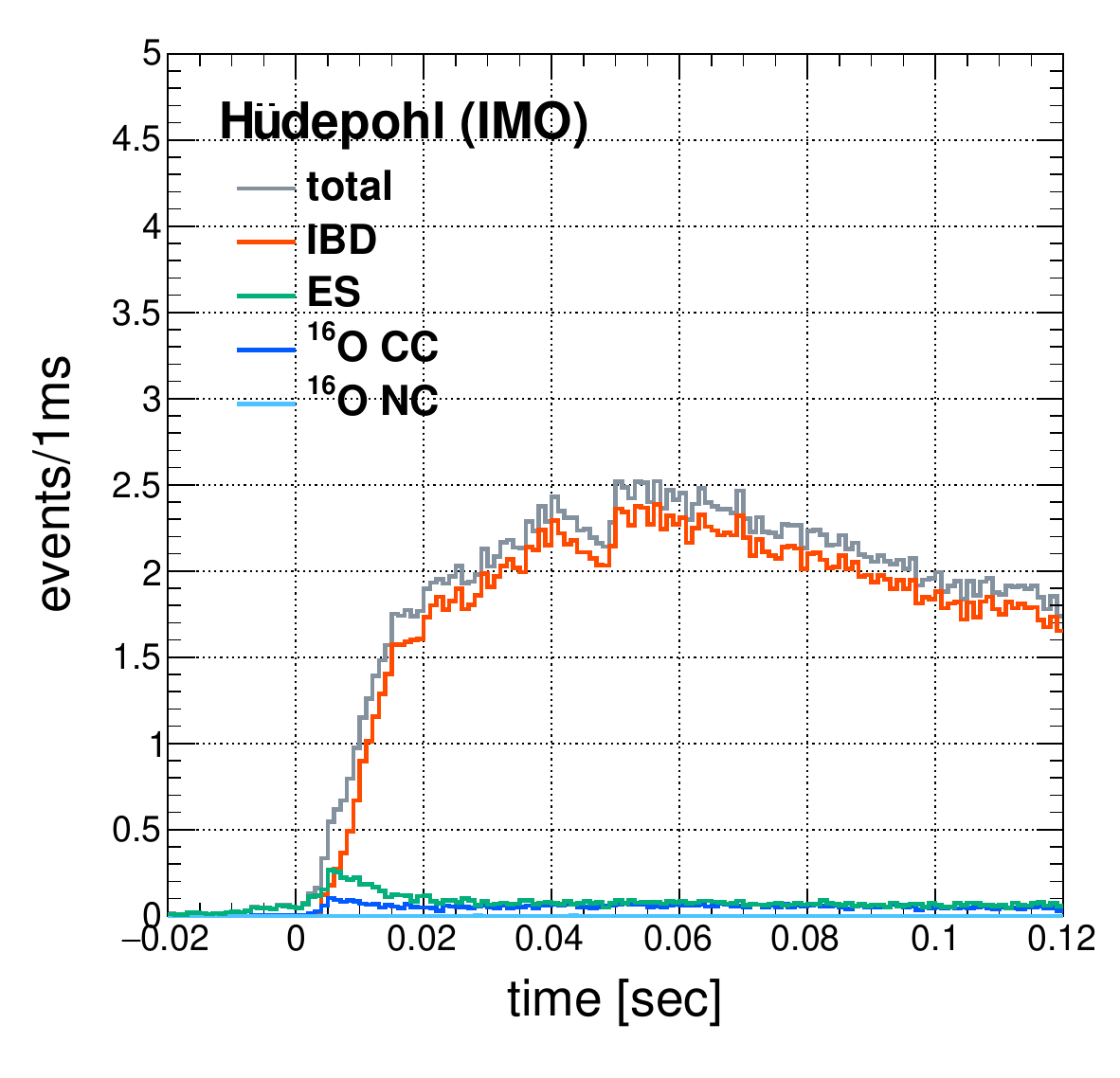}{0.33\textwidth}{}
    \fig{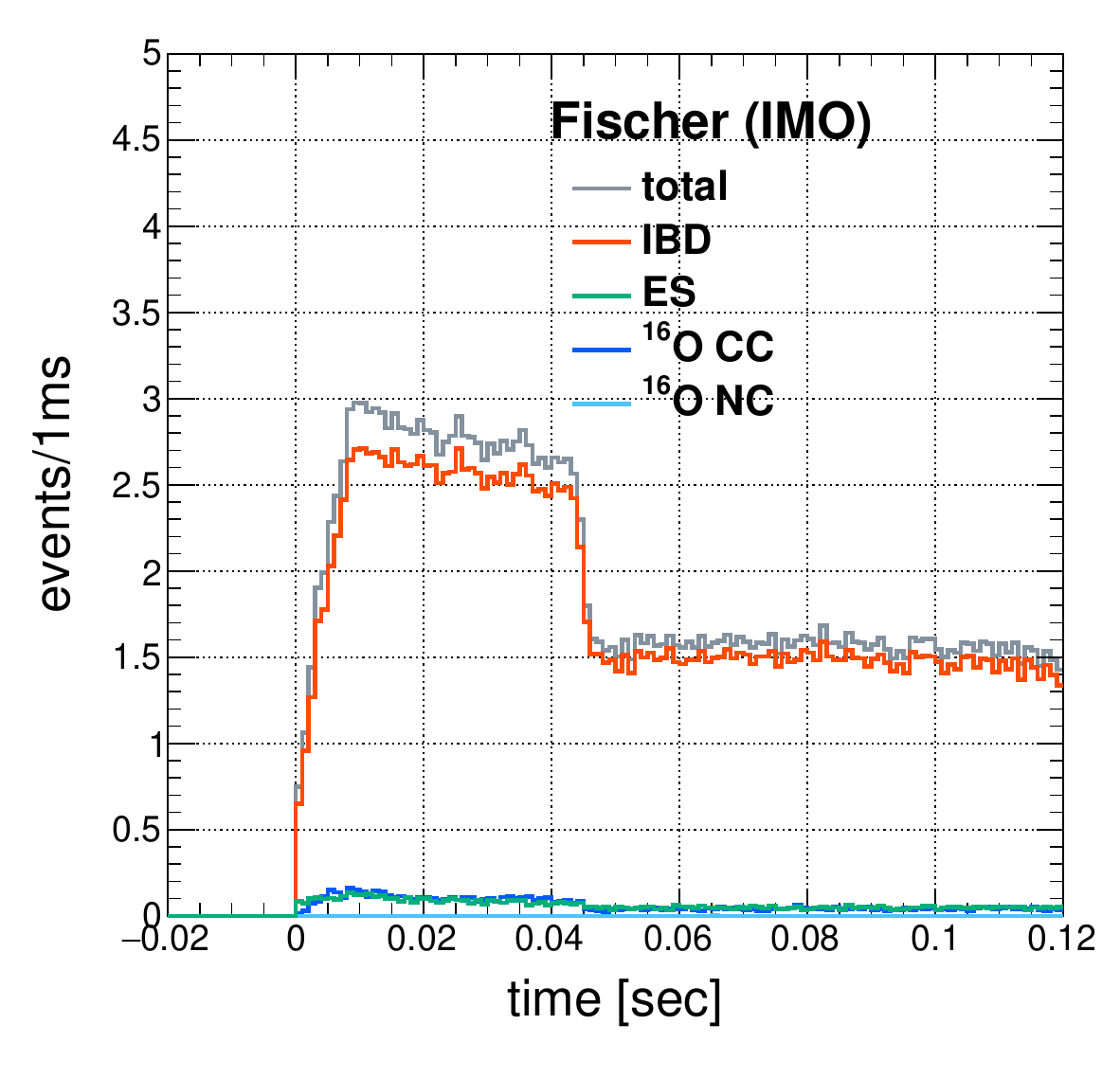}{0.33\textwidth}{}
    \fig{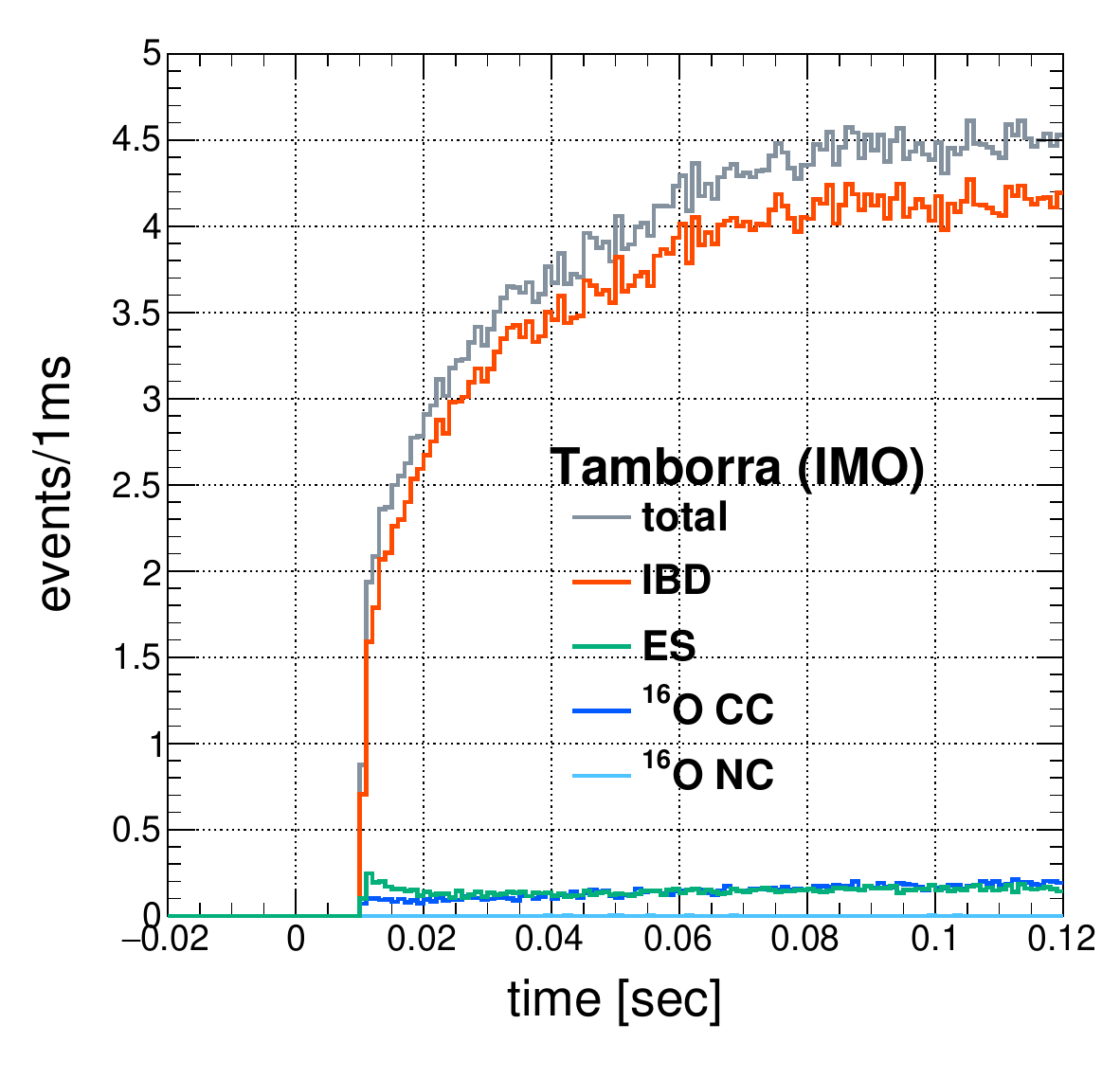}{0.33\textwidth}{}
}
\vspace{-0.5cm}
\caption{Comparison of time evolution up to 0.12~s among interactions for each model for an SN burst located at 10~kpc in the NMO scenario (top six panels) and the IMO scenario (bottom six panels).}
\label{fig:NMOandIMOTime100msInteractionsEachModel}
\end{figure}

\begin{figure}[htb!]
\gridline{
\fig{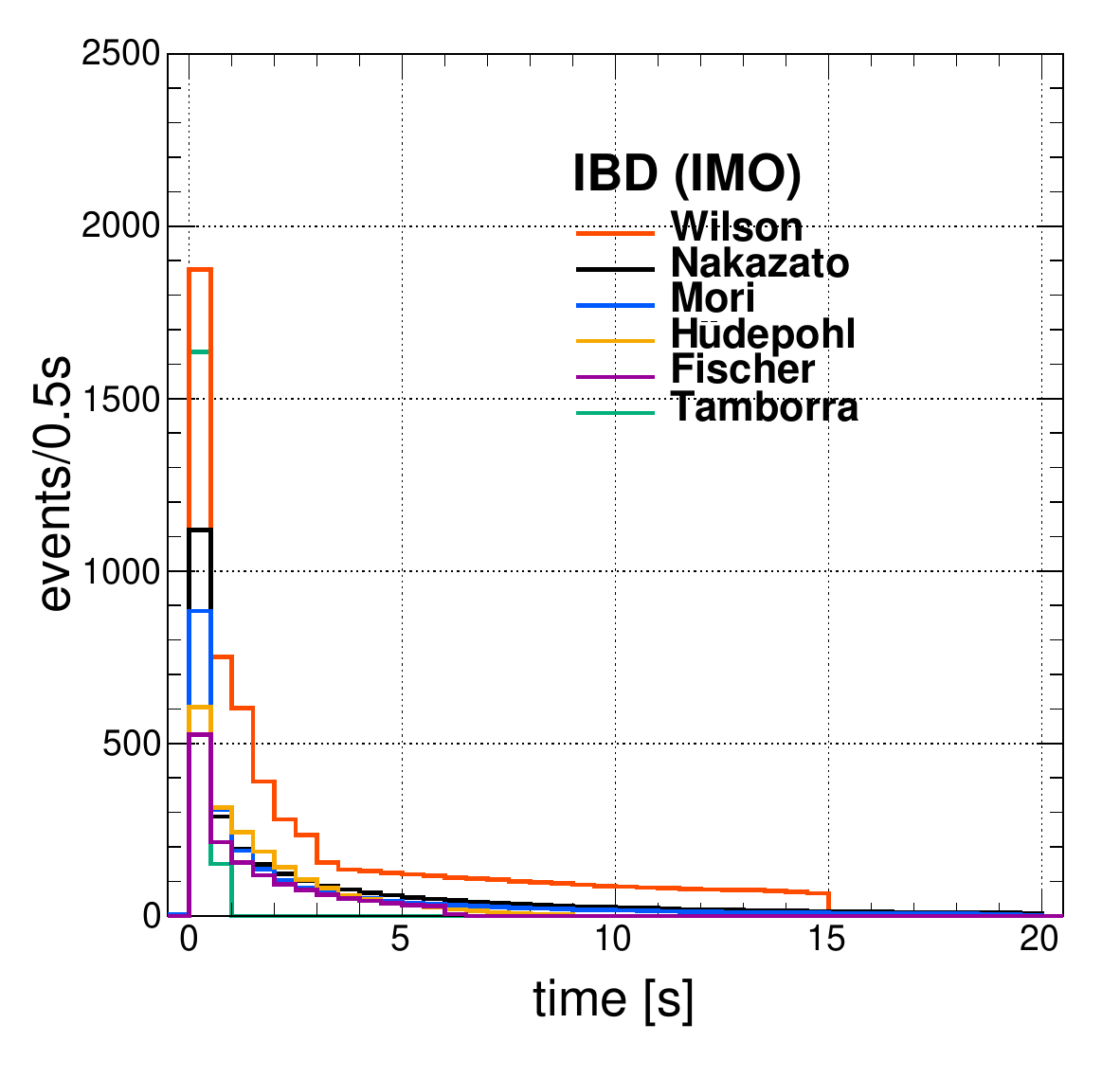}{0.3\textwidth}{(a) IBD, up to 20~s}
\fig{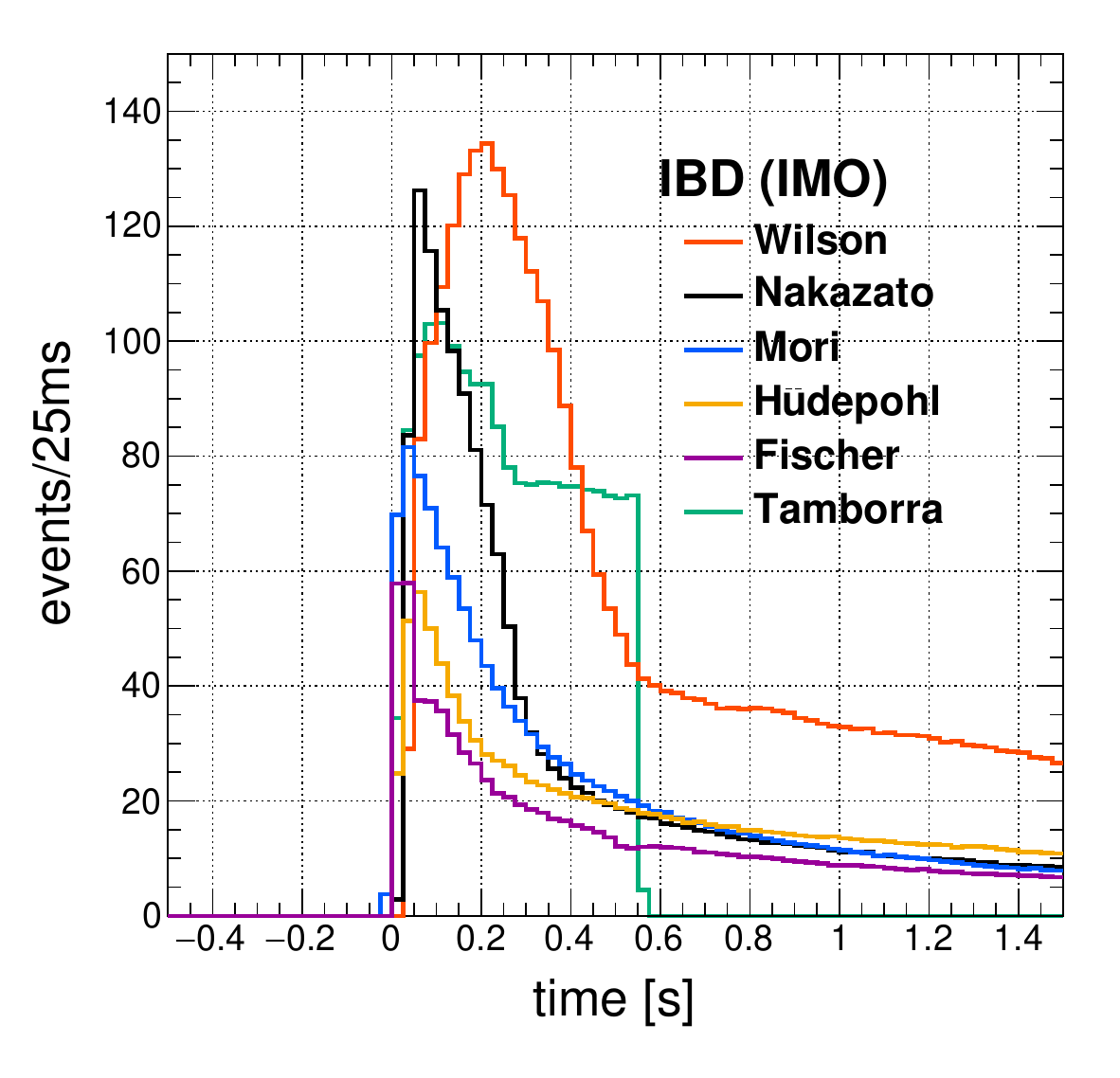}{0.3\textwidth}{(b) IBD, up to 1.5~s}
\fig{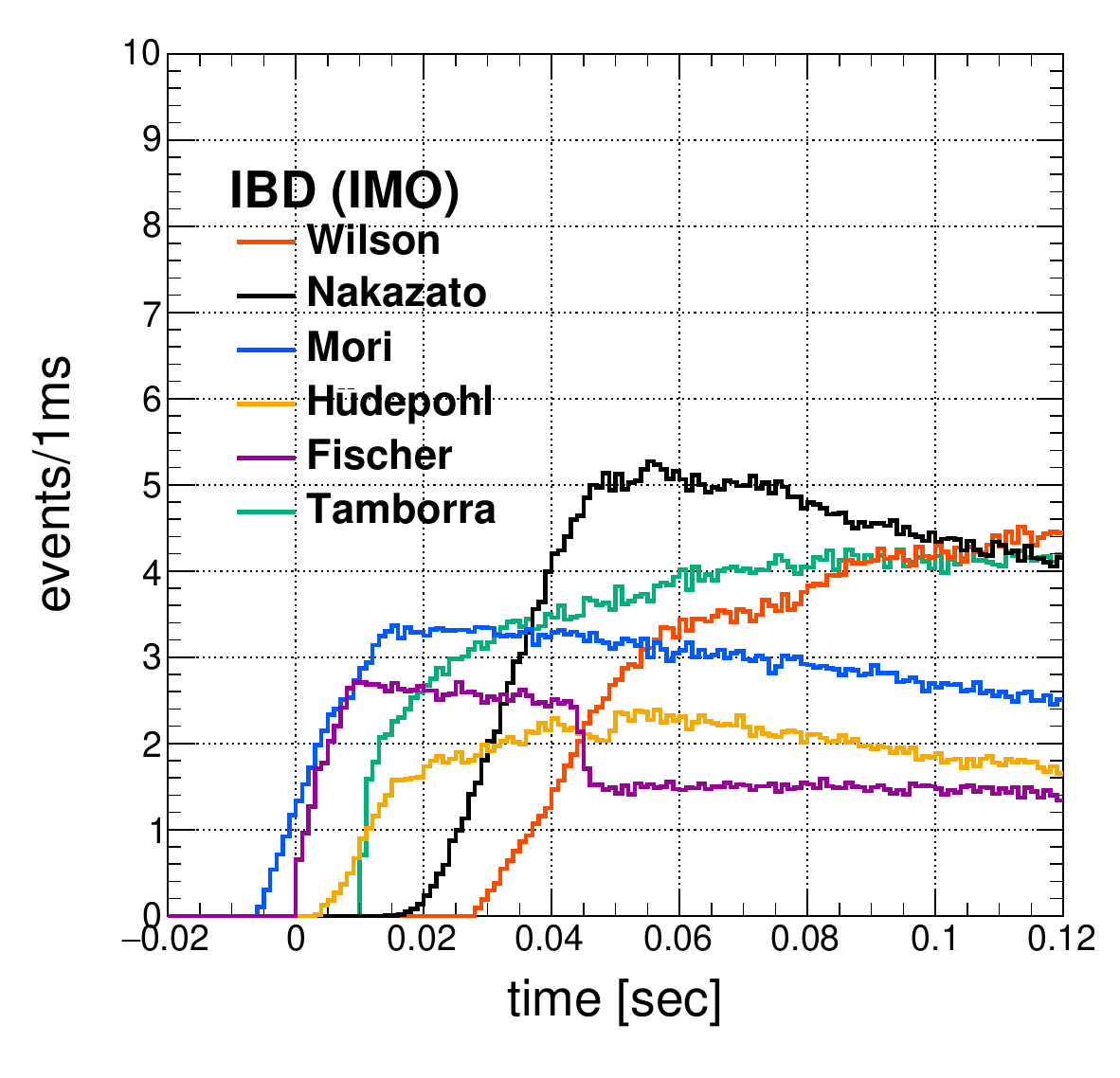}{0.3\textwidth}{(c) IBD, up to 0.12~s}
}
\vspace{-1.2cm}
\gridline{
\fig{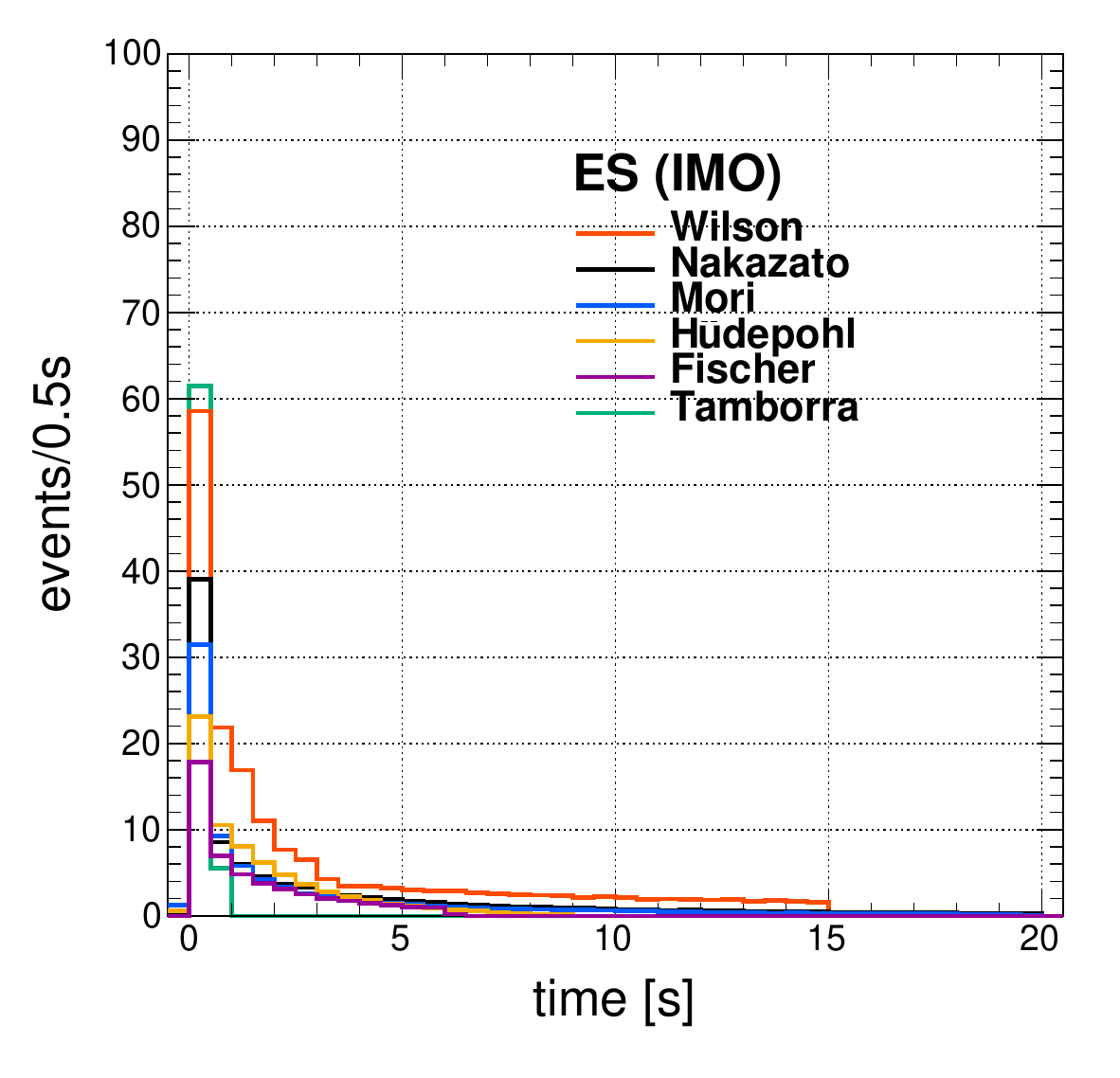}{0.3\textwidth}{(d) ES, up to 20~s}
\fig{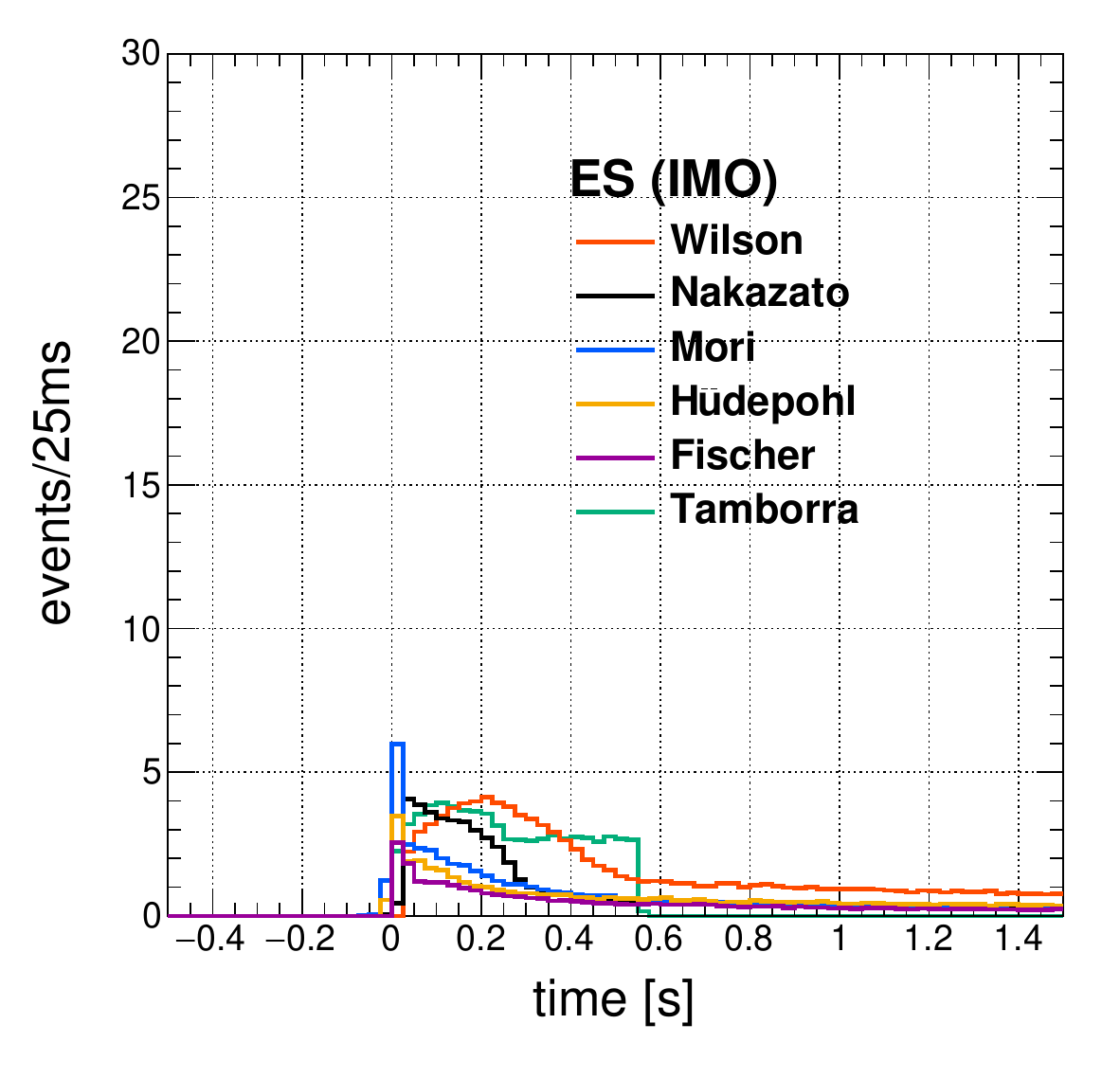}{0.3\textwidth}{(e) ES, up to 1.5~s}
\fig{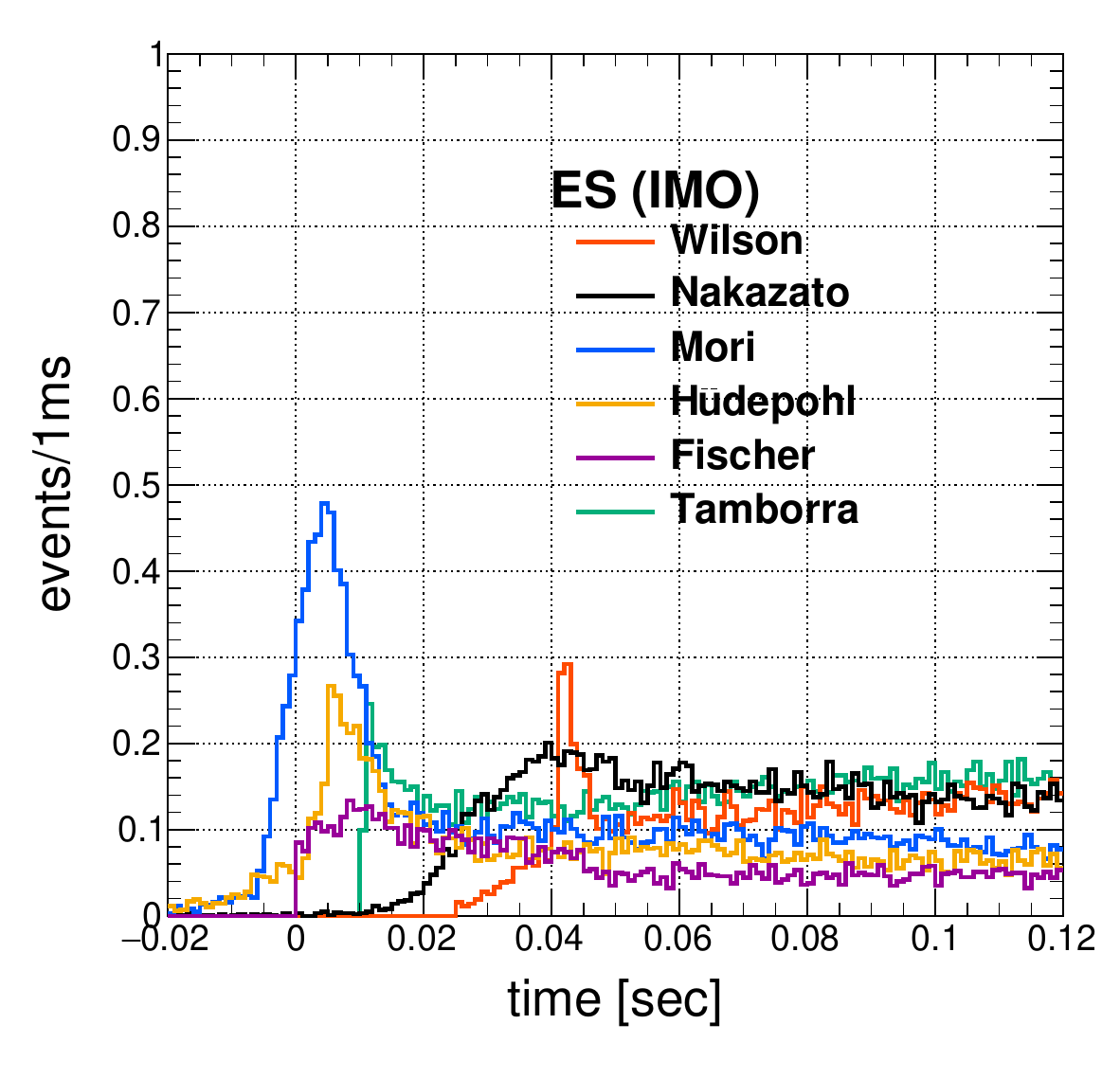}{0.3\textwidth}{(f) ES, up to 0.12~s}
}
\vspace{-1.2cm}
\gridline{
\fig{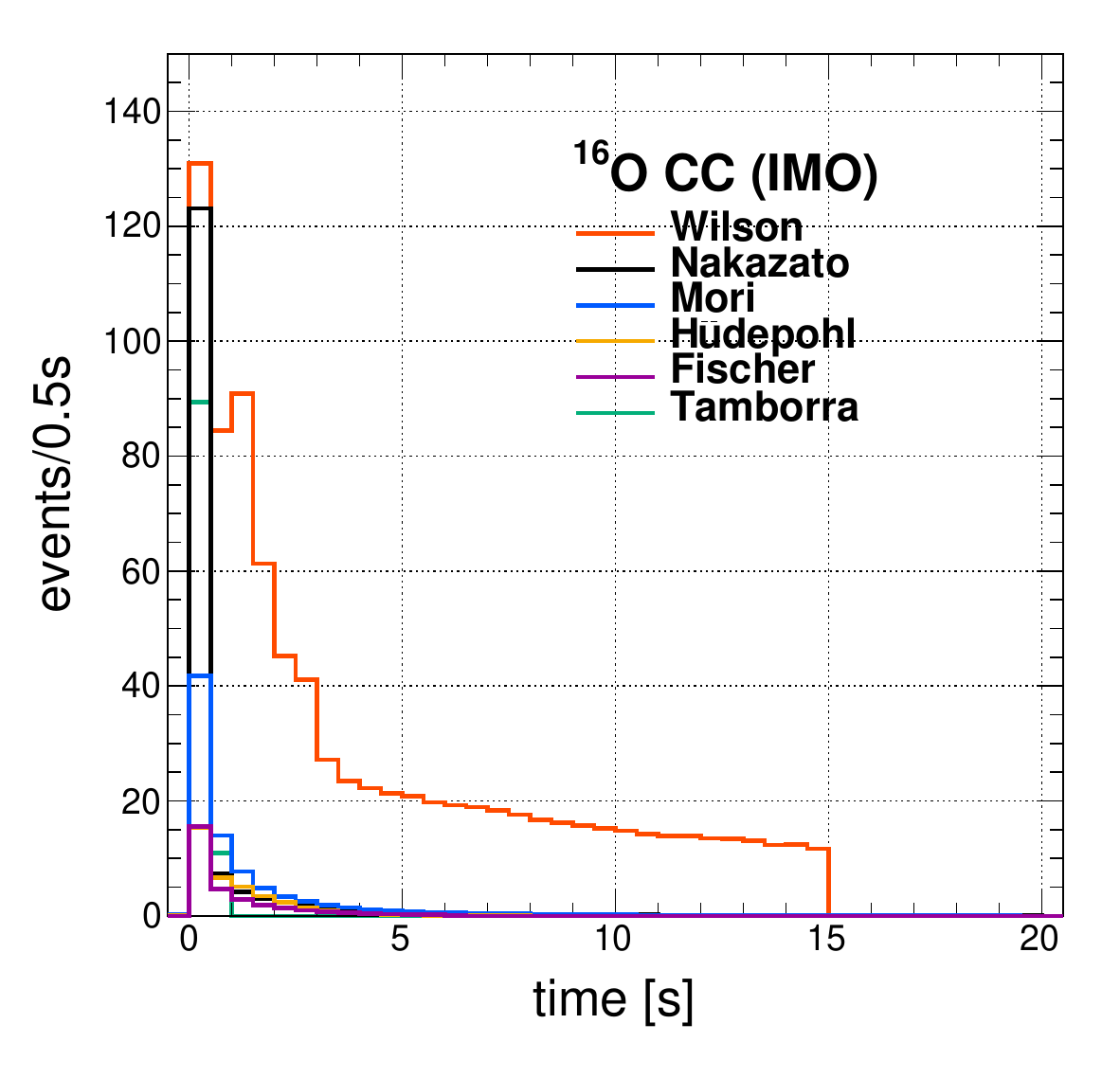}{0.3\textwidth}{(g) $^{16}$O~CC, up to 20~s}
\fig{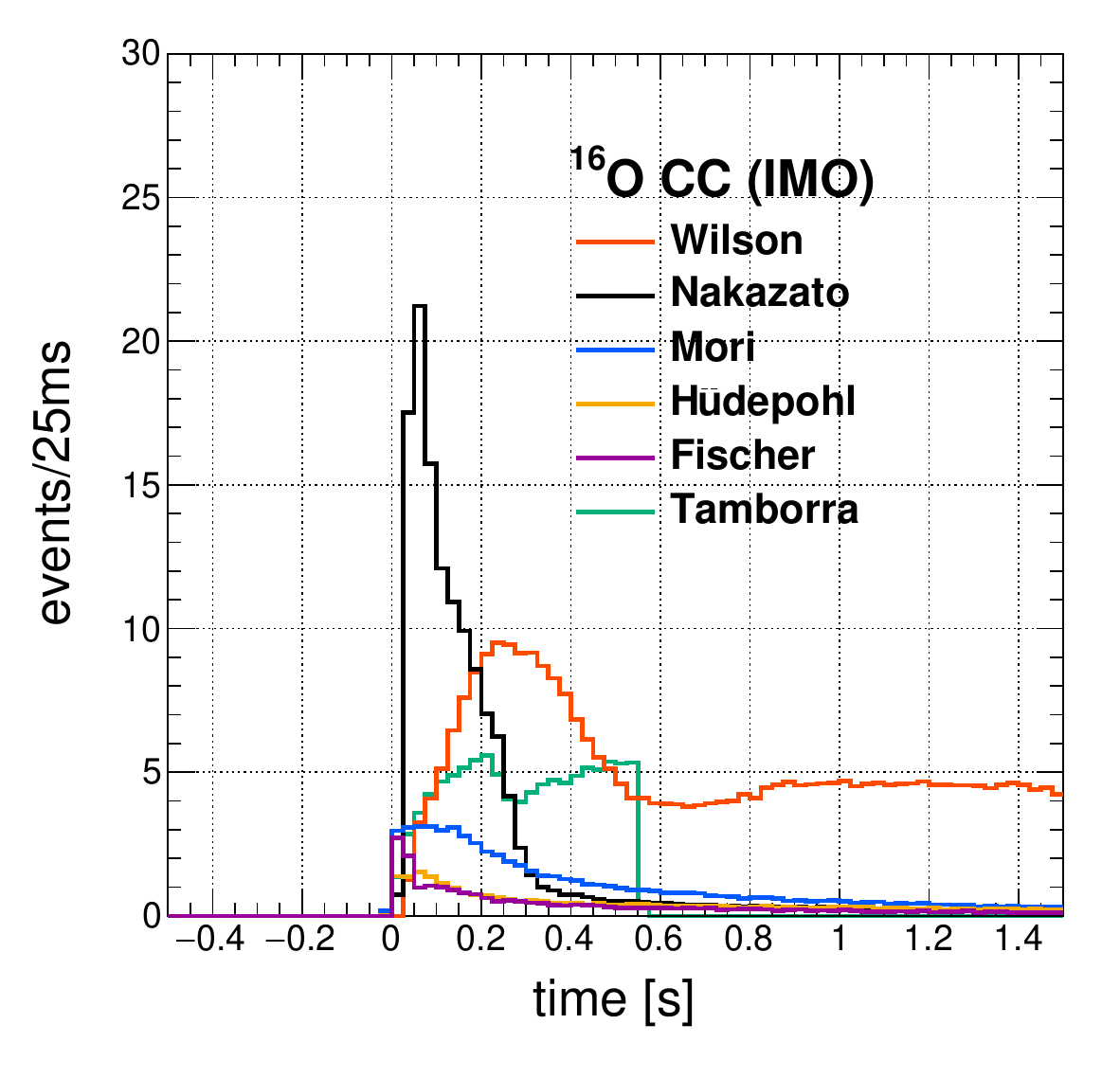}{0.3\textwidth}{(h) $^{16}$O~CC, up to 1.5~s}
\fig{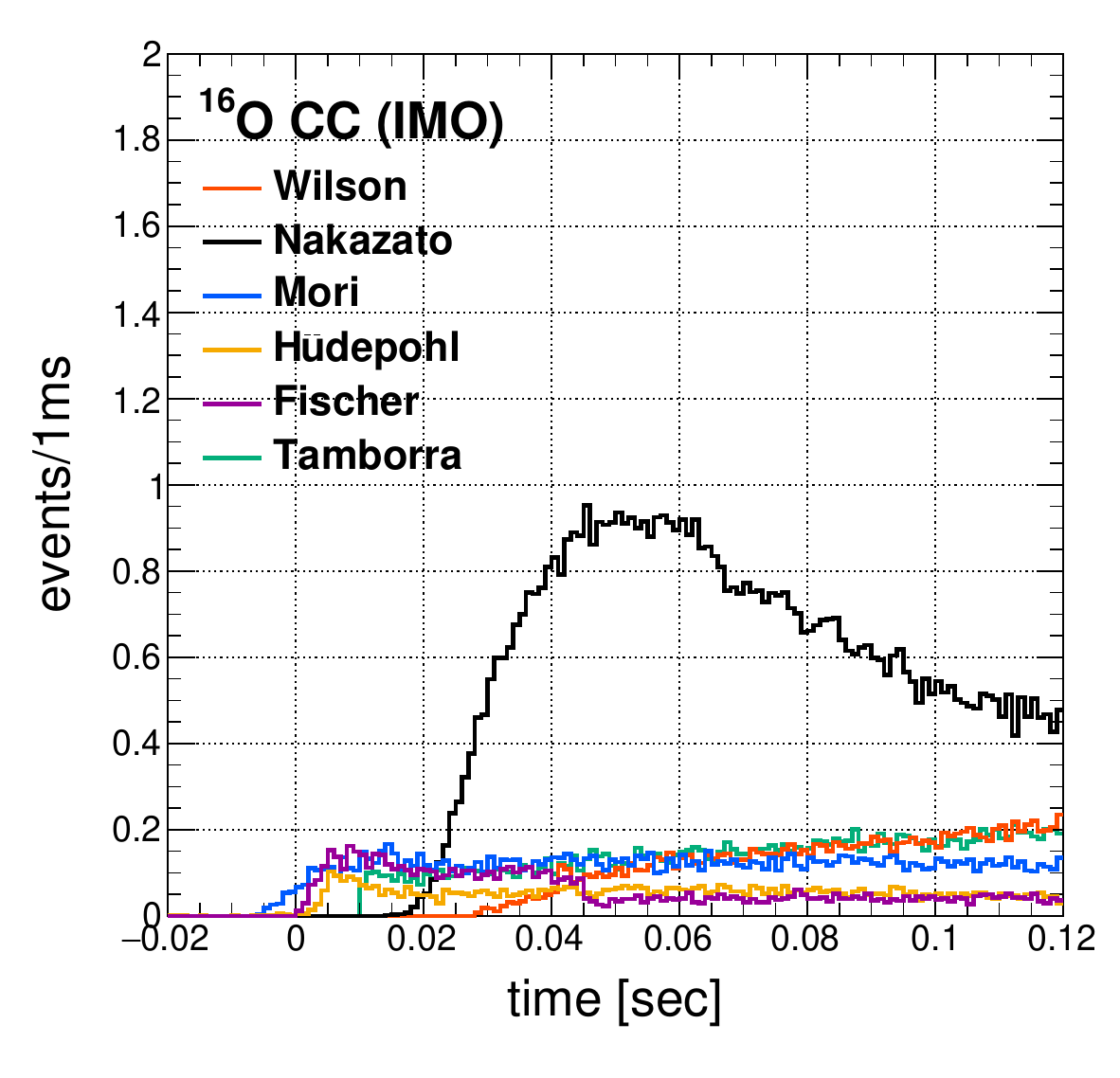}{0.3\textwidth}{(i) $^{16}$O~CC, up to 0.12~s}
}
\vspace{-1.2cm}
\gridline{
\fig{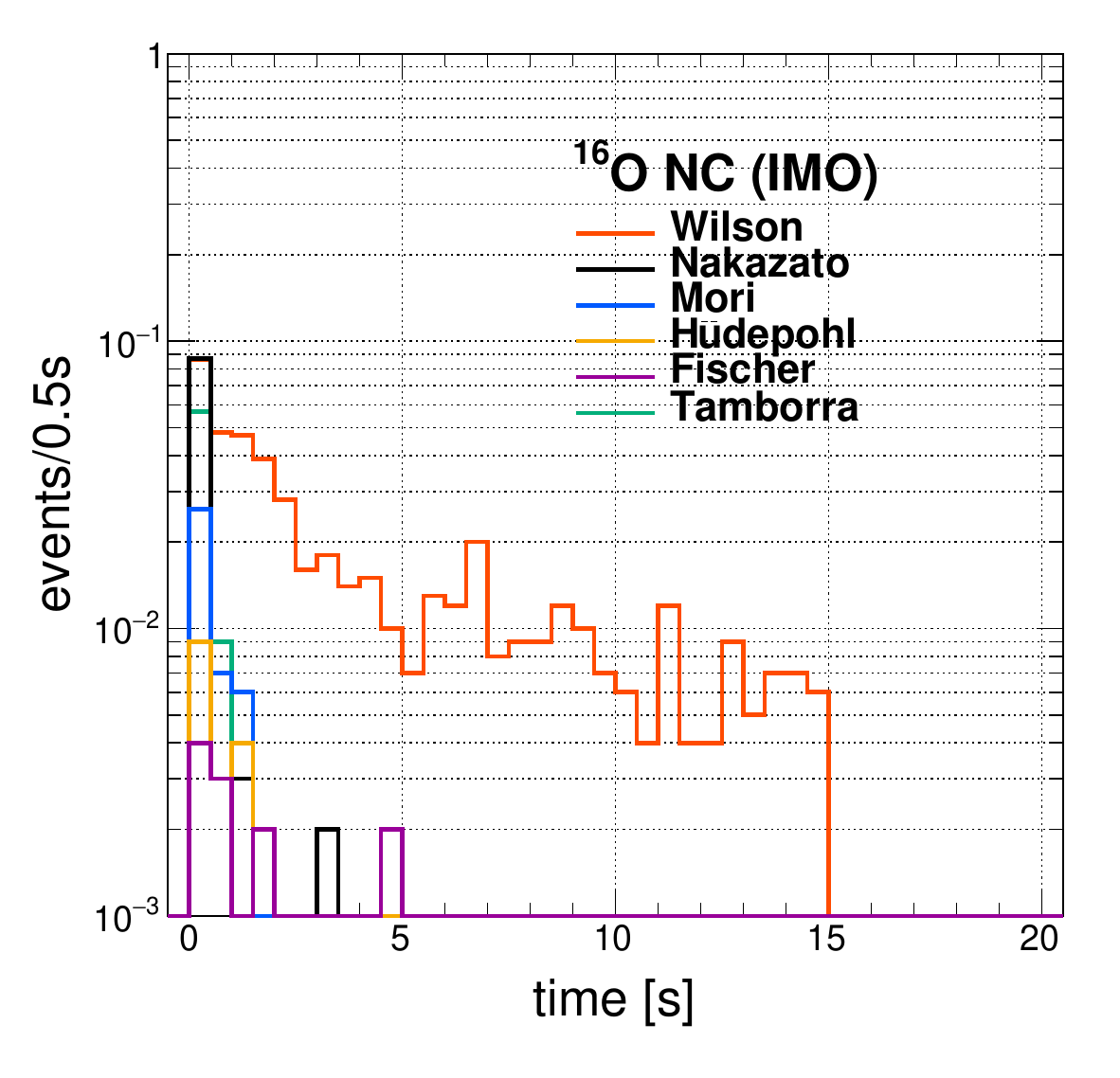}{0.3\textwidth}{}
\fig{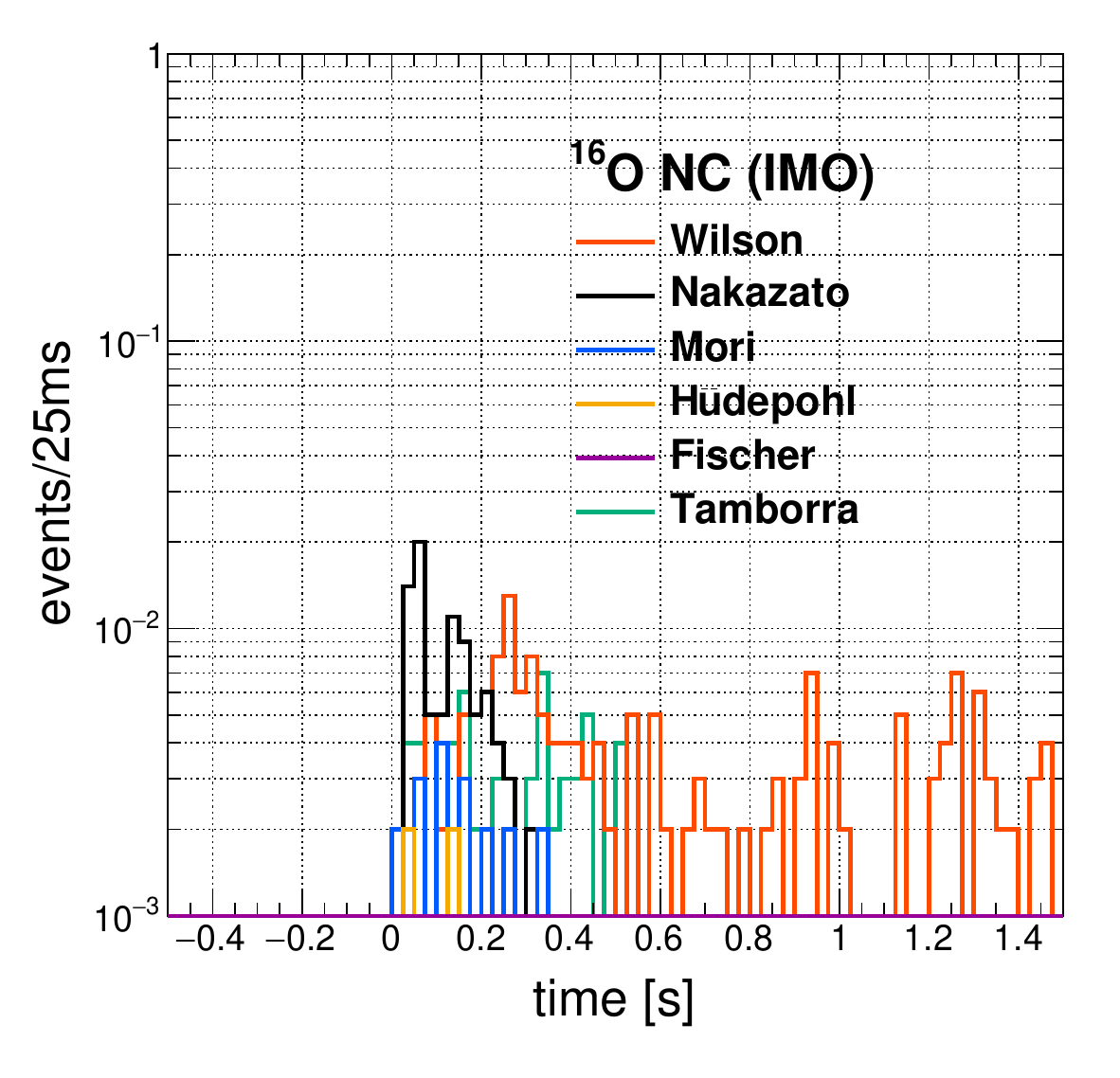}{0.3\textwidth}{}
\fig{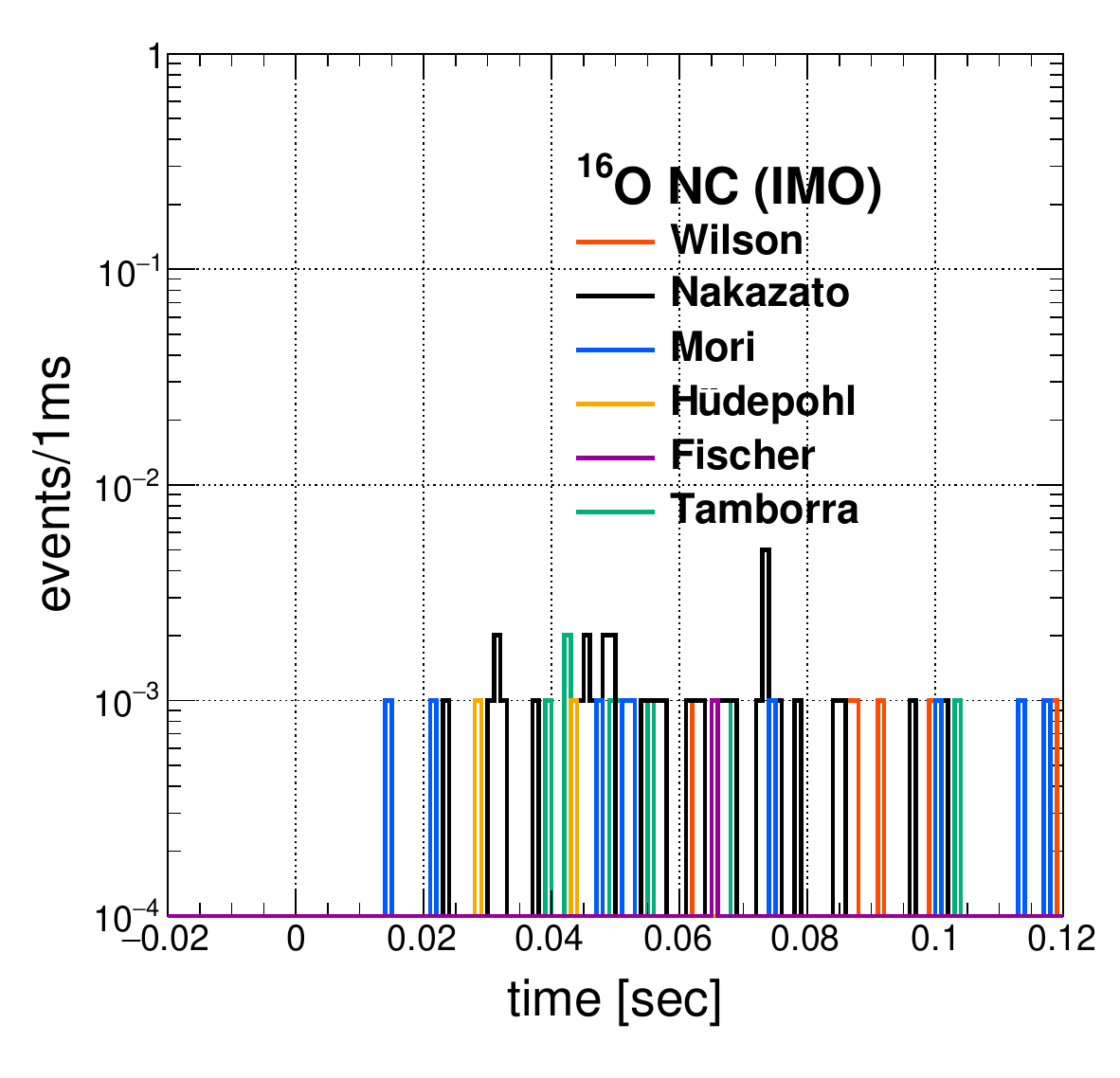}{0.3\textwidth}{}
}
\vspace{-0.5cm}
\caption{Comparison of time evolution among models for each interaction for an SN burst located at 10~kpc in neutrino oscillation with IMO.  The left, middle, and right columns show the time evolution up to 0.12~s, 1.5~s, and 20~s, respectively.  Each row represents IBD, ES, $^{16}$O~CC, and $^{16}$O~NC from top to bottom.}
\label{fig:IMOvarModelTimeInteractions}
\end{figure}

This appendix contains figures showing comparison of SK-Gd's response among models for SN burst located at 10~kpc with neutrino oscillation in the NMO and IMO scenarios, which could not be included in Section~\ref{sec:Results-truth}.
We note again that reconstructed events in the following figures correspond to events after reconstruction in SNWATCH in Figure~\ref{fig:SNWATCH_system} and after the time matching between the reconstructed time and the generated time by SKSNSim described in Section~\ref{sec:Results-truth}.
Figure~\ref{fig:NMOandIMOTimeWholeInteractionsEachModel} shows the comparison of time evolution of the number of events up to 20~s among interactions for each model in the NMO scenario (top six panels) and the IMO scenario (bottom six panels).
Figure~\ref{fig:NMOandIMOTime1sInteractionsEachModel} and Figure~\ref{fig:NMOandIMOTime100msInteractionsEachModel} show the same as Figure~\ref{fig:NMOandIMOTimeWholeInteractionsEachModel} but up to 1.5~s and 0.12~s, respectively.
From Figure~\ref{fig:NMOandIMOTime1sInteractionsEachModel}, we can see significant differences in the time structure of ES events (green) and $^{16}$O~CC events (blue);  In the Mori model (top right panel in Figure~\ref{fig:NMOandIMOTime1sInteractionsEachModel}), the ES events excess over $^{16}$O~CC events is observed at 0~s, which corresponds to the neutronization burst mentioned in Section~\ref{subsubsec:DifferenceAmongModels}.  The ES excess over $^{16}$O~CC is seen in all the models other than the Nakazato model.  In the Nakazato model, characterized by the large number of $^{16}$O~CC events within the first 0.25~s as shown in the top middle panel in Figure~\ref{fig:NMOandIMOTime1sInteractionsEachModel}, the yield of $^{16}$O~CC events exceeds the number of ES events.  The time structure of the Wilson model regarding $^{16}$O~CC is characterized by the long-lasting $^{16}$O~CC events excess over the ES events.  These excesses of $^{16}$O~CC events over ES events are caused by the larger cross section of $^{16}$O~CC interaction in higher energy regions (shown in Figure~\ref{fig:SNnuCrossSections}).  The Tamborra model has a similar characteristic to this; however, it differs from the other five models in the feature of an increasing number of events after the first neutrino emission peak; the number of events in the other models attenuates after the first neutrino emission peak.  
Figure~\ref{fig:IMOvarModelTimeInteractions} shows 
comparison of time evolution of events among models for each interaction covering three different time ranges in the IMO scenario, similar to Figure~\ref{fig:NMOvarModelTimeInteractions}.
The peaks due to the neutronization burst are stronger than those observed in Figure~\ref{fig:NMOvarModelTimeInteractions}; they are also observed in the H\"{u}depohl model (orange), slightly in the Nakazato model (black) and the Fischer model (purple), as well as for the Mori model (blue) and the Wilson model (red). 
Note that the reconstructed events in Figures~\ref{fig:NMOandIMOTimeWholeInteractionsEachModel}--\ref{fig:IMOvarModelTimeInteractions} satisfy the same event selection described in Section~\ref{subsec:EventReconstruction}.

\begin{figure}[htb!]
\gridline{
    \fig{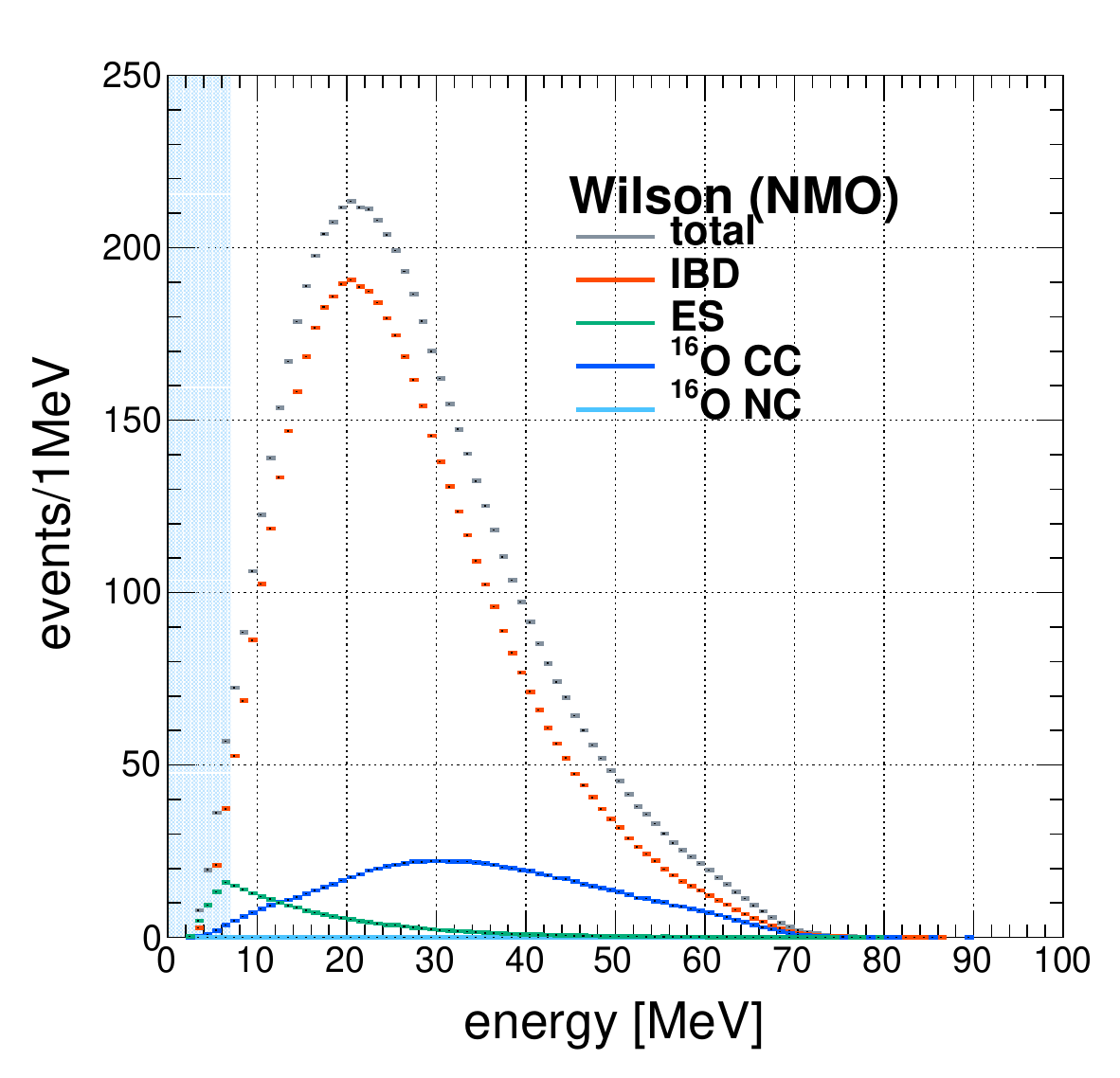}{0.31\textwidth}{(a) the Wilson model}
    \fig{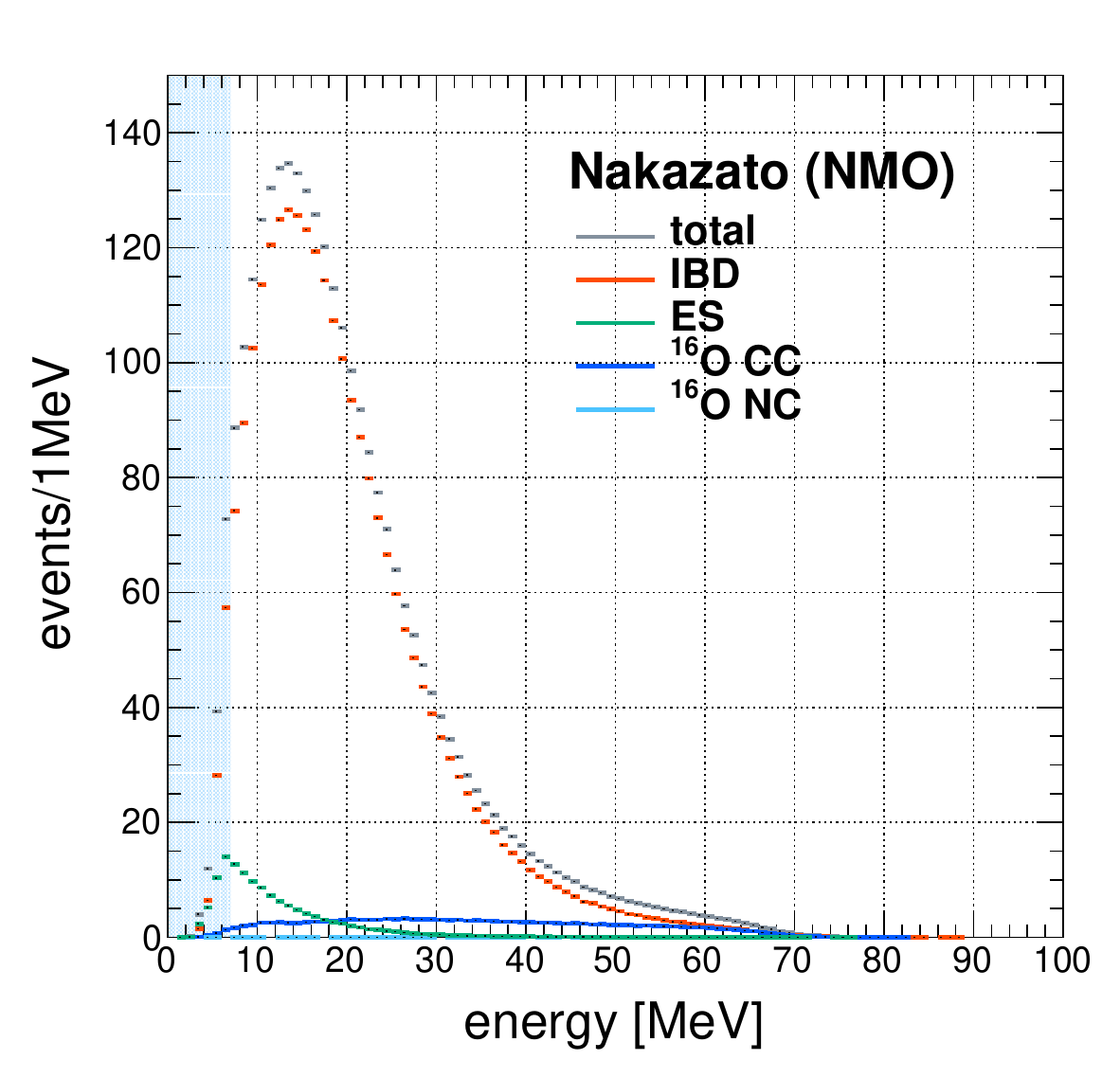}{0.31\textwidth}{(b) the Nakazato model}
    \fig{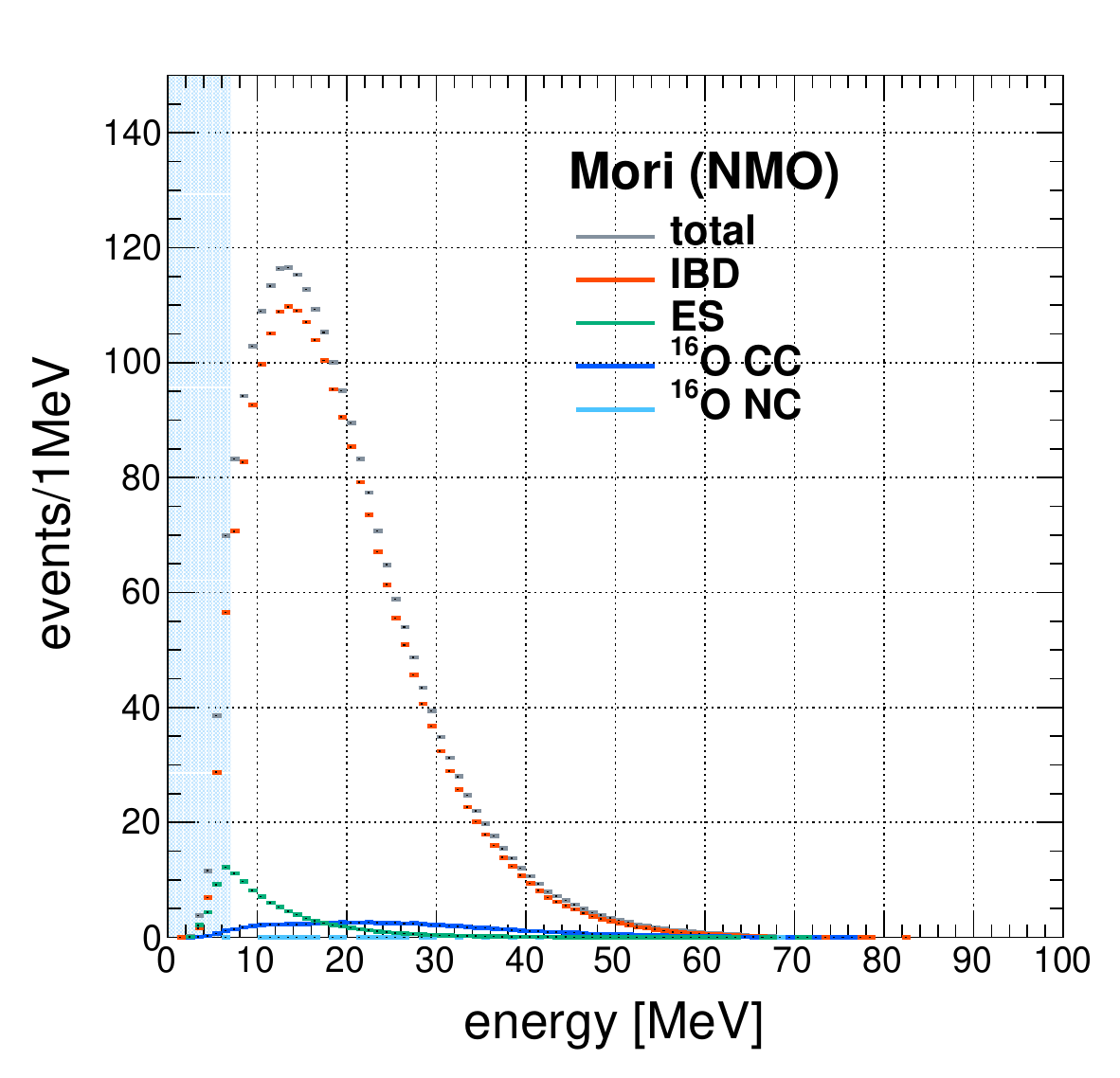}{0.31\textwidth}{(c) the Mori model}
}
\vspace{-1.14cm}
\gridline{
    \fig{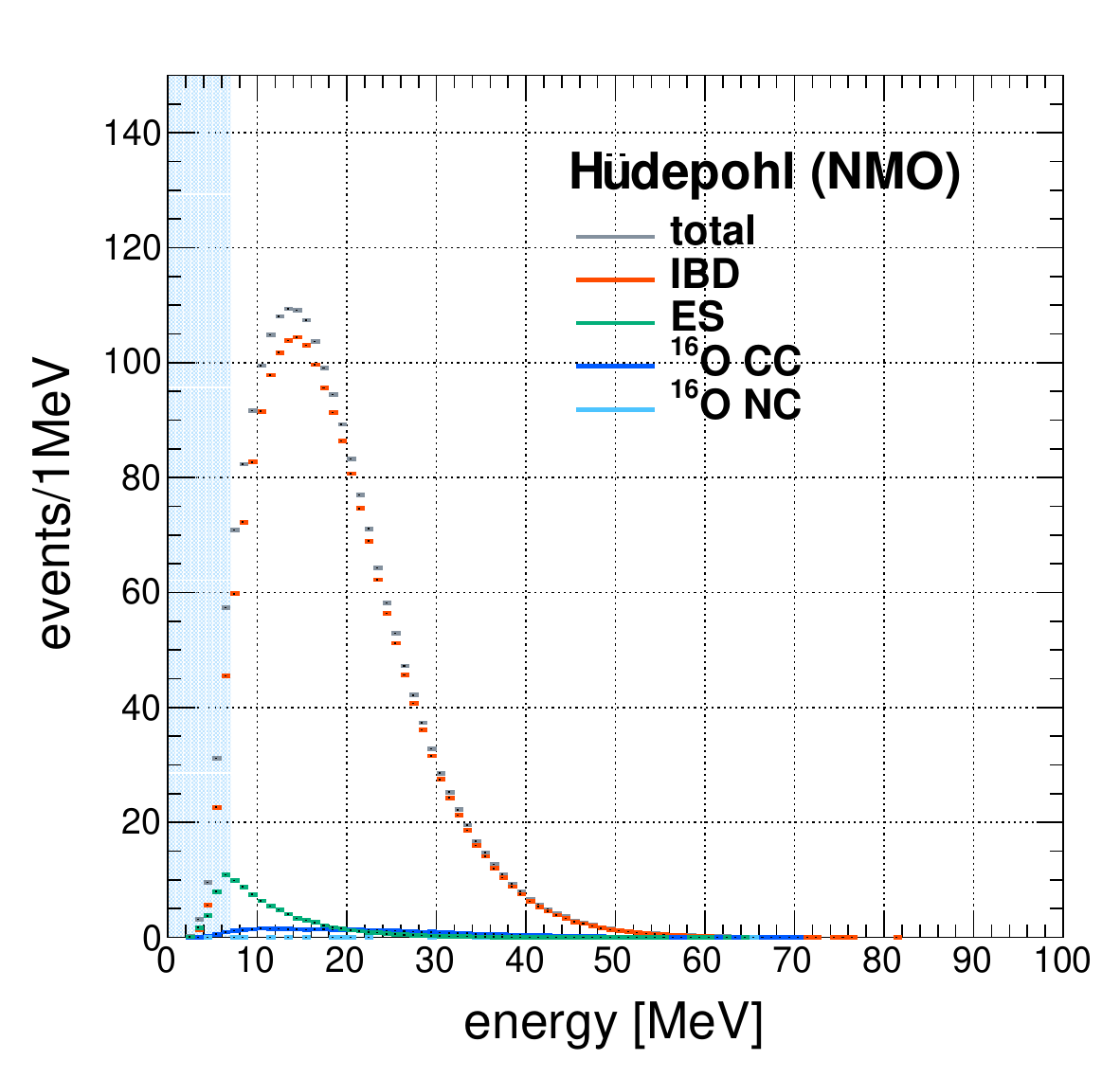}{0.31\textwidth}{(d) the H\"{u}depohl model}
    \fig{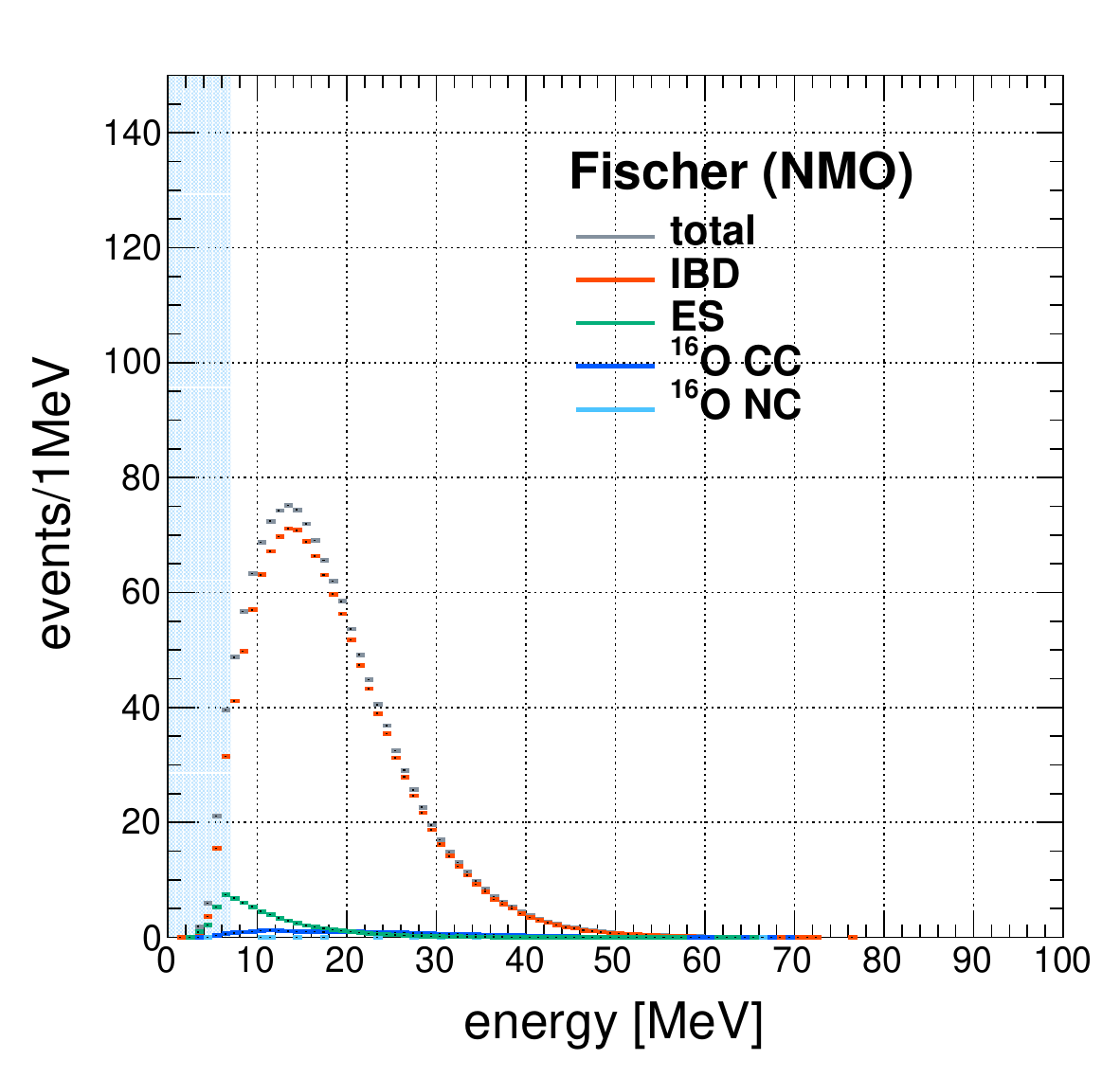}{0.31\textwidth}{(e) the Fischer model}
    \fig{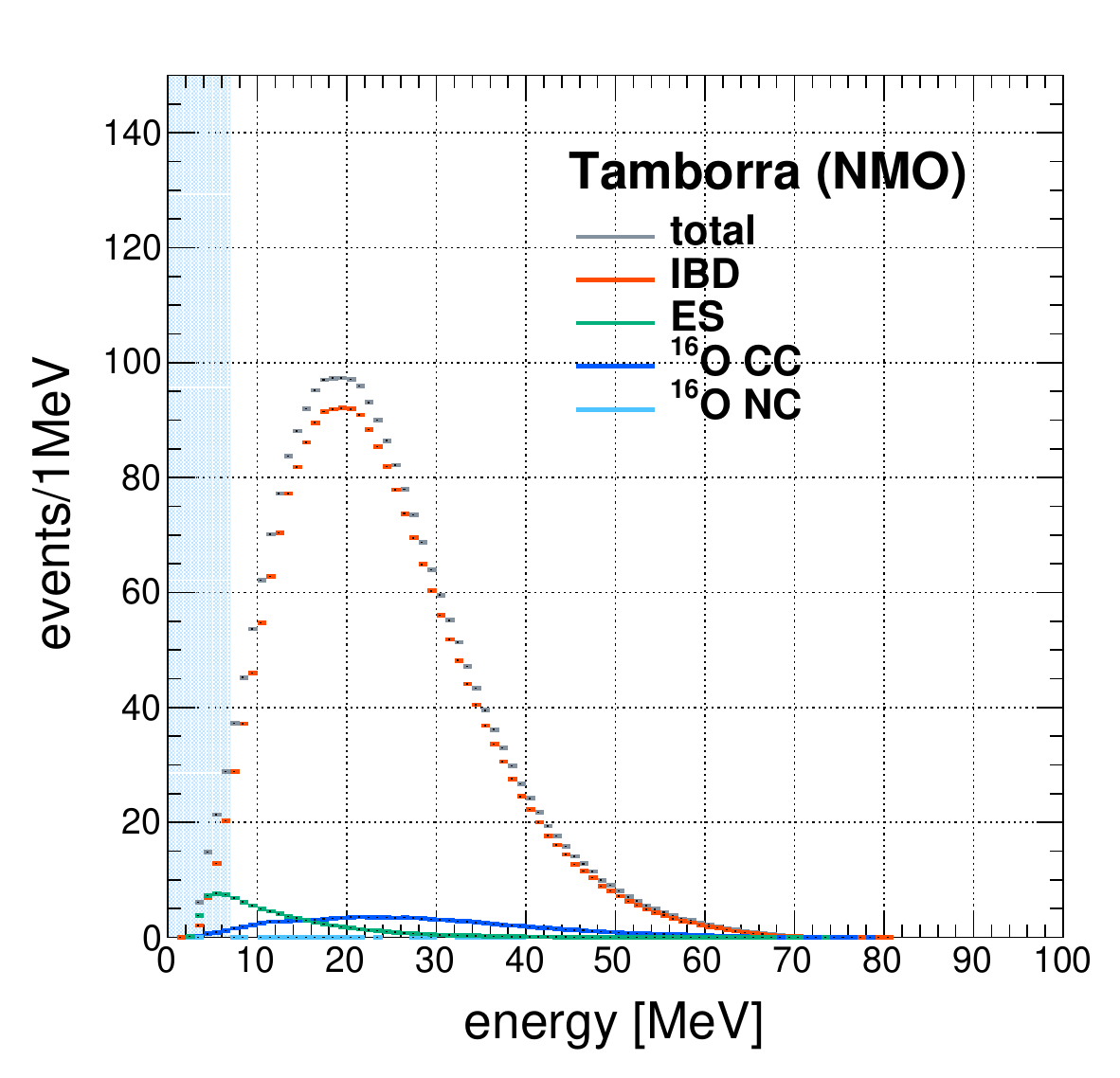}{0.31\textwidth}{(f) the Tamborra model}
}
\vspace{-1.14cm}
\gridline{
    \fig{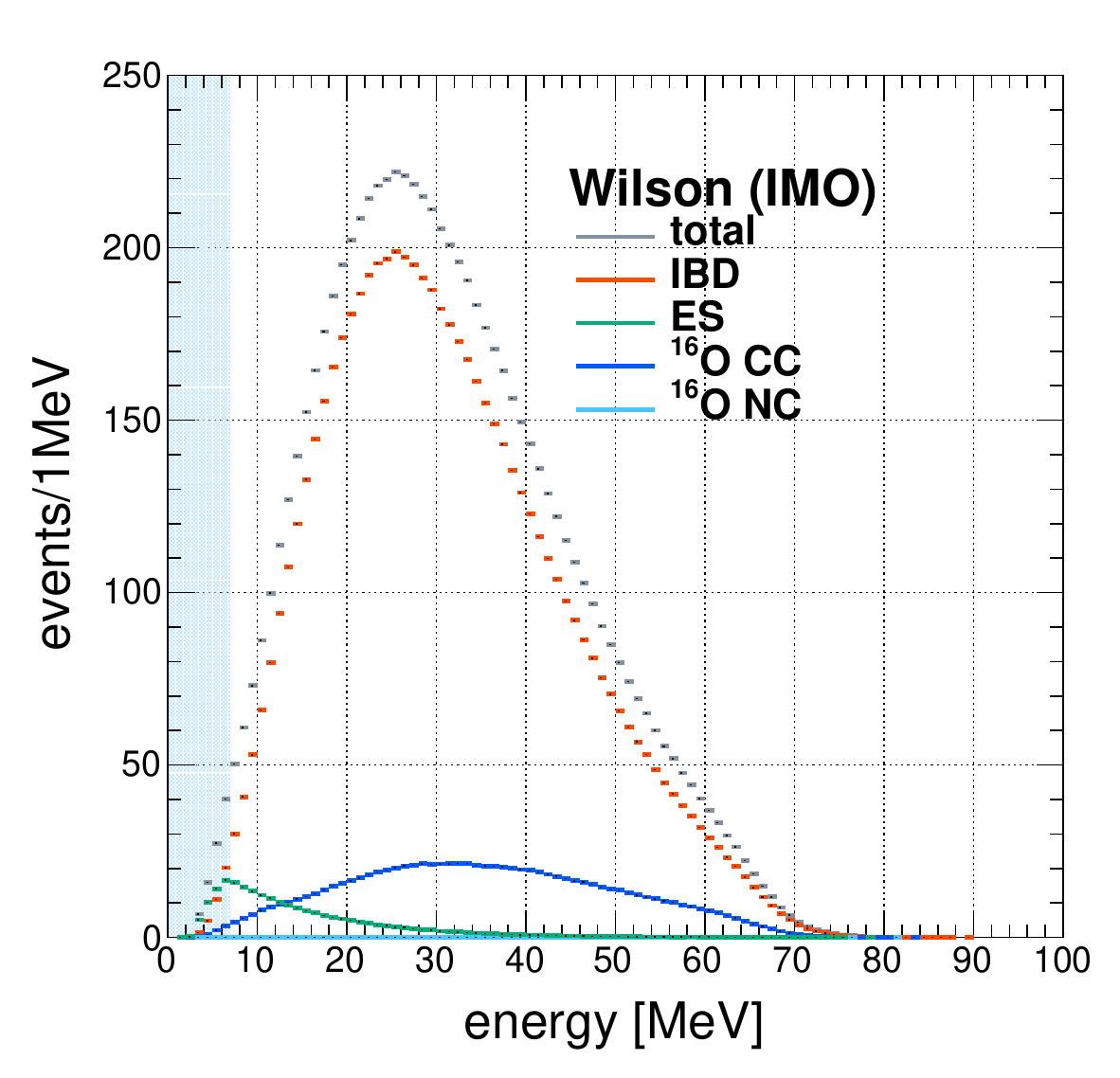}{0.31\textwidth}{(a) the Wilson model}
    \fig{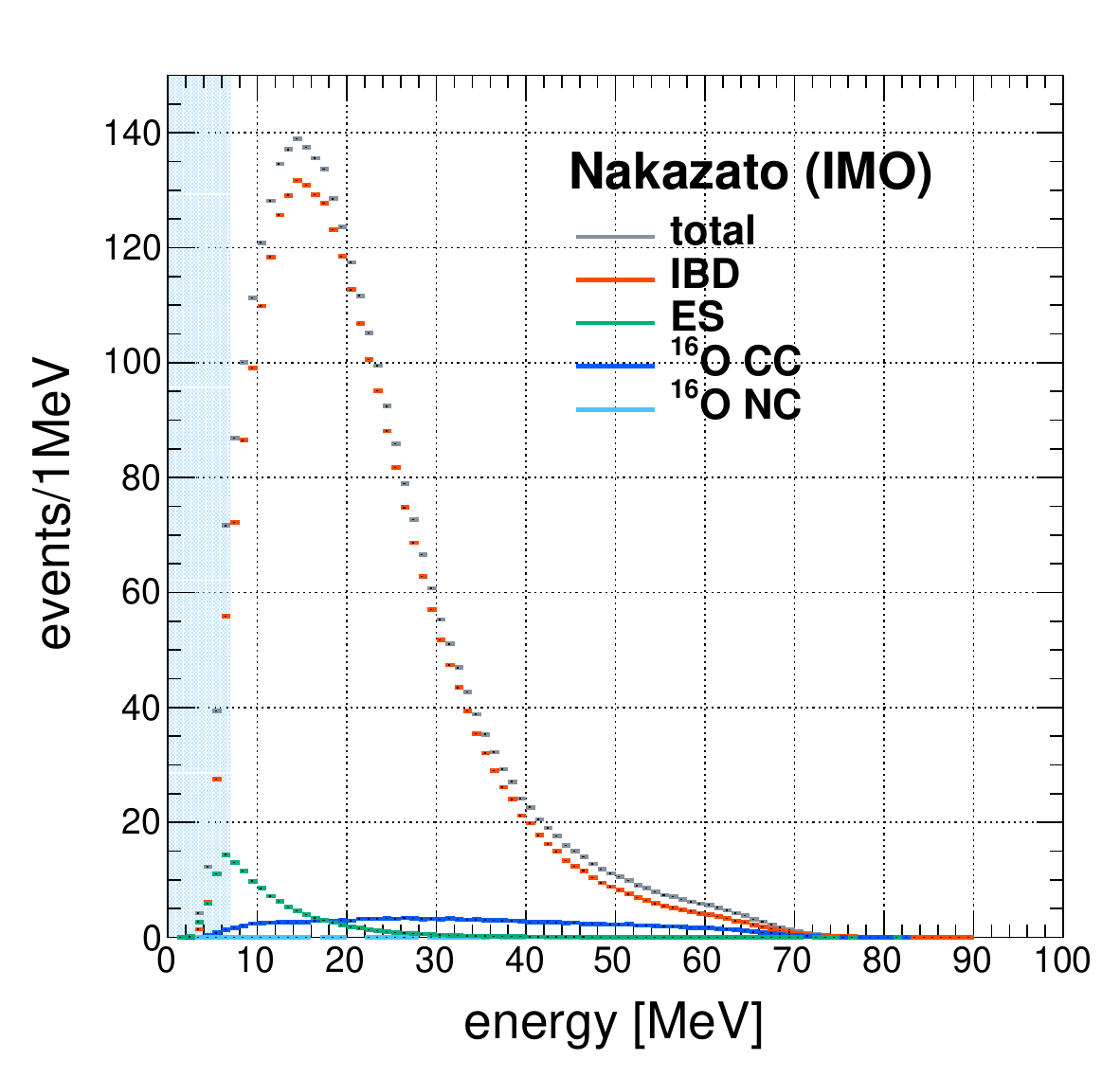}{0.31\textwidth}{(b) the Nakazato model}
    \fig{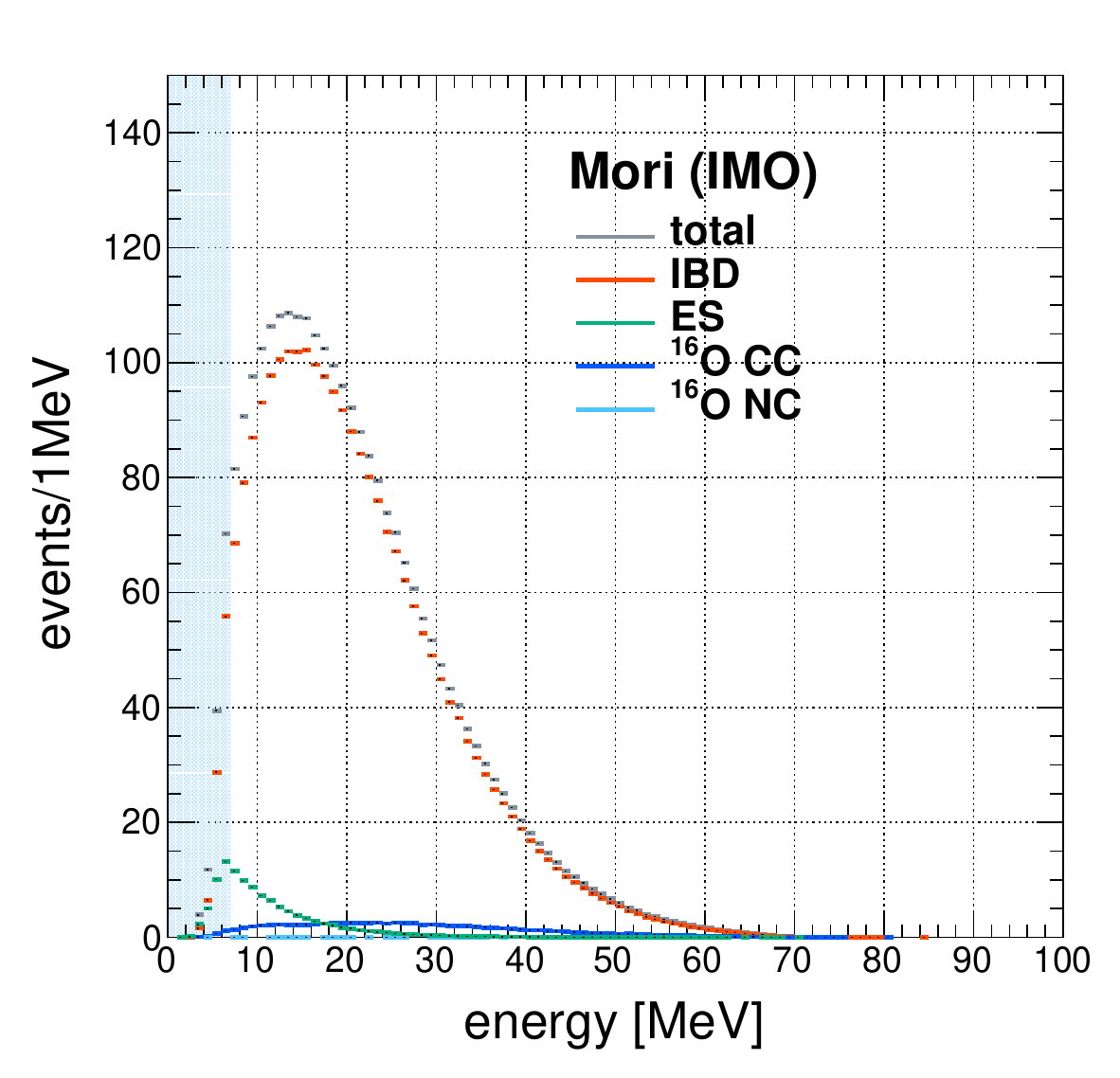}{0.31\textwidth}{(c) the Mori model}
}
\vspace{-1.14cm}
\gridline{
    \fig{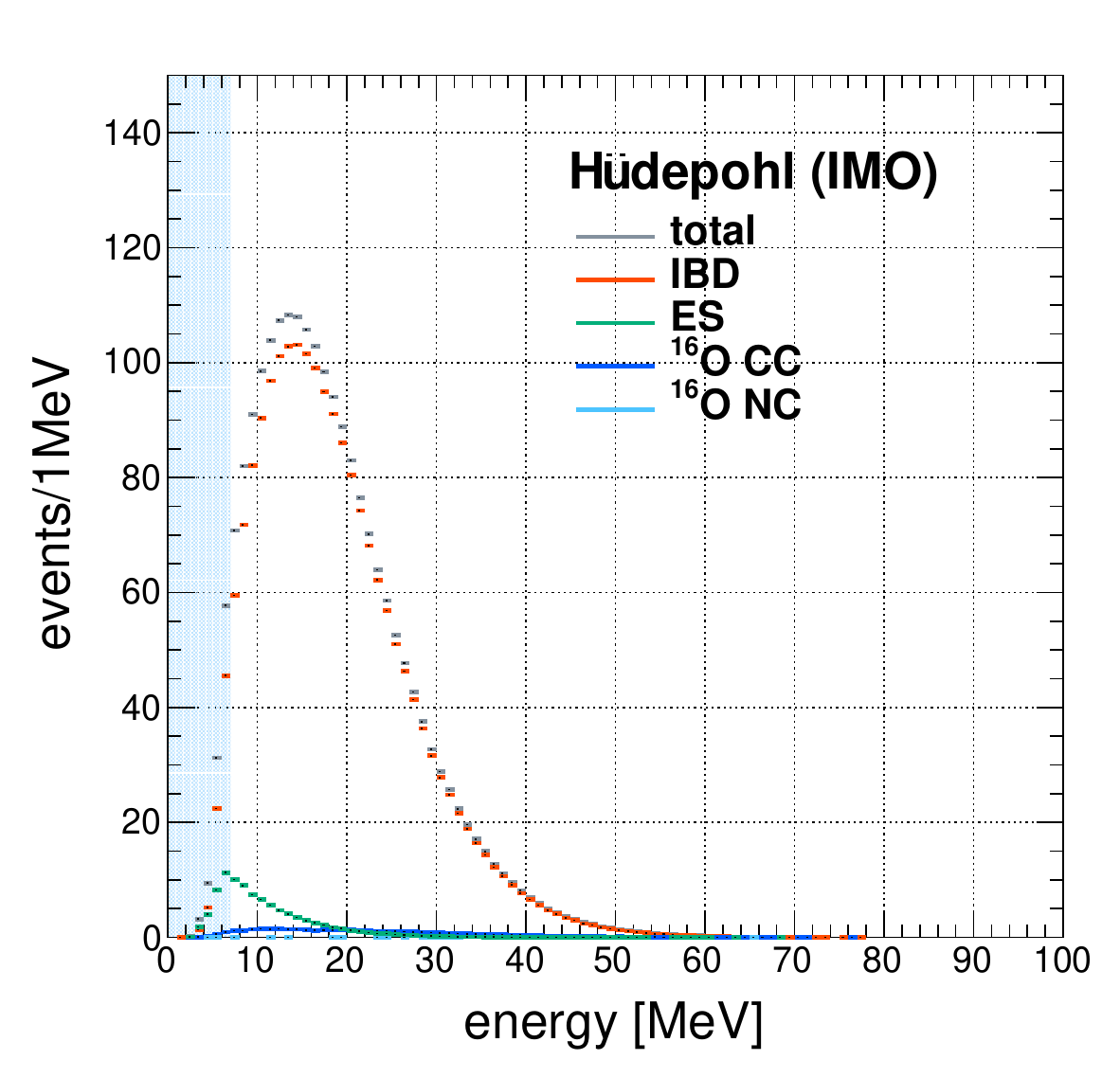}{0.31\textwidth}{}
    \fig{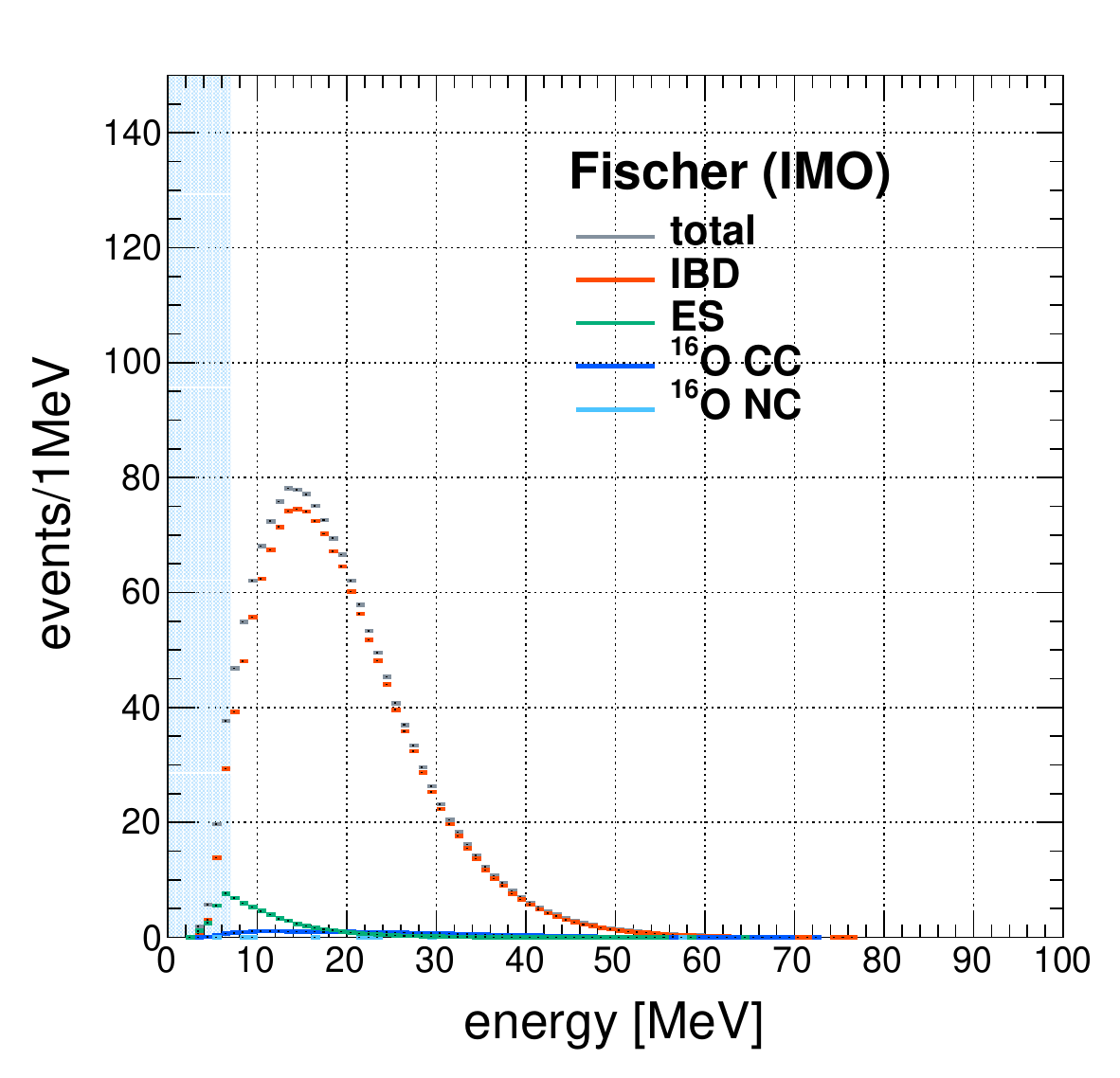}{0.31\textwidth}{}
    \fig{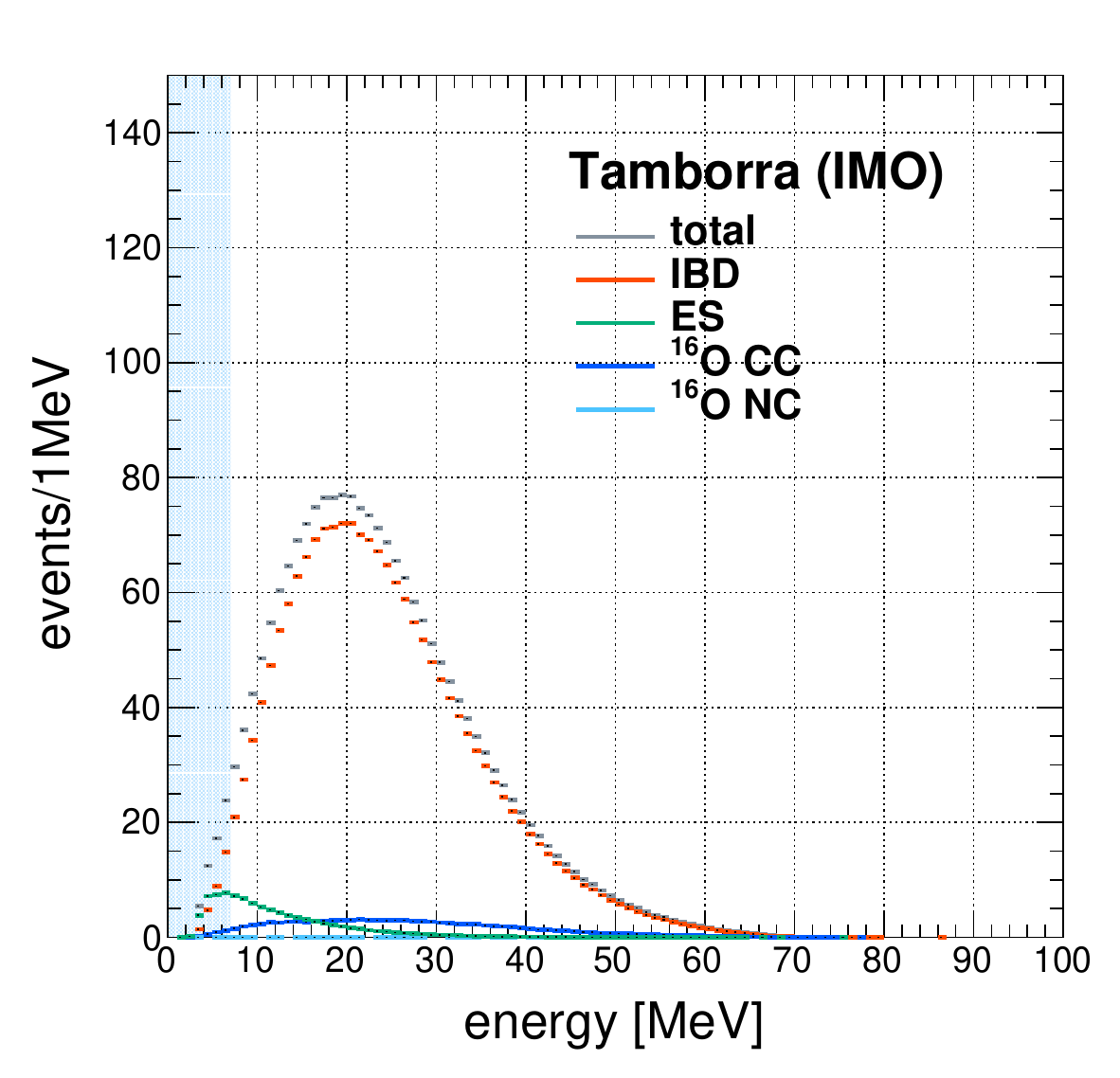}{0.31\textwidth}{}
}
\vspace{-0.5cm}
\caption{Comparison of energy spectra among interactions for each model in the IMO scenario. The energy region below the 7~MeV threshold for selecting ``prompt'' candidates is shaded in light blue.}
\label{fig:NMOandIMOEnergyInteractionsEachModel}
\end{figure}

\begin{figure}[htb!]
\gridline{\fig{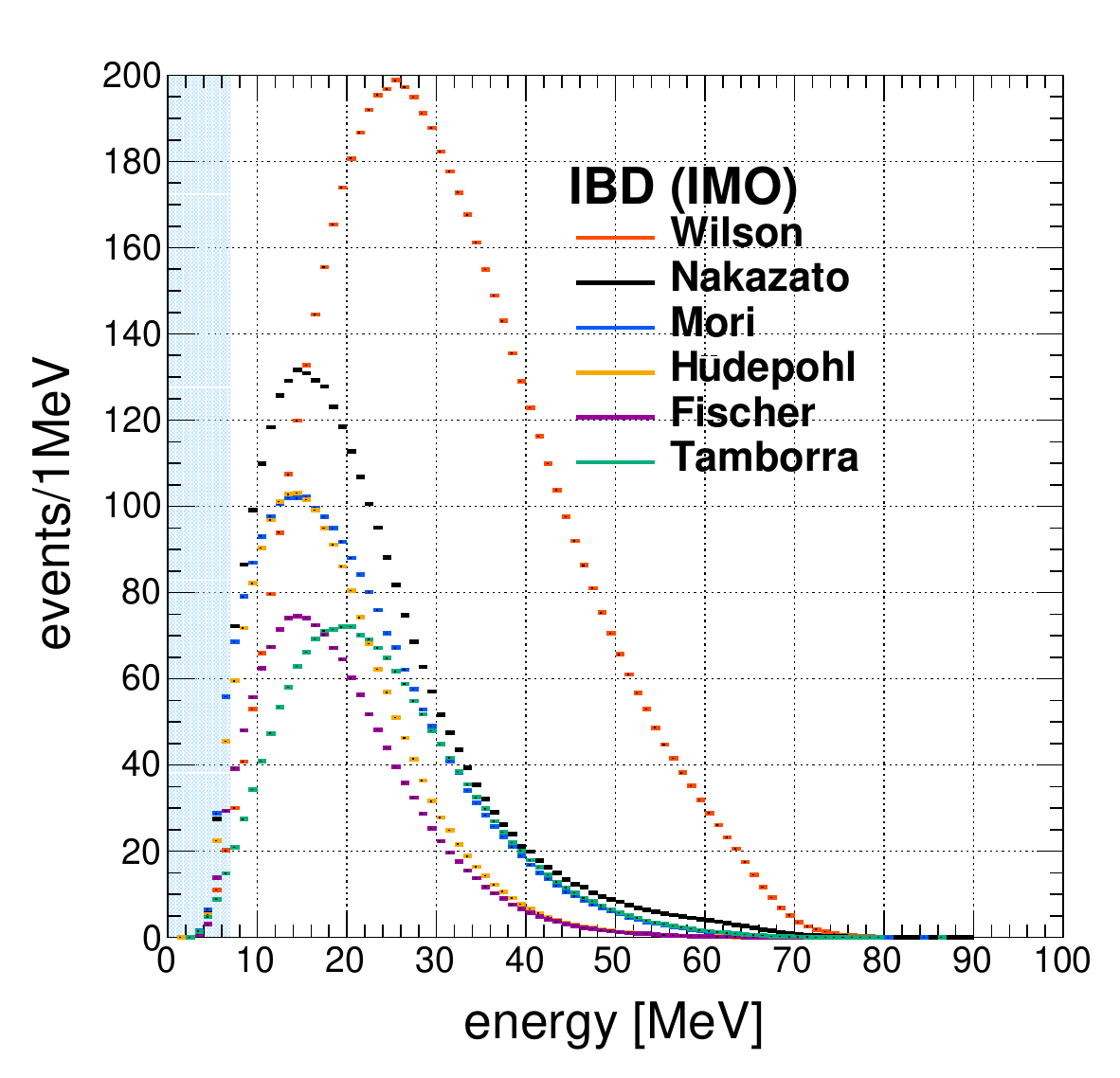}{0.35\textwidth}{(a) IBD}
          \fig{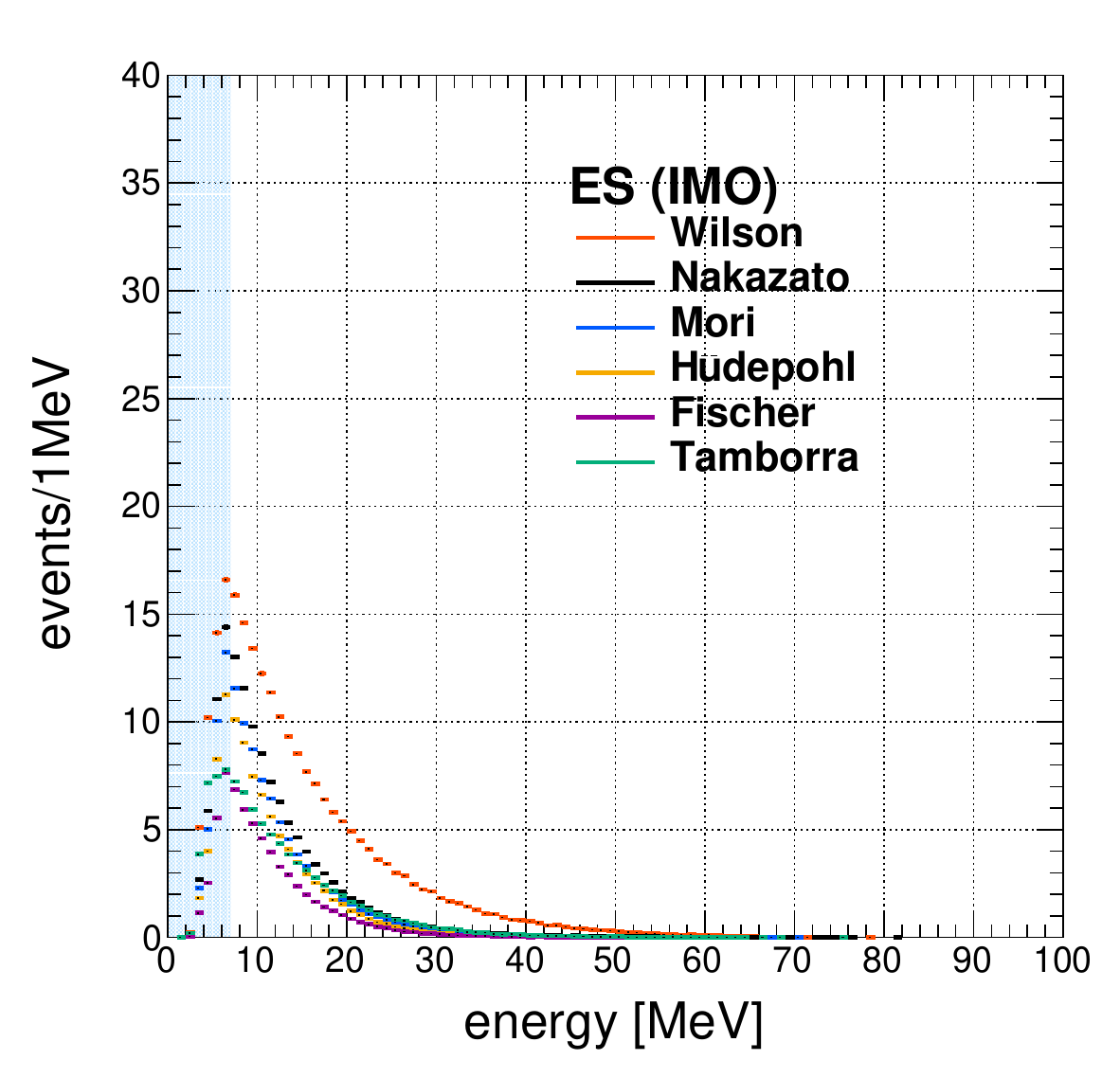}{0.35\textwidth}{(b) ES}}
\gridline{\fig{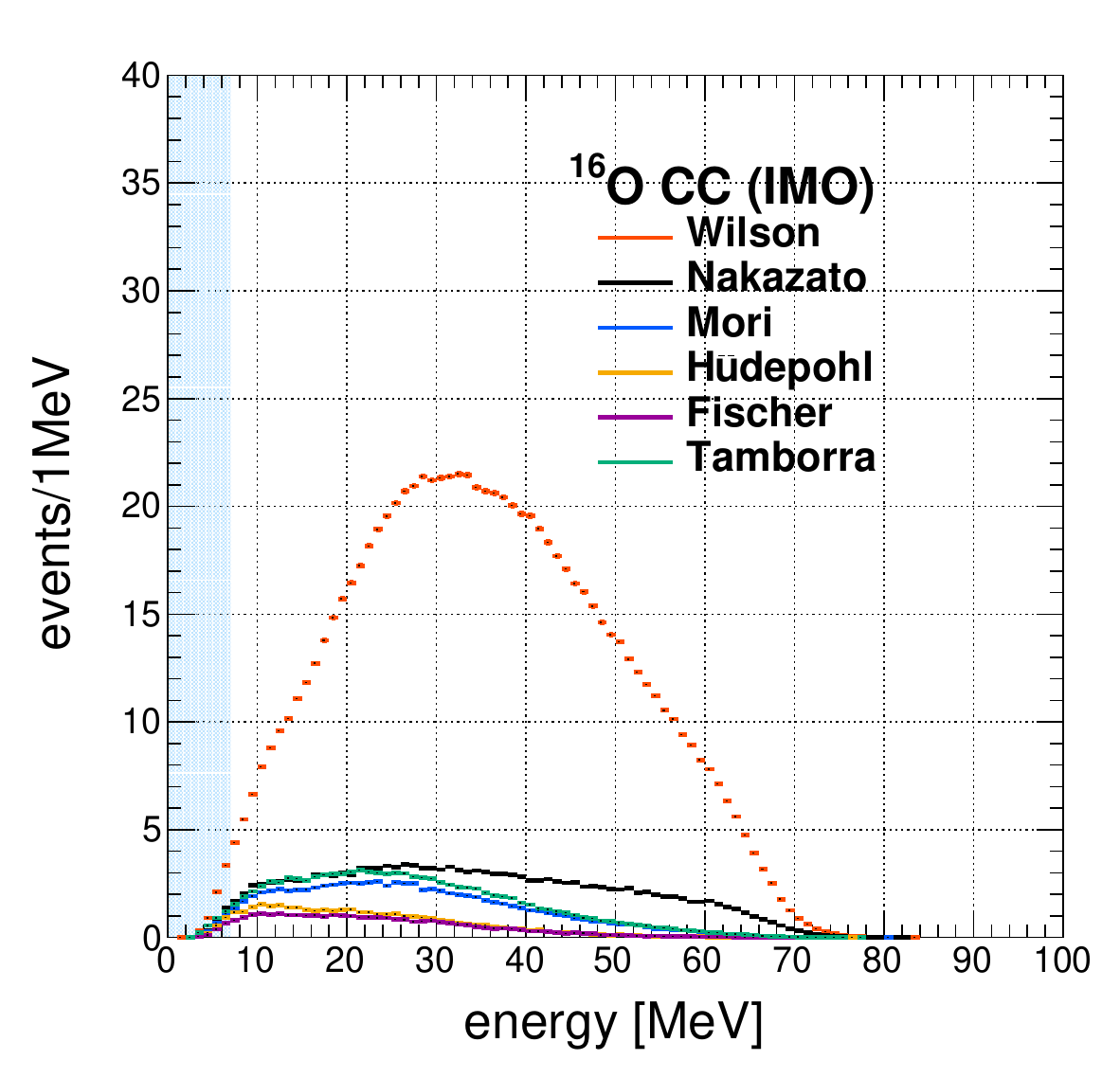}{0.35\textwidth}{(c) $^{16}$O~CC}
          \fig{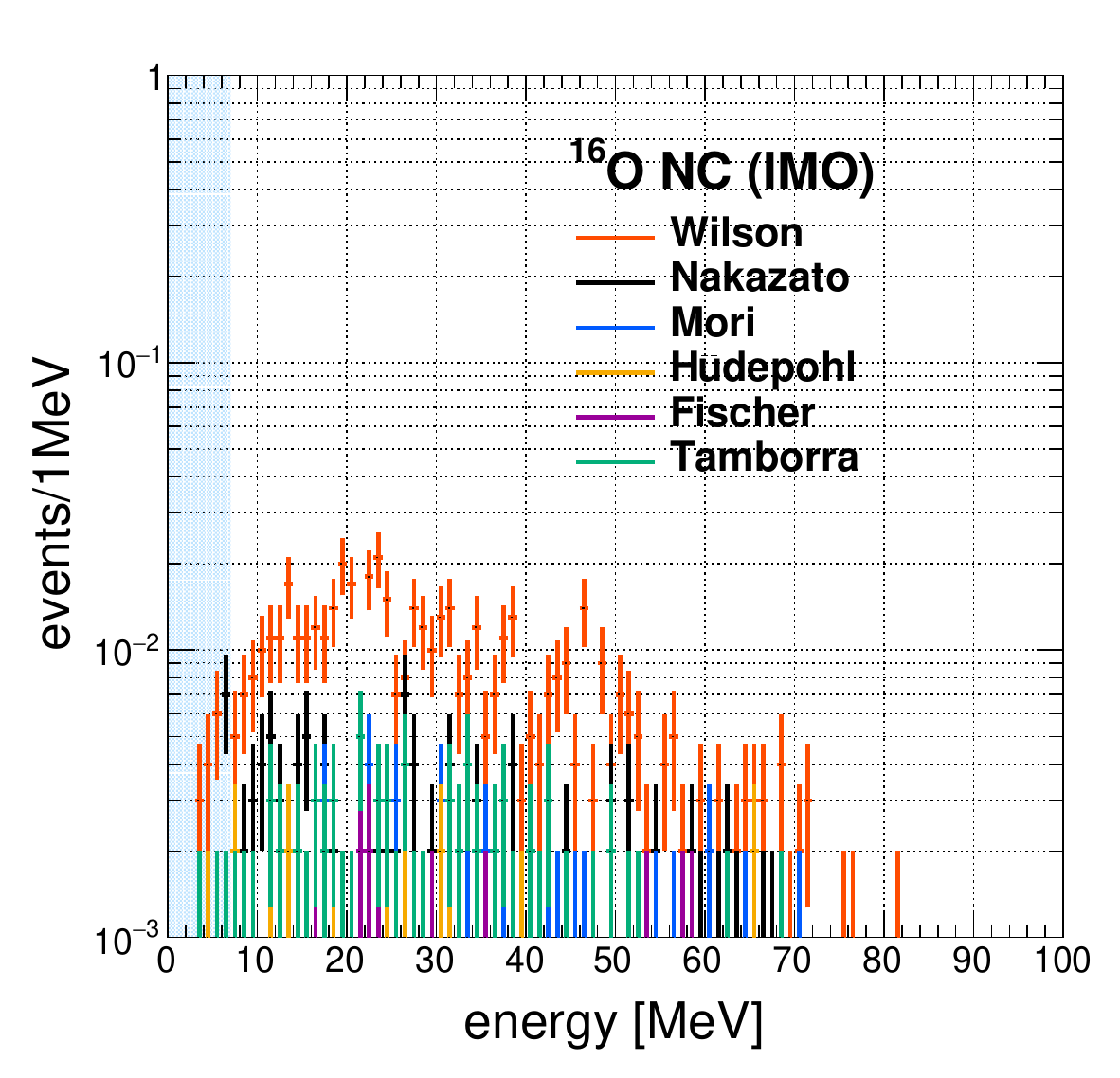}{0.35\textwidth}{(d) $^{16}$O~NC}}
\caption{Comparison of energy spectra among models for each interaction for an SN located at 10~kpc: (a)~IBD, (b)~ES, (c)~$^{16}$O~CC, and (d)~$^{16}$O~NC.  The energy region below the 7~MeV threshold for selecting ``prompt'' candidates is shaded in light blue.}
\label{fig:IMOvarModelEnergyInteractions}
\end{figure}

Figure~\ref{fig:NMOandIMOEnergyInteractionsEachModel} shows the energy spectra among interactions for each model in the NMO scenario (top six panels) and the IMO scenario (bottom six panels).  
It is obviously seen that the energy contribution from IBD events is dominant for all models.
Comparison of energy spectra among models in the IMO scenario is shown in Figure~\ref{fig:IMOvarModelEnergyInteractions}, similar to Figure~\ref{fig:NMOvarModelEnergyInteractions}.   
Note that the reconstructed events in Figures~\ref{fig:NMOandIMOEnergyInteractionsEachModel}--\ref{fig:IMOvarModelEnergyInteractions} satisfy the same event selection described in Section~\ref{subsec:EventReconstruction} except for the energy condition.

\begin{figure}[htb!]
\gridline{
    \fig{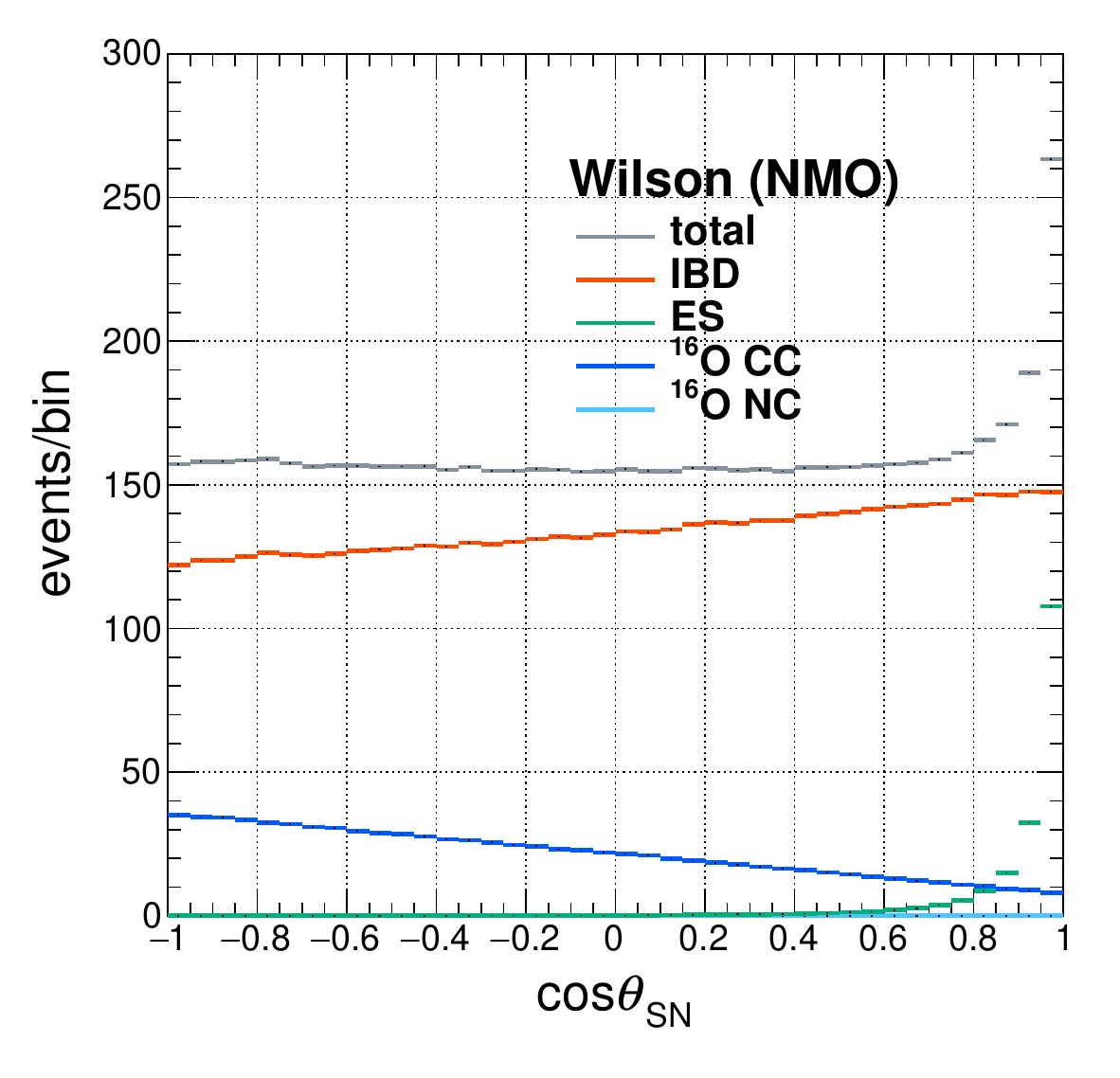}{0.33\textwidth}{(a) the Wilson model}
    \fig{Modification_23Dec_AngularDistribution_mtimePrompt_10kpc_NMO_recoCos_reactions_Nakazato.pdf}{0.33\textwidth}{(b) the Nakazato model}
    \fig{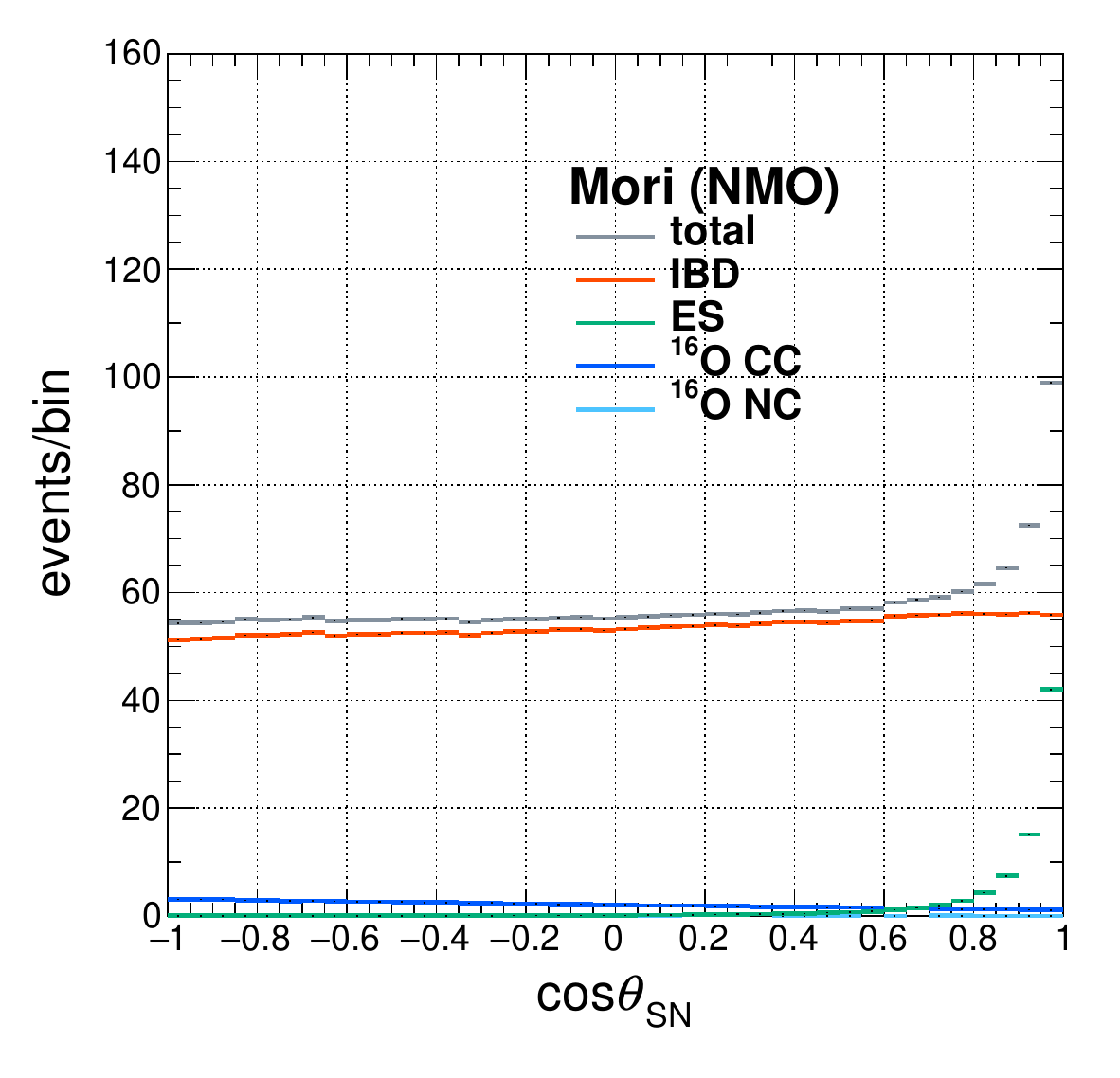}{0.33\textwidth}{(c) the Mori model}
}
\vspace{-1.2cm}
\gridline{
    \fig{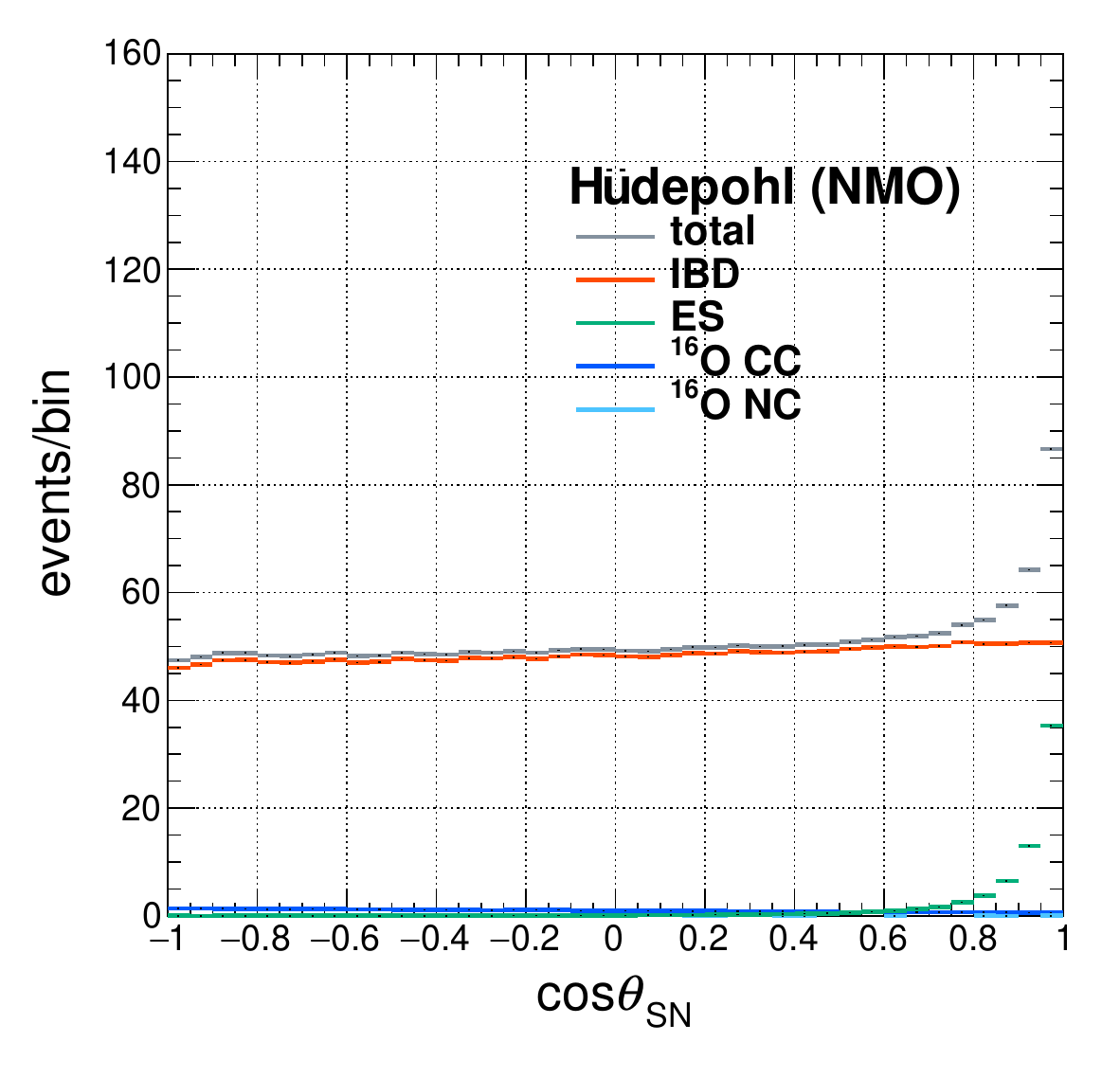}{0.33\textwidth}{(d) the H\"{u}depohl model}
    \fig{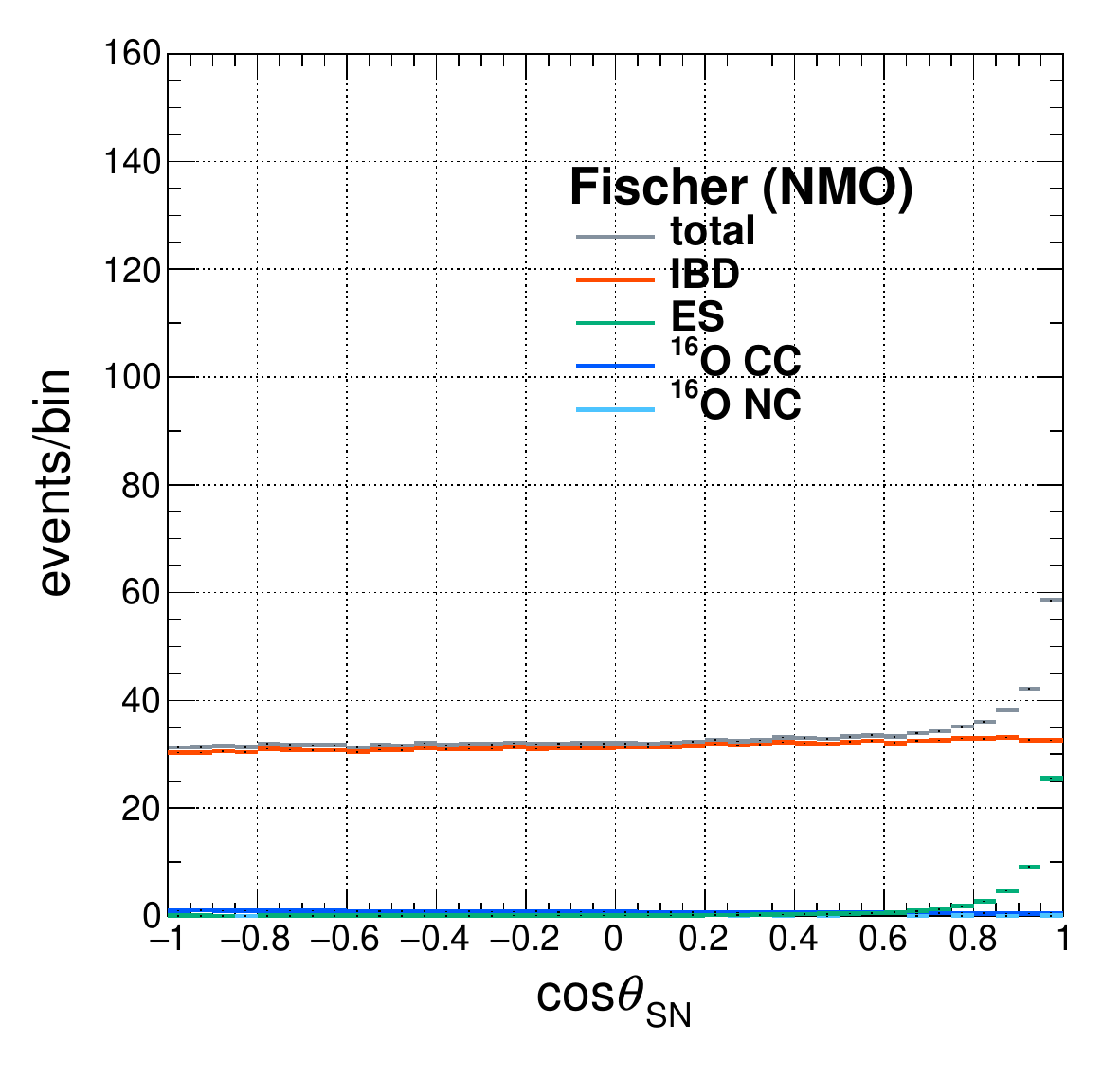}{0.33\textwidth}{(e) the Fischer model}
    \fig{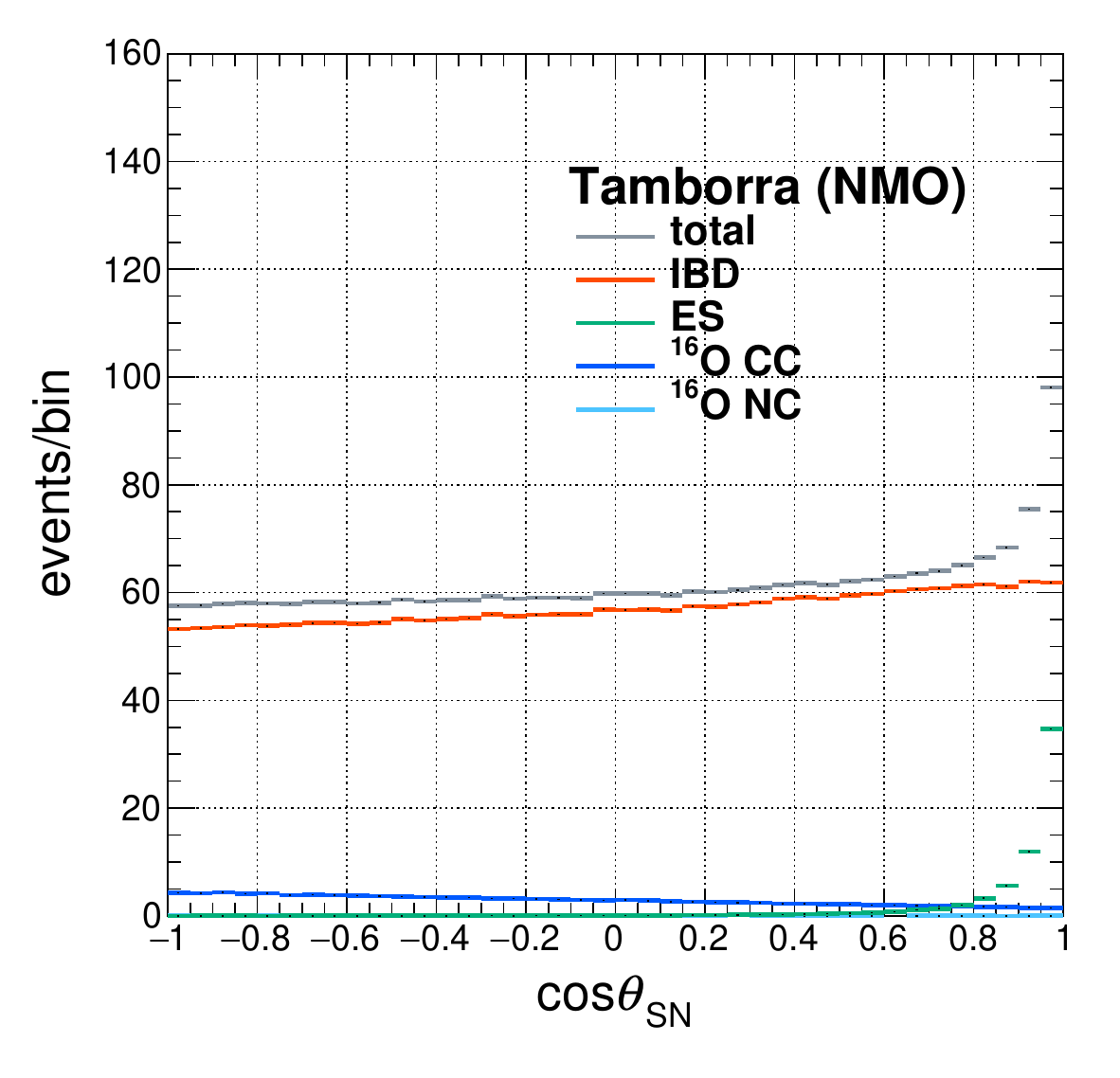}{0.33\textwidth}{(f) the Tamborra model}
}
\vspace{-1.2cm}
\gridline{
    \fig{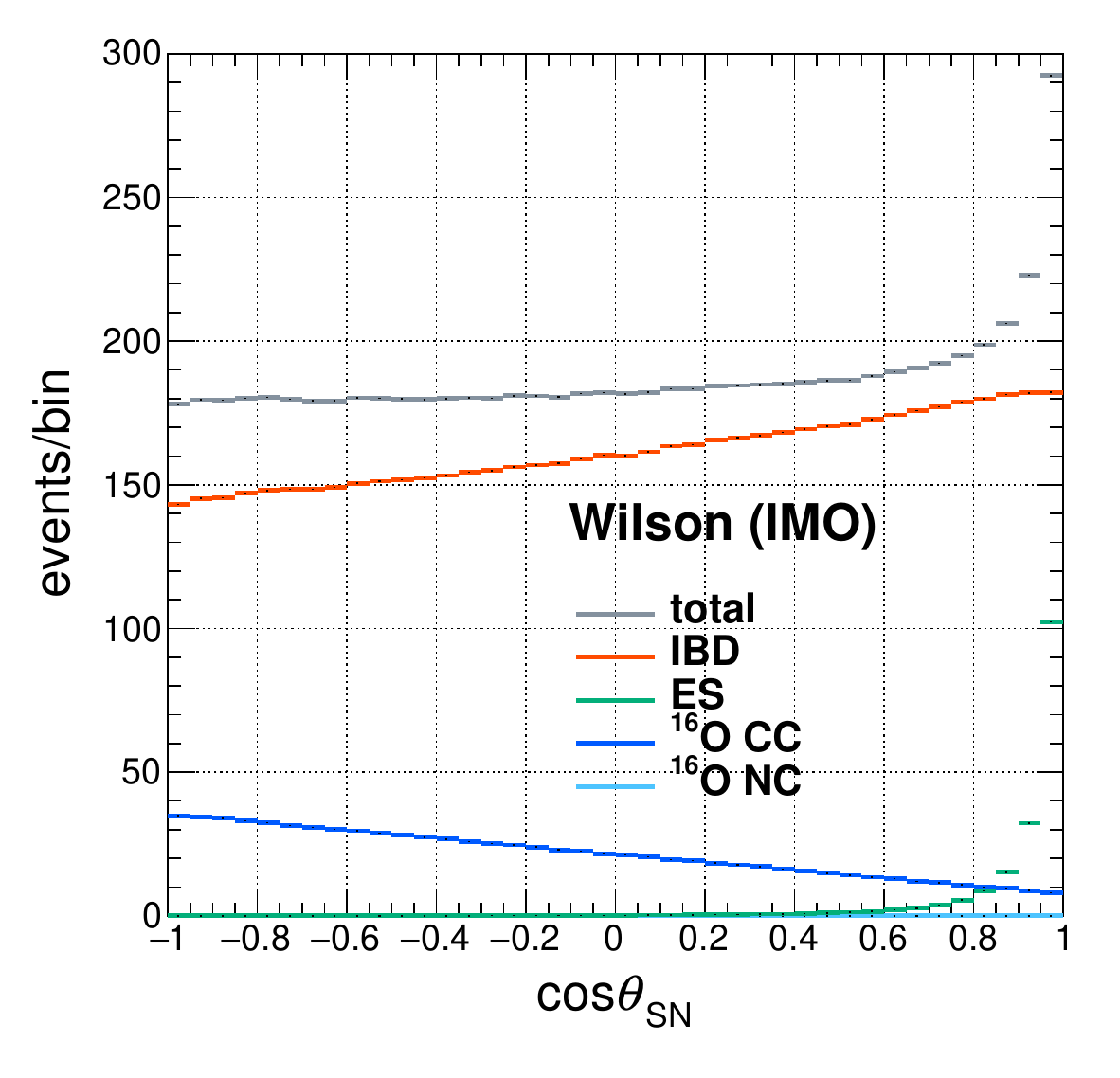}{0.33\textwidth}{}
    \fig{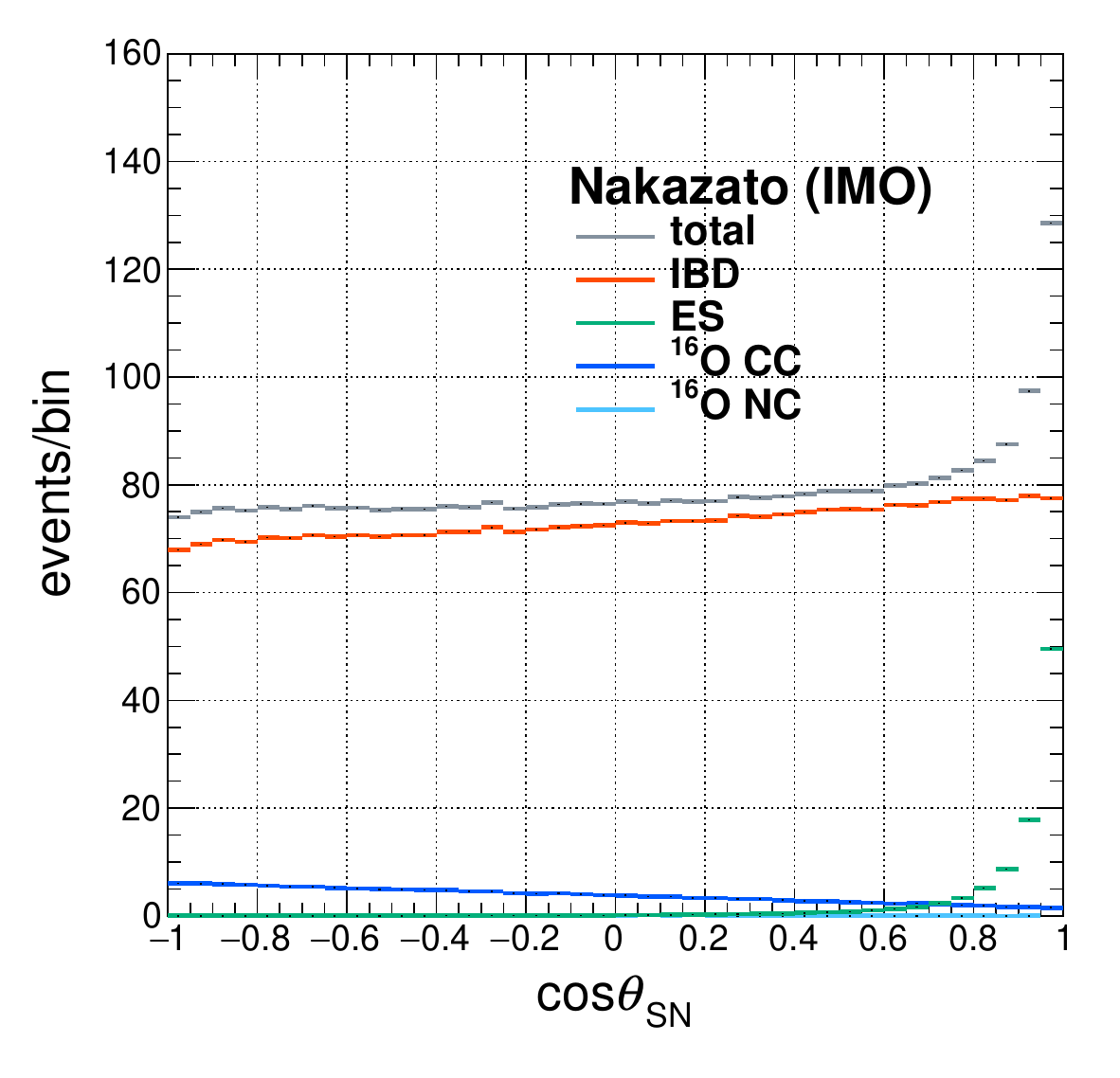}{0.33\textwidth}{}
    \fig{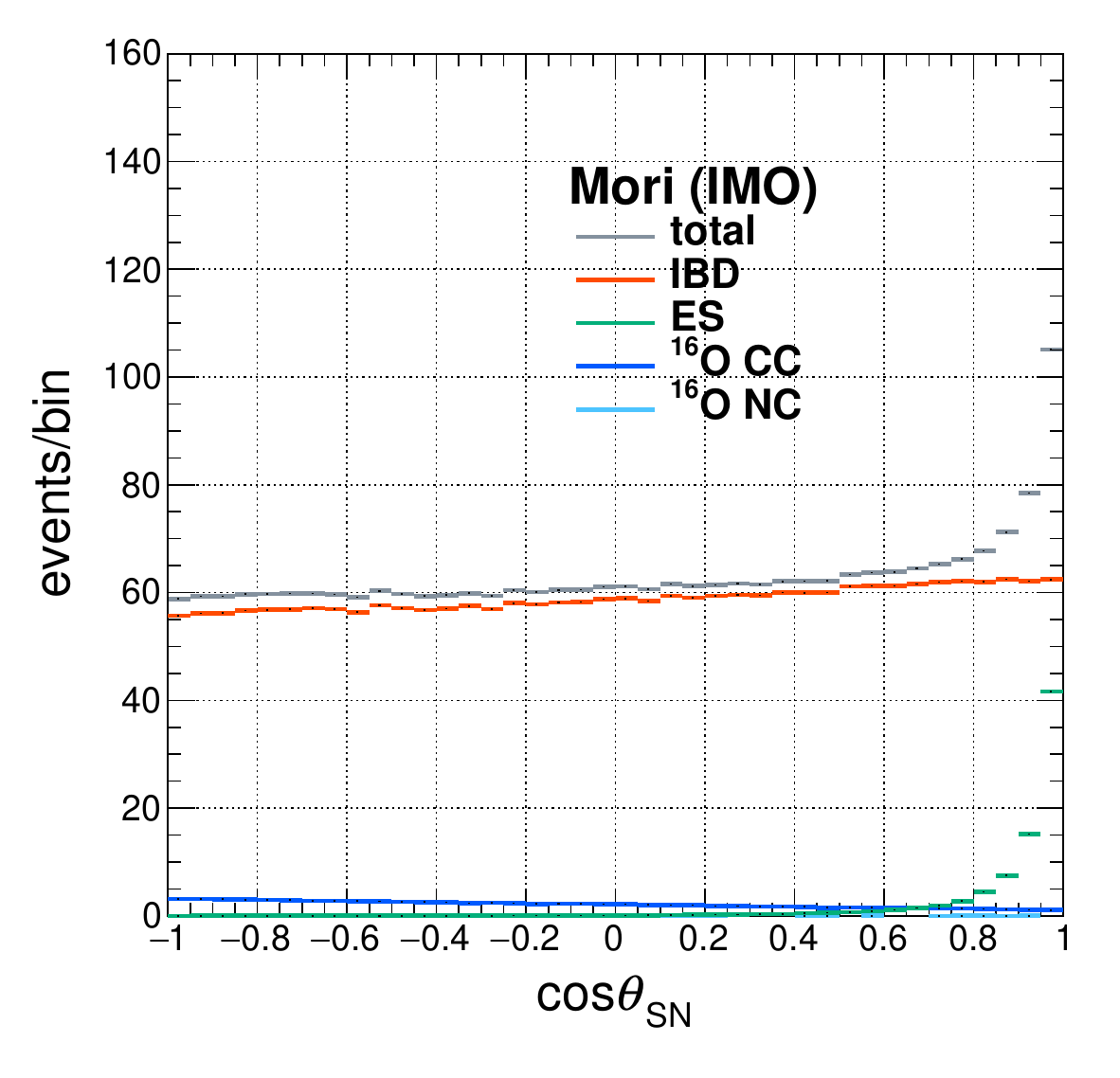}{0.33\textwidth}{}
}
\vspace{-1.2cm}
\gridline{
    \fig{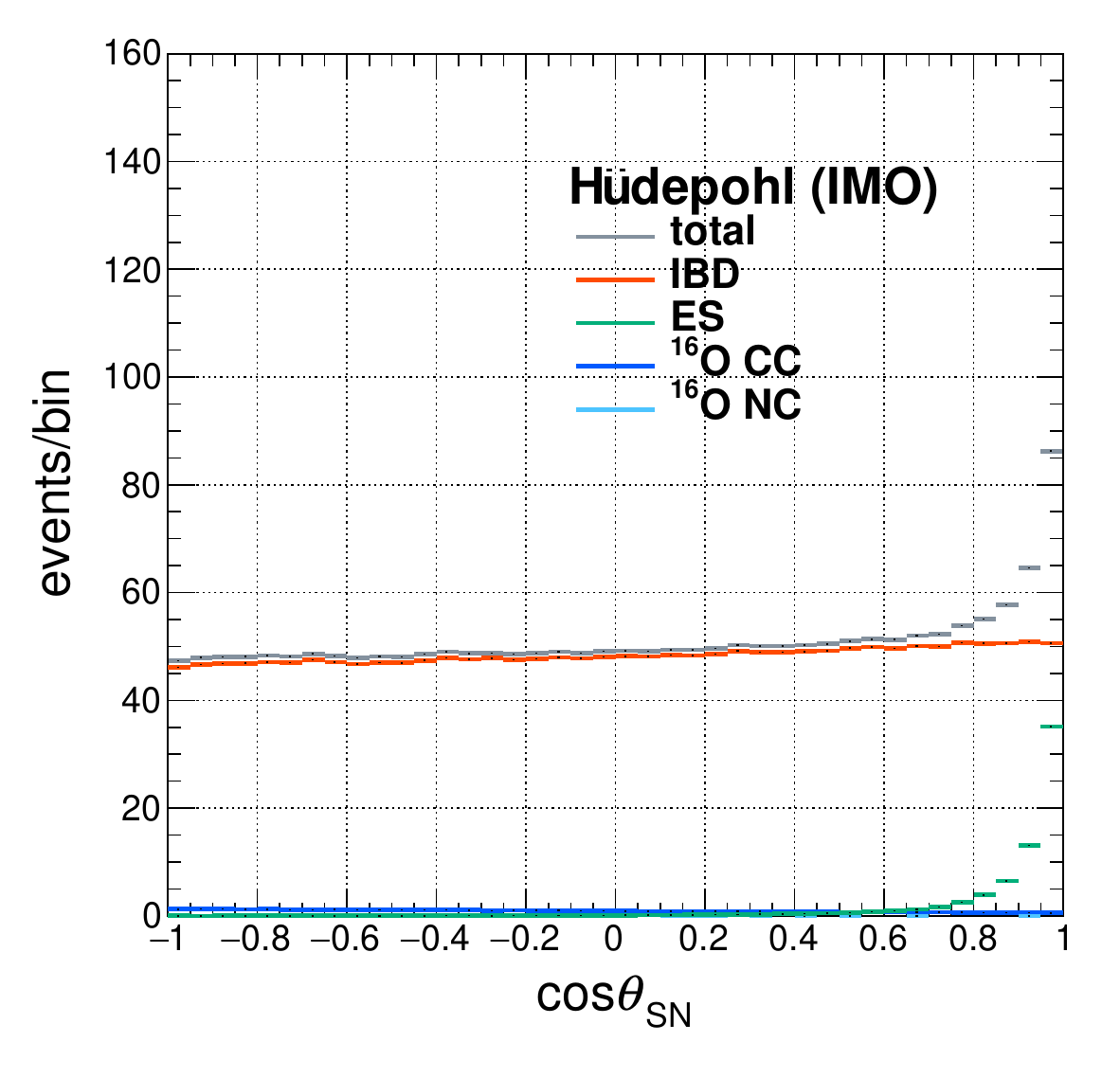}{0.33\textwidth}{}
    \fig{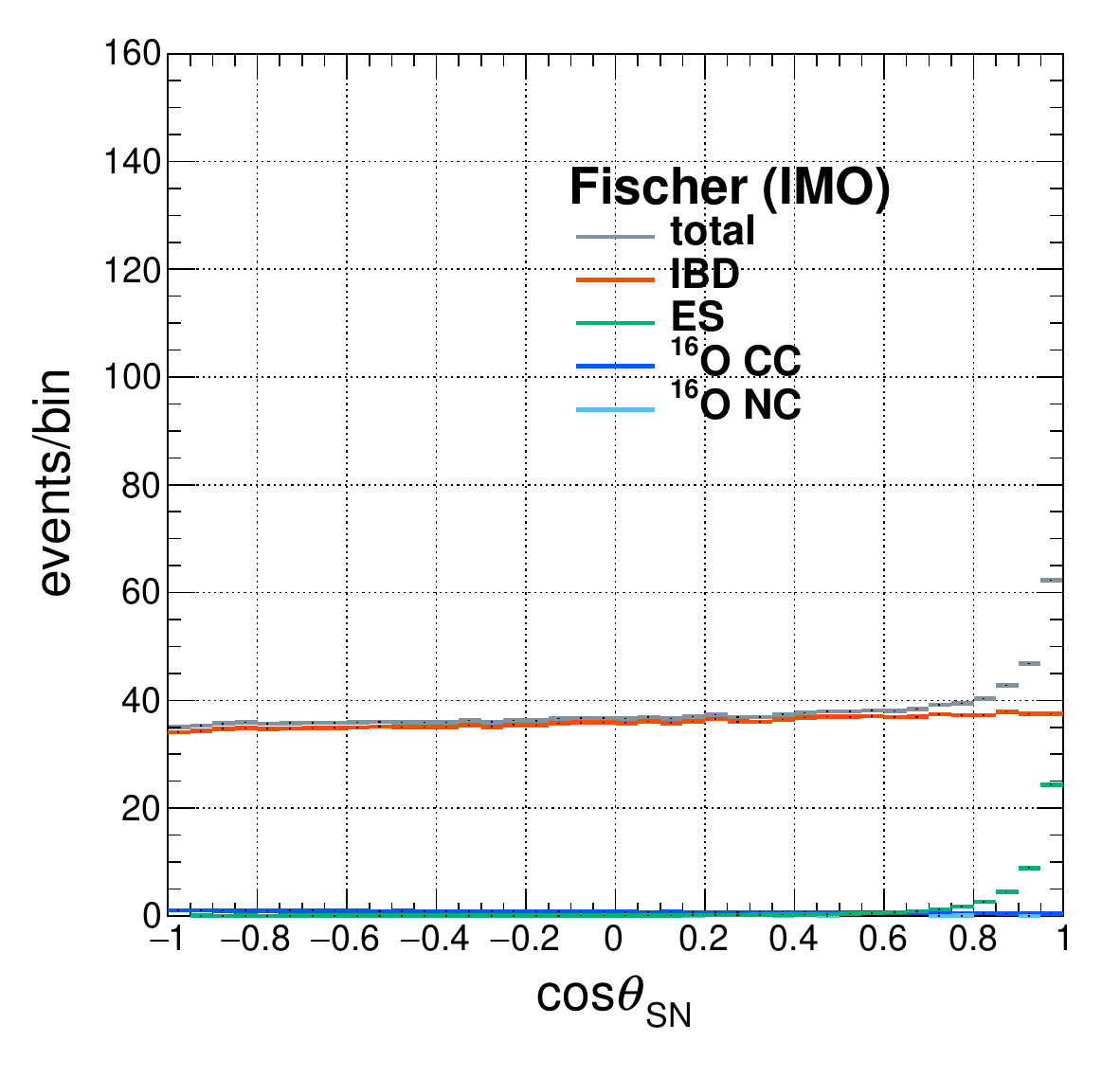}{0.33\textwidth}{}
    \fig{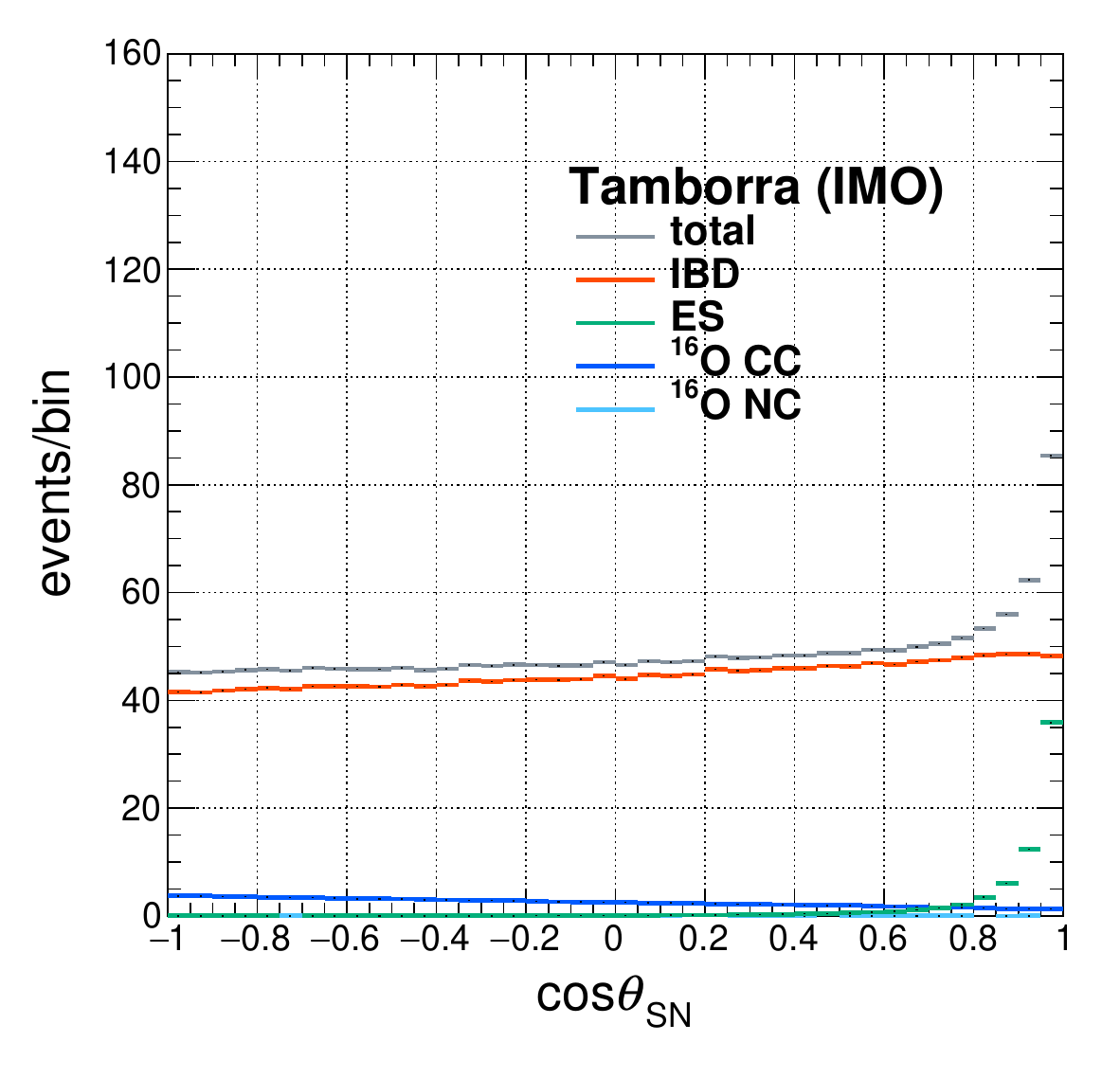}{0.33\textwidth}{}
}
\vspace{-0.5cm}
\caption{Comparison of $\cos\theta_\mathrm{SN}$ distribution among interactions for each model for an SN burst located at 10~kpc in the NMO scenario (top six panels) and the IMO scenario (bottom six panels).}
\label{fig:CosineInteractionsEachModelNMOandIMO}
\end{figure}

\begin{figure}[htb!]
\gridline{\fig{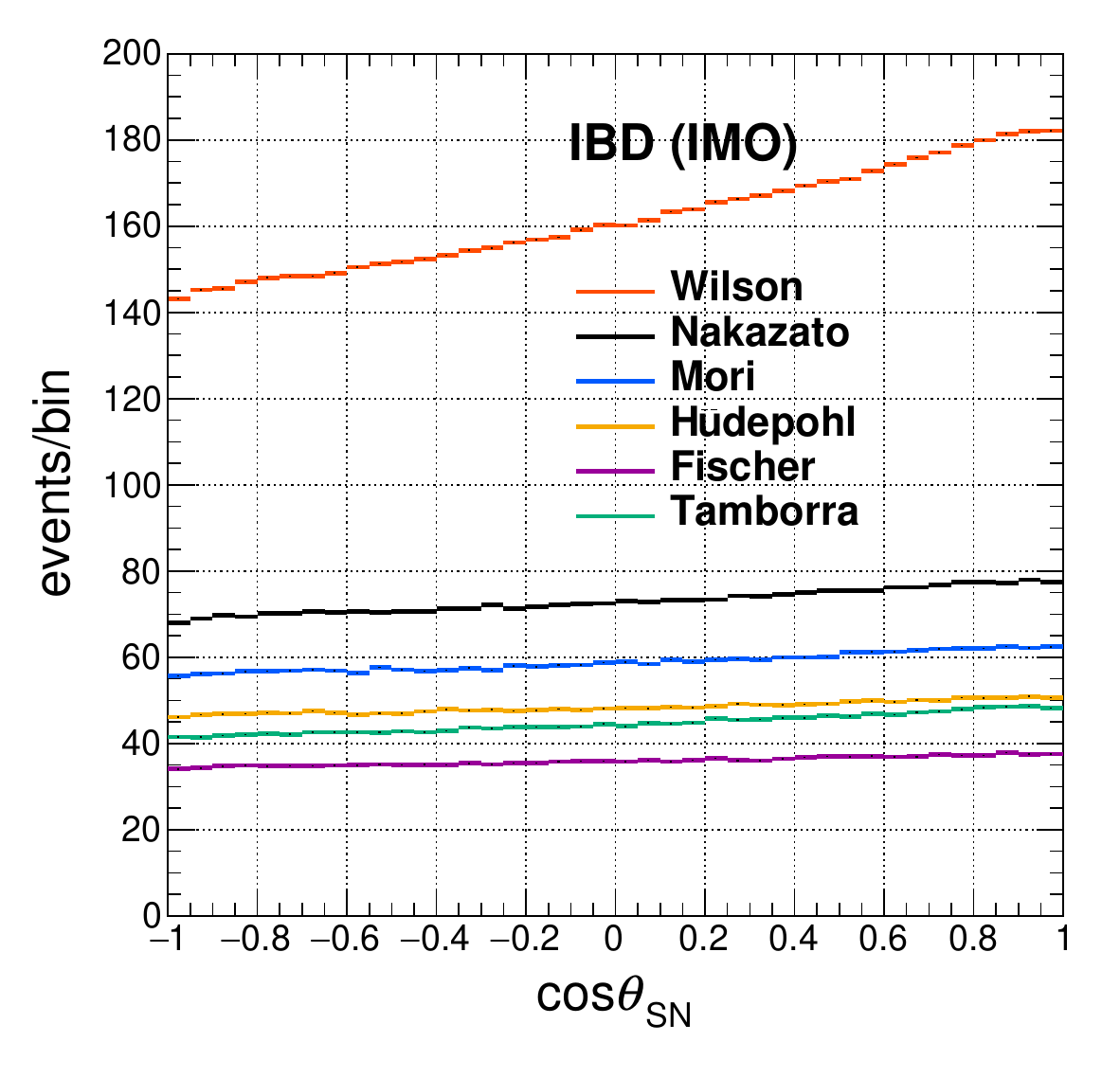}{0.35\textwidth}{(a) IBD}
          \fig{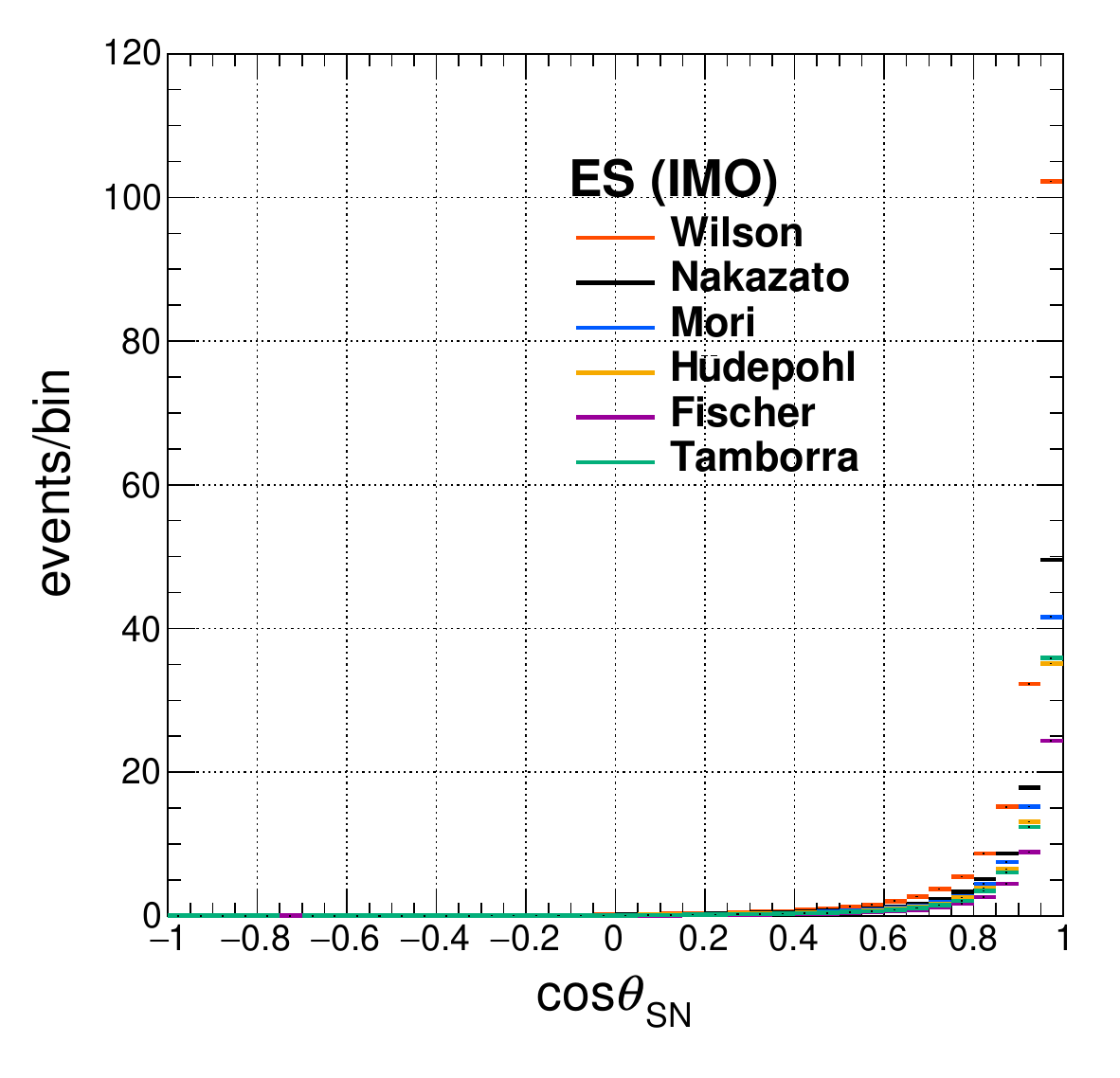}{0.35\textwidth}{(b) ES}}
\gridline{\fig{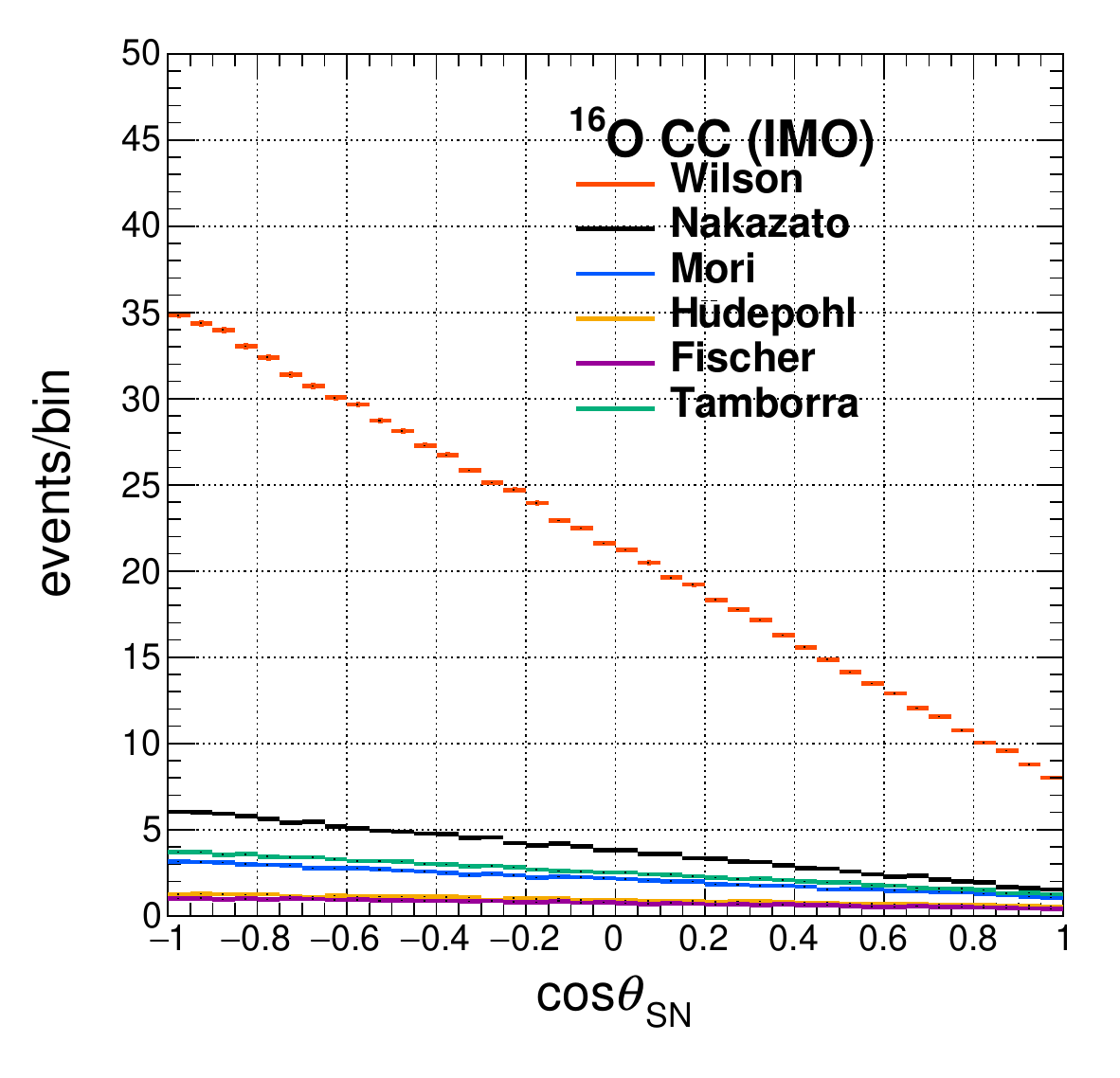}{0.35\textwidth}{(c) $^{16}$O~CC}
          \fig{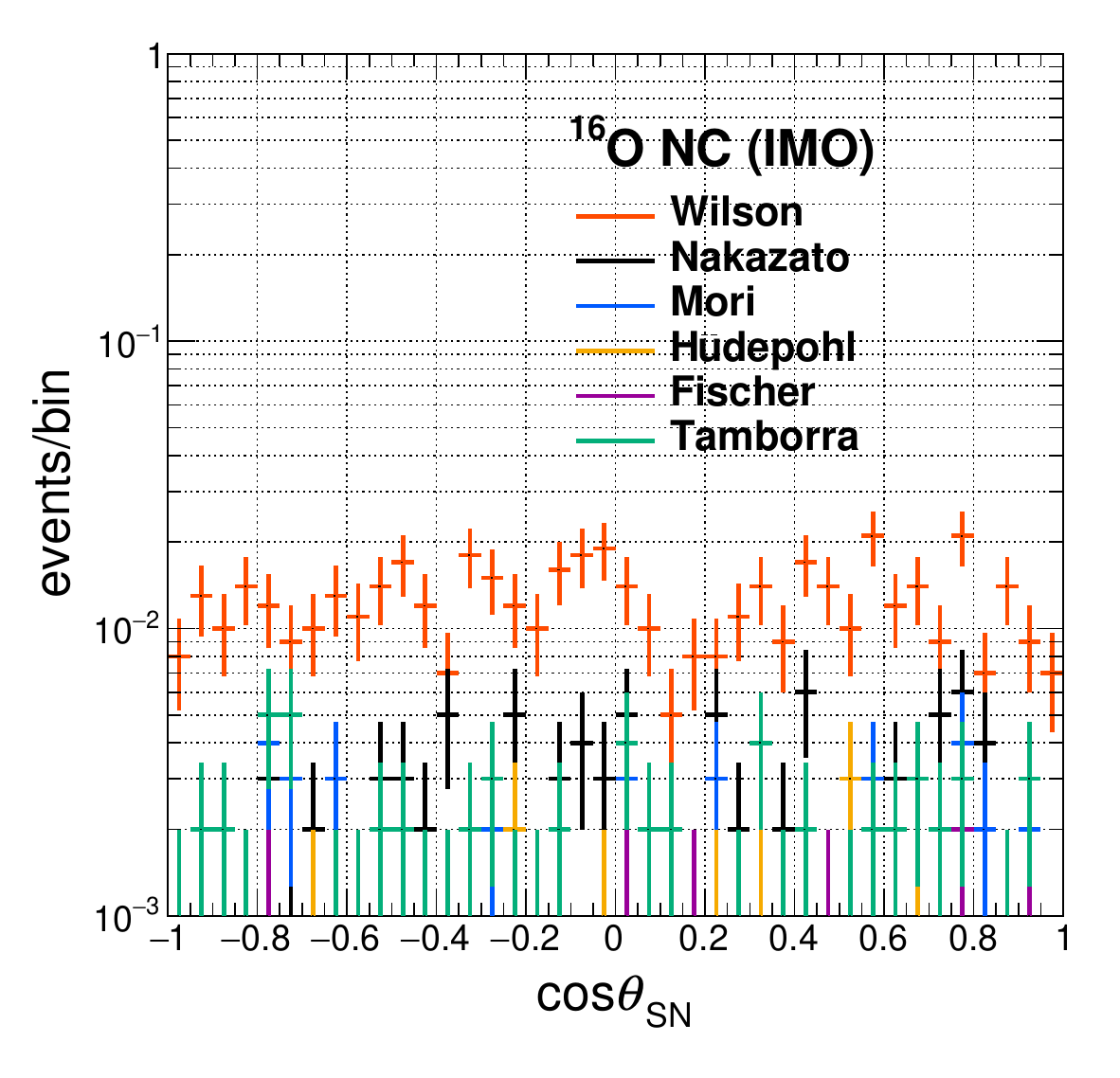}{0.35\textwidth}{(d) $^{16}$O~NC}}
\caption{Comparison of$\cos\theta_\mathrm{SN}$ distribution among models for each interaction for SN at 10~kpc: (a)~IBD, (b)~ES, (c)~$^{16}$O~CC, and (d)~$^{16}$O~NC.}
\label{fig:IMOvarModelCosineInteractions}
\end{figure}

Figure~\ref{fig:CosineInteractionsEachModelNMOandIMO} shows the $\cos\theta_\mathrm{SN}$ distribution of each interaction for each model for an SN burst located at 10~kpc in the NMO scenario (top six panels) and the IMO scenario (bottom six panels).  
Comparison of angular distribution of events among models for each interaction for an SN burst located at 10~kpc is shown in Figure~\ref{fig:IMOvarModelCosineInteractions}, similar to Figure~\ref{fig:NMOvarModelCosineInteractions}.
Note that the reconstructed events in Figures~\ref{fig:CosineInteractionsEachModelNMOandIMO}--\ref{fig:IMOvarModelCosineInteractions} satisfy the same event selection described in Section~\ref{subsec:EventReconstruction}.

\bibliography{sample631}{}
\bibliographystyle{aasjournal}



\end{document}